\documentclass[aps,amsmath,amssymb,showpacs,superscriptaddress,twocolumn,prb]{revtex4}

\pdfoutput=1

\usepackage{graphicx}
\usepackage{epsfig}
\usepackage{ulem}
\usepackage{amsmath}
\usepackage{amssymb}
\usepackage{physics}
\usepackage[usenames]{color}
\definecolor{orange}{rgb}{1,0.5,0}
\definecolor{violet}{rgb}{0.5,0,0.5}
\definecolor{mycolor}{rgb}{0.122, 0.435, 0.698}
\usepackage{xcolor}
\usepackage{mathtools}

\begin{document}

\title{Periodic polymers with increasing repetition unit: \\ 
Energy structure and carrier transfer.}

\date{\today}

\author{K. Lambropoulos}
\affiliation{Department of Physics, National and Kapodistrian University of Athens, Panepistimiopolis, Zografos, GR-15784, Athens, Greece}


\author{C. Vantaraki}
\affiliation{Department of Physics, National and Kapodistrian University of Athens, Panepistimiopolis, Zografos, GR-15784, Athens, Greece}

\author{P. Bilia}
\affiliation{Department of Physics, National and Kapodistrian University of Athens, Panepistimiopolis, Zografos, GR-15784, Athens, Greece}

\author{M. Mantela}
\affiliation{Department of Physics, National and Kapodistrian University of Athens, Panepistimiopolis, Zografos, GR-15784, Athens, Greece}

\author{C. Simserides}
\email{csimseri@phys.uoa.gr}
\homepage{http://users.uoa.gr/~csimseri/physics_of_nanostructures_and_biomaterials.html}
\affiliation{Department of Physics, National and Kapodistrian University of Athens, Panepistimiopolis, Zografos, GR-15784, Athens, Greece}






\begin{abstract}
We study the energy structure and the transfer of an extra electron or hole along periodic polymers made of $N$ monomers, with a repetition unit made of $P$ monomers, using a Tight-Binding wire model, where a site is a monomer (e.g., in DNA, a base pair), for $P$ even, and deal with two categories of such polymers:
made of the same monomer (GC..., GGCC..., etc) and made of different monomers (GA..., GGAA..., etc). 
We calculate the HOMO and LUMO eigenspectra, density of states and HOMO-LUMO gap and find some limiting properties these categories possess, as $P$ increases. We further examine the properties of the mean over time probability to find the carrier at each monomer. We introduce the \textit{weighted mean frequency} of each monomer and the \textit{total weighted mean frequency} of the whole polymer, as a measure of the overall transfer frequency content. We study the pure mean transfer rates. These rates can be increased by many orders of magnitude with appropriate sequence choice. Generally, homopolymers display the most efficient charge transfer. Finally, we compare the pure mean transfer rates with experimental transfer rates obtained by time-resolved spectroscopy.
\end{abstract}

\maketitle
\section{\label{sec:introduction} Introduction} 
A great part of the scientific community is interested in charge transfer and transport in biological systems: proteins~\cite{Page:2003,GrayWinkler:2010,Artes:2014,Kannan:2009}, enzymes~\cite{Moser:2010} and nucleic acids (DNA, RNA).
The electronic structure of nucleic acids and their charge transfer and transport properties are studied with the aim to understand their biological functions and their potential applications in nanotechnology.  The term \textit{transport} implies the use of electrodes between which electric voltage is applied, while the term \textit{transfer} means that the electron or hole, created e.g. by reduction or oxidation at a certain site, moves to more favorable sites. Although carrier transfer in DNA nearly vanishes after 10 to 20 nm~\cite{Simserides:2014, LChMKTS:2015,LChMKLTTS:2016}, DNA can be used as a molecular wire~\cite{Wohlgamuth:2013} for charge transport. Favoring geometries and base-pair sequences, the use of non-natural bases \cite{Fuentes-Cabrera:2007}, isomers and tautomers of the bases \cite{MMTS:2016}, the use of the triplet acceptor anthraquinone for hole injection~\cite{LewisWasielewski:2013} and so on, are being investigated. Structural fluctuations \cite{Zoli:2017} is another important factor which influences carrier movement through DNA~\cite{Gutierrez:2009,Gutierrez:2010,Siebbeles:1998,Woiczikowski:2009}. Charge transfer is relevant in DNA damage and repair \cite{Dandliker:1997,Rajski:2000,Giese:2006} and in discrimination between pathogenic and non-pathogenic mutations at an early stage \cite{Shih:2011}. Charge transport could probe DNA of different origin or organisms~\cite{Paez:2012}, mutations and diseases \cite{Shih:2008,Oliveira:2014}.
Charge transport in oxidatively damaged DNA under of structural fluctuations has also been investigated \cite{Lee:2012}. Also, electric charge oscillations govern the serum response factor - DNA recognition~\cite{Stepanek:2015}. 
Finally, guanine runs support delocalization over 4 - 5 guanine bases and resistance oscillations in such DNA segments have been observed theoretically and experimentally~\cite{Liu:2016}.

Today, after many years of research~\cite{FinkSchoenenberger:1999,Porath:2000,Yoo:2001,Xu:2004,Cohen:2005,Grozema:1999}, we realize that many factors (e.g. aqueousness, counterions, extraction process, electrodes, purity, substrate, structural fluctuations, geometry), influence carrier motion. Still, however, we must deeply understand how the base-pair sequence affects carrier motion.  This is one of the aims of the present work, for periodic sequences with increasing repetition unit.
In this work we use B-DNA as a prototype system, but the analysis can be projected to any similar one-dimensional polymer made of $N$ monomers. In the B-DNA case, a monomer is a base pair.
We also assume that the state or movement of an extra hole or electron in the polymer can be expressed through a combination of the HOMO (Highest Occupied Molecular Orbital) or LUMO (Lowest Unoccupied Molecular Orbital), respectively, of all monomers, cf. Eqs.~\eqref{TIP} and \eqref{TDP} below. This way, we define the HOMO regime and the LUMO regime.

Recent research shows that, certainly, carrier movement through B-DNA can be manipulated.
For example, the carrier transfer rate through DNA can be tuned by chemical modification\cite{KawaiMajimaBook:2015}. It was shown, using various natural and artificial nucleobases with different HOMO energies, that the hole transfer rate strongly depends on the difference between these HOMO energies~\cite{KawaiMajimaBook:2015}. The change in this difference from 0.78 to 0 eV resulted in an increase in the hole transfer rate by more than 3 orders of magnitude~ \cite{KawaiMajimaBook:2015}. 
In Ref.~\cite{LChMKLTTS:2016} we studied all possible B-DNA polymers of the form YXYX..., i.e.,
monomer polymers and dimer polymers with the Tight-Binding (TB) wire model used in the present article (TB I) and with a TB extended ladder model including diagonal hoppings (TB II). 
There are three types of such polymers (always mentioning only the $5'-3'$ base sequence of one strand): 
($\alpha'$) G... i.e. poly(dG)-poly(dC), A... i.e. poly(dA)-poly(dT), called here I1 
($\beta'$) GCGC..., CGCG..., ATAT..., TATA..., called here I2 and 
($\gamma'$) ACAC..., CACA..., TCTC..., CTCT..., AGAG..., GAGA..., TGTG..., GTGT.... called here D2, cf. Table~\ref{Table:TypesOfPolymers}.
We illustrated~\cite{LChMKLTTS:2016} that, generally, increasing the length of the polymer from 2 to 30 monomers results in a decrease of the \textit{pure} mean transfer rate by $\approx$ 2 orders of magnitude in type I1 polymers, while in types I2 and D2 polymers the decrease can be much more steep, $\approx$ from 3 to 7 orders of magnitude: cf. Figs. 20-22 of Ref.~\cite{LChMKLTTS:2016}. 

\textit{Ab initio} calculations \cite{YeJiang:2000, Barnett:2003,Artacho:2003,MehrezAnantram:2005,Voityuk:2008,Kubar:2008,TMLS:2017}
and model Hamiltonians~\cite{Simserides:2014, LChMKTS:2015,LChMKLTTS:2016,Cuniberti:2002,Roche-et-al:2003,Roche:2003,Yamada:2004,ApalkovChakraborty:2005,Klotsa:2005,Shih:2008,Yi:2003,Caetano:2005,Wang:2006} have been used to explore the variety of experimental results and the underlying charge transfer or transport mechanism.
The former are currently limited to very  short segments, while the latter allow to address systems of realistic length~\cite{Albuquerque:2014,Albuquerque:2005,Cuniberti:2007}.
Here we study rather long sequences, hence we adopt the latter approach.
The Tight-Binding approximation has been used in DNA for decades. It was realized that since the DNA bases have delocalized $\pi$ electrons, these will interact and this interaction will not be negligible \cite{LadikAppel:1964}.

In Section ~\ref{sec:theory} we delineate the basic theory behind the time-independent (Subsection~\ref{subsec:TIproblem}) and the time-dependent (Subsection~\ref{subsec:TDproblem}) problem.
In Section~\ref{sec:Results} we discuss our results for 
polymers made of the same monomer and 
polymers made of different monomers.
In Section~\ref{sec:Conclusion} we state our conclusions.

\section{\label{sec:theory} Theory} 
The present work is a part of a series of papers discussing the effect of the most important intrinsic factor, i.e. the base sequence, that affects charge transfer in DNA. We use some physical quantities introduced before~\cite{Simserides:2014, LChMKTS:2015, LChMKLTTS:2016}, while introducing some new ones, such as the \textit{weighted mean frequency} (WMF) and the total weighted mean frequency (TWMF), that can act as measures of the frequency content of the extra carrier transfer. We use a simple wire model, where the site is a monomer. 
The model can potentially be enriched by adding to the Hamiltonian terms such as the electron-phonon coupling, the spin-orbit coupling, reservoirs that resemble the environment, external fields, etc, but here we focus on the understanding of the base sequence. The introduction of many more --and of largely unknown value-- parameters, would cast shade on the influence the factor we study has on charge transfer.
We call $\mu$ the monomer index, $\mu = 1,2,\dots,N$. 
For B-DNA, we mention only the $5'-3'$ base sequence of one strand.
For example, we denote two successive monomers by YX, meaning that 
the base pair X-X$_{\text{compl}}$ is separated and twisted by 3.4 {\AA} and $36^{\circ}$, respectively, relatively to the base pair Y-Y$_{\text{compl}}$, around the B-DNA growth axis. 
X$_{\text{compl}}$ (Y$_{\text{compl}}$) is the complementary base of X (Y).


\subsection{\label{subsec:TIproblem} Stationary States - Time-independent problem} 

The TB \textit{wire} model Hamiltonian can be written as
\begin{equation} \label{WireHamiltonian}
\hat{H}_\text{W} =
\sum_{\mu=1}^{N} E_\mu \ketbra{\mu}{\mu} +
\bigg( \sum_{\mu=1}^{N-1} t_{\mu,\mu+1} \ketbra{\mu}{\mu+1} + h.c. \bigg).
\end{equation}
$E_\mu$ is the on-site energy of the $\mu$-th monomer, and $t_{\mu,\lambda} = t_{\lambda,\mu}^*$ is the hopping integral between monomers $\mu$ and $\lambda$. 
The state of a polymer can be expressed as
\begin{equation} \label{TIP}
\ket{\mathcal{P}} = \sum_{\mu=1}^{N} v_\mu \ket{\mu}.
\end{equation}
Substituting Eqs. \eqref{WireHamiltonian} and \eqref{TIP} to the time-independent Schr\"odinger equation
\begin{equation} \label{TISchr}
\hat{H} \ket{\mathcal{P}} = E \ket{\mathcal{P}},
\end{equation}
we arrive to a system of $N$ coupled equations
\begin{equation} \label{TIsystembp}
E_\mu v_{\mu} + t_{\mu, \mu+1} v_{\mu+1} + t_{\mu, \mu-1} v_{\mu-1} = E v_{\mu}, 
\end{equation}
which is equivalent  to the eigenvalue-eigenvector problem
\begin{equation} \label{eigenvproblem}
H\vec{v} = E \vec{v},
\end{equation}
where $H$ is the hamiltonian matrix of order $N$, composed of the TB parameters $E_\mu$ and $t_{\mu,\lambda}$, and $\vec{v}$ is the vector matrix composed of the coefficients $v_\mu$. The diagonalization of $H$ leads to the determination of the eigenenergy spectrum (\textit{eigenspectrum}), $\{E_k\}$, $k =1,2,\dots,N$, for which we suppose that $E_1<E_2<\dots<E_{N}$, as well as to the determination of the occupation probabilities for each eigenstate, $\abs{v_{\mu k}}^2$, where $v_{\mu k}$ is the $\mu$-th component of the $k$-th eigenvector. $\{v_{\mu k}\}$ are normalized, and their linear independence is checked in all cases.

Having determined the eigenspectrum, we can compute the density of states (DOS), generally given by
\begin{equation} \label{DOS}
g(E) = \sum_{k=1}^{N} \delta(E-E_k).
\end{equation}
Changing the view of a polymer from one (e.g. top) to the other (e.g. bottom) side of the growth axis, reflects the hamiltonian matrix $H$ of the polymer on its main antidiagonal. This reflected Hamiltonian, $H^{\text{equiv}}$, describes the \textit{equivalent polymer}~\cite{LChMKLTTS:2016}. 

\subsection{\label{subsec:TDproblem} Time-dependent problem} 
To describe the spatiotemporal evolution of an extra carrier (hole/electron), inserted or created (e.g. by oxidation/reduction) in a particular monomer of the polymer, we consider the state of the polymer as
\begin{equation} \label{TDP}
\ket{\mathcal{P}(t)} = \sum_{\mu=1}^{N}C_\mu(t) \ket{\mu},
\end{equation}
where $\abs{C_\mu(t)}^2$ is the probability to find the carrier at the $\mu$-th monomer at time $t$. Substituting Eqs. \eqref{WireHamiltonian} and \eqref{TDP} to the time-dependent Schr\"odinger equation
\begin{equation} \label{TDSchr}
i\hbar \pdv{t}\ket{\mathcal{P}(t)} = \hat{H} \ket{\mathcal{P}(t)},
\end{equation}
we arrive at a system of $N$ coupled differential equations 
\begin{equation} \label{TDsystembp}
i\hbar \dv{C_\mu}{t} = E_\mu C_\mu + t_{\mu, \mu+1} C_{\mu+1} + t_{\mu, \mu-1} C_{\mu-1}.
\end{equation}
Eq. \eqref{TDsystembp} is equivalent  to a 1$^\text{st}$ order matrix differential equation of the form
\begin{equation} \label{Matrixdiffeq}
\dot{\vec{C}}(t) = -\frac{i}{\hbar}H \vec{C}(t),
\end{equation}
where $\vec{C}(t)$ is a vector matrix composed of the coefficients $C_\mu(t), \;\; \mu = 1, 2, \dots, N$. Eq.~\eqref{Matrixdiffeq} can be solved with the eigenvalue method, i.e., by looking for solutions of the form $\vec{C}(t) = \vec{v} e^{-\frac{i}{\hbar}Et} \Rightarrow \dot{\vec{C}}(t) = -\frac{i}{\hbar}E \vec{v}e^{-\frac{i}{\hbar}Et}$. Hence, Eq.~\eqref{Matrixdiffeq} leads to the eigenvalue problem of Eq.~\eqref{eigenvproblem}, that is, $H\vec{v} = E \vec{v}$. Having determined the eigenvalues and eigenvectors of $H$, the general solution of Eq.~\eqref{Matrixdiffeq} is
\begin{equation} \label{generalsolution}
\vec{C}(t) = \sum_{k=1}^{N} c_k \vec{v}_k e^{-\frac{i}{\hbar}E_kt},
\end{equation}
where the coefficients $c_k$ are determined from the initial conditions.
In particular, if we define the $N \times N$ eigenvector matrix $V$, with elements $v_{\mu k}$, then it can be shown that the vector matrix $\vec{c}$, composed of the coefficients $c_k, \;\; k = 1, 2, \dots, N$, is given by the expression
\begin{equation} \label{c_kdef}
\vec{c} = V^T\vec{C}(0).
\end{equation}
Suppose that initially the extra carrier is placed at the $\lambda$-th monomer, i.e., $C_\lambda(0) = 1$, $C_\mu(0) = 0, \forall \mu \neq \lambda$. Then,
\begin{equation} \label{cmatrix}
\vec{c} = \begin{bmatrix}
v_{\lambda 1}\\\vdots\\v_{\lambda k}\\\vdots\\v_{\lambda N}
\end{bmatrix}.
\end{equation}
In other words, the coefficients $c_k$ are given by the row of the eigenvector matrix which corresponds to the monomer the carrier is initially placed at. 

From Eq.~\eqref{generalsolution} it follows that the probability to find the extra carrier at the $\mu$-th monomer is
\begin{equation} \label{probabilities}
\abs{C_\mu(t)}^2 = \sum_{k=1}^{N} c_k^2v_{\mu k}^2 + 2 \sum_{k=1}^{N} \sum_{\substack{k'=1 \\k'<k}}^{N} c_k c_{k'} v_{\mu k} v_{\mu k'} \cos(2\pi f_{kk'}t),
\end{equation}
where
\begin{equation} \label{fandT}
f_{kk'} = \frac{1}{T_{kk'}} = \frac{E_k-E_{k'}}{h}, \; \forall k > k',
\end{equation}
are the frequencies ($f_{kk'}$) or periods ($T_{kk'}$) involved in charge transfer. If $m$ is the number of discrete eigenenergies, then, the number of different $f_{kk'}$ or $T_{kk'}$ involved in carrier transfer is $S= {m \choose 2} = \frac{m!}{2!(m-2)!} = \frac{m(m-1)}{2}$.
If there are no degenerate eigenenergies (which holds for all cases studied here, but e.g. does not hold for \textit{cyclic} I1 polymers~\cite{LChMKTS:2015}), then $m = N$. If eigenenergies are symmetric relative to some central value, then, $S$ decreases (there exist degenerate $f_{kk'}$ or $T_{kk'}$). Specifically, in that case, $S = \frac{m^2}{4}$, for even $m$ and $S = \frac{m^2-1}{4}$ for odd $m$.

From Eq.~\eqref{probabilities} it follows that the mean over time probability to find the extra carrier at the $\mu$-th monomer is
\begin{equation} \label{meanprobabilities}
\expval{\abs{C_\mu(t)}^2} = \sum_{k=1}^{N} c_k^2v_{\mu k}^2.
\end{equation}

Furthermore, from Eq.~\eqref{probabilities} it can be shown that the one-sided Fourier amplitude spectrum that corresponds to the probability $\abs{C_\mu(t)}^2$ is given by
\begin{equation} \label{Fourierspectra}
\resizebox{\hsize}{!}{$\displaystyle
\abs{\mathcal{F}_\mu(f)} = \sum_{k=1}^{N} c_k^2v_{\mu k}^2 \delta(f) + 2 \sum_{k=1}^{N} \sum_{\substack{k'=1 \\ k'<k}}^{N} \abs{c_k c_{k'} v_{\mu k} v_{\mu k'}} \delta(f-f_{kk'}).$}
\end{equation}
Hence, the Fourier amplitude of frequency $f_{kk'}$  is $2 |c_k v_{\mu k} c_{k'} v_{\mu k'}|$. We can further define the WMF of monomer $\mu$ as
\begin{equation} \label{Eq:WMFdef}
f_{WM}^\mu = \frac{\displaystyle\sum_{k=1}^{N} \sum_{\substack{k'=1 \\k'<k}}^{N} |c_k v_{\mu k} c_{k'} v_{\mu k'}| f_{kk'}}{
\displaystyle \sum_{k=1}^{N} \sum_{\substack{k'=1 \\k'<k}}^{N} |c_k v_{\mu k} c_{k'} v_{\mu k'}|}.
\end{equation}
WMF expresses the mean frequency content of the extra carrier oscillation at monomer $\mu$. Having determined the WMF for all monomers, we can now obtain a measure of the overall frequency content of carrier oscillations in the polymer: Since $f_{WM}^\mu$ is the weighted mean frequency of monomer $\mu$ and $\left\langle|{C_\mu(t)|^2}\right\rangle$ is the mean probability of finding the extra carrier at monomer $\mu$, we define the TWMF as
\begin{equation} \label{Eq:TWMFdef}
f_{TWM} = \sum_{\mu=1}^{N} f_{WM}^\mu \left\langle|C_\mu(t)|^2\right\rangle.
\end{equation}

A quantity that evaluates simultaneously the magnitude of charge transfer and the time scale of the phenomenon, is the \textit{pure} mean transfer rate~\cite{Simserides:2014}
\begin{equation} \label{pmtr}
k_{\lambda \mu} = \frac{\expval{\abs{C_\mu(t)}^2}}{t_{\lambda \mu}}.
\end{equation}
$t_{\lambda \mu}$ is the \textit{mean transfer time}, i.e.,
having placed the carrier initially at monomer $\lambda$,
the time it takes for the probability to find the extra carrier at monomer $\mu$,
$\abs{C_\mu(t)}^2$, to become equal to its mean value, $\expval{\abs{C_\mu(t)}^2}$, for the first time. 
For the \textit{pure} mean transfer rates it holds
$ k_{\lambda \mu} = k_{\mu \lambda}=k_{(N-\lambda+1)(N-\mu+1)}^{\text{equiv}} = k_{(N-\mu+1)(N-\lambda+1)}^{\text{equiv}}$.

\section{\label{sec:Results} Results} 
One could think of many types of periodic polymers, 
some of which are shown synoptically in Table~\ref{Table:TypesOfPolymers}. 
We just give an example of the sequence, e.g., 
for type I4 we give the example GGCC..., but there are obviously other similar sequences: CCGG..., AATT..., TTAA.... $P$ is the number of monomers in the repetition unit, e.g., for type I4, $P = 4$. In this article, we illustrate our results using the sequence examples of Table~\ref{Table:TypesOfPolymers}. Similar conclusions hold, obviously, for all other members of the same type.

\begin{table}[h!]
\caption{The types of polymers mentioned in this work. I (D) denotes polymers made of the identical (different)  monomers. $P$ is the number of monomers in the repetition unit. We only mention the $5'-3'$ base sequence.}
\label{Table:TypesOfPolymers}
\begin{tabular}{|c|c|c|} \hline
(I,D)$P$ & sequence example        \\ \hline \hline
I1       & G...   or   A...        \\ \hline
I2       & GC...                   \\ \hline
I3       & GGC...                  \\ \hline
I4       & GGCC...                 \\ \hline
I6       & GGGCCC...               \\ \hline
I8       & GGGGCCCC...             \\ \hline
I10      & GGGGGCCCCC...           \\ \hline
I20      & GGGGGGGGGGCCCCCCCCCC... \\ \hline \hline
D2       & GA...                   \\ \hline
D4       & GGAA...                 \\ \hline
D6       & GGGAAA...               \\ \hline
D8       & GGGGAAAA...             \\ \hline
D10      & GGGGGAAAAA...           \\ \hline
D20      & GGGGGGGGGGAAAAAAAAAA... \\ \hline
\end{tabular}
\end{table}

The TB parameters for B-DNA are the same as in Refs.~\cite{Simserides:2014,LChMKTS:2015,LChMKLTTS:2016}, unless otherwise stated. The HOMO/LUMO hopping integrals are given in Table~\ref{Table:HoppingIntegrals}.
The HOMO/LUMO base-pair on-site energies are~\cite{HKS:2010-2011} $E_\textrm{G-C} = -8.0/-$4.5 eV, 
$E_\textrm{A-T} = -8.3/-$4.9 eV. More details about the parameter choice can be found in Refs.~\cite{Simserides:2014, HKS:2010-2011}.

\begin{table}[h!]
\caption{The HOMO/LUMO hopping integrals $t_{\mu,\lambda}$, in meV,  
between successive base pairs $\mu,\lambda$ ~\cite{Simserides:2014}.}
\label{Table:HoppingIntegrals}
\begin{tabular}{|c|c|c|c|c|} \hline
$\mu,\lambda$& $t_{\mu,\lambda}$    && $\mu,\lambda$   &  $t_{\mu,\lambda}$ \\ \hline
GG $\equiv$ CC & $-$100/20          &&  AA $\equiv$ TT &  $-$20/$-$29       \\ \hline
GC             &   10/$-$10         &&  CG             &  $-$50/$-$8        \\ \hline
AT             &   35/0.5           &&  TA             &   50/2             \\ \hline
CT $\equiv$ AG &  $-$30/3           &&  TC $\equiv$ GA &  $-$110/$-$1       \\ \hline
CA $\equiv$ TG &  $-$10/17          &&  AC $\equiv$ GT &   10/32            \\ \hline
\end{tabular}
\end{table}

\subsection{\label{subsec:EigenDOSGap} Eigenspectra, Density of States, Energy Gap} 
\begin{figure*} [h!]\vspace{-0.8cm}
\includegraphics[width=0.45\textwidth]{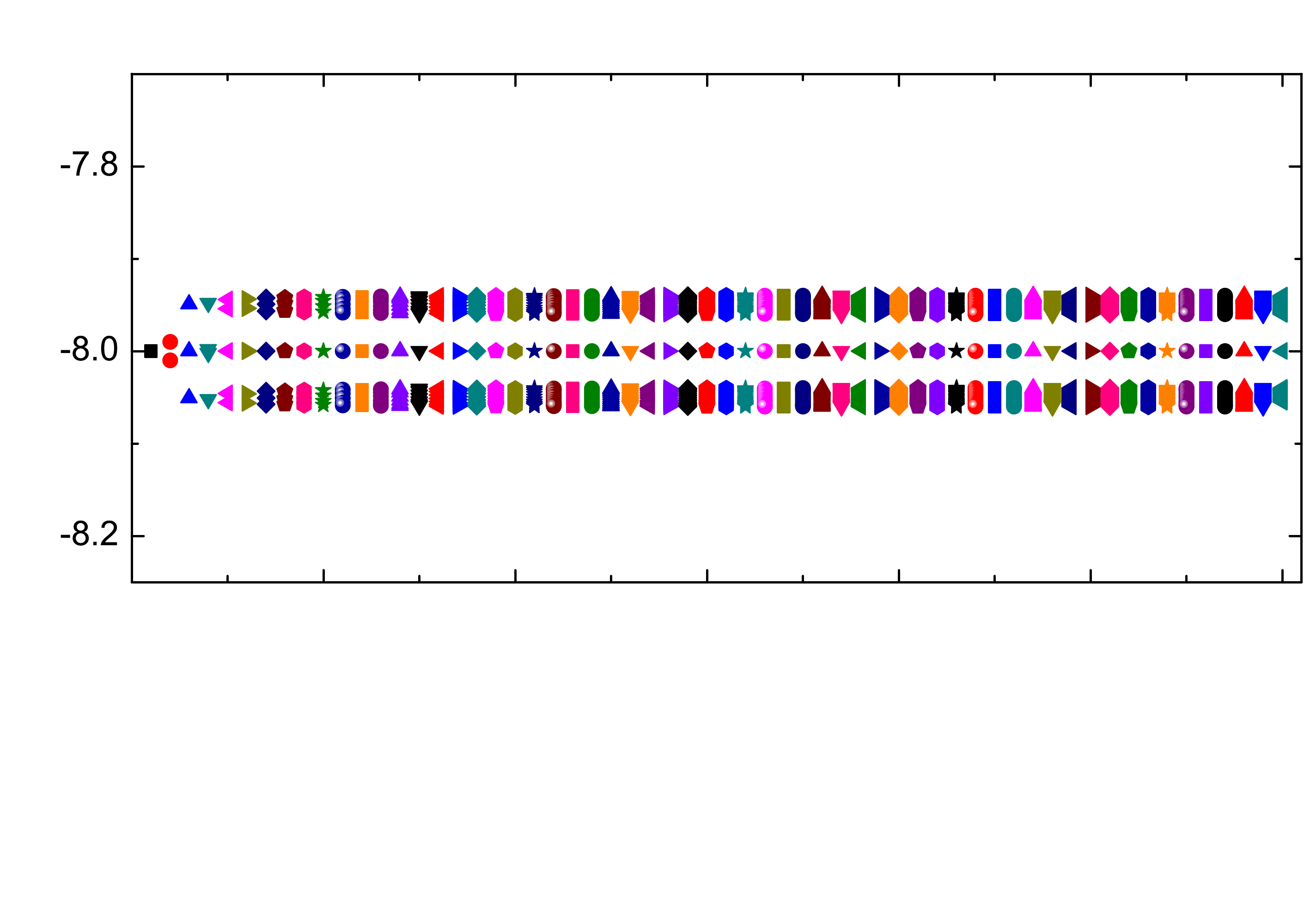}   
\includegraphics[width=0.45\textwidth]{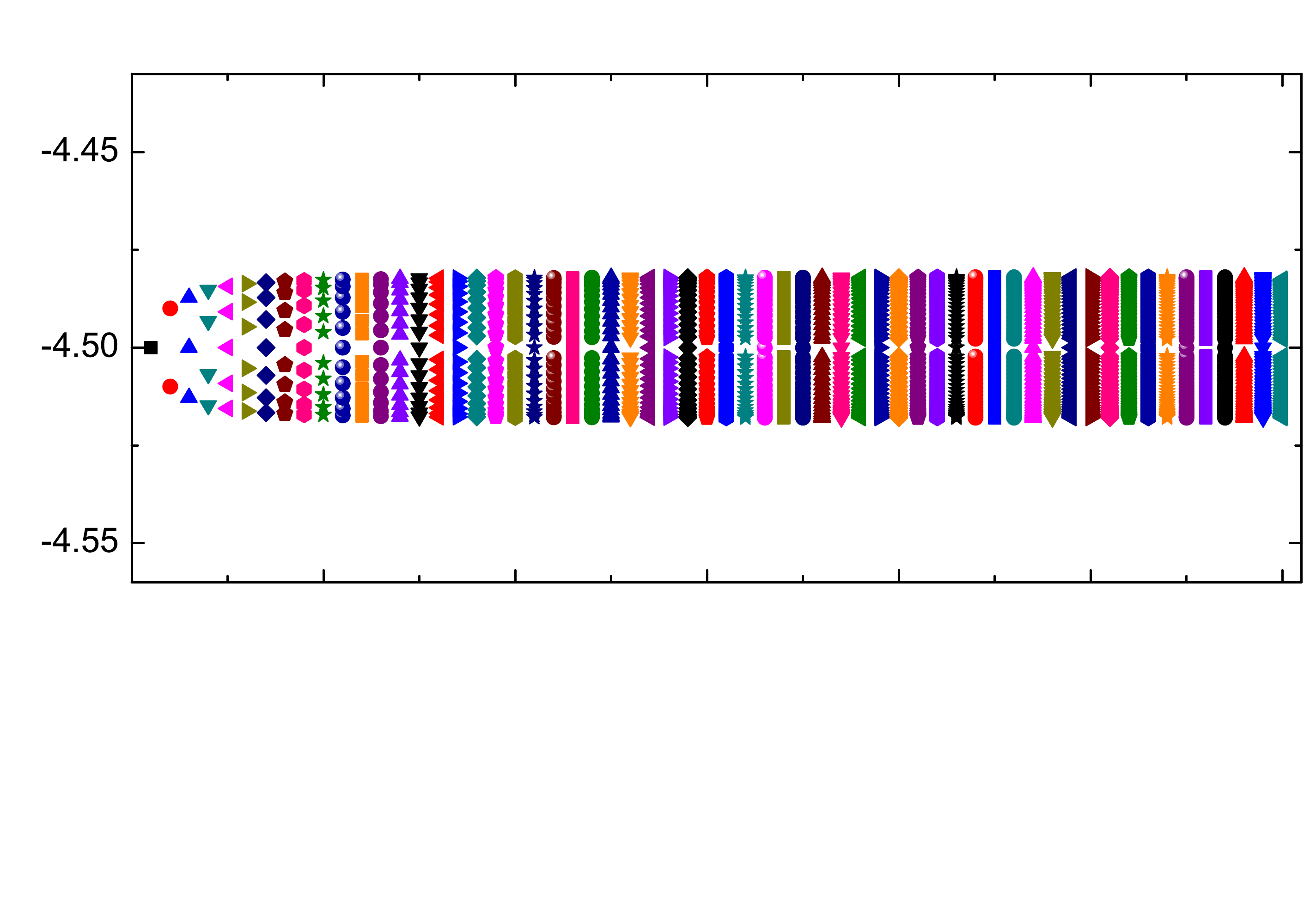} \\ \vspace{-2.5cm}
\includegraphics[width=0.45\textwidth]{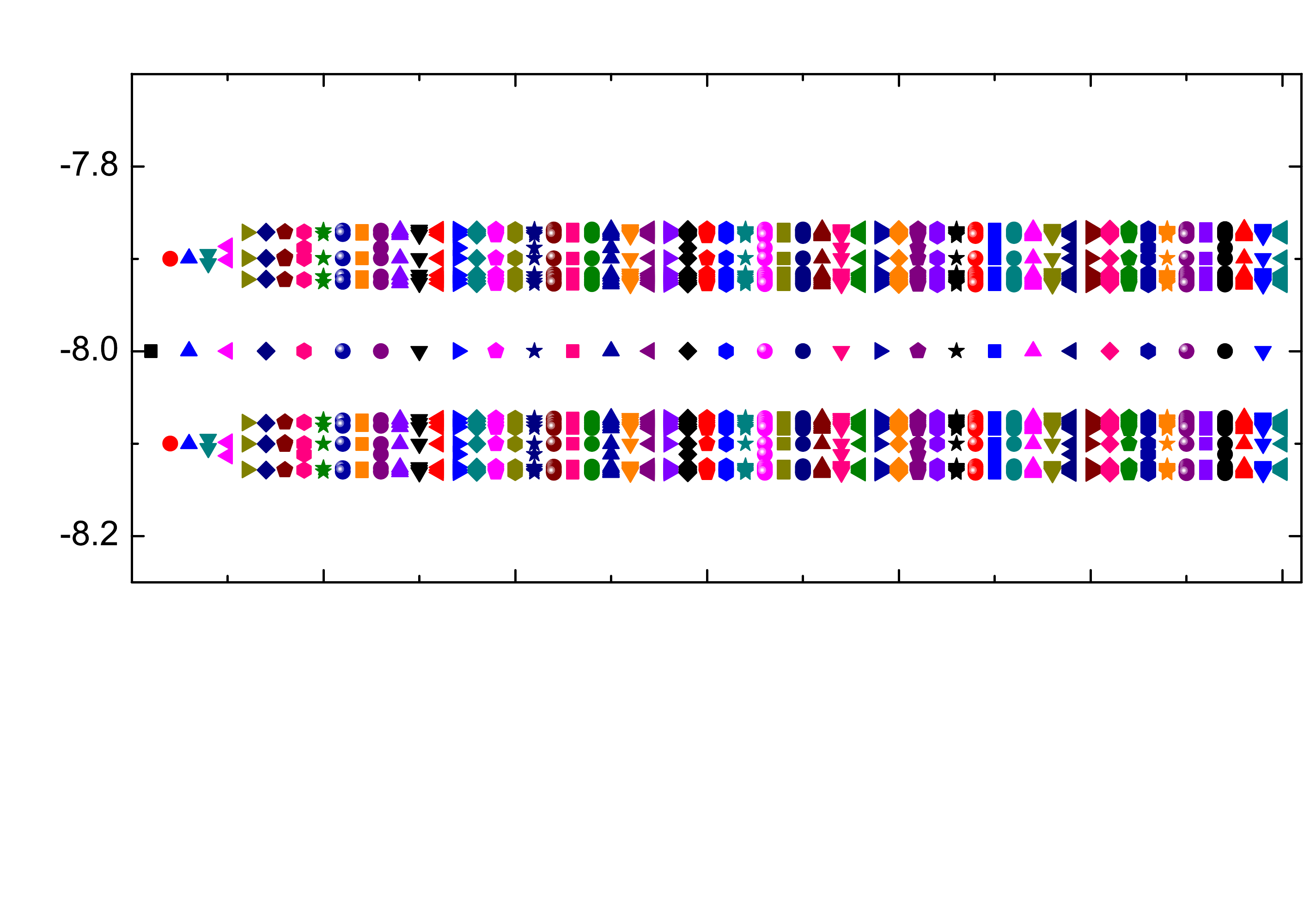}   
\includegraphics[width=0.45\textwidth]{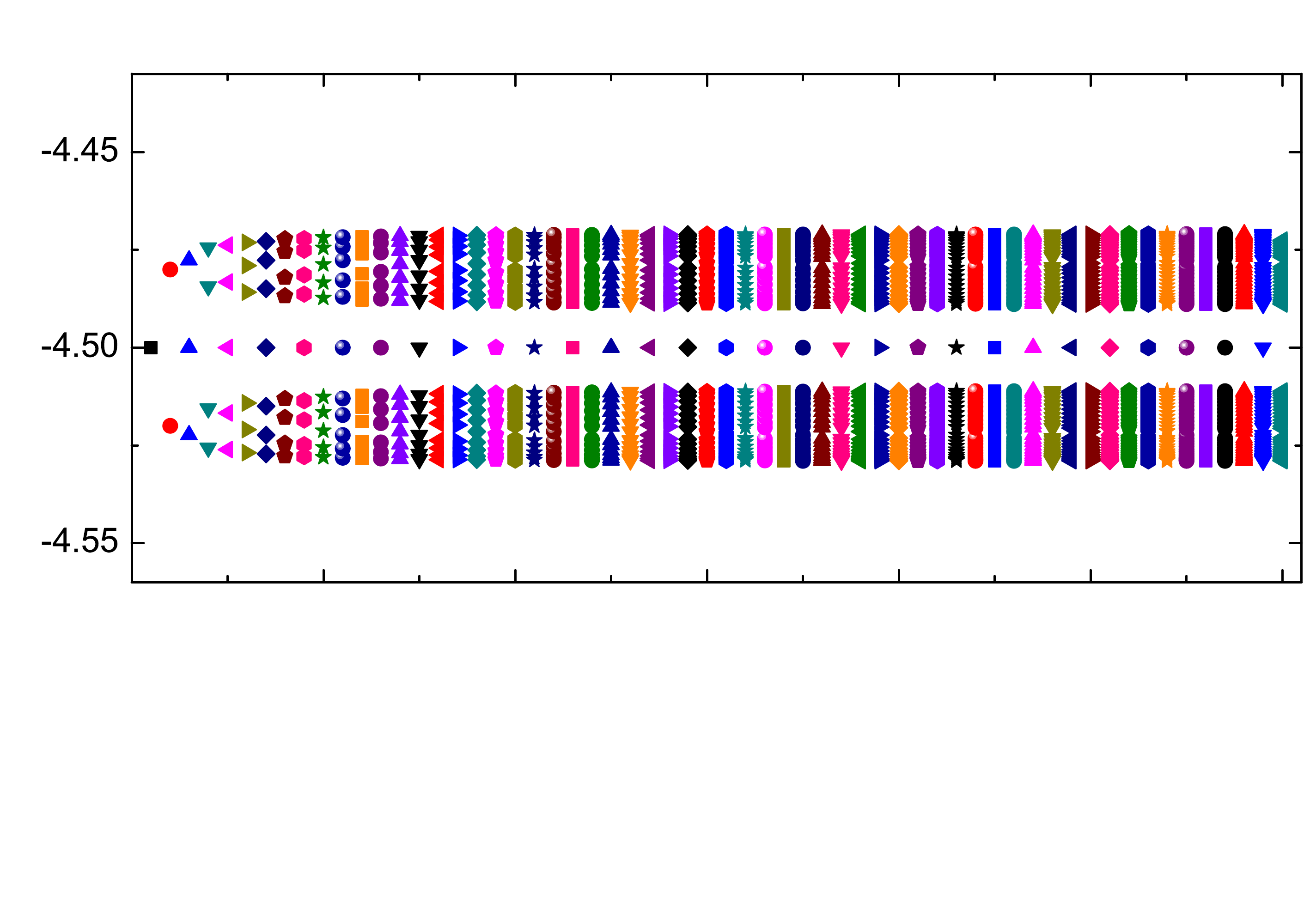} \\ \vspace{-2.5cm}
\includegraphics[width=0.45\textwidth]{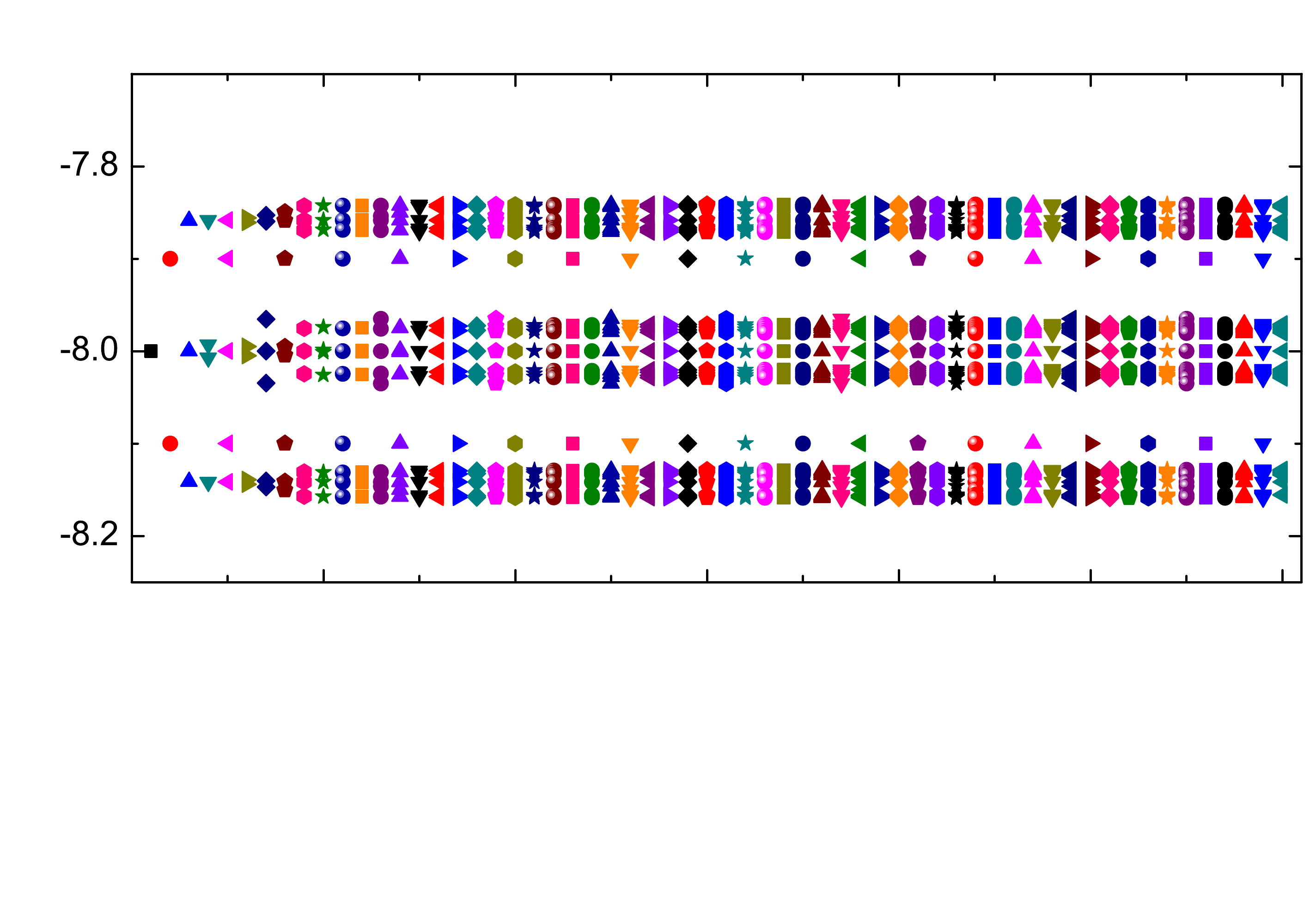}   
\includegraphics[width=0.45\textwidth]{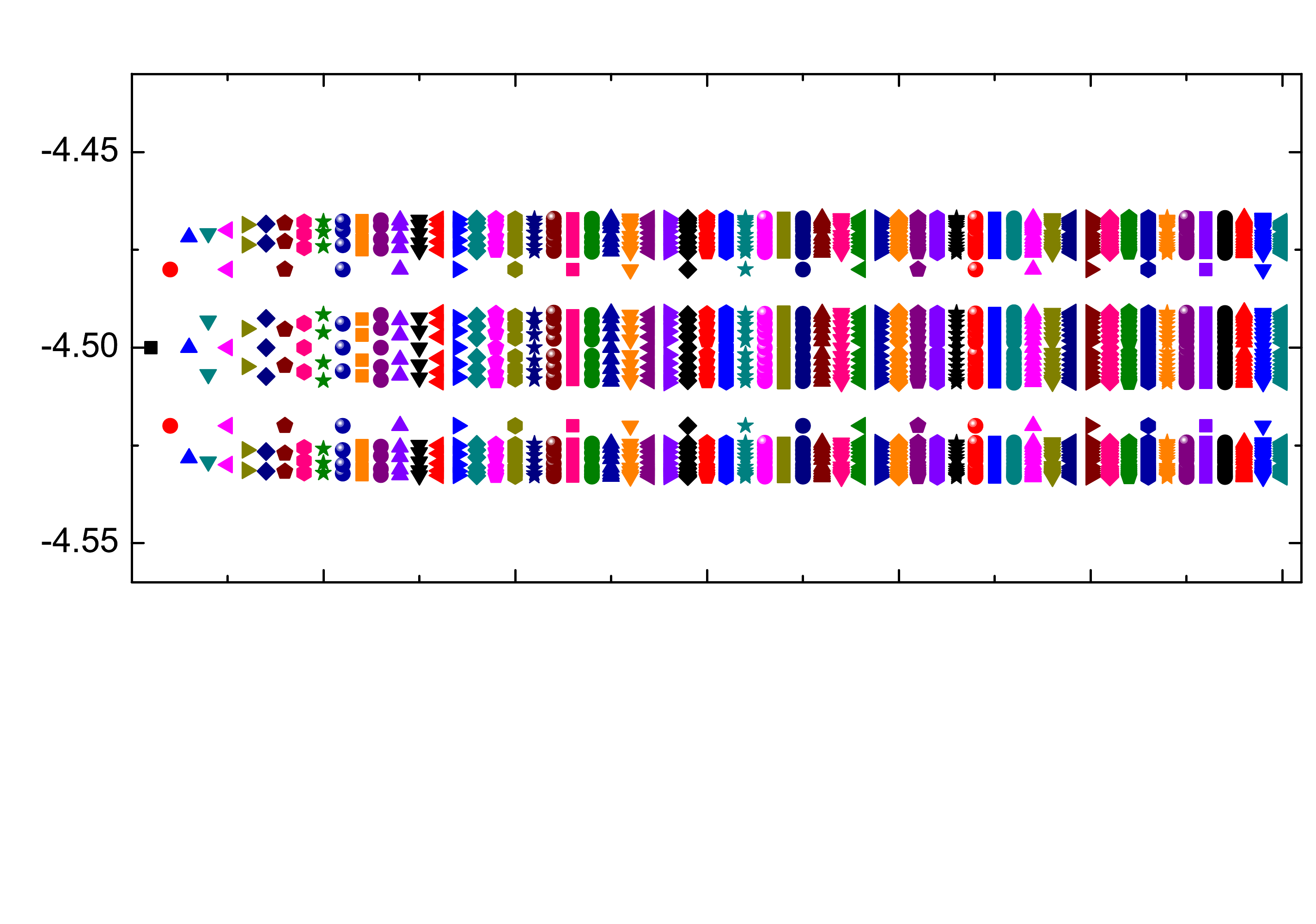} \\ \vspace{-2.5cm}
\includegraphics[width=0.45\textwidth]{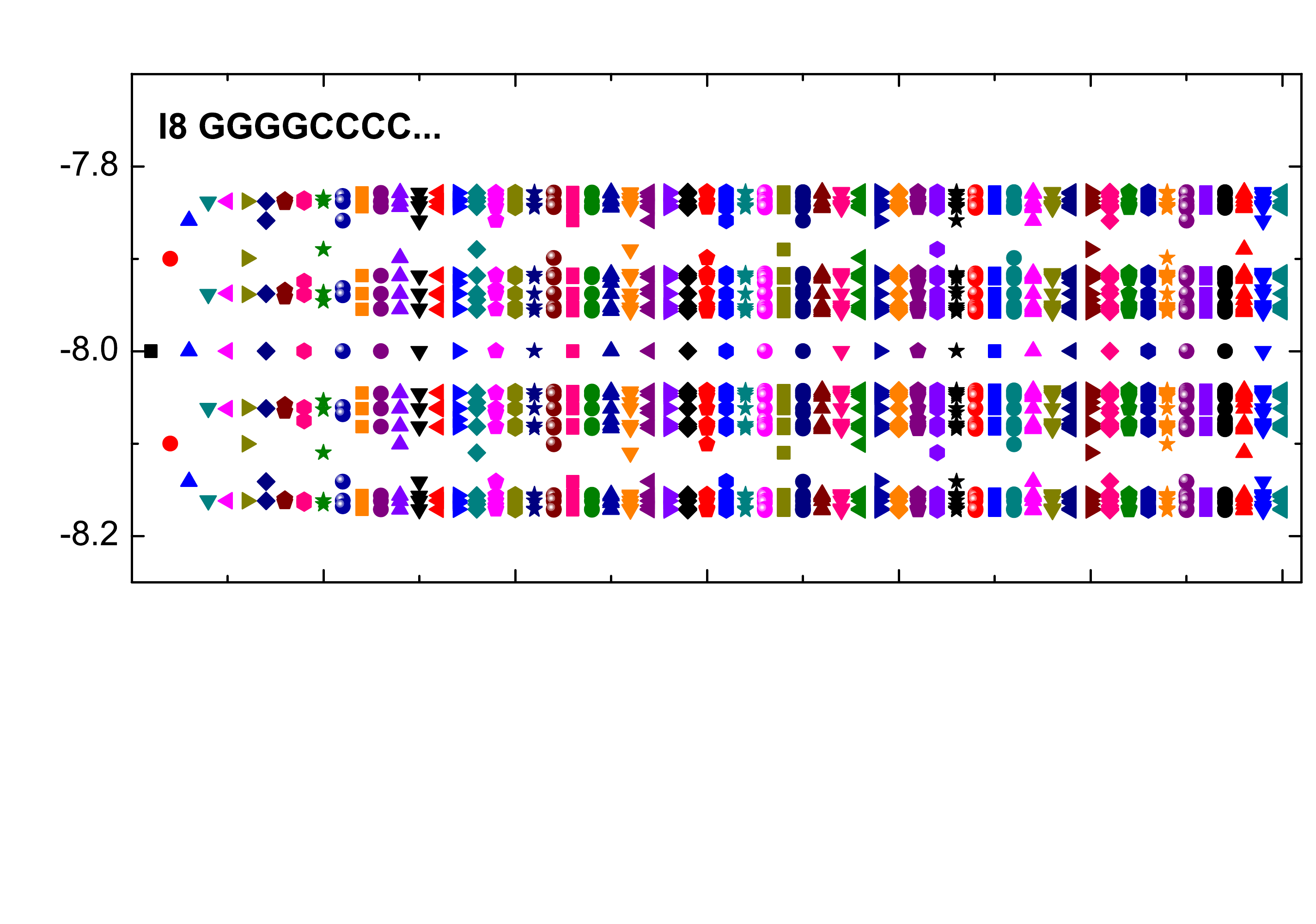}    
\includegraphics[width=0.45\textwidth]{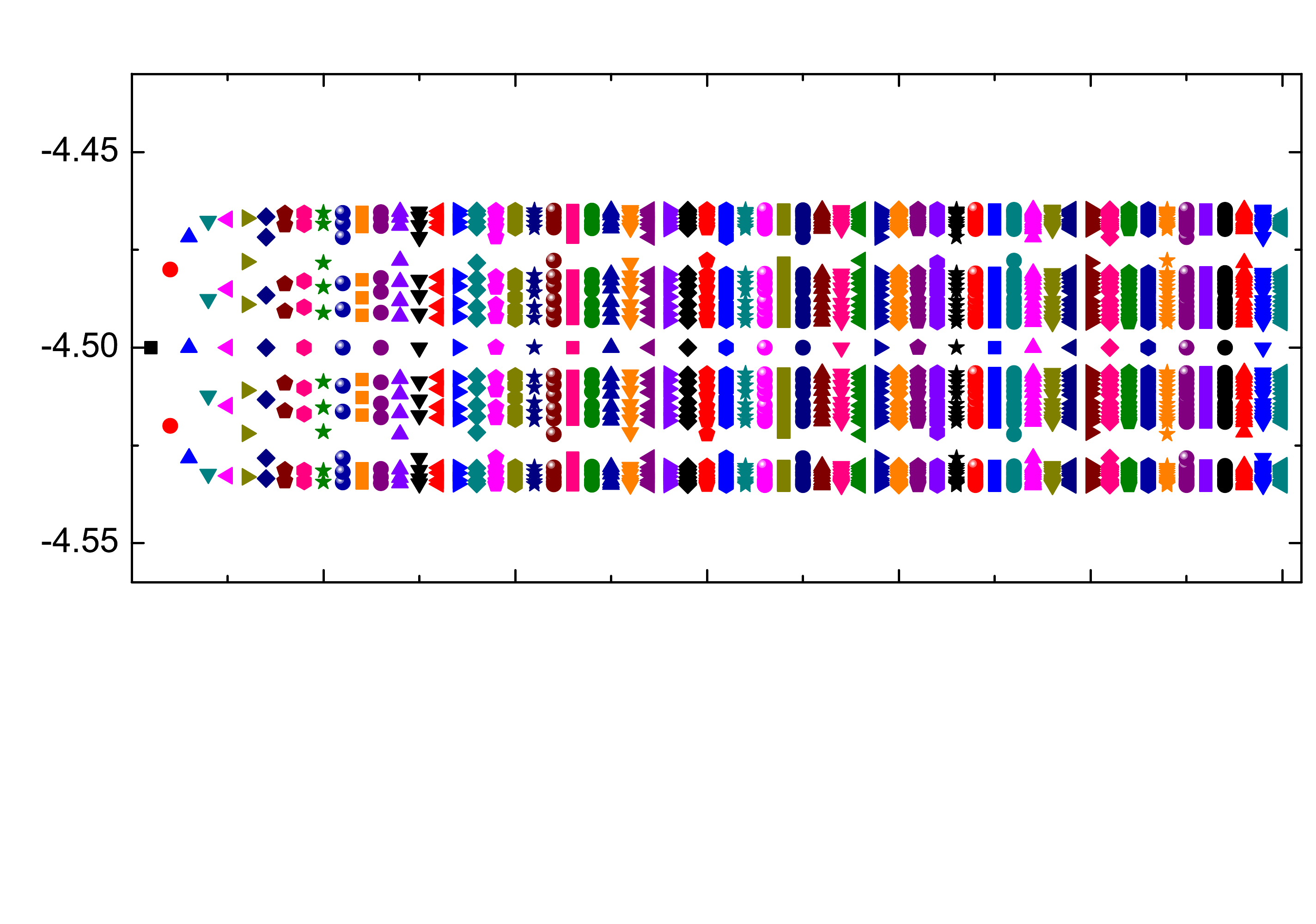} \\ \vspace{-2.5cm}
\includegraphics[width=0.45\textwidth]{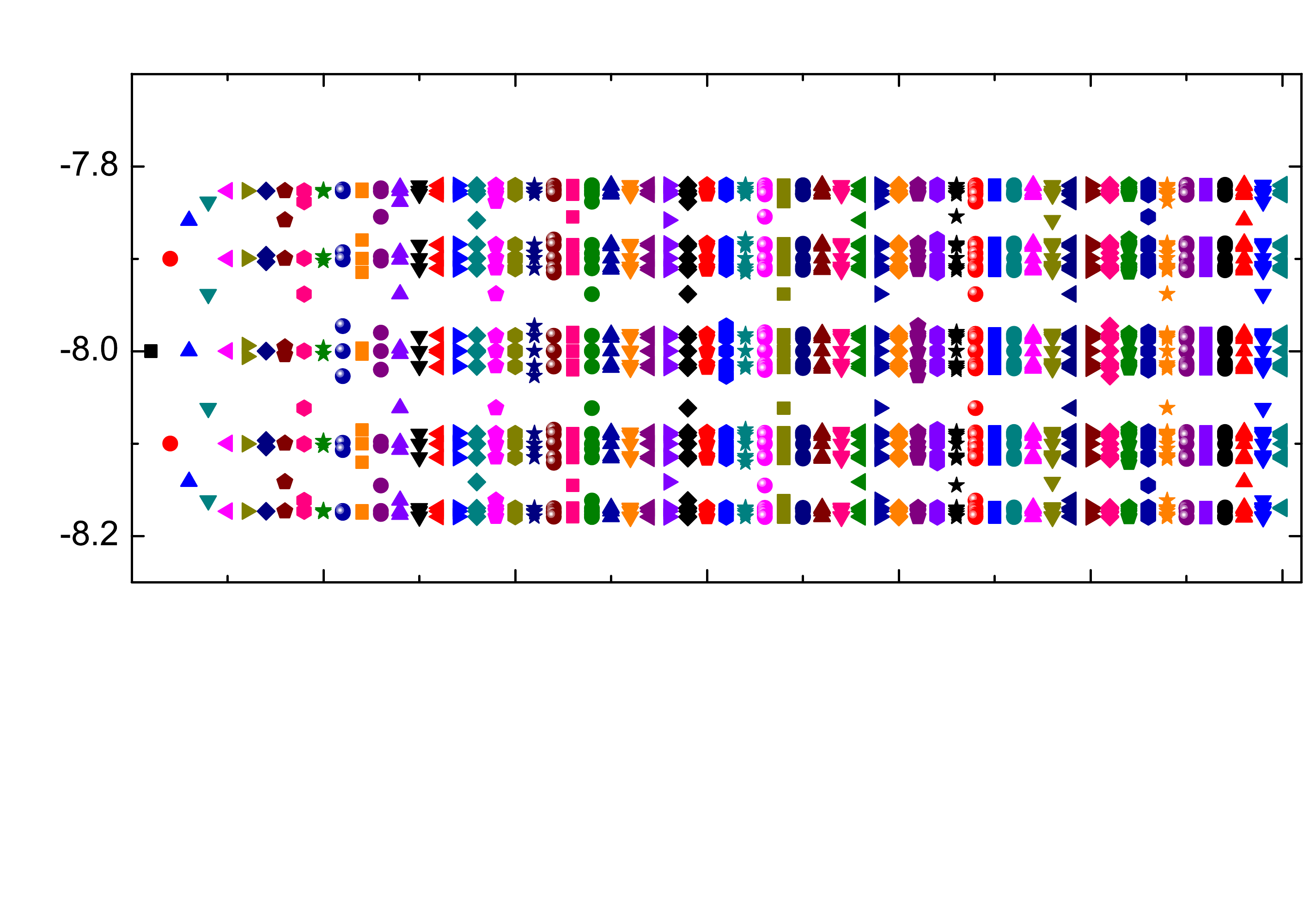}   
\includegraphics[width=0.45\textwidth]{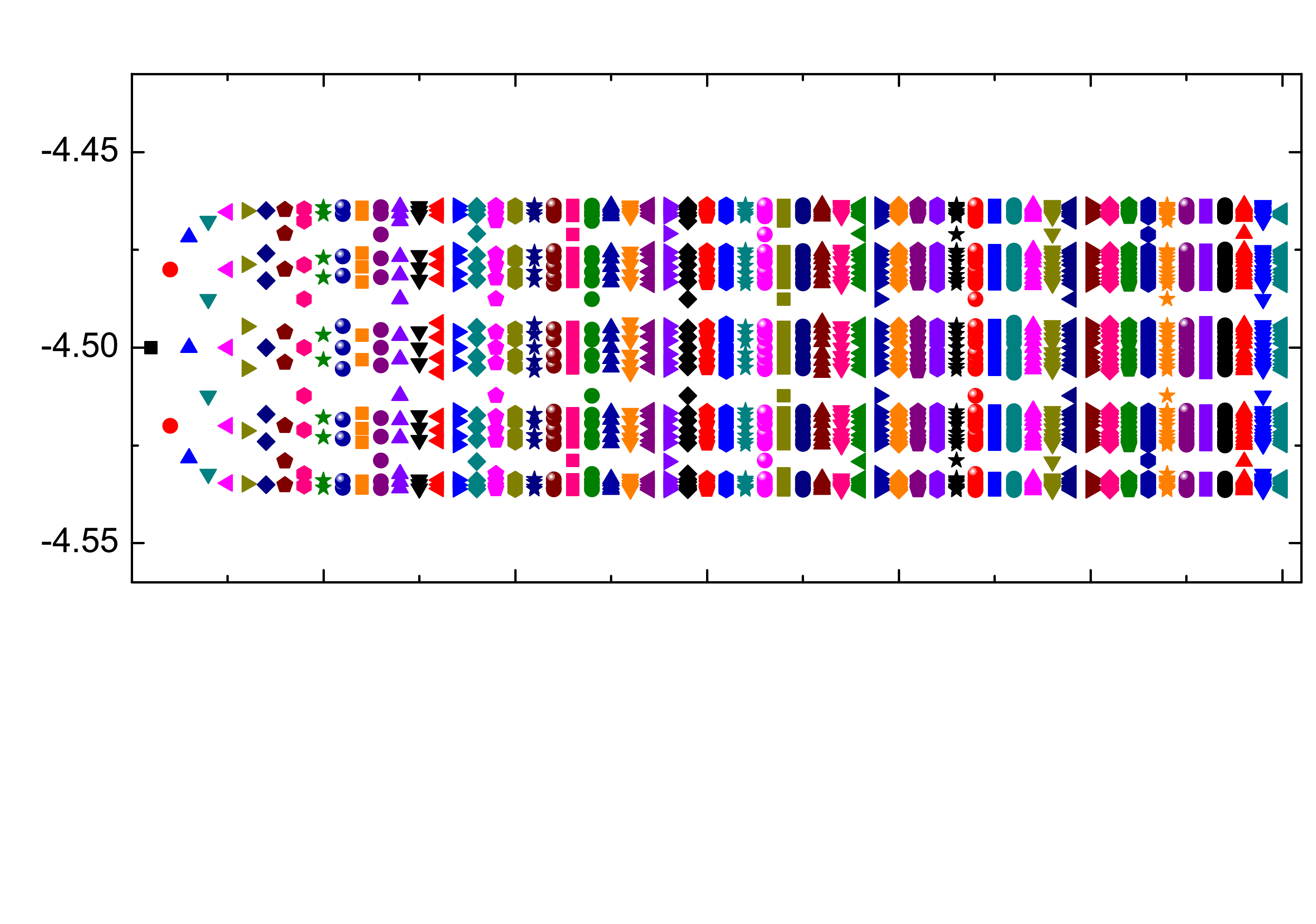} \\ \vspace{-2.5cm}
\includegraphics[width=0.45\textwidth]{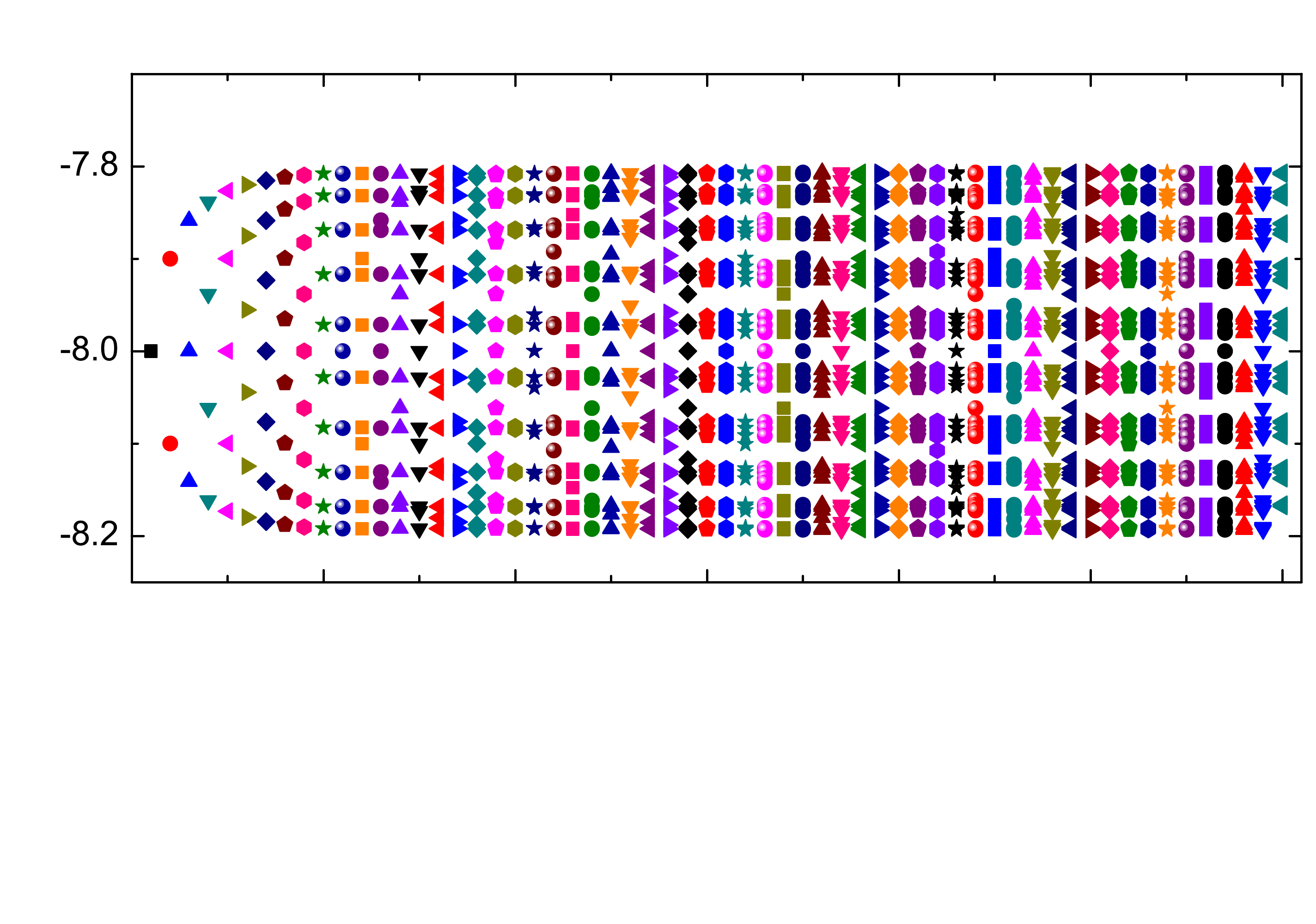}   
\includegraphics[width=0.45\textwidth]{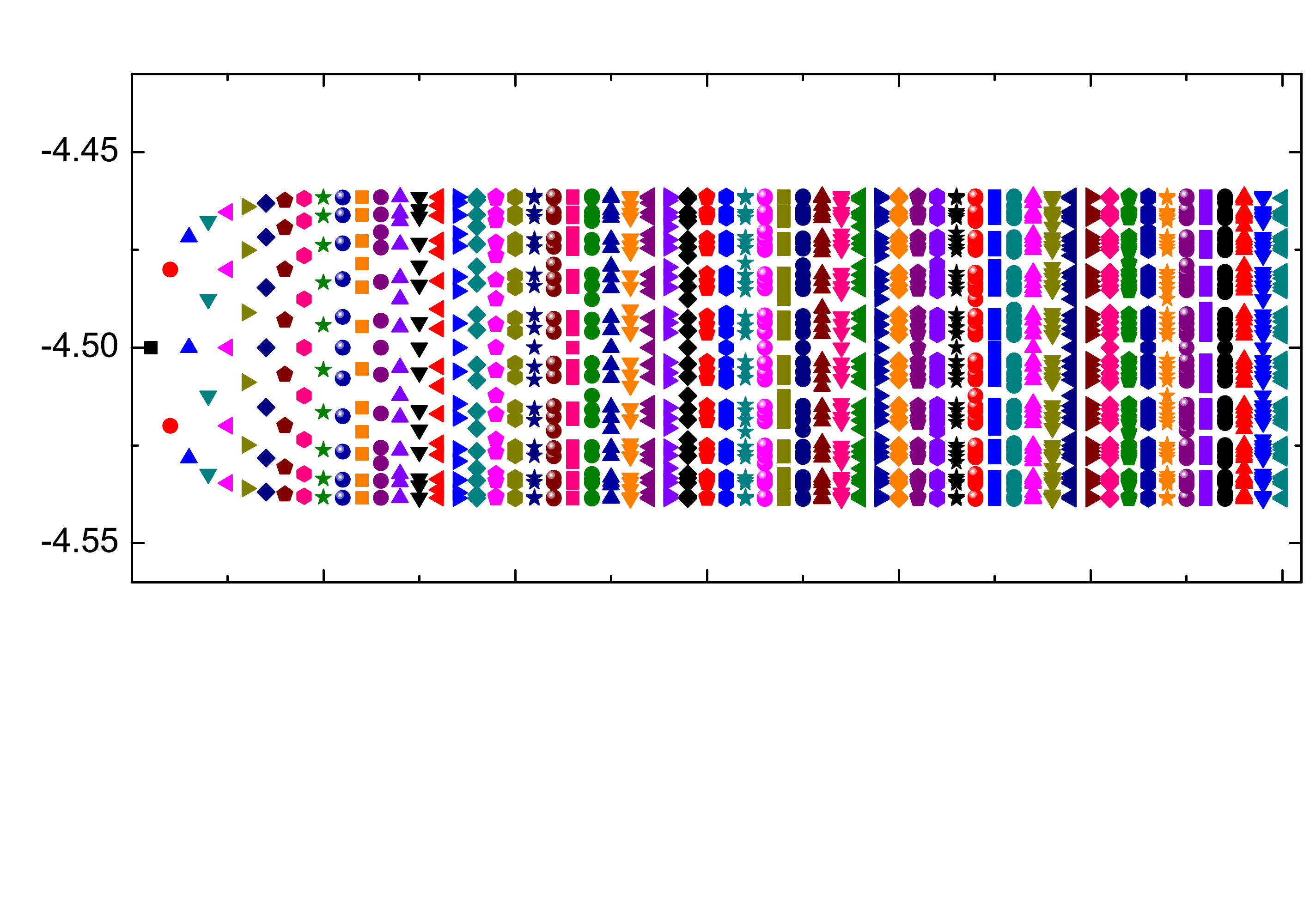} \\ \vspace{-2.5cm}
\includegraphics[width=0.45\textwidth]{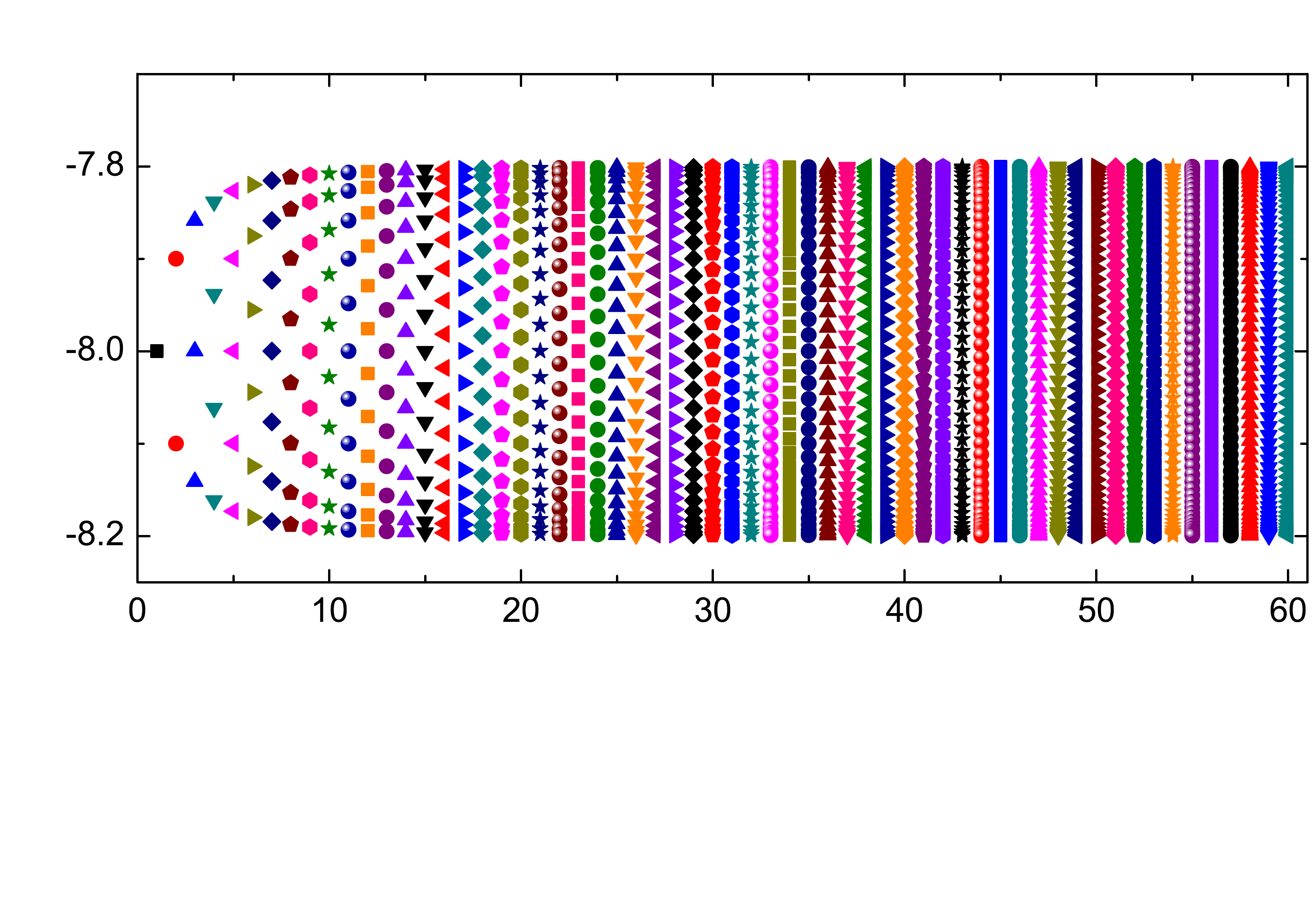}   
\includegraphics[width=0.45\textwidth]{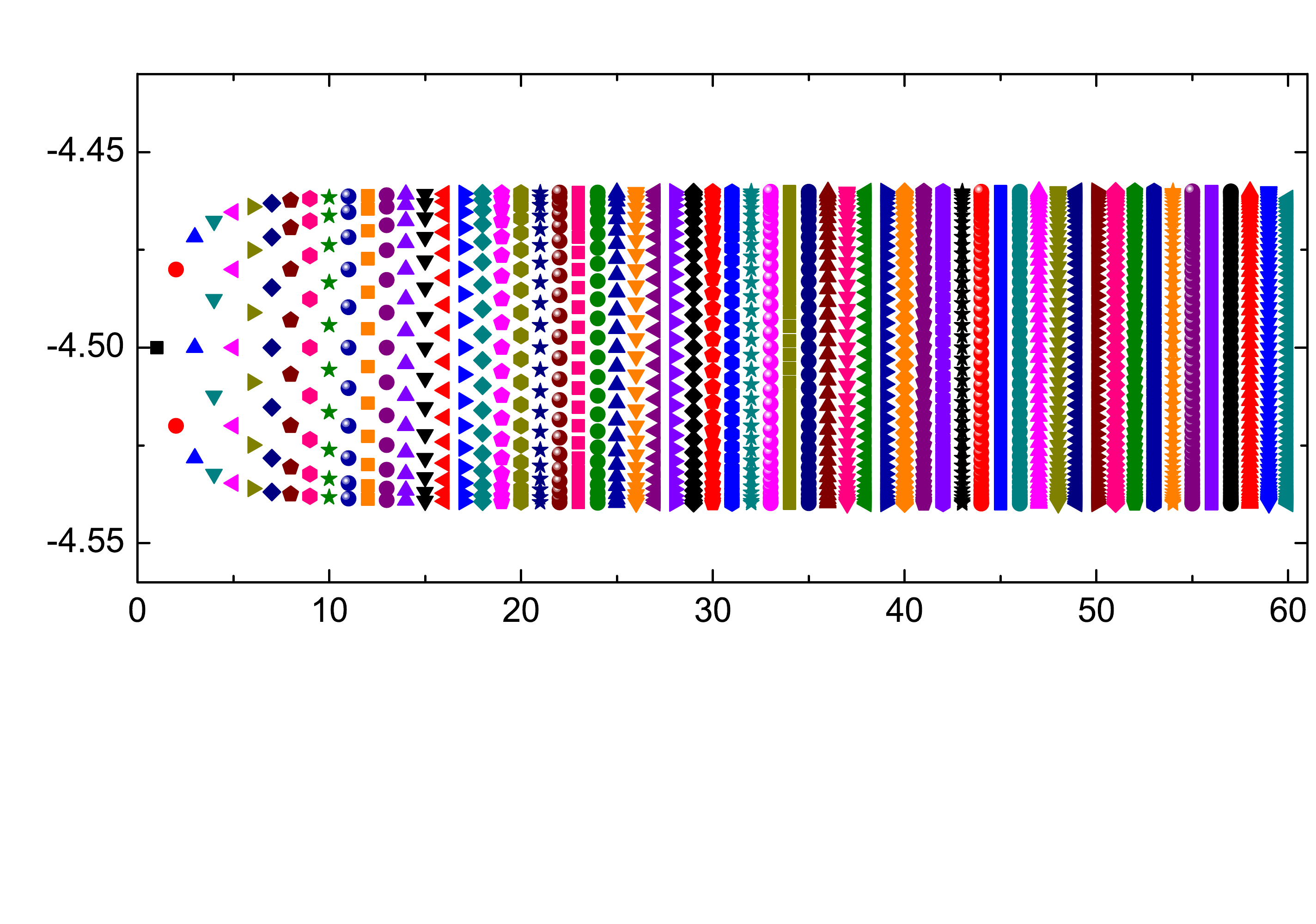}  \vspace{-1.7cm}
\caption{Eigenspectra of I polymers. HOMO regime (left) and LUMO regime (right).}
\label{fig:Eigenspectra-gaps-ChV}
\end{figure*}

\begin{figure*} [h!]\vspace{-0.8cm}
\includegraphics[width=0.45\textwidth]{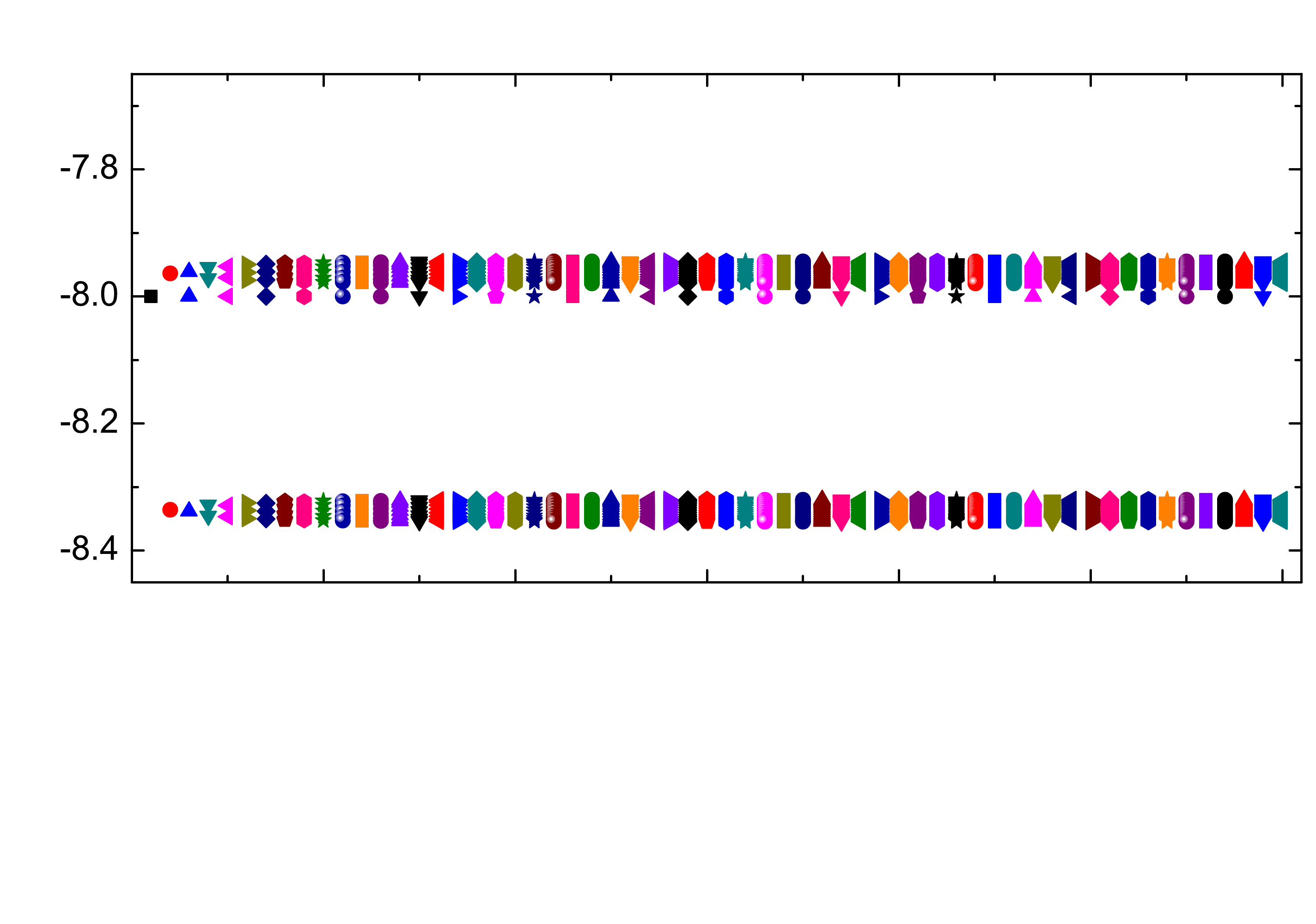}   
\includegraphics[width=0.45\textwidth]{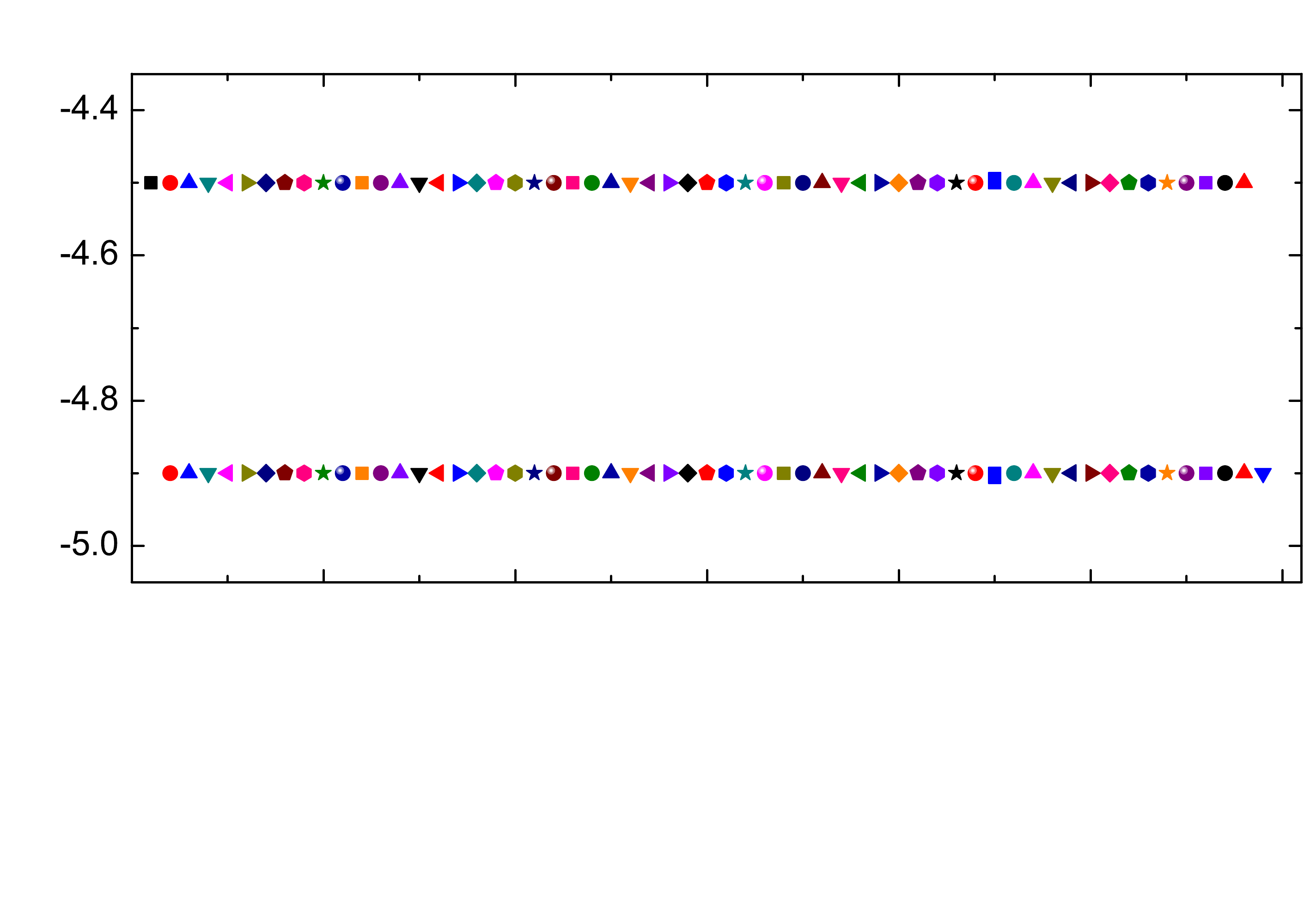} \\ \vspace{-2.5cm}
\includegraphics[width=0.45\textwidth]{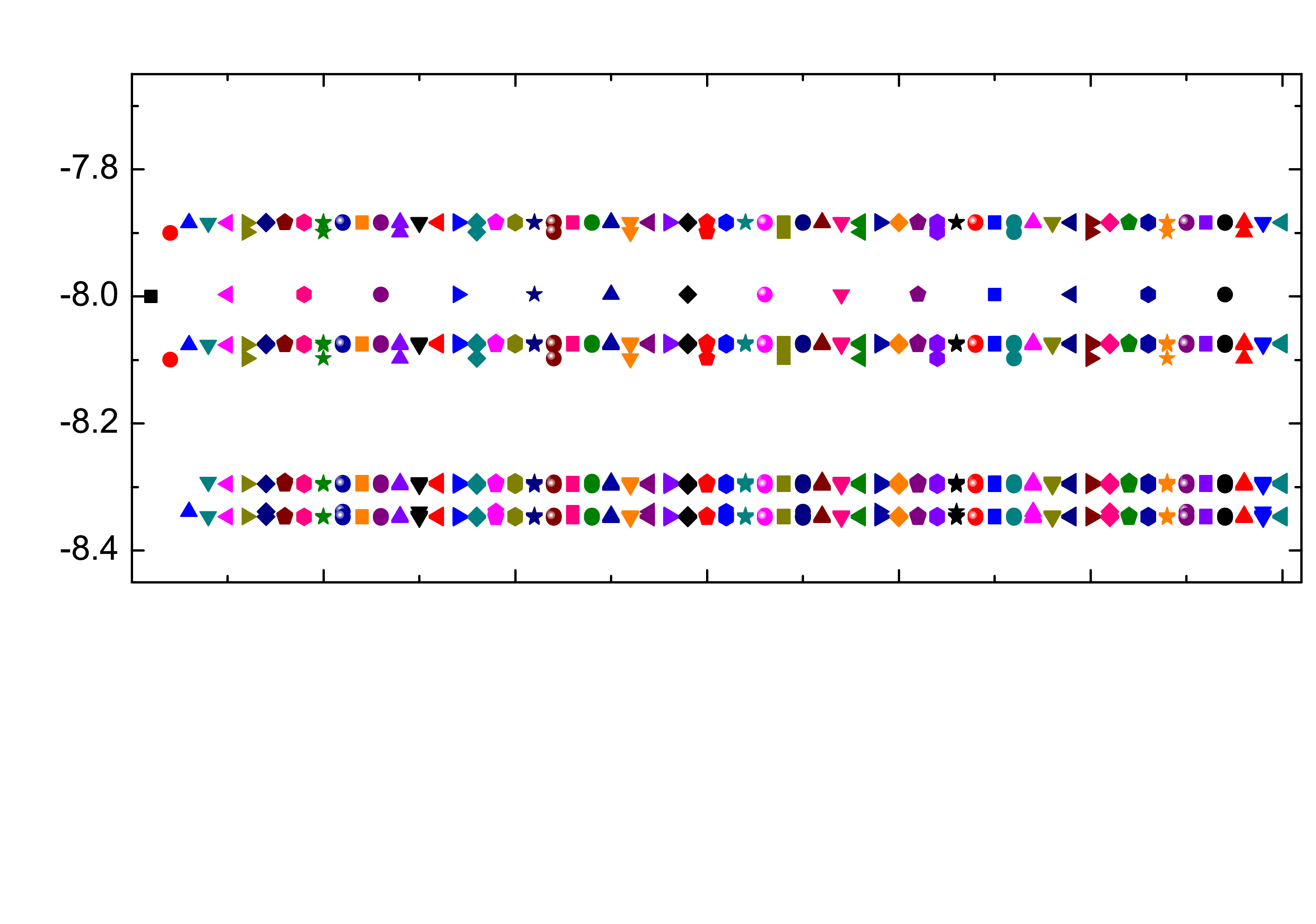}   
\includegraphics[width=0.45\textwidth]{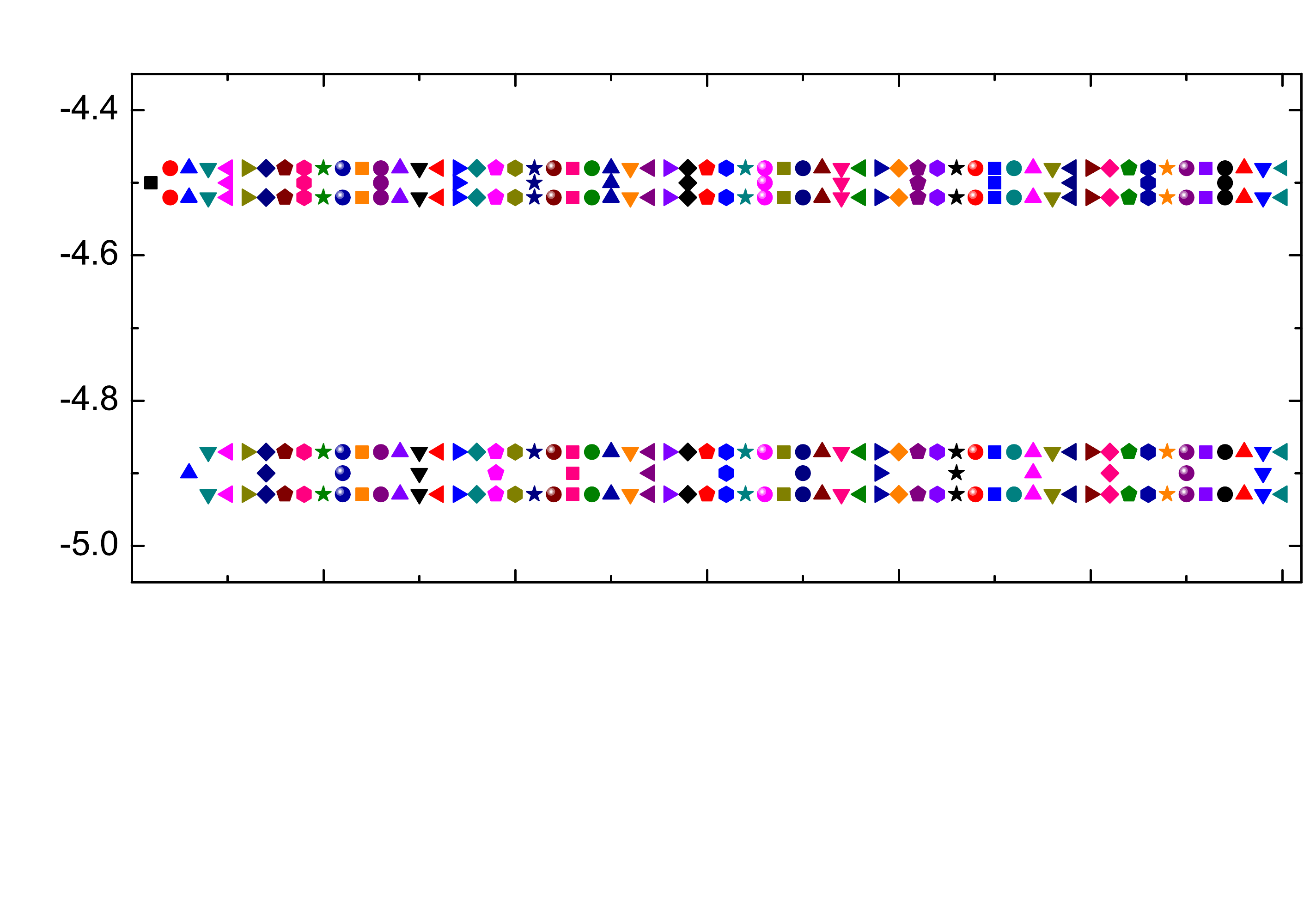} \\ \vspace{-2.5cm}
\includegraphics[width=0.45\textwidth]{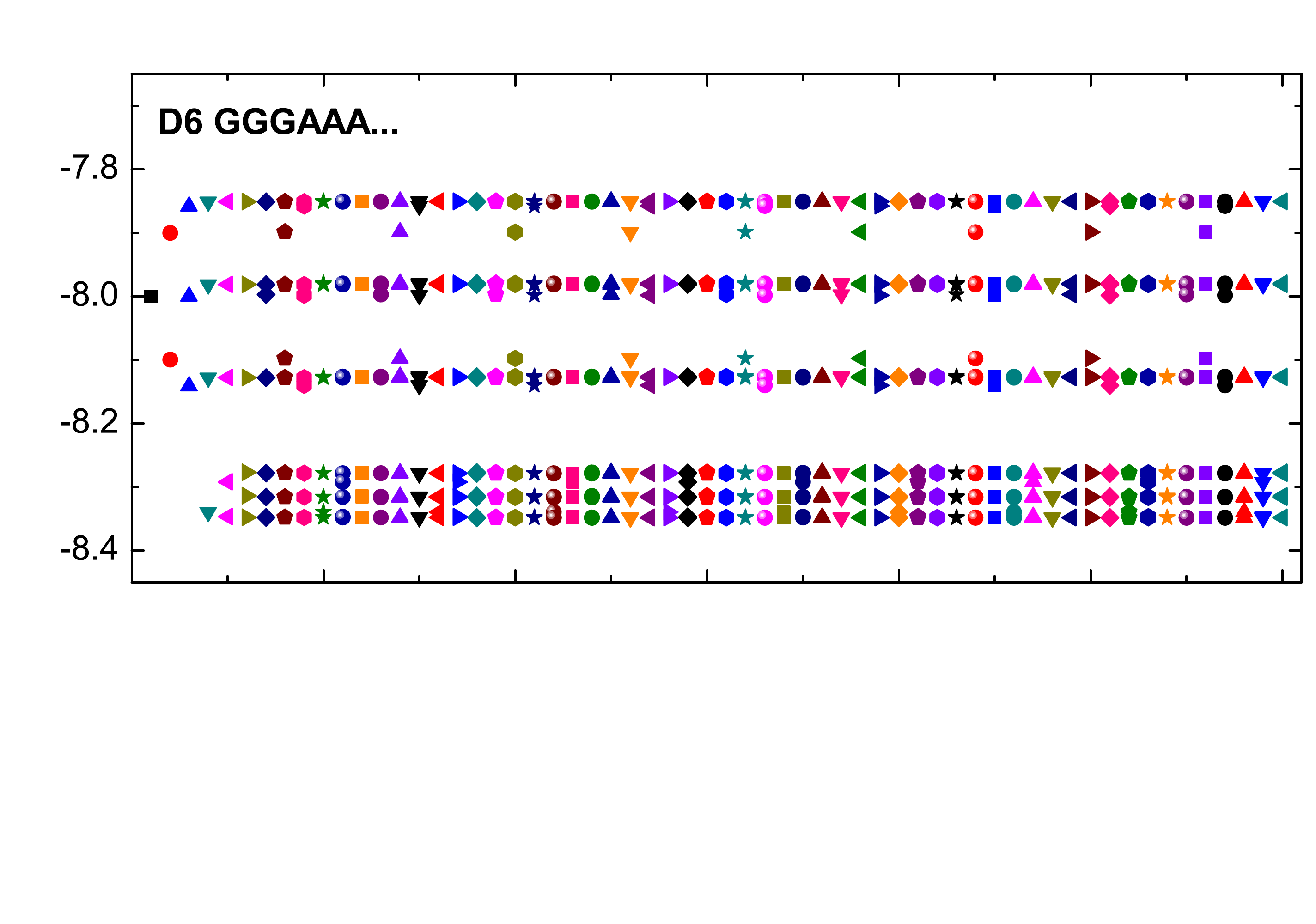}   
\includegraphics[width=0.45\textwidth]{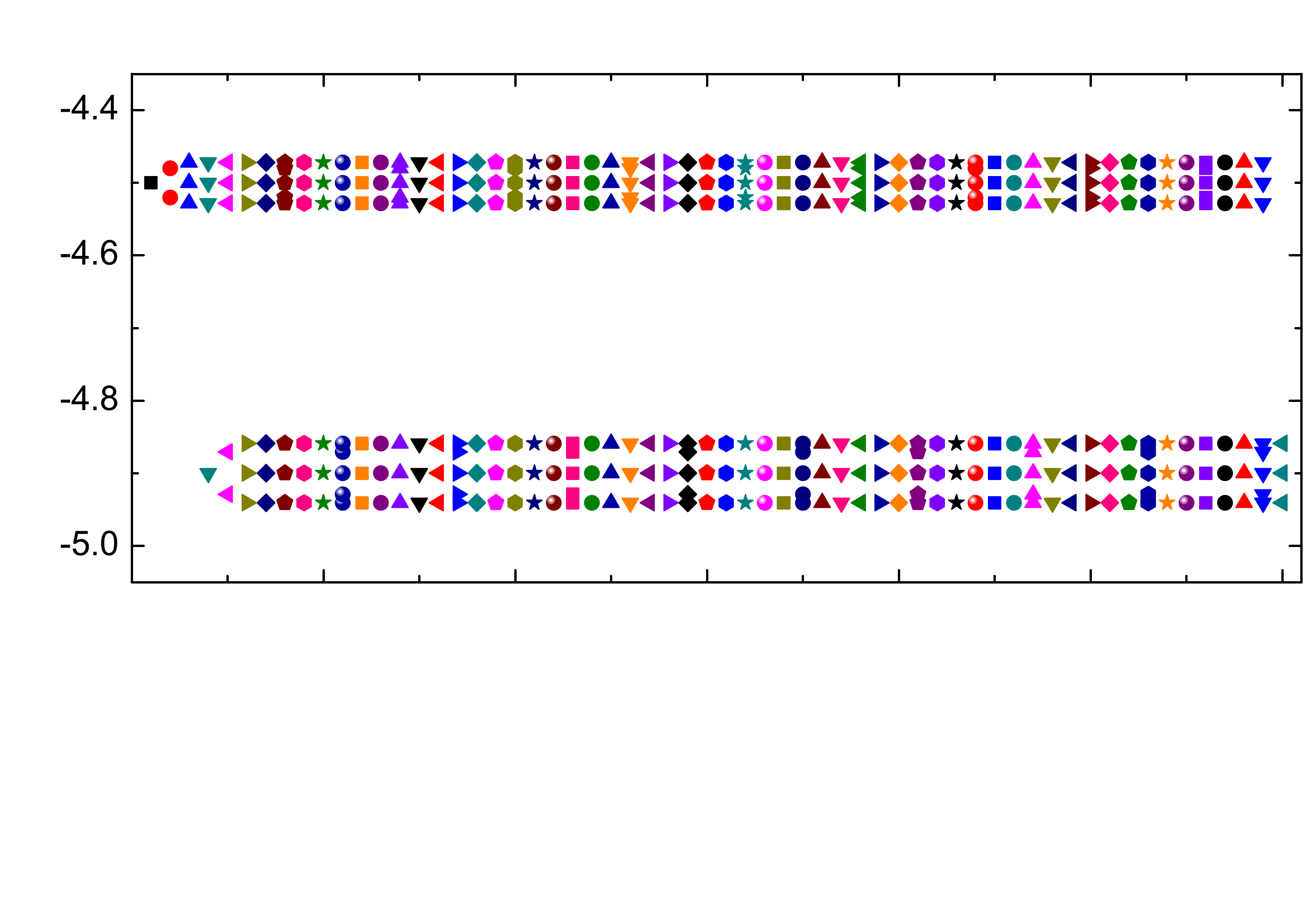} \\ \vspace{-2.5cm}
\includegraphics[width=0.45\textwidth]{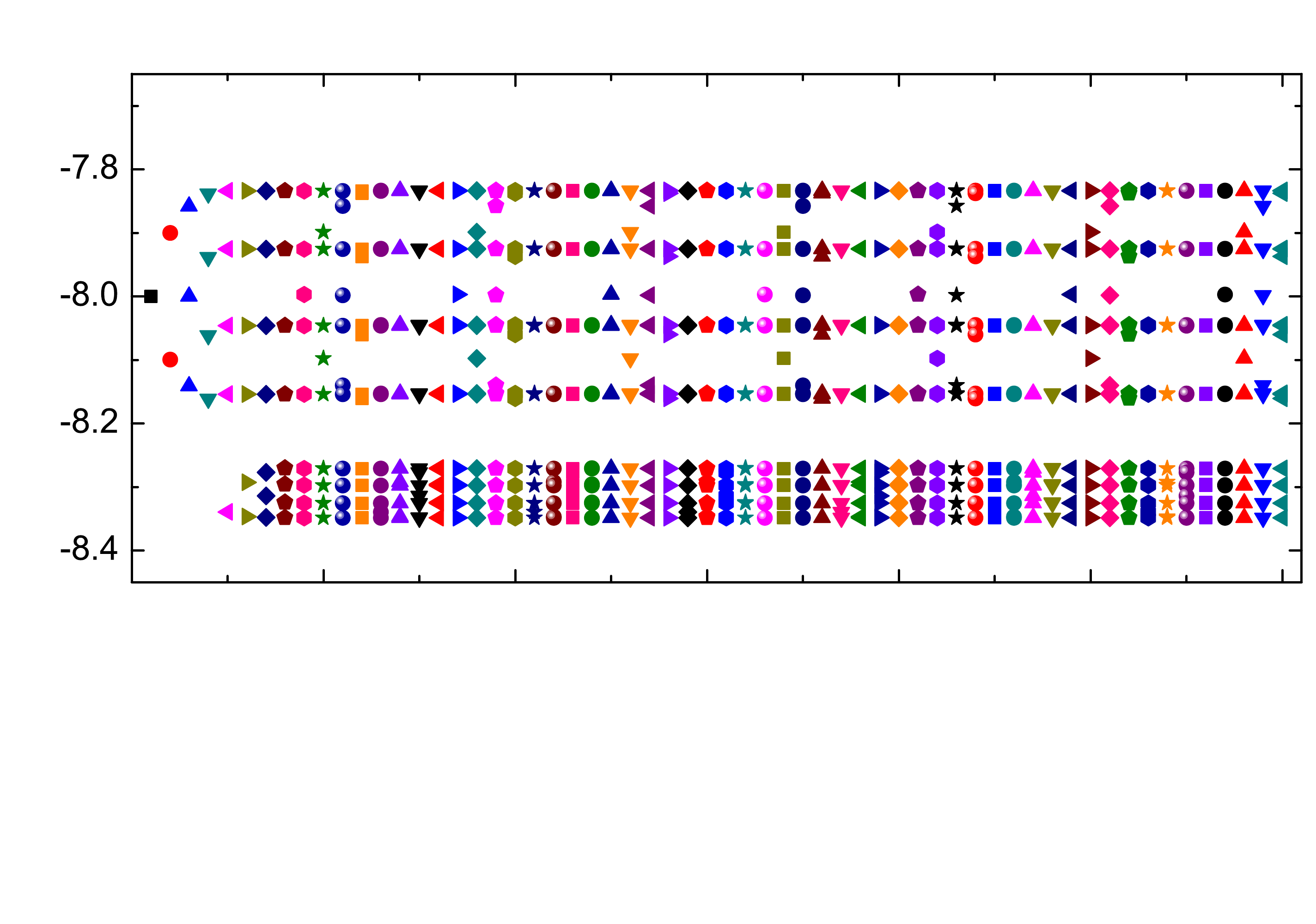}    
\includegraphics[width=0.45\textwidth]{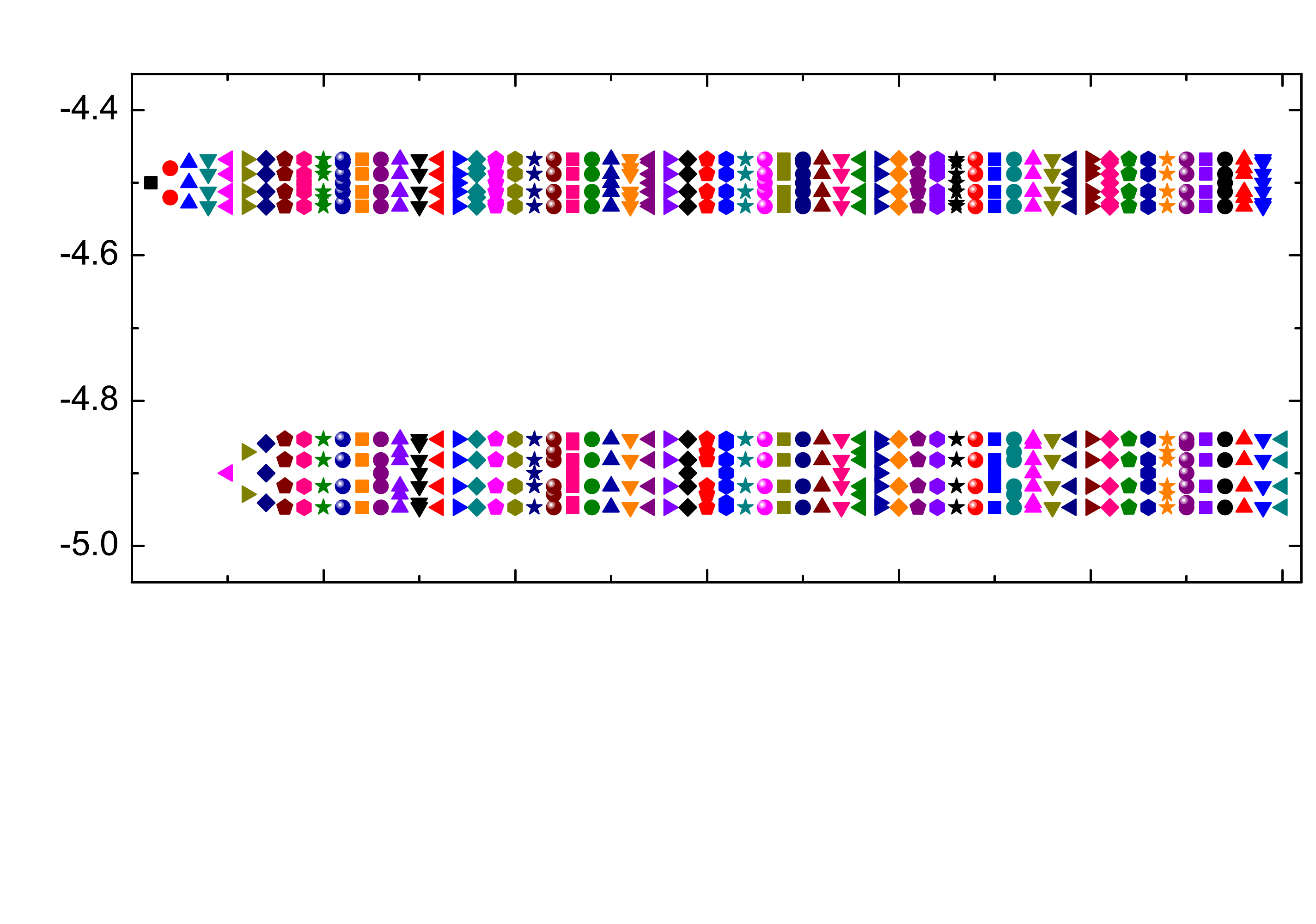} \\ \vspace{-2.5cm}
\includegraphics[width=0.45\textwidth]{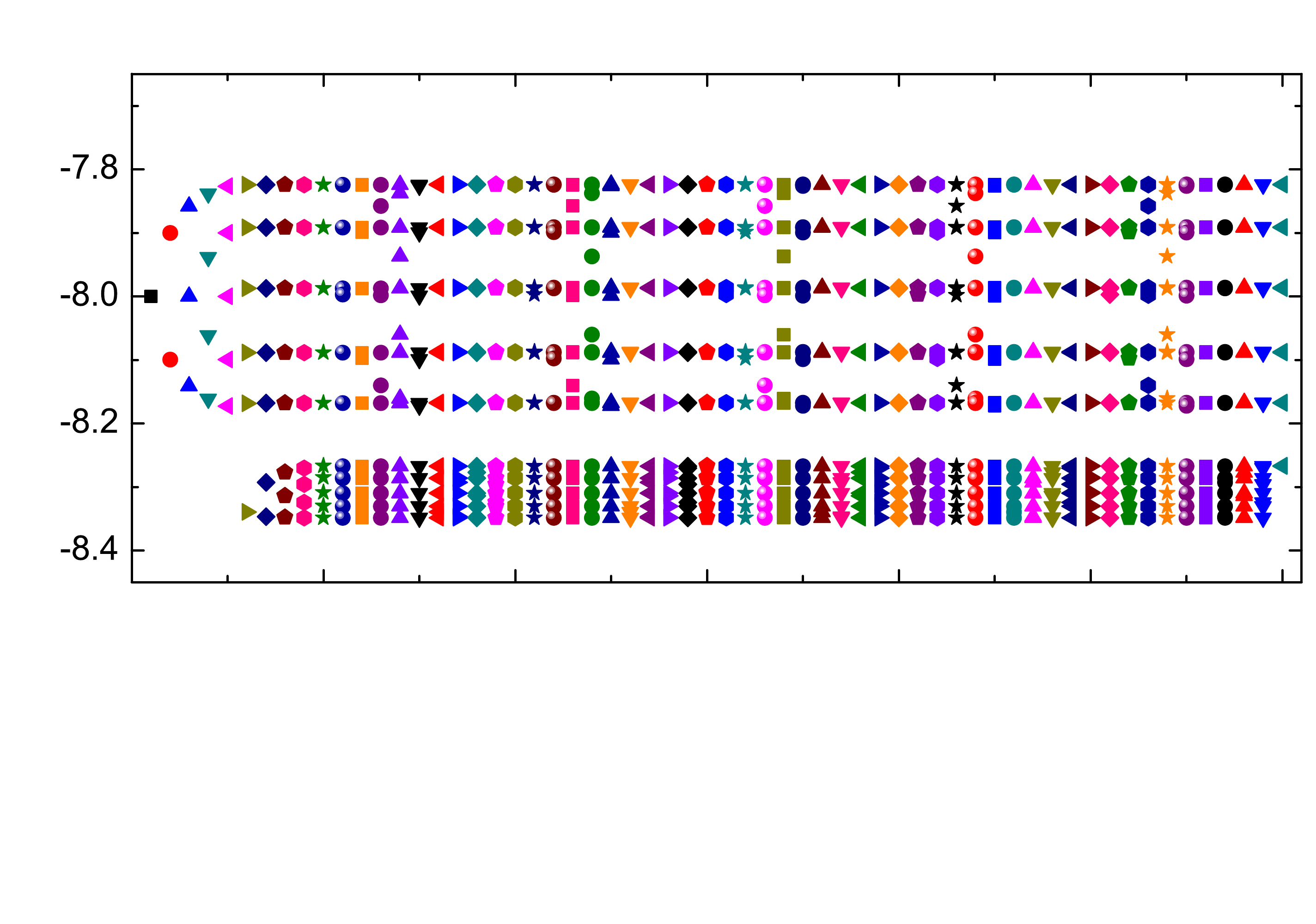}   
\includegraphics[width=0.45\textwidth]{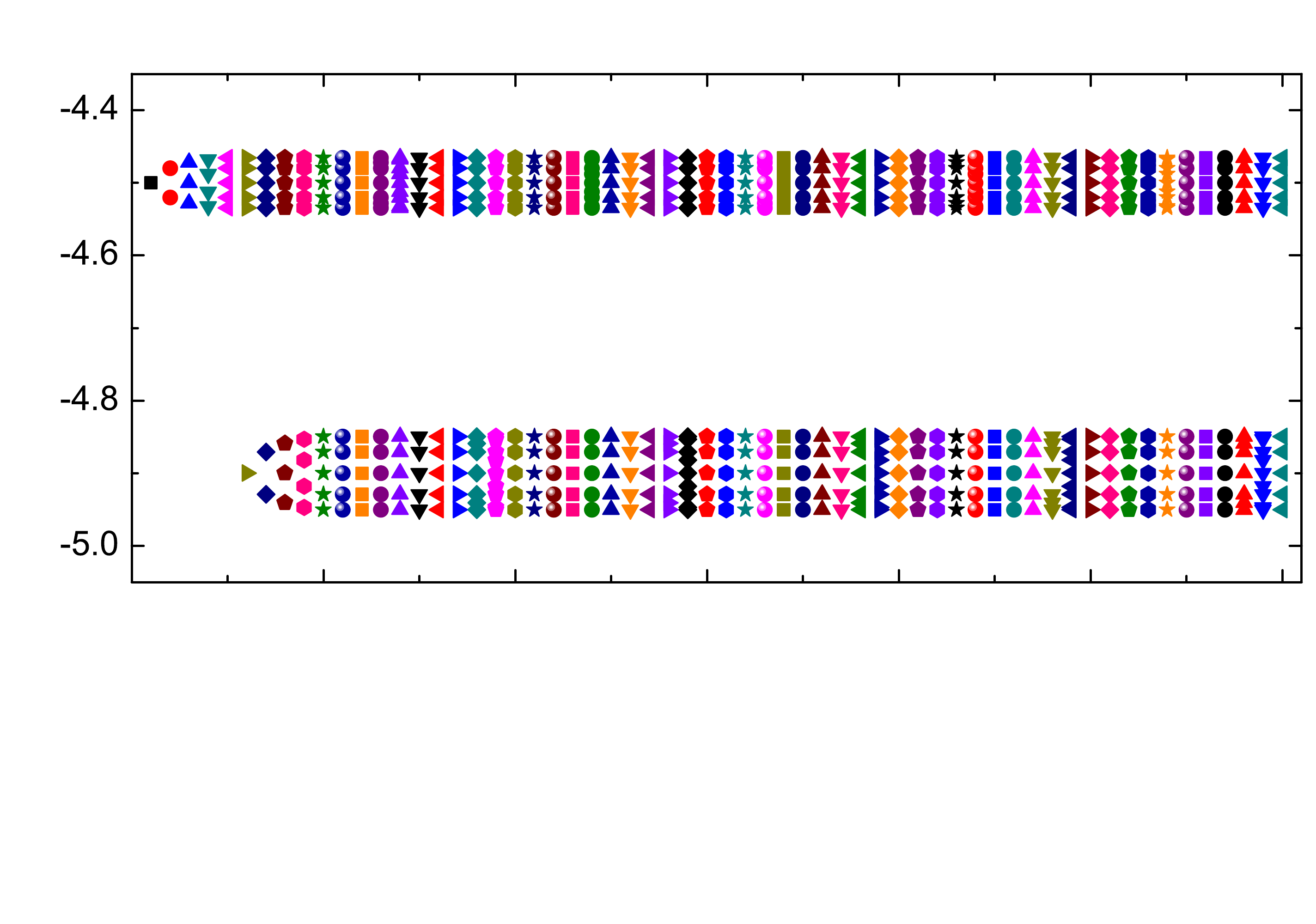} \\ \vspace{-2.5cm}
\includegraphics[width=0.45\textwidth]{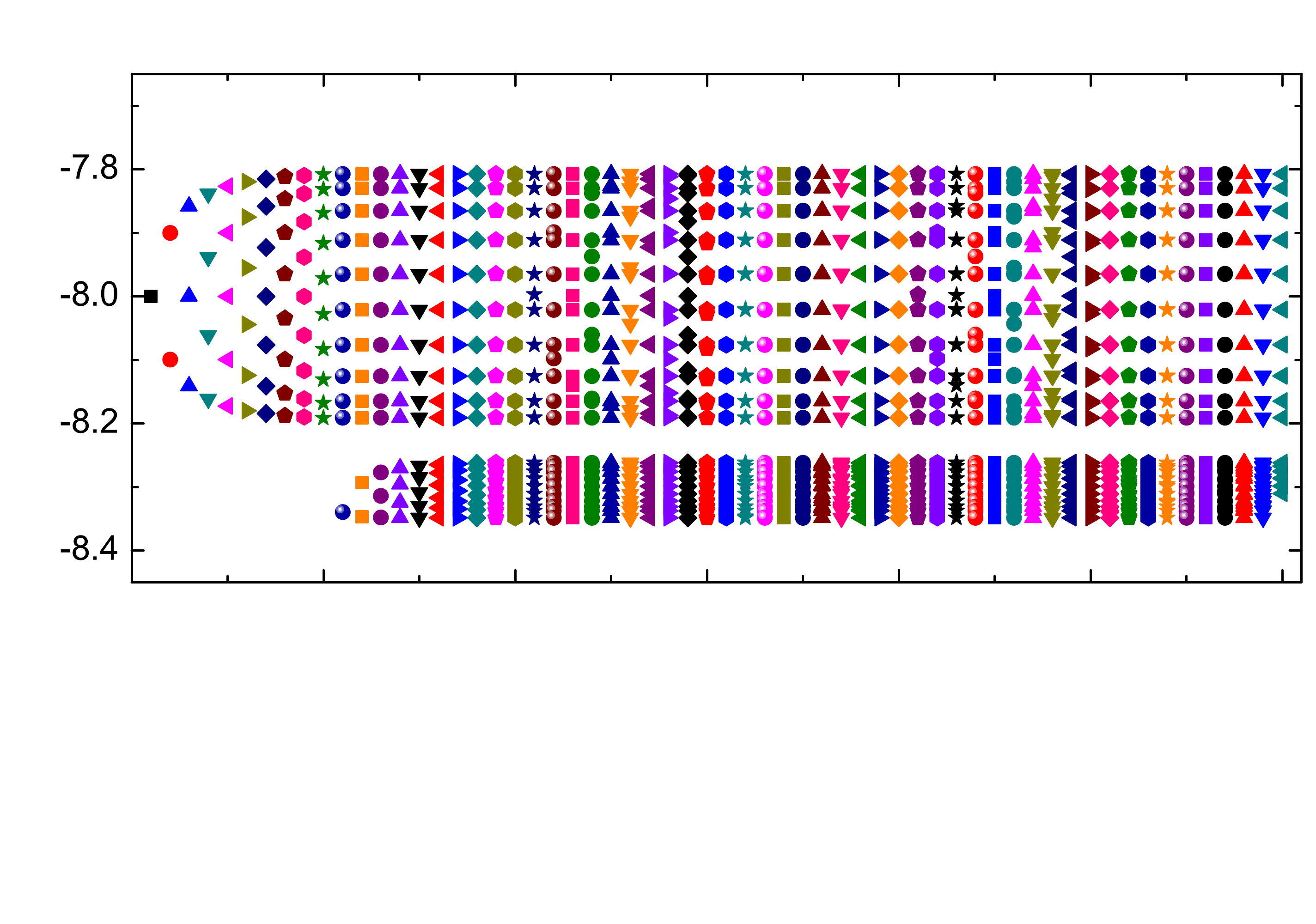}   
\includegraphics[width=0.45\textwidth]{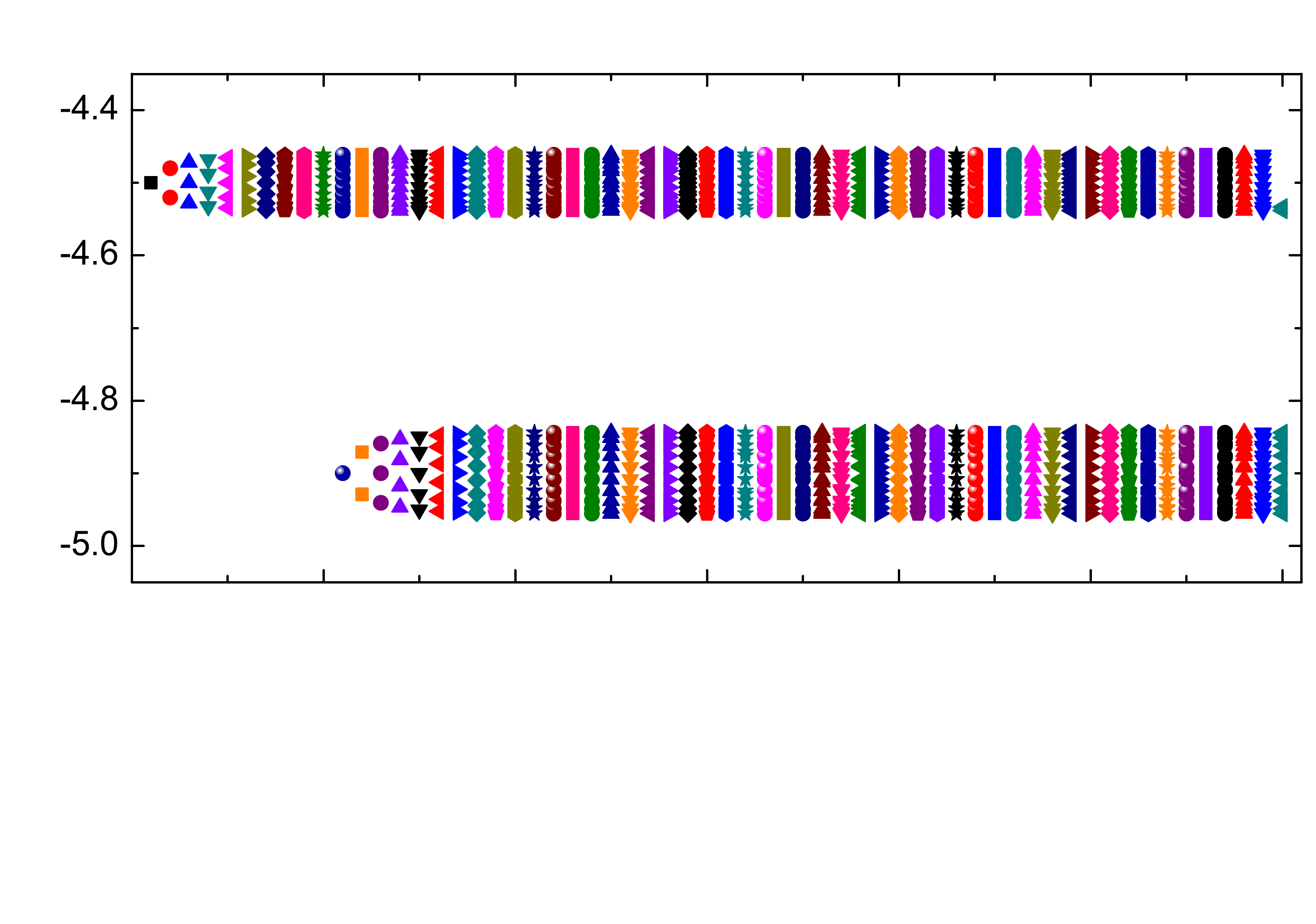} \\ \vspace{-2.5cm}
\includegraphics[width=0.45\textwidth]{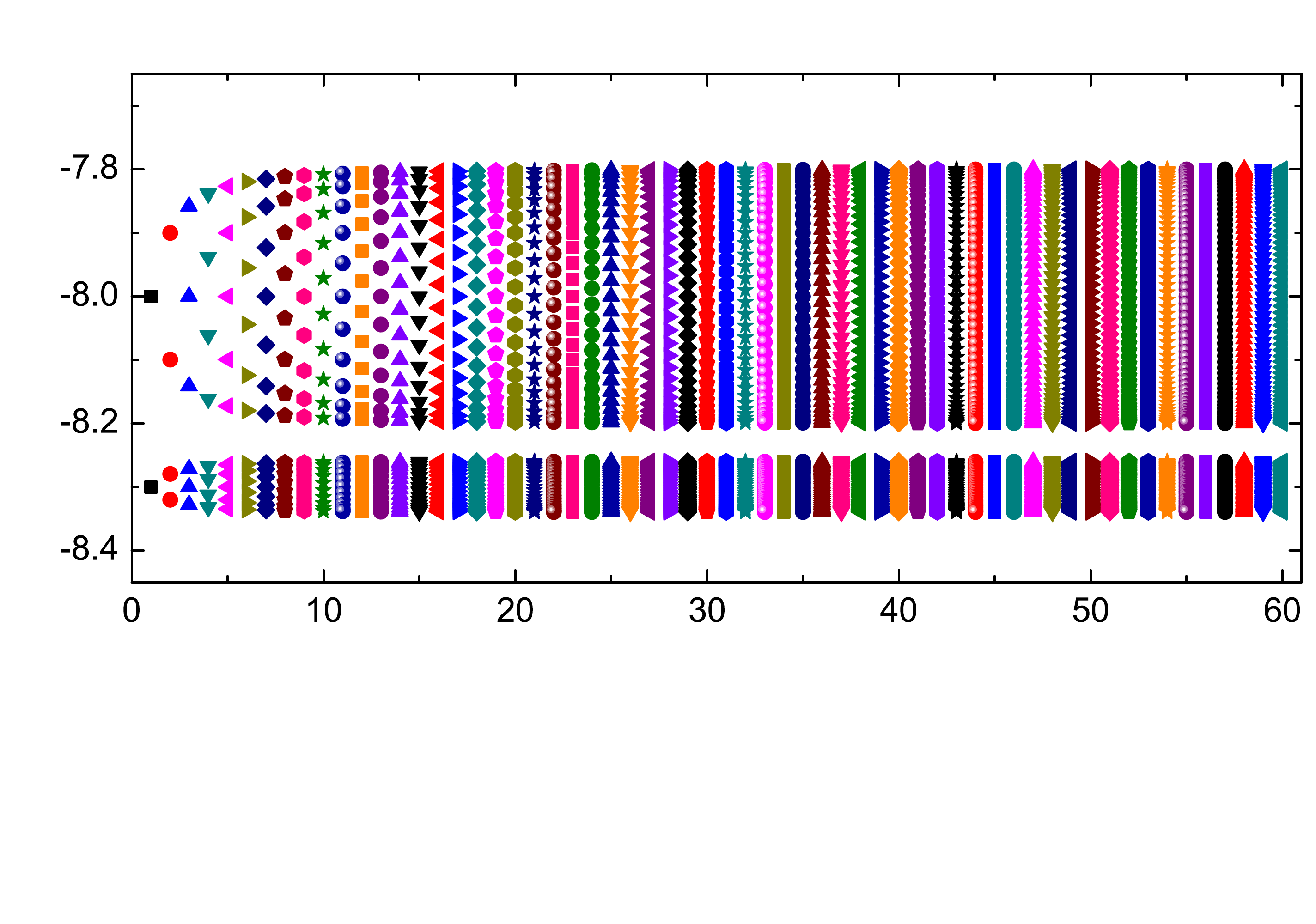}
\includegraphics[width=0.45\textwidth]{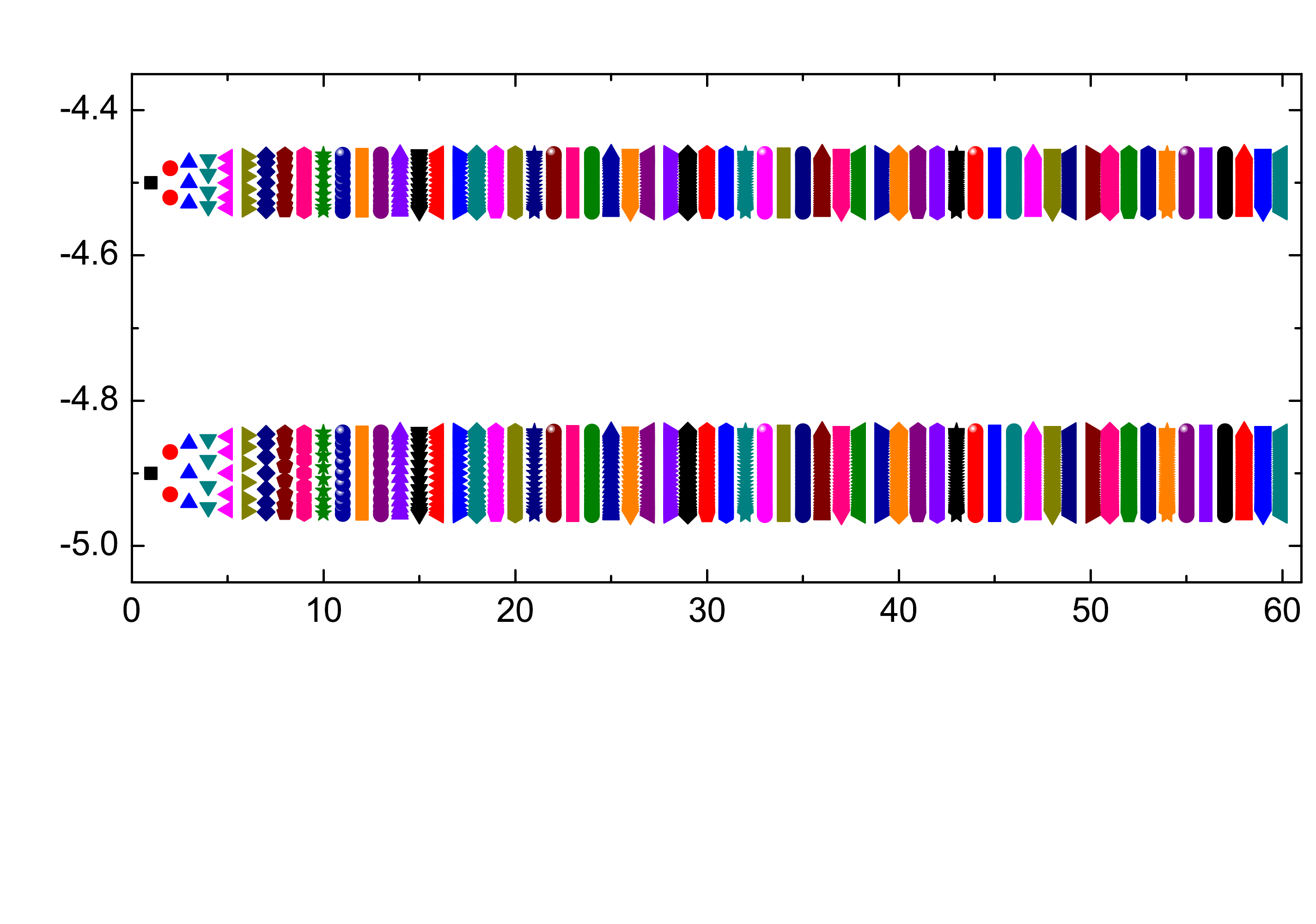}  \vspace{-1.7cm}
\caption{Eigenspectra of D polymers as well as of I1 (G...) and I1 (A...) polymers plotted together. HOMO regime (left) and LUMO regime (right).}
\label{fig:Eigenspectra-gaps-PMp}
\end{figure*}

\begin{figure*} [h!]
\includegraphics[width=0.95\textwidth]{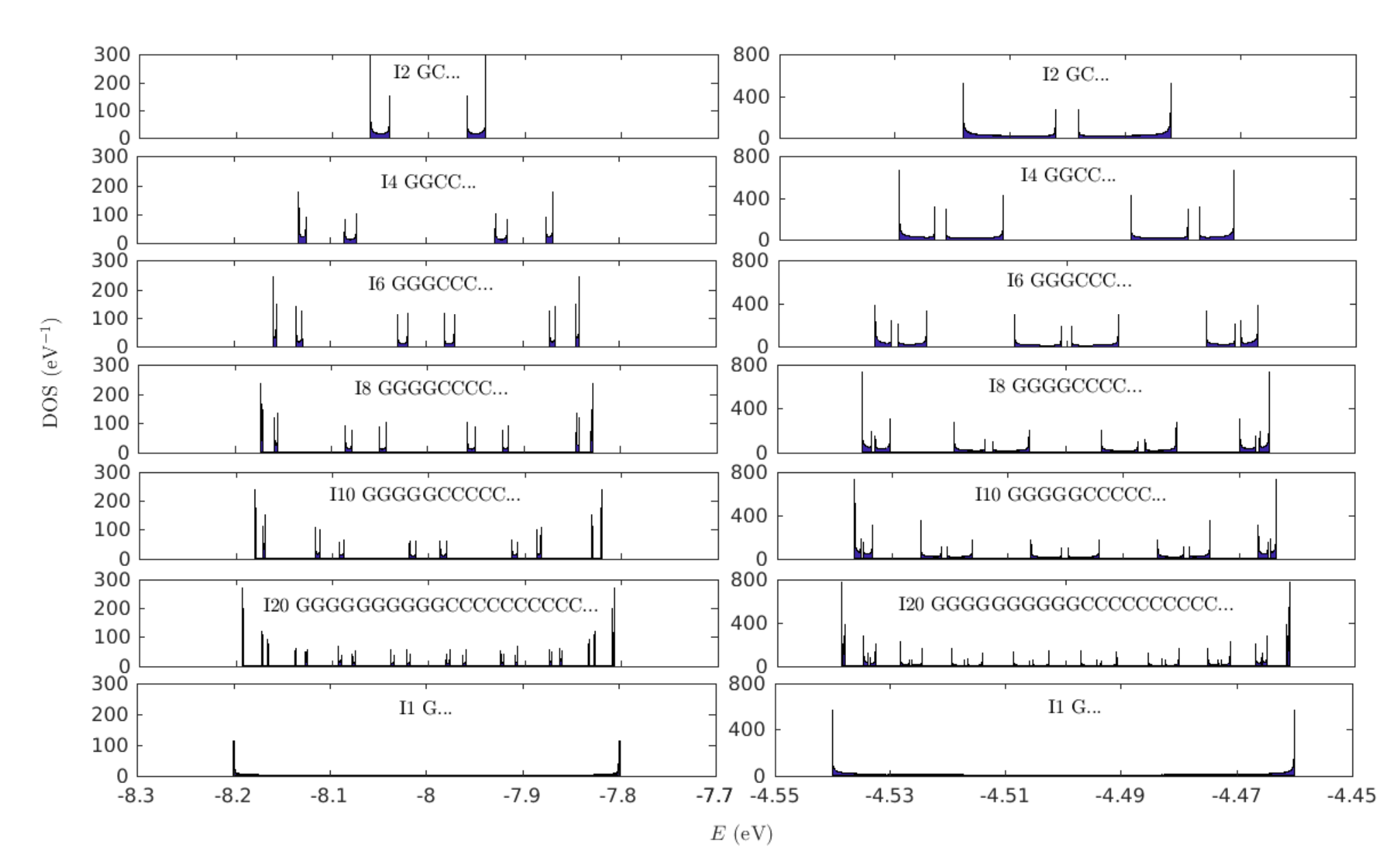}  \vspace{-0.5cm}
\caption{Density of states of I polymers for the HOMO (left) and the LUMO (right) regime.} 
\label{fig:DOS-ChV}
\end{figure*}

\begin{figure*} [h!]
\includegraphics[width=0.95\textwidth]{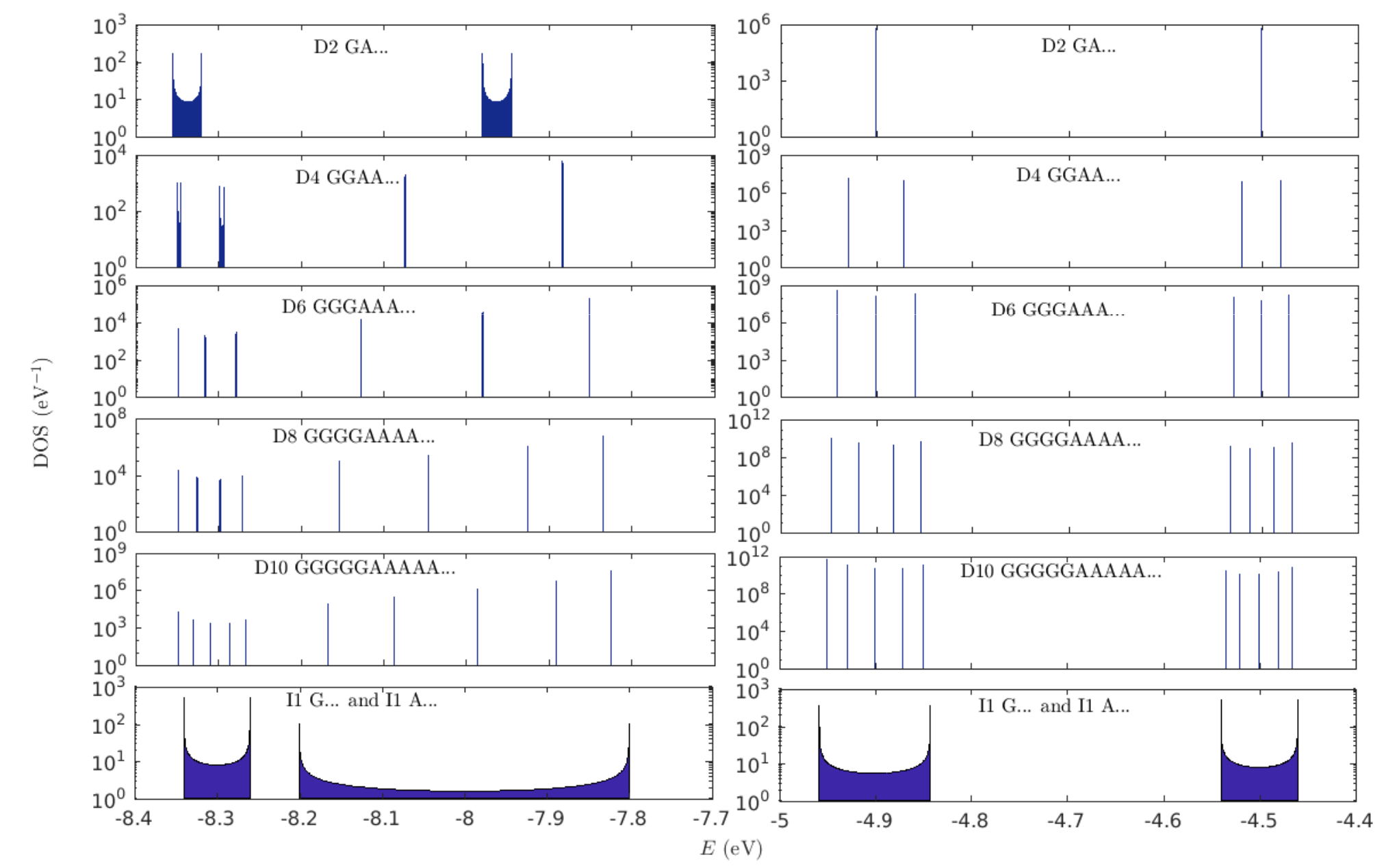}  \vspace{-0.5cm}
\caption{Density of states of D polymers as well as of I1 (G...) and I1 (A...) polymers plotted together, for the HOMO (left) and the LUMO (right) regime.}
\label{fig:DOS-PMp}
\end{figure*}

In Figs.~\ref{fig:Eigenspectra-gaps-ChV}~and~\ref{fig:Eigenspectra-gaps-PMp}, 
we show the HOMO and LUMO eigenspectra of 
[I2, I4, I6, I8, I10, I20 and I1 polymers] and 
[D2, D4, D6, D8, D10, D20 and I1 (G...), I1 (A...) polymers], and 
in Figs.~\ref{fig:DOS-ChV}~and~\ref{fig:DOS-PMp} we plot the corresponding DOS.
The HOMO and LUMO bands of each polymer consist of $P$ subbands, e.g., 
for I6 or D6 polymers, the number of subbands is 6. 
Some eigenenergies protrude periodically from the subbands at certain relationships between $N$ and $P$. 
At the limits of subbands, Van Hove singularities occur. 
The subbands are separated by small energy gaps, which, increasing $P$, decrease. 

For polymers made of identical monomers (cf. Figs.~\ref{fig:Eigenspectra-gaps-ChV}-\ref{fig:DOS-ChV}), 
all eigenvalues are symmetric around the monomer on-site energy and for $N$ odd the trivial eigenvalue, equal to the monomer on-site energy, exists. Increasing $P$, the eigenspectra tend to the eigenspectra of I1 polymers, and  
the DOS tends to the DOS of I1 polymers. 

For polymers made of different monomers (cf. Figs.~\ref{fig:Eigenspectra-gaps-PMp}-\ref{fig:DOS-PMp}), 
increasing $P$, the eigenenergies gather around the two monomer on-site energies. 
Increasing $P$, the eigenspectra gather within the limits defined by the union of eigenspectra of I1 (G...) and I2 (A...) polymers.  
In Fig.~\ref{fig:DOS-ChV}, increasing $P$, the subbands become narrower, but they are wide enough (e.g. for I10 of a few meV) so that their DOS minima remain low enough; therefore we don't have to change the vertical presentation scale. In Fig.~\ref{fig:DOS-PMp}, the already very narrow subbands, increasing $P$, become even narrower (e.g for D10 two to nine orders of magnitude narrower than for I10), which drives the DOS minima in each subband much higher; therefore, to depict the DOS, we have to increase the vertical presentation scale.
Finally, we notice that apart the eigenvalues for $N=Pn$ can be obtained analytically and recursively with the help of the Chebyshev polynomials of the second kind, via the Transfer Matrix Method~\cite{LS:2018}.

The energy gap of a monomer is the difference between its LUMO and HOMO levels. The energy gap of a polymer is the difference between 
the lowest level of the LUMO regime and the highest level of the HOMO regime,  
because we assume that the orbitals --one per site-- which contribute to the HOMO (LUMO) band are occupied (empty), since in both possible monomers there is an even number of $p_z$ electrons contributing to the $\pi$ stack~\cite{HKS:2010-2011}.
The energy gaps of all possible I1, I2, D2 polymers, for TB I and TB II, can be found in Ref.~\cite{LChMKLTTS:2016}. 

\begin{figure} [h!]
\includegraphics[width=0.45\textwidth]{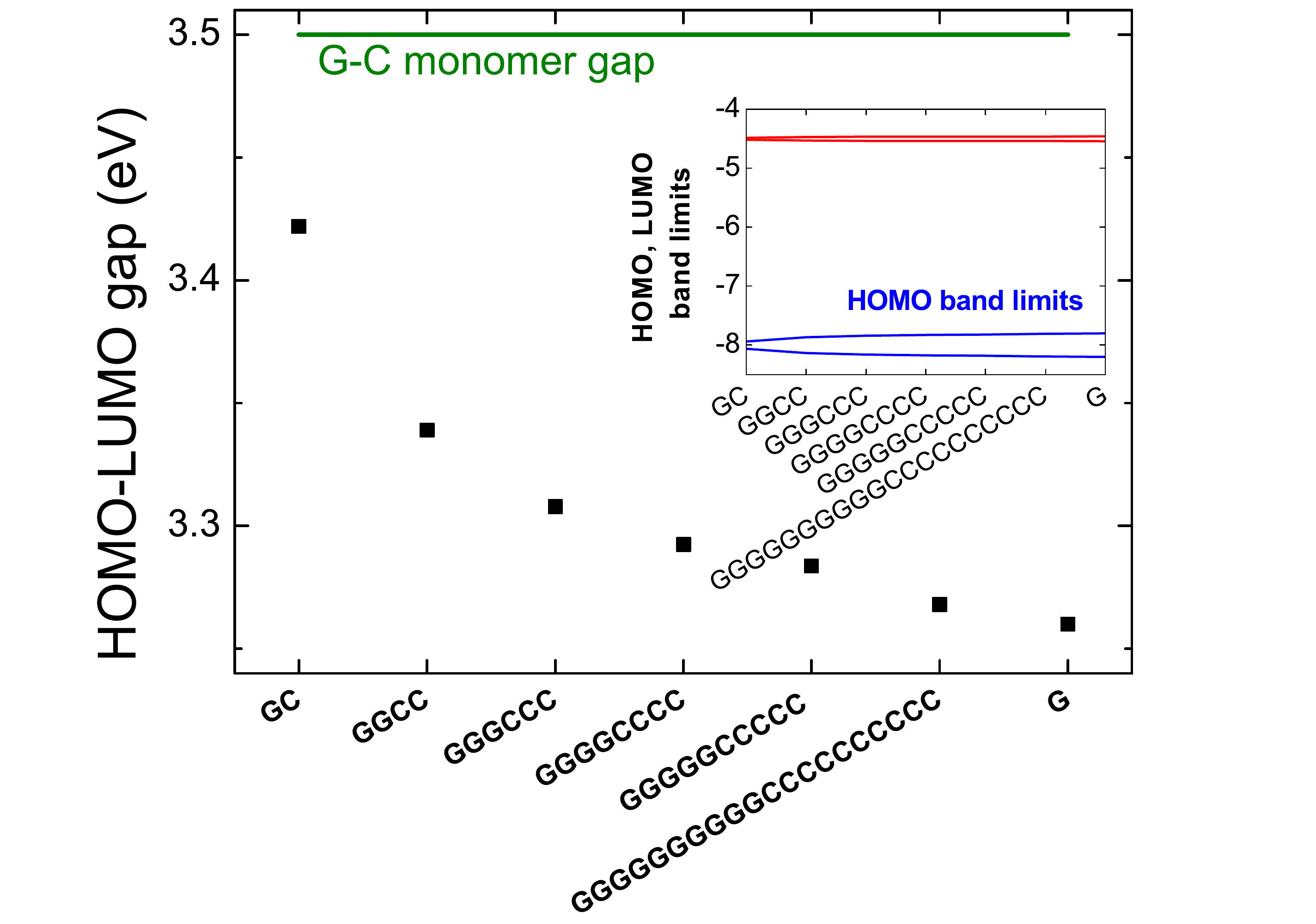}
\includegraphics[width=0.45\textwidth]{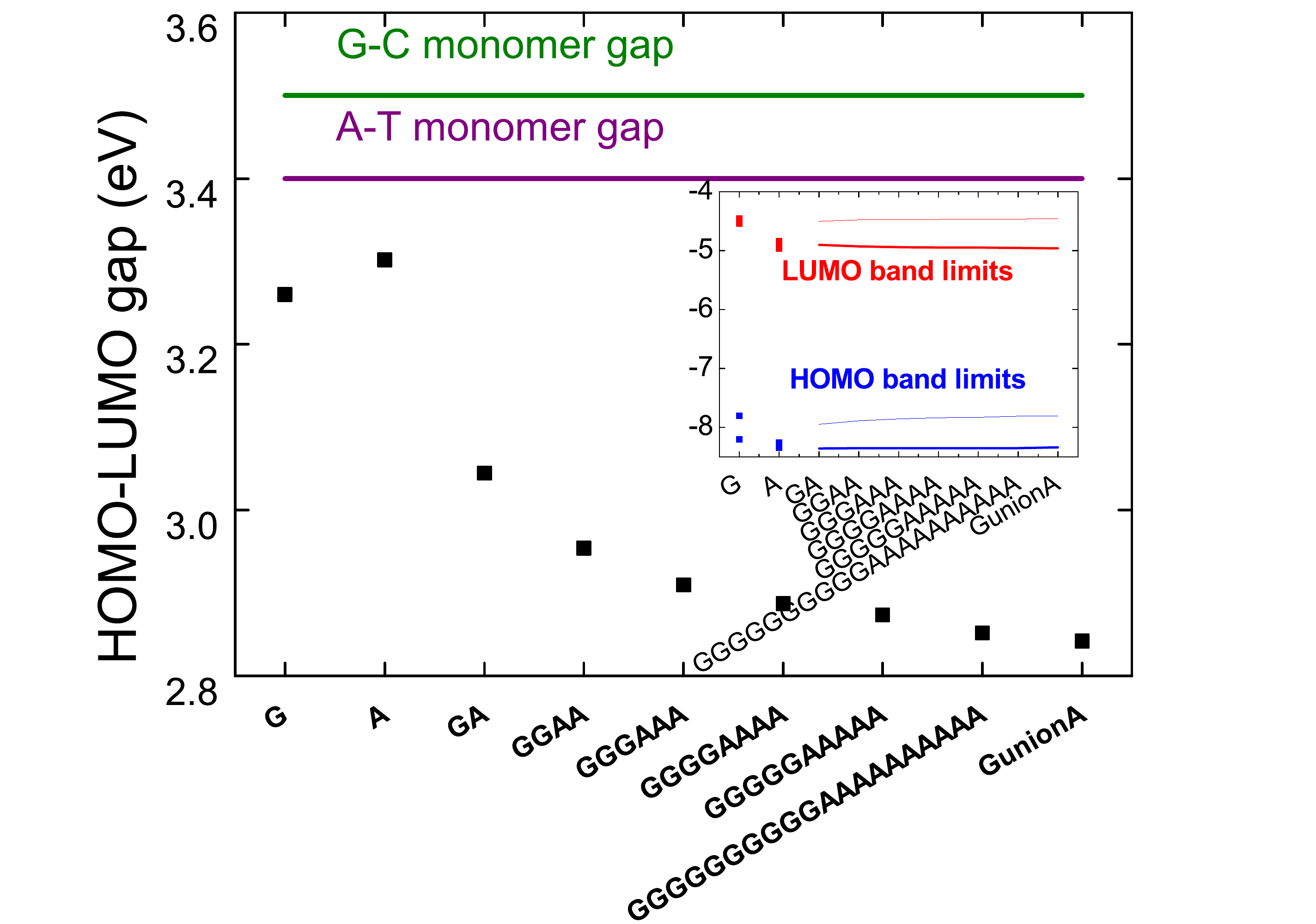}
\caption{Energy gaps. Upper panel: I2 (GC...), I4 (GGCC...), I6 (GGGCCC...), I8 (GGGGCCCC...), I10  (GGGGGCCCCC...), I20 (GGGGGGGGGGCCCCCCCCCC...) polymers, as well as I1 (G...) polymers. The horizontal green line at 3.5 eV shows the energy gap of the monomer (G-C base pair). Inset: HOMO and LUMO band upper and lower limits. Other variants of I polymers follow the same trend, e.g. the gap of I3 (GGC...) is $\approx$ 3.36 eV.
Lower Panel: D2 (GA...), D4 (GGAA...), D6 (GGGAAA...), D8 (GGGGAAAA...), D10 (GGGGGAAAAA...), D20 (GGGGGGGGGGAAAAAAAAAA...) polymers, as well as the union of I1 (G...) and I1 (A...) polymers. The horizontal green line at 3.5 eV (purple line at 3.4 eV) shows the energy gap of the G-C (A-T) base pair. Inset: HOMO and LUMO band upper and lower limits.}
\label{fig:gaps}
\end{figure}

At the large-$N$ limit, increasing $P$, the gaps of I2, I4, I6, ... polymers approach the gap of I1 polymer (cf. upper panel of Fig.~\ref{fig:gaps}). 
Indeed, increasing the repetition unit in the mode GC, GGCC, GGGCCC, ..., finally results in a G...GC...C polymer which is almost G... with just a switch from G to C at the middle of the polymer. 
Hence, at the large-$N$ limit, the energy gap of I1 polymers is the smallest of these series of polymers. 
For the same reason, increasing $P$, the eigenspectra and the DOS of I2, I4, I6, ... polymers tend to the eigenspectra and the DOS of I1 polymers (cf. Figs.~\ref{fig:Eigenspectra-gaps-ChV}, \ref{fig:DOS-ChV}).


\clearpage

At the large-$N$ limit, increasing $P$, the gaps of D2, D4, D6, ... polymers approach the gap of the union of I1 (G...) and I1 (A...) polymers (cf. lower panel of Fig.~\ref{fig:gaps}), which is $\approx$ 0.5 eV lower than the gaps of the relevant homopolymers (G..., A...). Increasing the repetition unit in the mode GA..., GGAA..., GGGAAA... and so on, finally results in a G...GA...A polymer which is \textit{energetically} almost a union of separated G...  and A... polymers. This happens due to the large difference of G-C and A-T on-site energies in comparison with the $t_\textrm{GA}$ hopping integral. 
Increasing $P$, the lowering of the energy gap in the case of D polymers [$\approx$ 0.6 (0.7) eV relative to the A-T (G-C) monomer gap] is much bigger than in the case of I polymers [$\approx$ 0.25 eV relative to the G-C monomer gap].

\subsection{\label{subsec:M(ot)P} Mean over time Probabilities} 
\begin{figure} [h]
	\centering
	\includegraphics[width=0.9\columnwidth]{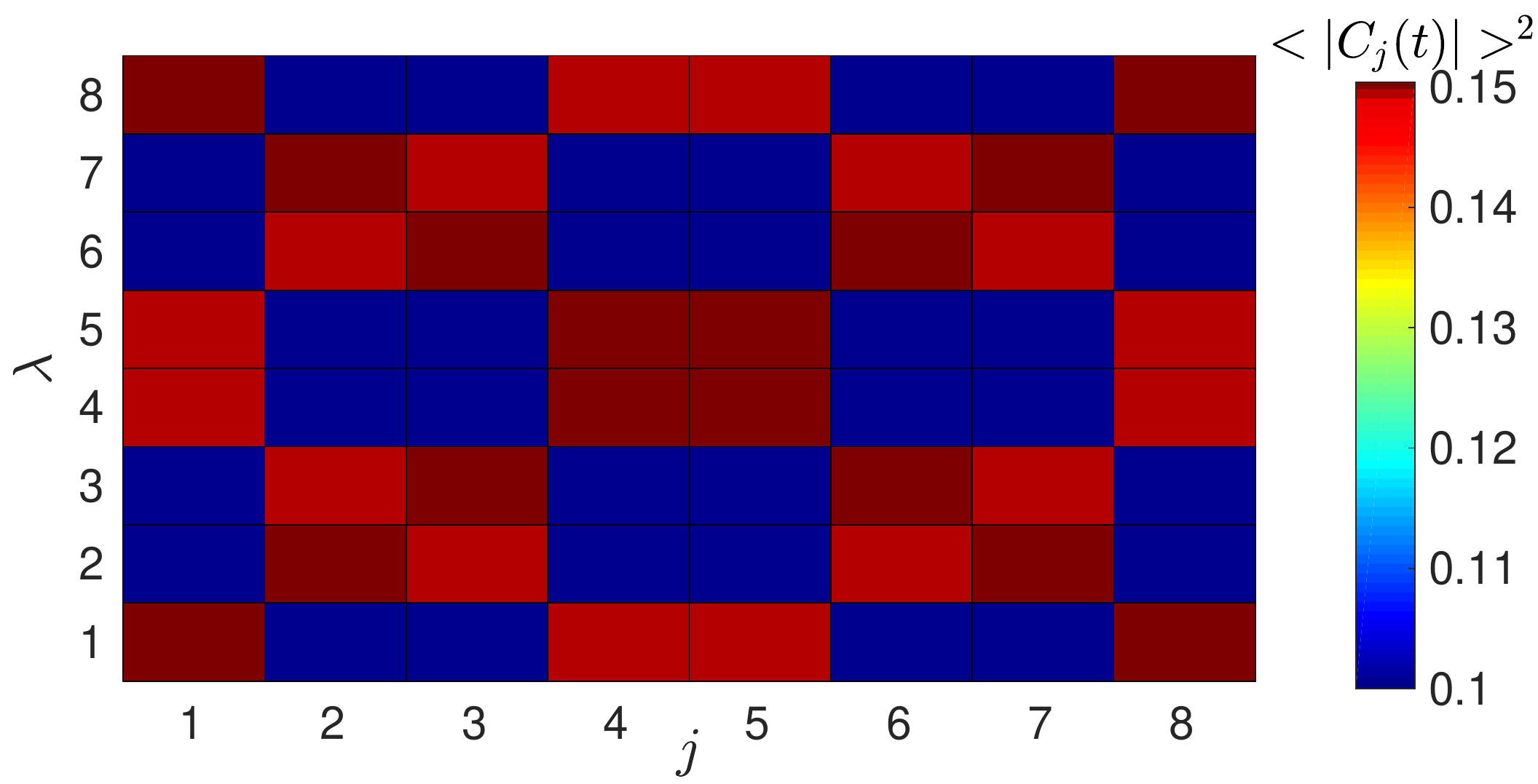}
	\caption{Mean (over time) probabilities to find an extra hole in a GGGGCCCC polymer for all possible initial placements.}
	\label{fig:GGGGCCCCmeanprobs}
\end{figure}

\begin{figure} [h]
	\centering
	\includegraphics[width=0.5\columnwidth]{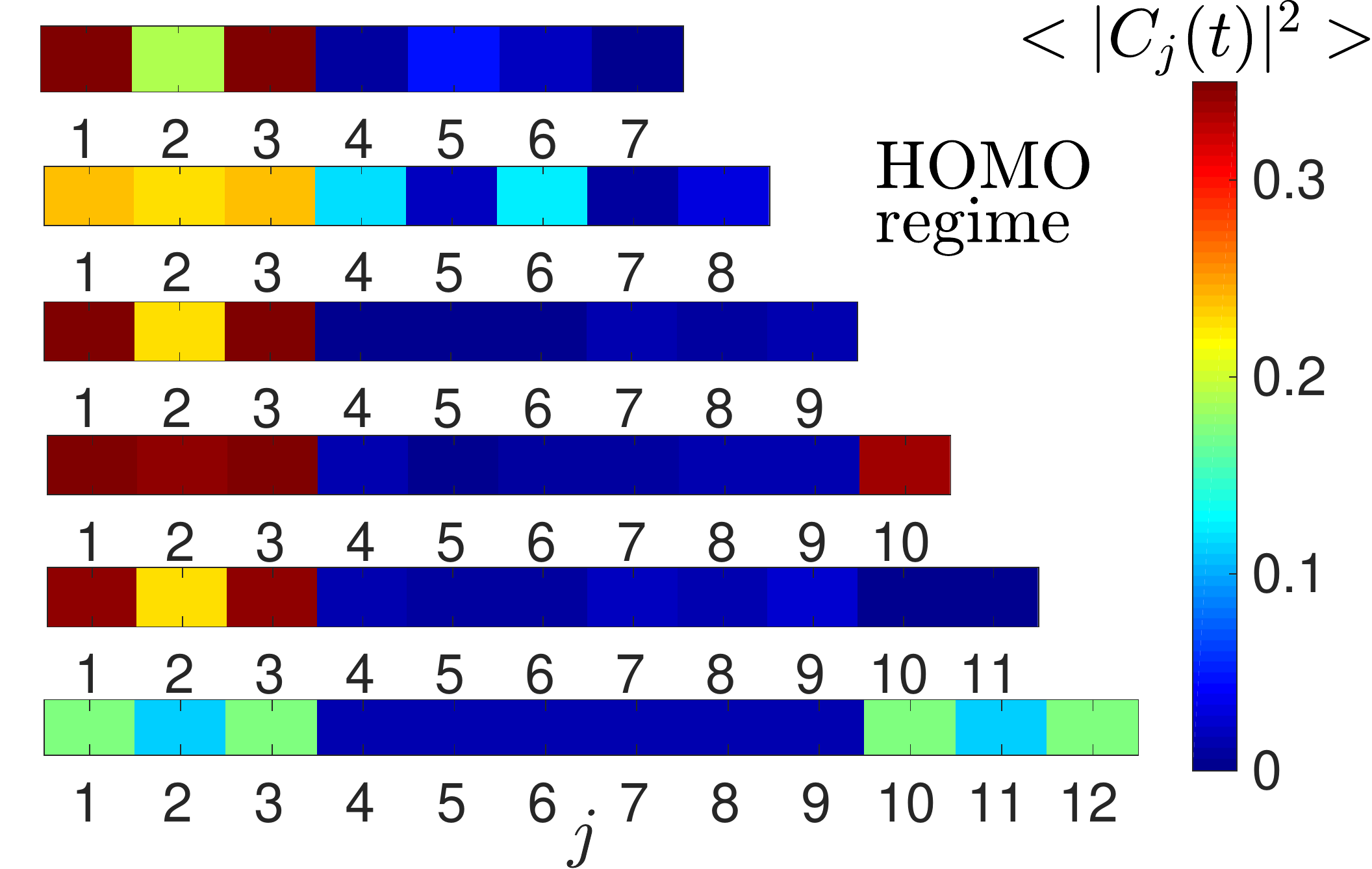}\includegraphics[width=0.5\columnwidth]{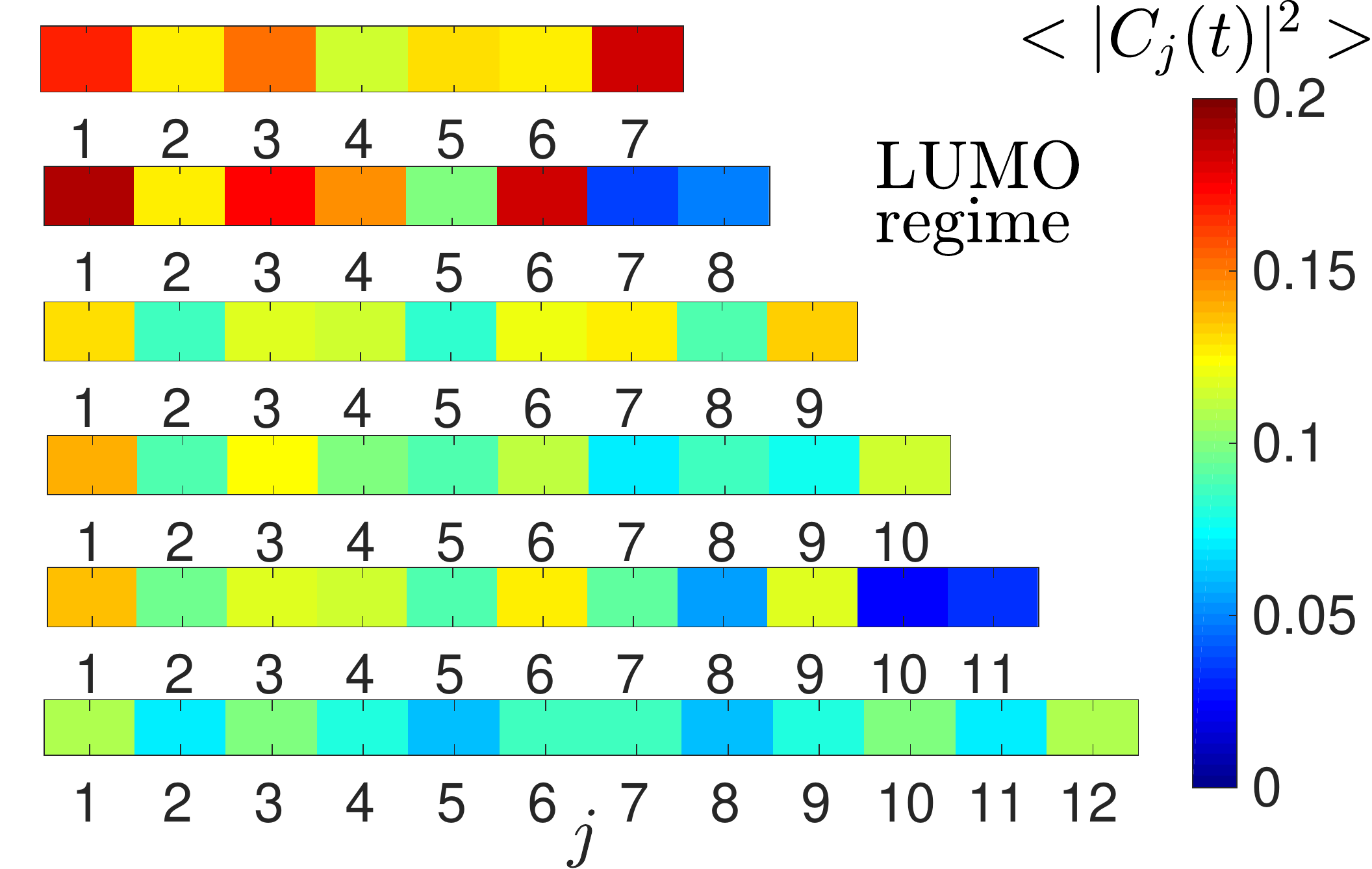}
	\caption{Mean (over time) probabilities to find an extra hole (left) and electron (right), initially placed at the first monomer, in a GGGCCC... polymer ($P=6$) made up of $N = P + \tau$, $\tau =1,\dots,P$ monomers.}
	\label{fig:GGGCCCmeanprobs}
\end{figure}

The main aspects of our results for the mean (over time) probabilities for I2, I4, ... polymers are summarized in Figs.~\ref{fig:GGGGCCCCmeanprobs} and~\ref{fig:GGGCCCmeanprobs} for some example cases.
For $N$ equal to natural multiples of $P$ ($N = Pn, n \in \mathcal{N}^*$), \textit{palindromicity} holds, i.e., the probabilities are palindromic. 
This is due to the fact that for $N = Pn$, the hamiltonian matrices of these polymers are palindromic, i.e. reading them from top left to bottom right and vice versa gives the same result. The palindromicity for $N = nP$ is shown in Fig.~\ref{fig:GGGGCCCCmeanprobs}, for an example I8 (GGGGCCCC) polymer, for all possible initial placements of an extra hole. It is evident that palindromicity holds for all initial conditions. Hence, in these polymer cases, the appropriate choice of the monomer the carrier is injected to, can lead to enhanced presence at specific sites at its other end, leading to more efficient transfer.
For $N \neq Pn$, palindromicity is lost. This is shown in Fig.~\ref{fig:GGGCCCmeanprobs}, for an example case of an I6 (GGGCCC) polymer.
In the HOMO regime, all studied polymers with $N \neq Pn$, show increased mean (over time) probabilities at the $\frac{P}{2}$ initial monomers. For example, for type I6 polymers, for $N \neq 6n$, we have increased probabilities at the first, second and third monomer (left panel of Fig.~\ref{fig:GGGCCCmeanprobs}).
This property is so evident in the HOMO regime due to the magnitude of the hopping integrals, cf. Table~\ref{Table:HoppingIntegrals}:
$t_\textrm{GG}$ is the greater of all, and $t_\textrm{GC}$ is much smaller than $t_\textrm{CG}$. In the LUMO regime, this property cannot be clearly seen, because $t_\textrm{GG}$ is the greater of all, but $t_\textrm{CG}$ and $t_\textrm{GC}$ have similar values . For the same reason, in the LUMO regime, for $N = Pn +\frac{P}{2}$, we have an almost palindromic behavior (see e.g. the right panel of Fig.~\ref{fig:GGGCCCmeanprobs}).
For I1 polymers and initial placement of the carrier at the first monomer,
the mean over time probability to find the carrier 
at the first or at the last monomer is $\psi$ and 
at any other monomer is $\chi$, ~\cite{LChMKTS:2015} where 
\begin{equation}\label{psichi-typea-allmostallcase}
\psi=\frac{3}{2(N+1)}, \quad \chi=\frac{1}{N+1}.
\end{equation}
Increasing $P$, the relevant probabilities of I2, I4, I6, ... polymers tend to the I1 probabilities $\psi$ and $\chi$ of Eq.~\eqref{psichi-typea-allmostallcase}. 

\begin{figure} [h]
	\centering
	\includegraphics[width=0.9\columnwidth]{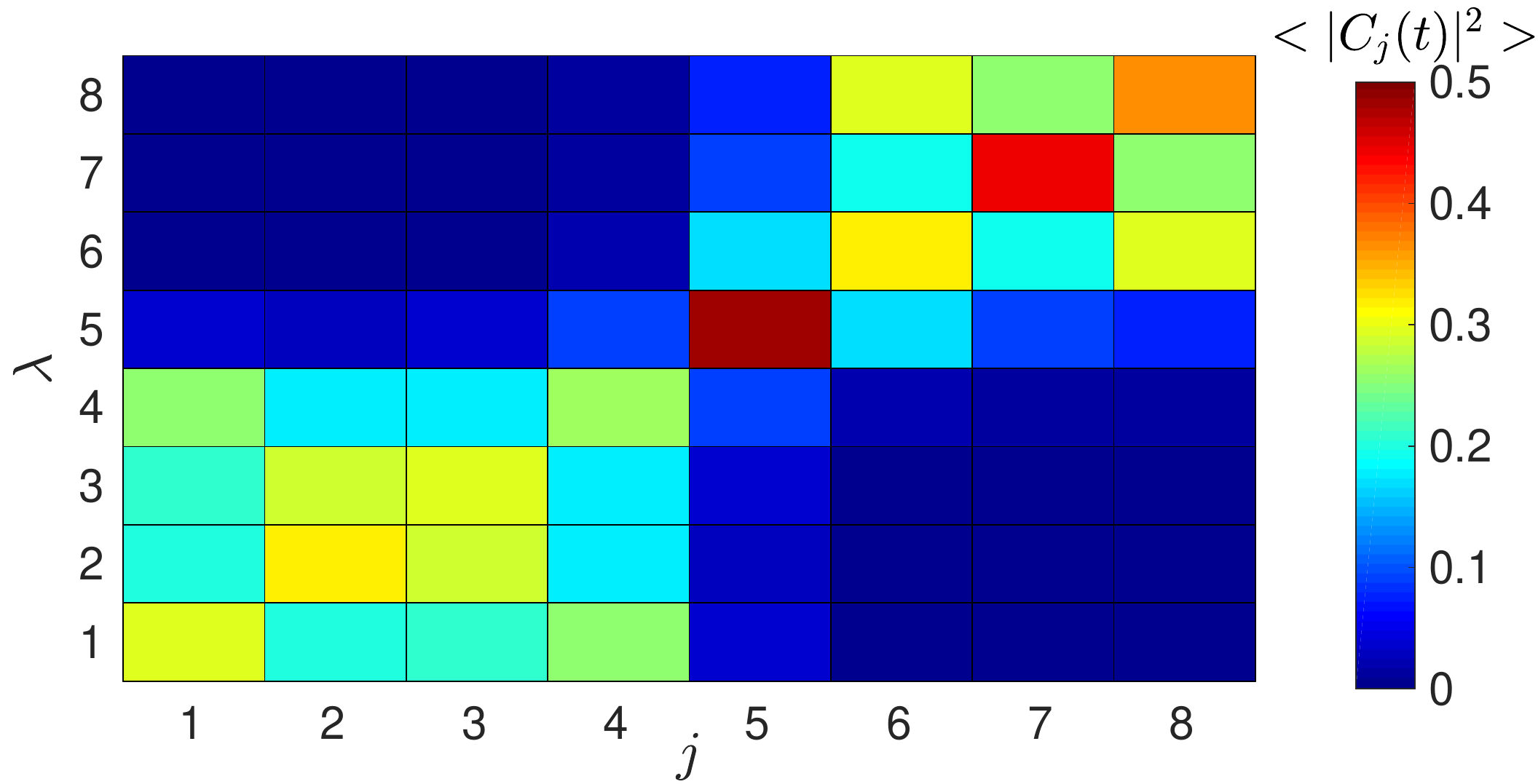}
	\caption{Mean (over time) probabilities to find an extra hole in a GGGGAAAA polymer for all possible initial placements.}
	\label{fig:GGGGAAAAmeanprobs}
\end{figure}

The main features of our results for the mean (over time) probabilities for D2, D4, ... polymers are summarized in Fig.~\ref{fig:GGGGAAAAmeanprobs} for the example case of an I8 (GGGGAAAA) polymer, for all possible initial placements of an extra hole. A basic observation for polymers made of different monomers is that if we initially place the carrier at a G-C monomer the probability to find it at an A-T monomer is small, and vice versa.

Detailed numerical results displaying all the above mentioned features, having placed the hole or electron initially at the first monomer, for $N = P + \tau$, $\tau = 0, 1, \dots, P-1$, can be found in the Supplementary Material, and specifically in Figs.~\ref{fig:ProbabilitiesHL-I2}-\ref{fig:ProbabilitiesHL-I10} for polymers made up of identical monomers, and in Figs.~\ref{fig:ProbabilitiesHL-D2}-\ref{fig:ProbabilitiesHL-D10}, for polymers made up of different monomers.

\subsection{\label{subsec:Frequencies} Frequency Content} 
The Fourier spectra of the time-dependent probability to find the carrier at each monomer, are, generally, in the THz regime. A general remark is that when the dominant frequencies i.e. those with the greater Fourier amplitudes are smaller (bigger), the carrier transfer 
is slower (faster). 

For $N = Pn$, $n\in\mathcal{N}$, for I1, I2, I4, I6, ... polymers, the Fourier spectra of the time-dependent probability to find an extra carrier at the various monomers, either for the HOMO or the LUMO regime, are palindromic, i.e., they are identical for the $\mu$-th and $(N-\mu+1)$-th monomer. This stems from the palindromicity characterizing the hamiltonian matrices for $N = Pn$, $n\in\mathcal{N}$. The Fourier spectra of the probability to find an extra carrier at the first and at the last monomer, having placed it initially at the first monomer, for I and D polymers, for the HOMO and the LUMO regime, for $N = P + \tau$, $\tau = 0, 1, \dots, P-1$, as well as similar diagrams for greater $N$, can be found in Refs.~\cite{Vantaraki:2017,Bilia:2018}. 
Since for $N \neq Pn$, $n\in\mathcal{N}$ the hamiltonian matrices are not palindromic, the Fourier spectra are also not palindromic. Preliminary analysis of the frequency content of I1, I2, D2, I3, I4 and I6 polymers, for TB I and TB II, including the Fourier spectra, the WMFs and the TWMF as a function of $N$ can be found in Ref.~\cite{LMS-PIERS:2017}, with TB parameters taken from Ref.~\cite{HKS:2010-2011}.

\begin{figure}[!h]
\includegraphics[width=0.4\textwidth]{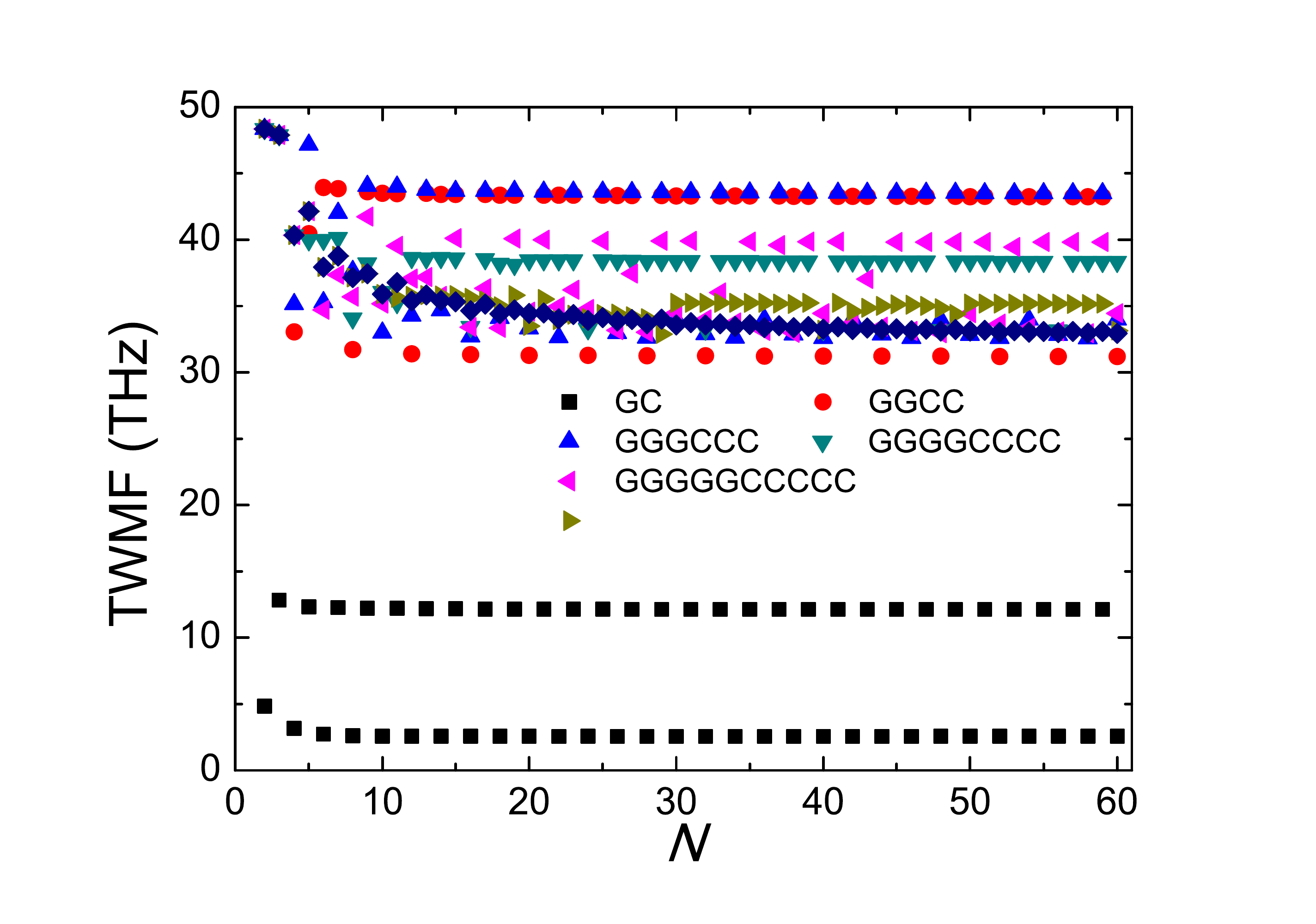}
\includegraphics[width=0.4\textwidth]{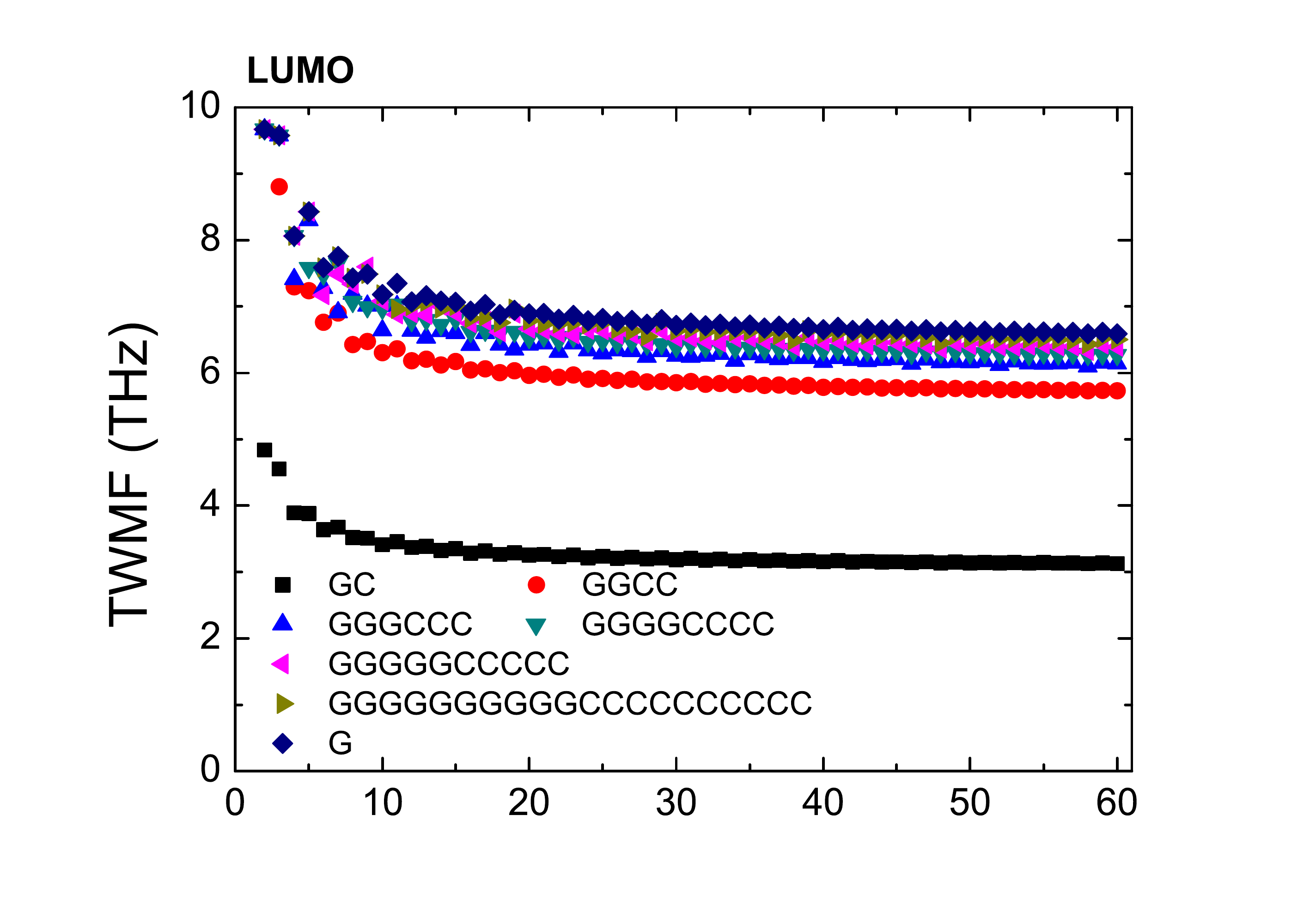}
\caption{Total Weighted Mean Frequency (TWMF) as a function of the number of monomers $N$ in the polymer, having placed the carrier initially at the first monomer, for   
I1 (G...), I2 (GC...), I4 (GGCC...), I6 (GGGCCC...), I8 (GGGGCCCC...), I10 (GGGGGCCCCC...), and
I20 (GGGGGGGGGGCCCCCCCCCC...) polymers, for the HOMO and the LUMO regime.}
\label{fig:TWMFHLI}
\end{figure}

Next, we focus on the TWMF as a function of $N$ for various  types of polymers made of identical monomers (cf. Fig.~\ref{fig:TWMFHLI}).
In I2 (GC...) polymers, only two hopping integrals are involved: $t_{\textrm{GC}}$, $t_{\textrm{CG}}$. 
In I4 (GGCC...), I6 (GGGCCC...), ... polymers, three hopping integrals are involved: $t_{\textrm{GG}}, t_{\textrm{GC}}, t_{\textrm{CG}}$.
This is the reason that in the limit of large $N$, the TWMF for I2 polymers tends to a different frequency region than for I4, I6, ... polymers. 
For I4, I6, ... polymers, increasing $P$, the role of $t_{\textrm{GG}}$ gradually increases, hence, this series of polymers has as a limit I1 (G...) polymers, where only one 
hopping integral is involved: $t_{\textrm{GG}}$. 
In particular, the TWMF of I4, I6, ... polymers, in the limit of large $N$, tends to the TWMF of I1 polymers. 

\begin{figure}[!h]
\includegraphics[width=0.4\textwidth]{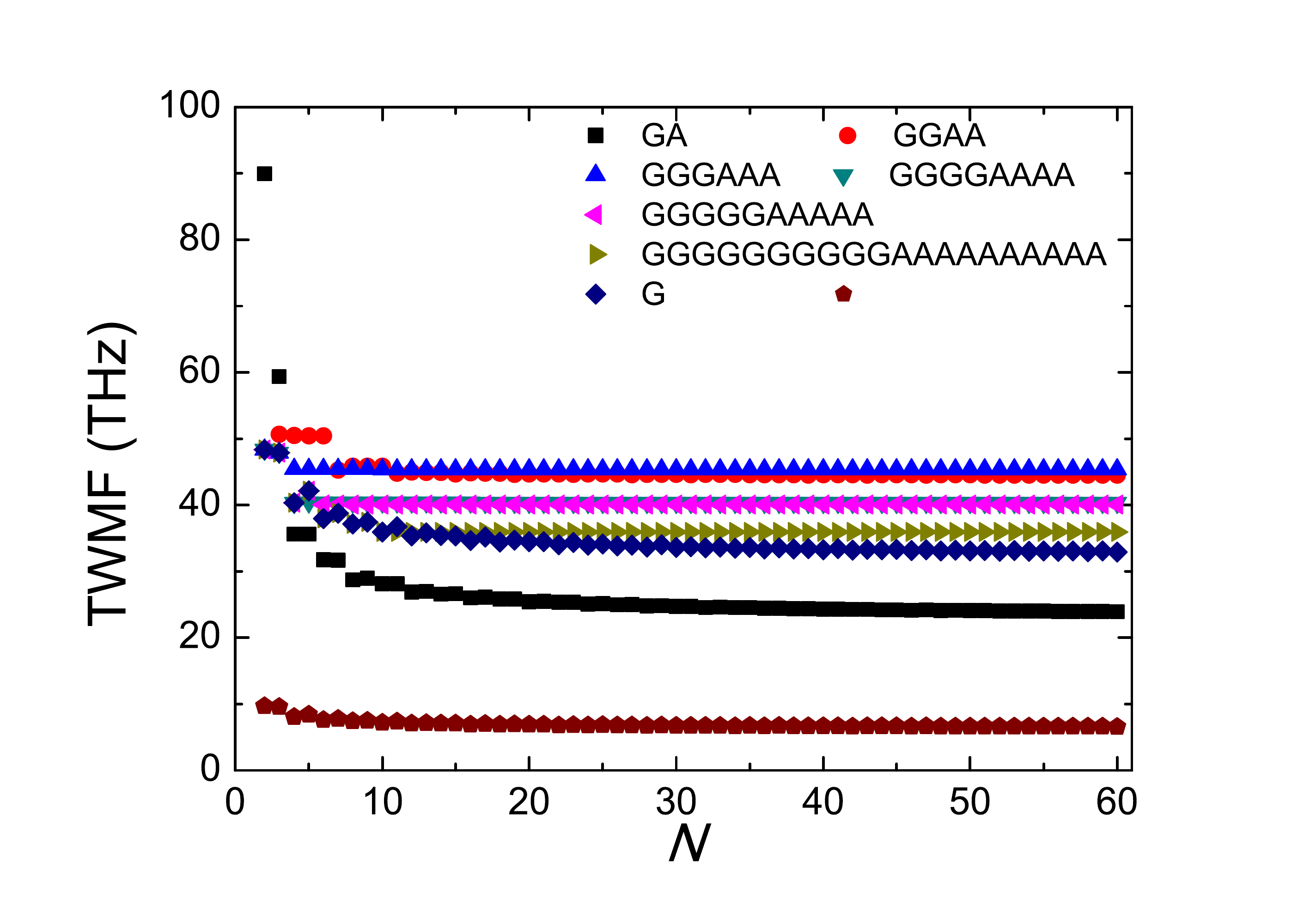}
\includegraphics[width=0.4\textwidth]{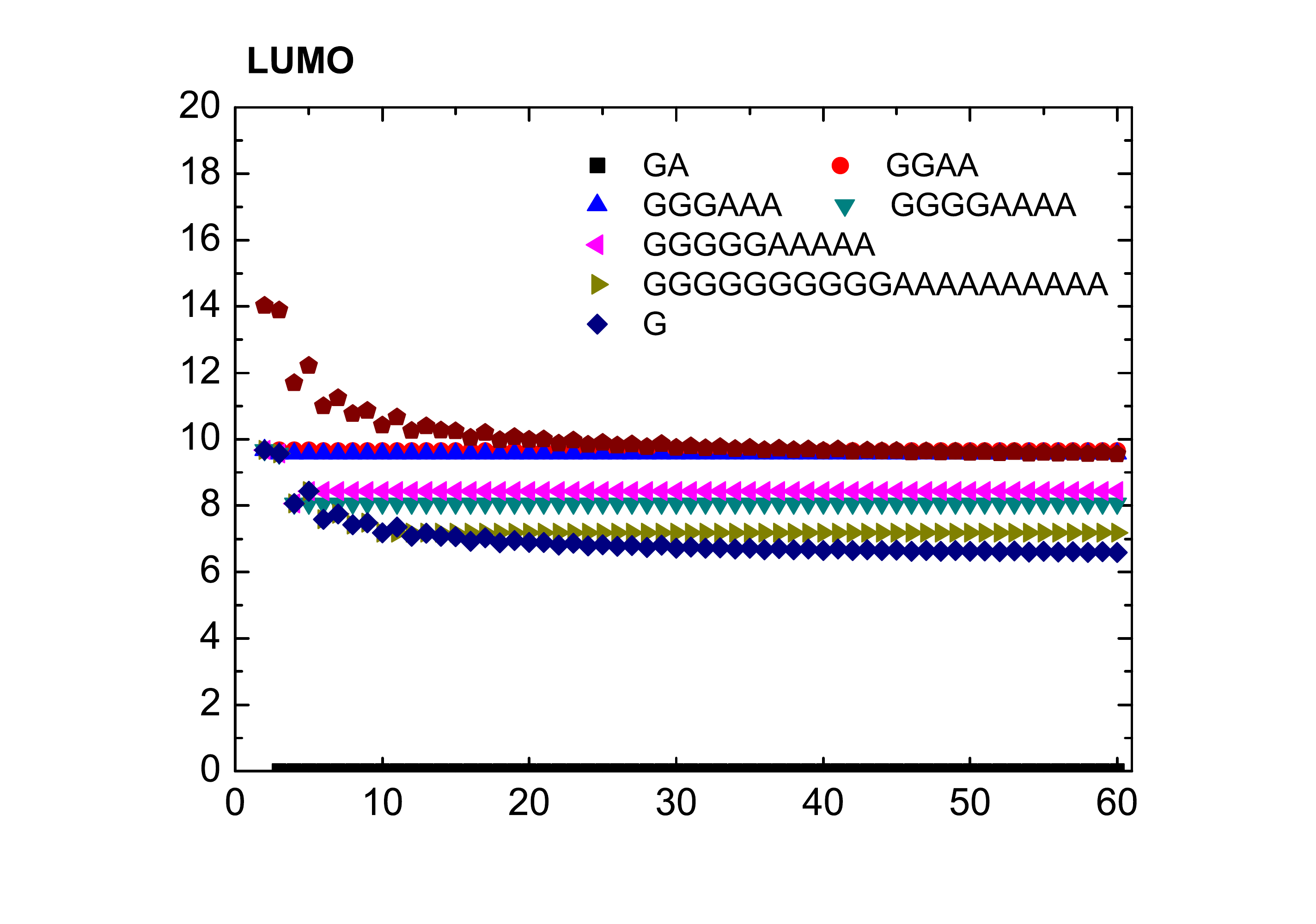}
\caption{Total Weighted Mean Frequency (TWMF) as a function of the number of monomers $N$ in the polymer, having placed the carrier initially at the first monomer, for   
I1 (G...), I1 (A...), D2 (GA...), D4 (GGAA...), D6 (GGGAAA...), D8 (GGGGAAAA...), D10 (GGGGGAAAAA...), and
D20 (GGGGGGGGGGAAAAAAAAAA...) polymers, for the HOMO and the LUMO regime.}
\label{fig:TWMFHLD}
\end{figure}

Next, we focus on the TWMF as a function of $N$ for various types of polymers made of different monomers (cf. Fig.~\ref{fig:TWMFHLD}). In D2 (GA...) polymers, only two hopping integrals are involved: $t_{\textrm{GA}}$, $t_{\textrm{AG}}$. In D4 (GGAA...), D6 (GGGAAA...), ... polymers, four hopping integrals are involved: $t_{\textrm{GG}}, t_{\textrm{GA}}, t_{\textrm{AG}}, t_{\textrm{AA}}$.
This is the reason that in the limit of large $N$, the TWMF for D2 polymers tends to a different frequency region than for D4, D6, ... polymers.
For D4, D6, ... polymers, increasing $P$, the role of $t_{\textrm{GG}}$ gradually increases; the same happens with the role of $t_{\textrm{AA}}$. However, if we place the carrier initially at the first G-C monomer, the probability to find it at any A-T monomer is very small (e.g., cf. Fig.~\ref{fig:GGGGAAAAmeanprobs}). 
Hence, for initial placement of the carrier at a G-C monomer (like in Fig.~\ref{fig:TWMFHLD}), 
the TWMF of D4, D6, ... polymers, in the limit of large $N$, tends to the TWMF of I1 (G...) polymers, where only one hopping integral is involved: $t_{\textrm{GG}}$. 

The frequencies involved in charge transfer are given by Eq.~\eqref{fandT}.
Hence, the maximum frequency is determined by the maximum difference of eigenenergies, i.e., by the upper and lower limits of the HOMO or LUMO band. Since increasing $P$, the eigenspectra of I2, I4, I6, ... polymers tend to the eigenspectra of I1 polymers, the maximum frequencies of these polymers also tend to the maximum frequency of I1 polymers.
Since increasing $P$, the eigenspectra of D2, D4, D6, ... polymers tend to the eigenspectra of the union of I1 (G...) and I1 (A...) polymers, the maximum frequencies of these polymers also tend to the maximum frequency of the union of I1 (G...) and I1 (A...) polymers. Numerical results regarding the behavior of maximum frequency can be found in Fig.~\ref{fig:maxfN60HL} in the Supplementary Material.

\subsection{\label{subsec:k} \textit{Pure} Mean Transfer Rates} 

\begin{figure} [h!]
\includegraphics[width=0.45\textwidth]{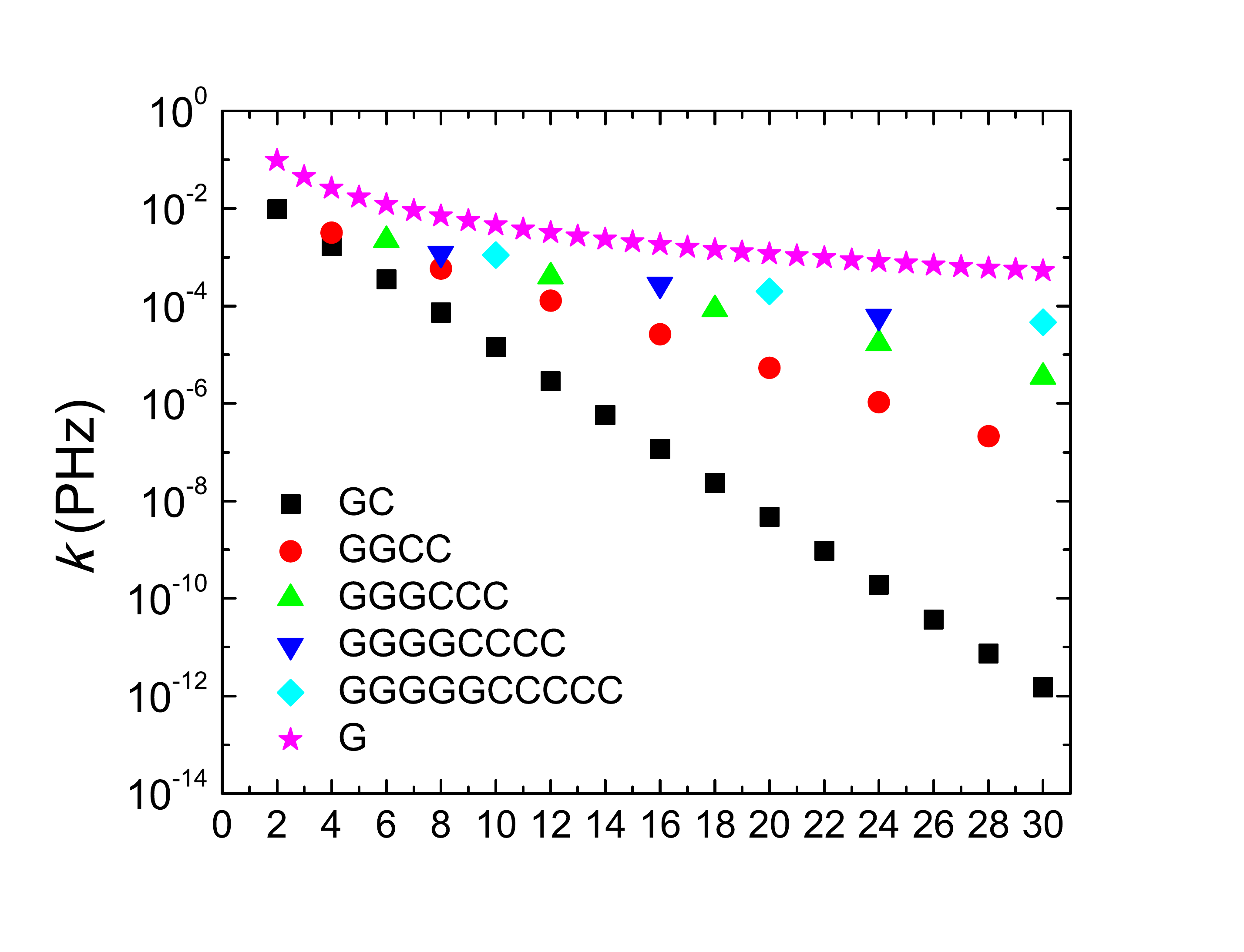}
\includegraphics[width=0.45\textwidth]{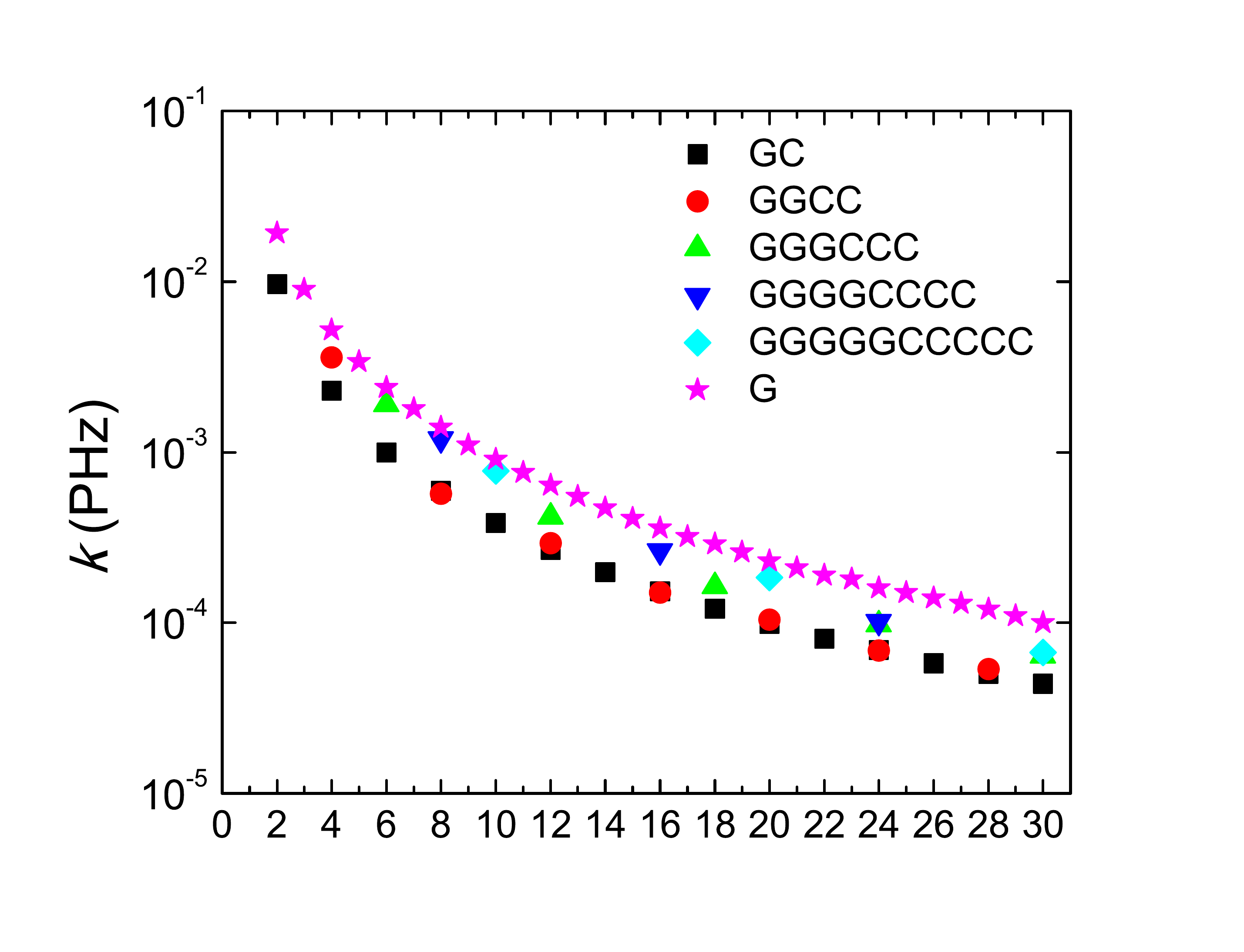}
\caption{\textit{Pure} mean transfer rate $k$ of type I1 (G...), I2 (GC...), I4 (GGCC...), I6  (GGGCCC...), I8 (GGGGCCCC...) and I10 (GGGGGCCCCC...) polymers, as function of the number of monomers $N$ in the polymer, for $N$ equal to natural multiples of their $P$, for HOMO (upper panel) and LUMO (lower panel).}
\label{fig:kofN-ChV}
\end{figure}

Next, we study the \textit{pure} mean transfer rates, $k_{1,N}$, from the first to the last monomer. For simplicity, we drop the indices. An impressive case where appropriate sequence choice can increase $k$ by many orders of magnitude is shown in Fig.~\ref{fig:kofN-ChV}. We depict $k(N)$ for $N=nP$, either for HOMO or for LUMO, for  type I1 (G...), I2 (GC...), I4 (GGCC...), I6 (GGGCCC...), I8 (GGGGCCCC...) and I10 (GGGGGCCCCC...) polymers. These polymers are palindromic (cf. Sec.~\ref{subsec:M(ot)P}), hence there is enhanced presence of the extra carrier at the last monomer. Results for any $N$ i.e. $N = Pn$ or $N \neq Pn$, $n\in\mathcal{N}^*$, can be found elsewhere \cite{Vantaraki:2017}.

In all cases, $k(N)$ is a decreasing function. The electron $k$ range is many orders of magnitude narrower than the hole $k$ range, due to the much smaller difference between the hopping integrals ($t_\textrm{GG}$, $t_\textrm{GC}$, $t_\textrm{CG}$) involved (cf. Table~\ref{Table:HoppingIntegrals}). 

In Fig.~\ref{fig:kofN-ChV} we observe that for $N = Pn$, starting from type I2 (GC...) polymers and increasing $P$, i.e. for types I4 (GGCC...), I6 (GGGCCC...), ... polymers, $k$ takes increasingly larger values. In other words, the degree of transfer difficulty is greater for type I2 (GC...) polymers and decreases gradually for types I4, I6, ... I10 polymers. And so it will be if we still increase $P$ taking similar types of polymers. However, $k(N)$ has an upper limit which is $k(N)$ of type I1 polymers. The latter polymers are structurally simpler (more precisely, they have the simplest possible structure), a fact that favors charge transfer along them, so their transfer rates are higher than those of the other polymer types. As $P$ increases, the influence of $t_{GC}$ and $t_{CG}$ becomes less significant, hence this upper limit appears.

Generally, for an extra electron (LUMO), $\ln\!k(\ln\!N)$ is approximately linear, of the form $\ln\!k = \ln\!k_0 - \eta \ln\!N$, a relation that generally does not hold for an extra hole (HOMO). However, for type I1  (G...) polymers the above mentioned linear relation holds both for HOMO and LUMO. For HOMO, generally $k(d)$, where $d = (N-1)\times 3.4$ {\AA} is the charge transfer distance, is approximately of the form $\ln\!k = \ln\!k_0 - \beta d$, a relation that does not generally hold for an extra electron (LUMO).

To gain further insight, we have performed the exponential fits $k = k_0 e^{-\beta d}$ and $k = A + k_0 e^{-\beta d}$ as well as the power-law fit $k = k_{0}' N^{-\eta}$. In all cases we studied $k$ from the first to the last monomer, under the condition $N=nP$, $N < 40$. We observe that for I2 (GC...), I4 (GGCC...), I6 (GGGCCC...), I8 (GGGGCCCC...), I10 (GGGGGCCCCC...) polymers the HOMO regime is better characterized by exponential fits and the LUMO regime by power-law fits. For type I1 (G...) polymers the power-law fits are better both for the HOMO and the LUMO regimes. his fact can also be easily seen in Fig.~\ref{fig:kofN-ChV}, where in the HOMO regime all I2, I4, I6, I8, I10 polymers show a linear relation between $\ln\!k$ and $d$, while, type I1 polymers do not satisfy this linear relation. In the LUMO regime, none of I2, I4, I6, I8, I10 polymers satisfies a linear relation between $\ln\!k$ and $d$. In the LUMO regime, all I2, I4, I6, I8, I10 polymers as well as I1 polymers satisfy an almost linear $\ln\!k$ - $\ln\!N$ relation.

$\ln\!k(\ln\!N)$ and $\ln\!k(d)$, as well as detailed fit results are shown in Figs.~\ref{fig:kofN-ChV-SM}-\ref{fig:betaeta-ChV} in the Supplementary Material. In the HOMO regime for I2, I4, I6, I8, I10 polymers, 
the exponent $\beta$ is characterized by small errors, while 
the exponent $\eta$ by much larger errors. 
For I1 polymers, the exponent $\eta$ shows minimum error, while 
the exponent $\beta$ much larger error. 
On the contrary, in the LUMO regime, all polymers show very small errors for the exponent $\eta$ and much larger for the exponent $\beta$. This shows that the exponential fits are better for the HOMO regime for I2, I4, I6, I8, I10 polymers, while, the power-law fits are better for I1 polymers both for the HOMO and the LUMO regimes and for I2, I4, I6, I8, I10 polymers for the LUMO regime.
Increasing $P$, in the series I2, I4, I6, I8, I10 polymers,
the exponent $\beta$ decreases, i.e. the fall of $k(d)$ is less steep as $P$ increases.
The $\eta$ exponent values for  I2, I4, I6, I8, I10 polymers are similar with its value for I1 polymers.
Hence, we observed, increasing $P$, a convergence of the mean transfer rates to those of type I1 polymers.

\begin{figure} [h!]
\includegraphics[width=0.45\textwidth]{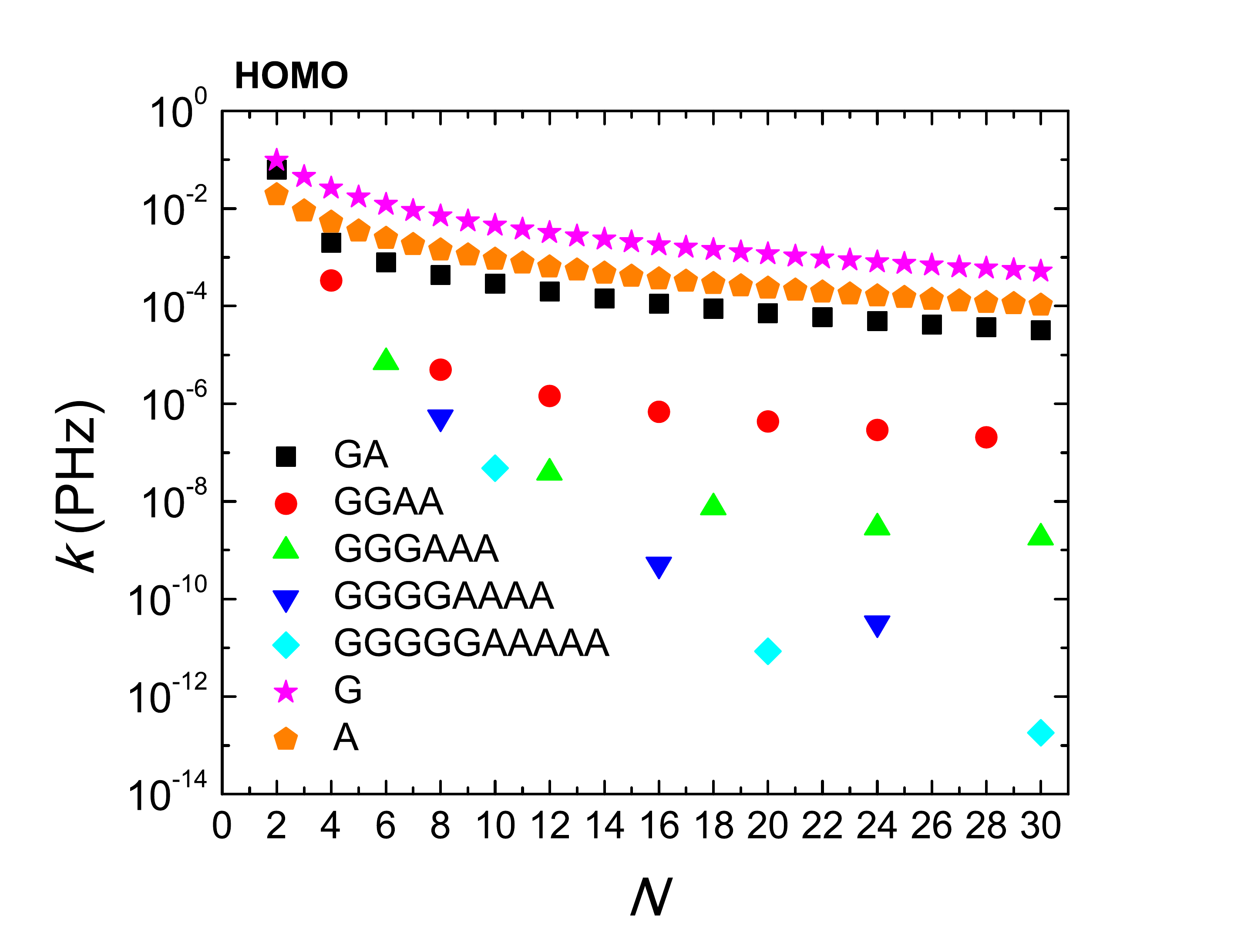}
\includegraphics[width=0.45\textwidth]{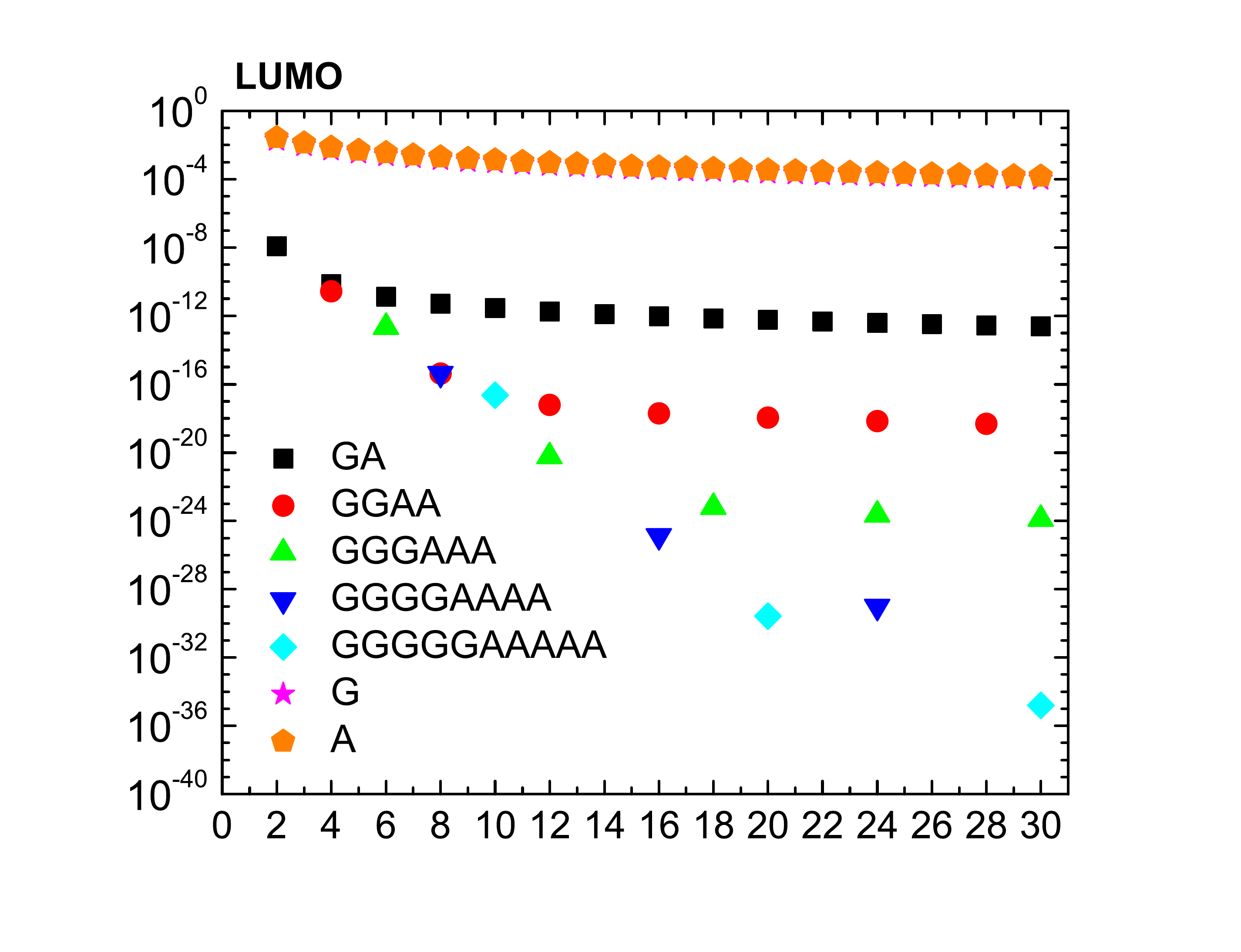}
\caption{\textit{Pure} mean transfer rate $k$ of type I1 (G...), I1 (A...), D2 (GA...), D4 (GGAA...), D6  (GGGAAA...), D8 (GGGGAAAA...) and D10 (GGGGGAAAAA...) polymers, as function of the number of monomers $N$ in the polymer, 
for $N$ equal to natural multiples of their $P$, for HOMO (upper panel) and LUMO (lower panel).}
\label{fig:kofN-PMp}
\end{figure}

For polymers made of different monomers $k(N)$, is depicted in Fig.~\ref{fig:kofN-PMp}. We observe that, while
for type I1 polymers (G... and A...) $k$ drops $\approx$ by only 2 to 3 orders of magnitude, increasing $N$ from 2 to 30, as the number of A in the repetition unit increases, $k(N)$ drops dramatically by many more orders of magnitude. Again, this behavior shows that the \textit{pure} mean transfer rate can be increased by many orders of magnitude by appropriate choice of the repetition unit. $\ln\!k(\ln\!N)$ and $\ln\!k(d)$ are shown in Fig.~\ref{fig:kofN-PMp-SM} in the Supplementary Material. 

Finally, a comparison of $k$ for all possible I1, I2 and D2 polymers is presented in Figs.~\ref{fig:kofN-TBI}-\ref{fig:k29or30perk3or2-TBI} in the Supplementary Material.
	
All in all, our results suggest that type I1 polymers are the best for electron or hole transfer.

\subsection{\label{subsec:KEXP} Transfer Rates in Experiments} 
Comparison with the experiment in terms of transfer rates is not as straightforward as it may seem from a simplistic first view, because the easiness of charge transfer is usually measured experimentally via the quantification of relevant products. For example, when the hole is not transferred we obtain the product PN;  when the hole is transferred we obtain the product PY. The concentration of PN and PY can be measured by an indirect complicated method like polyacrylamide gel electrophoresis and piperidine treatment \cite{Meggers:1998,Giese:2001}. Being so complicated, this method, although it has revealed important aspects of hole transfer such as the sequence dependence of the relative hole transfer efficiency, it does not provide the kinetics of hole transfer in DNA~\cite{KawaiMajima:2013}. 
Generally, there is no proof that the concentrations of PN and PY are exactly proportional to the degree of charge transfer, although, generally, greater charge transfer means greater concentration of PY.  
Quantum mechanically, only a percentage of the carrier passes through the bridge connecting the carrier donor to the carrier acceptor. PN and PY will depend on the speed as well as on the percentage of carrier transfer. However, the concentration of PY is not strictly proportional to the amount of carrier transfer and it is not strictly inversely proportional to the time of transfer. Moreover, since quantum mechanically, only a percentage of the carrier passes through the bridge, the definition of the time of transfer is problematic.
Also, usually it is implied that the production of PY is much faster than the carrier transfer, although this might not be always the case \cite{Lakhno:2004}. Then, only the relative behavior of the theoretically calculated transfer rates and the experimentally measured transfer rates has some meaning. This was, e.g. realized in Ref.~\cite{Lakhno:2004}, where the theoretically determined transfer rates had to be divided by a factor of $8.9 \times 10^{10}$ Hz, to be compared with the experimentally determined transfer rates of Ref.~\cite{Meggers:1998}. 

Our point of view is different, since the quantity we use, the \textit{pure} mean transfer rate~\cite{Simserides:2014}, given by Eq.~\ref{pmtr}, uses simultaneously the magnitude of charge transfer and the time scale of the phenomenon. Additionally, DNA is a dynamical structure, i.e. the geometry is not fixed. Hence, the TB parameters any TB model uses have to be utilized with care. Large variations of the TB parameters are expected in a real situations and also, large variations of the TB parameters have been obtained by different theoretical methods by different authors, cf. e.g. Ref.~\cite{Simserides:2014} and references therein.

A more direct experimental approach is to use time-resolved spectroscopy like transient absorption to observe the products of charge transfer~\cite{Lewis:1997,Wan:2000,KawaiMajima:2013}.

The hole transfer kinetics of various short DNA segments have been experimentally investigated in Ref.~\cite{KawaiMajima:2013} with time-resolved spectroscopy~\cite{KawaiMajima:2013}.
Table~\ref{Table:KawaiMajima} provides a comparison of the $\beta$ (\AA$^{-1}$) factors between the experiment of Ref.~\cite{KawaiMajima:2013}, where the transfer rate is fitted into $K = K_{0} \mathrm{e}^{- \beta d}$ and the TB wire model calculations, where the \textit{pure} mean transfer rate is fitted into $k=k_{0} \mathrm{e}^{- \beta d}$.  
\begin{table}[h!]
\caption{Comparison of $\beta$ (\AA$^{-1}$) between the experiment of Ref.~\cite{KawaiMajima:2013}, where the transfer rate is fitted into $K = K_{0} \mathrm{e}^{- \beta d}$ and the TB wire calculations, where the \textit{pure} mean transfer rate is fitted into $k=k_{0} \mathrm{e}^{- \beta d}$. $d$ is the charge transfer distance. 
(S) denotes the TB parametrization of Ref.~\cite{Simserides:2014}.
(M) denotes the TB parametrization of Ref.~\cite{Simserides:2014} modified by putting 
$t_\textrm{GG} \to 2 \; t_\textrm{GG}$, 
$t_\textrm{AG} \to 0.15 \; t_\textrm{AG}$, 
$t_\textrm{AA} \to 2 \; t_\textrm{AA}$, 
$t_\textrm{AC} \to t_\textrm{AC}/3$.}
\label{Table:KawaiMajima}
\begin{tabular}{|l|c|c|c|} \hline
sequence               & $\beta$ exp & $\beta$ (S) &  $\beta$ (M)  \\ \hline
G(A)$_n$G, $n=$ 0,1,2  & 1.6         & 1.14        &  1.64         \\ \hline
G(A)$_n$G, $n=$ 2,3    & 0.6         & 0.64        &  0.61         \\ \hline
G(T)$_n$G, $n=$ 1,2,3  & 0.6         & 0.80        &  0.59         \\ \hline
G(A)$_n$C, $n=$ 0,1    & 1.5         & 0.86        &  1.51         \\ \hline
G(A)$_n$C, $n=$ 1,2    & 1.0         & 1.41        &  1.07         \\ \hline
\end{tabular}
\end{table}

Transient absorption measurements where used in Refs.~\cite{Lewis:1997,Wan:2000} and 
$\beta$ values $\approx$ 0.5 - 0.7 \AA$^{-1}$, where reported,  
for exponential fits of the measured transfer rates $K = K_{0} \mathrm{e}^{- \beta d}$, 
where $d$ is the charge transfer distance.
In the experiment of Ref.~\cite{Wan:2000}, in sequences Ap(A)$_n$G, $n=0, 1,2,3$, where Ap denotes 2-aminopurine, $\beta = 0.57$ \AA$^{-1}$. 
Our simple TB wire model for the sequence A(A)$_n$G, $n=0, 1,2,3$,  
using the parametrization of Ref.~\cite{Simserides:2014} gives $\beta = 0.44$ \AA$^{-1}$; 
using the parametrization of Ref.~\cite{HKS:2010-2011} gives $\beta = 0.52$ \AA$^{-1}$.
The experiments of Refs.~\cite{Lewis:1997,Wan:2000} give transfer rates $K$ whose order of magnitude is close to our theoretically determined \textit{pure} mean transfer rates $k$. 

The measurement, using transient absorption spectra, of distance- and temperature-dependent rate constants for charge separation in capped hairpins in which a stilbene hole acceptor and hole donor were separated by sequences A$_3$G$_n$, $n$ in the range 1 - 19, were reported in Ref.~\cite{Conron:2010}. The measured transfer rates $K$ are of the orders $10^{-6}$ to $10^{-8}$ PHz and an exponential fit of the form $K = K_{0} \mathrm{e}^{- \beta d}$, gives $\beta \approx 0.07$ \AA$^{-1}$.
If we use the simple TB wire model with the TB parametrization of Ref.~\cite{Simserides:2014}, modified by putting
$t_\textrm{GG} \to 1.35 \; t_\textrm{GG}$, 
$t_\textrm{AG} \to 0.15 \; t_\textrm{AG}$, 
$t_\textrm{AA} \to 1.35 \; t_\textrm{AA}$, 
to fit the \textit{pure} mean transfer rate into $k=k_{0} \mathrm{e}^{- \beta d}$,  
we obtain $k$ of similar magnitude with $K$ and $\beta \approx 0.05$ \AA$^{-1}$. 

In summary, the simple TB wire model can grasp qualitatively the experimental behavior.

\section{\label{sec:Conclusion} Conclusion} 
We comparatively studied the energy structure and the transfer of an extra carrier, electron or hole, along $N$-monomer periodic polymers (made of the same monomer, i.e.  I1, I2, I4, I6, I8, I10, I20, as well as made of different monomers, i.e. D2, D4, D6, D8, D10, D20), using the TB wire model. The number of monomers in the repetition unit is $P$. We determined various physical quantities: the HOMO and LUMO eigenspectra and density of states, the HOMO-LUMO gap, the mean over time probability to find the carrier at each monomer, the frequency content of carrier transfer and the \textit{pure} mean transfer rate. To express clearly the frequency content, using the Fourier spectra, we defined two new physical quantities: the weighted mean frequency of each monomer and the total weighted mean frequency of the whole polymer.
 
For periodic polymers made of identical monomers (I), 
the eigenenergies are always symmetric relative to the monomer on-site energy. 
Increasing $P$,
we have witnessed convergence of types I2, I4, I6, ... polymers, to  type I1 polymers, in terms of eigenspectra, density of states, energy gaps, mean over time probabilities to find the carrier at the first and last monomers, frequency content (total weighted mean frequency), and \textit{pure} mean transfer rates, i.e. for all the properties we studied. In other words, increasing $P$, the physical properties of I2, I4, I6, ... polymers tend to those of the relevant homopolymer. The homopolymer has the smallest HOMO-LUMO gap. Generally, for homopolymers, the magnitude of $k$ is larger and the fall of $k(N)$ is less steep. As $P$ increases, the influence of $t_{GC}$ and $t_{CG}$ becomes less significant, hence $k$ of the homopolymer acts as an upper limit. Moreover, we have ascertained palindromicity of physical properties such as the mean (over time) probabilities and the Fourier spectra, when the number of monomers is a natural multiple of the $P$.

For polymers made of different monomers (D), the eigenenergies gather around the two monomers' on-site energies. Increasing $P$, the gap decreases, converging to the gap of the union of the two relevant homopolymers, which is $\approx$ 0.5 eV lower than the gaps of the relevant homopolymers. As far as the mean probabilities are regarded, if we initially place the carrier at a G-C monomer, the probability to find it at an A-T monomer is small, and vice versa. Increasing $P$, $k$ (from the first to the last monomer) falls dramatically.

Some further general remarks: For both I and D polymers, the frequency content of carrier transfer (in terms of the TWMF) lies within the THz regime. For both I and D polymers, although $k(N)$ is a decreasing function, it can be increased, for the same $N$, by many orders of magnitude with appropriate sequence choice.  
The homopolymers (e.g. G... and A...), i.e. the structurally simplest cases, display higher pure mean transfer rates, hence they are more efficient in terms of electron and hole transfer. As far as comparison with experiments is concerned, the TB parameters any TB model uses have to be utilized with care, since large variations are expected in real situations. Transient absorption spectroscopy experiments give transfer rates whose order of magnitude is close to our theoretically determined pure mean transfer rates. Also, from a qualitative point of view, in terms of comparison of the inverse decay length of the transfer rates, TB can grasp the experimental  behavior.

Closing this article, we mention that, although here we used parameters relevant to B-DNA, this analysis holds for other periodic polymers of similar types, other than B-DNA.

\noindent \textbf{Acknowledgements}
K. Lambropoulos wishes to thank the General Secretariat of
Research and Technology (GSRT) and the Hellenic Foundation for
Research and Innovation (HFRI) for a PhD research scholarship. \\


\clearpage

\appendix
\renewcommand\thefigure{\thesection.\arabic{figure}}    
\setcounter{figure}{0}    

\begin{widetext}

\section{\label{apA} Supplementary Material} 

\begin{figure*}[!h]
\includegraphics[width=0.4\textwidth]{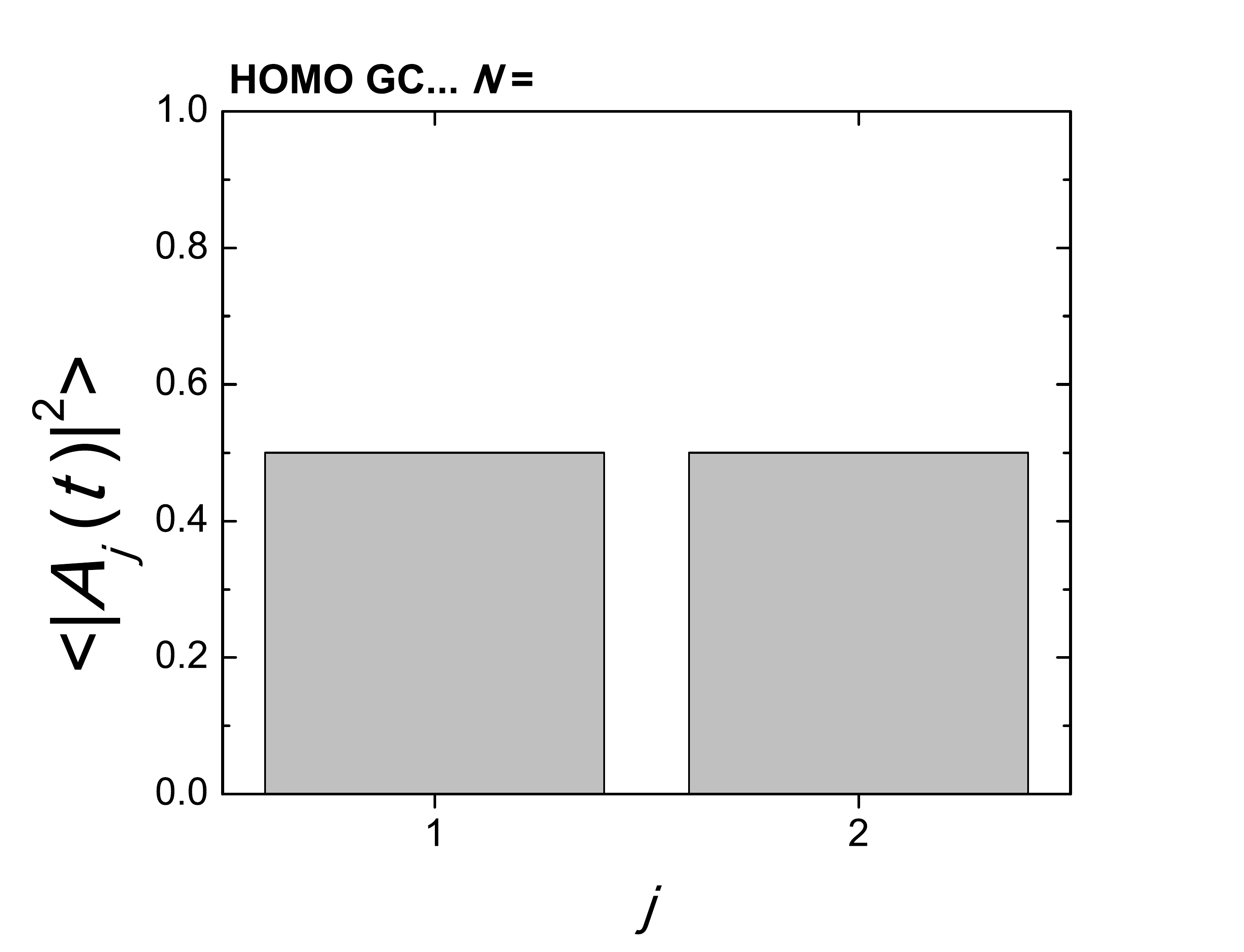}
\includegraphics[width=0.4\textwidth]{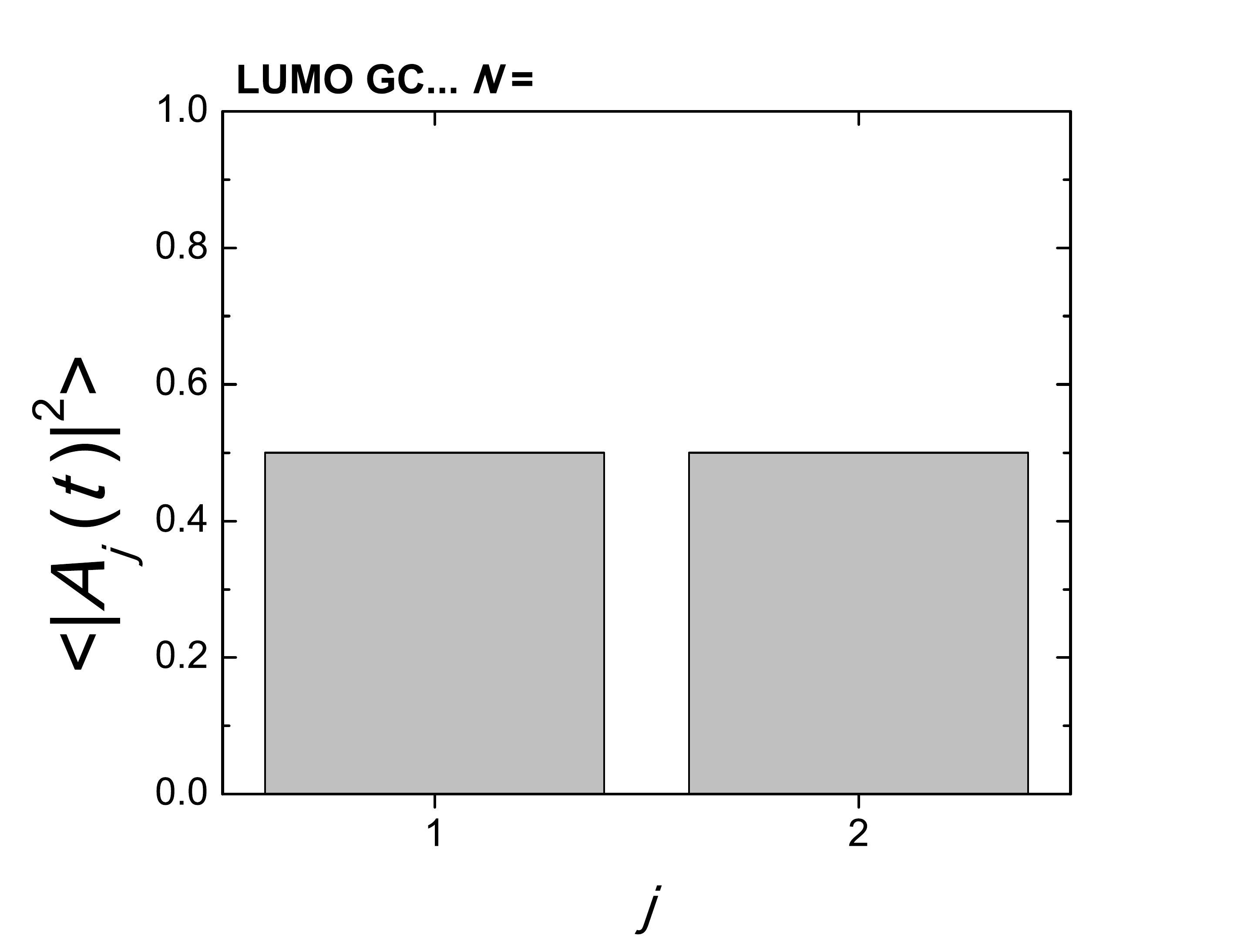} 
\includegraphics[width=0.4\textwidth]{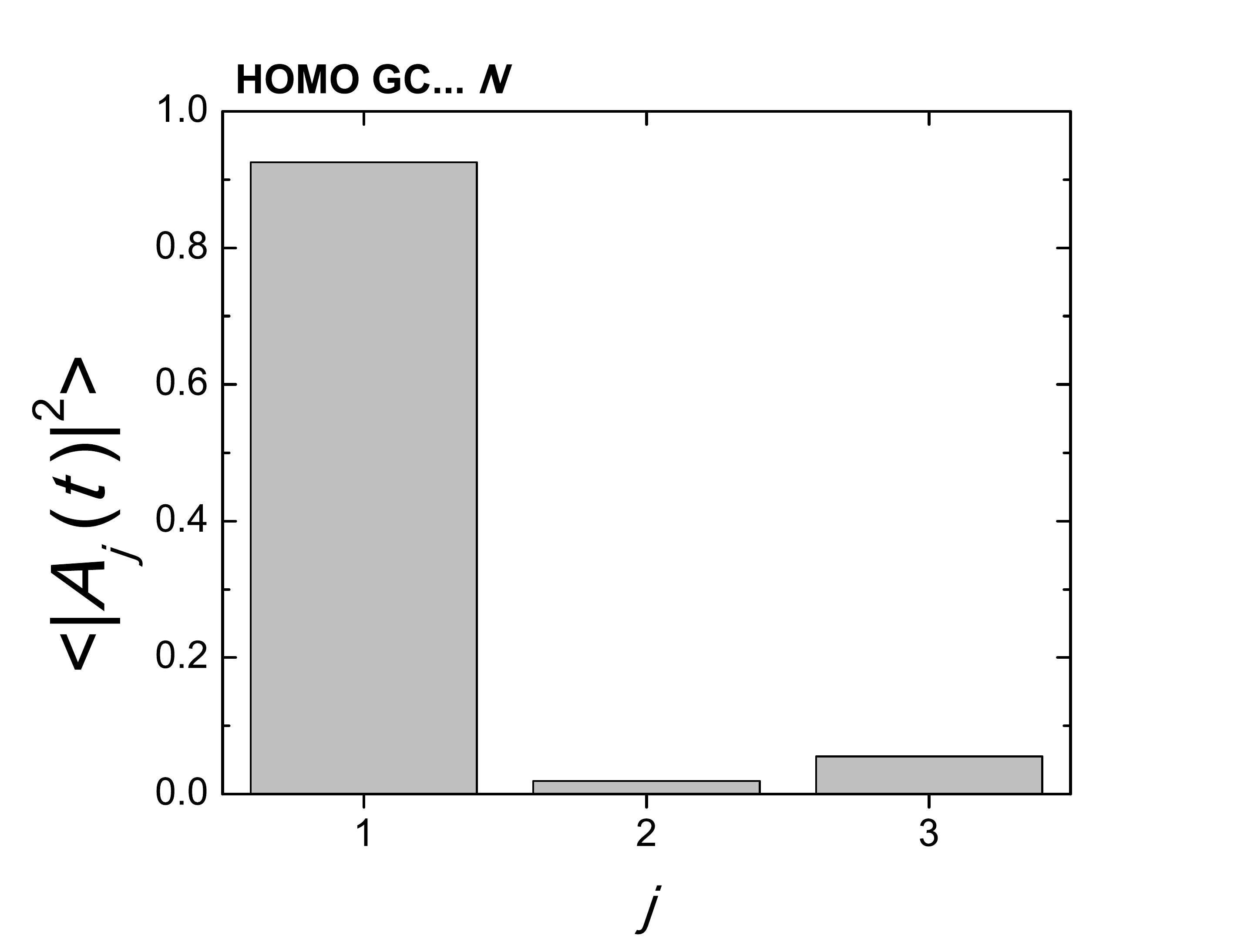}
\includegraphics[width=0.4\textwidth]{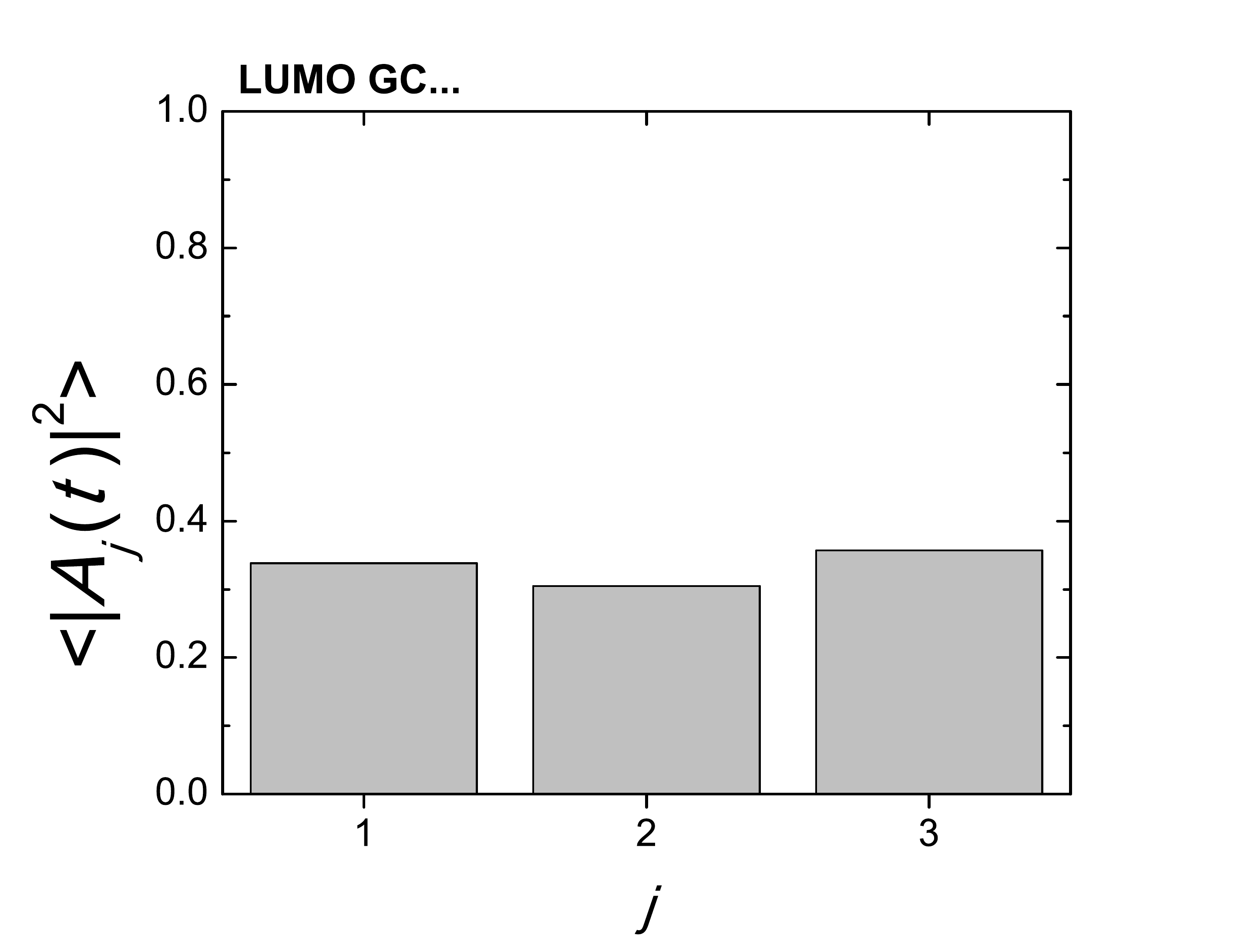}
\caption{Mean (over time) probabilities to find the extra carrier at each monomer $j$, having placed it initially at the first monomer, for I2 (GC...) polymers, for the HOMO (left) and the LUMO (right). $N = P + \tau$, $\tau = 0, 1, \dots, P-1$.}
\label{fig:ProbabilitiesHL-I2}
\end{figure*}

\begin{figure*}[!h]
\includegraphics[width=0.4\textwidth]{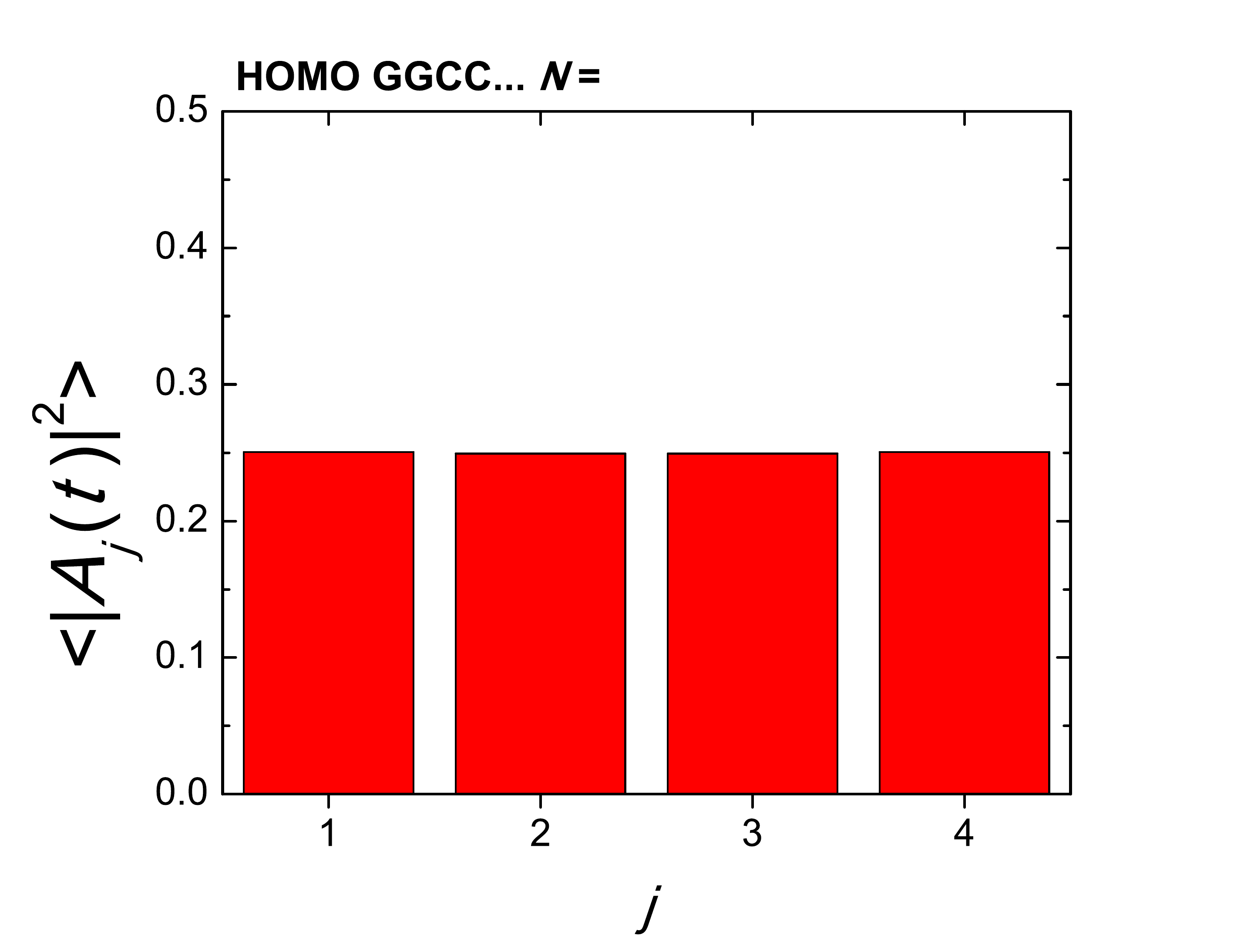}
\includegraphics[width=0.4\textwidth]{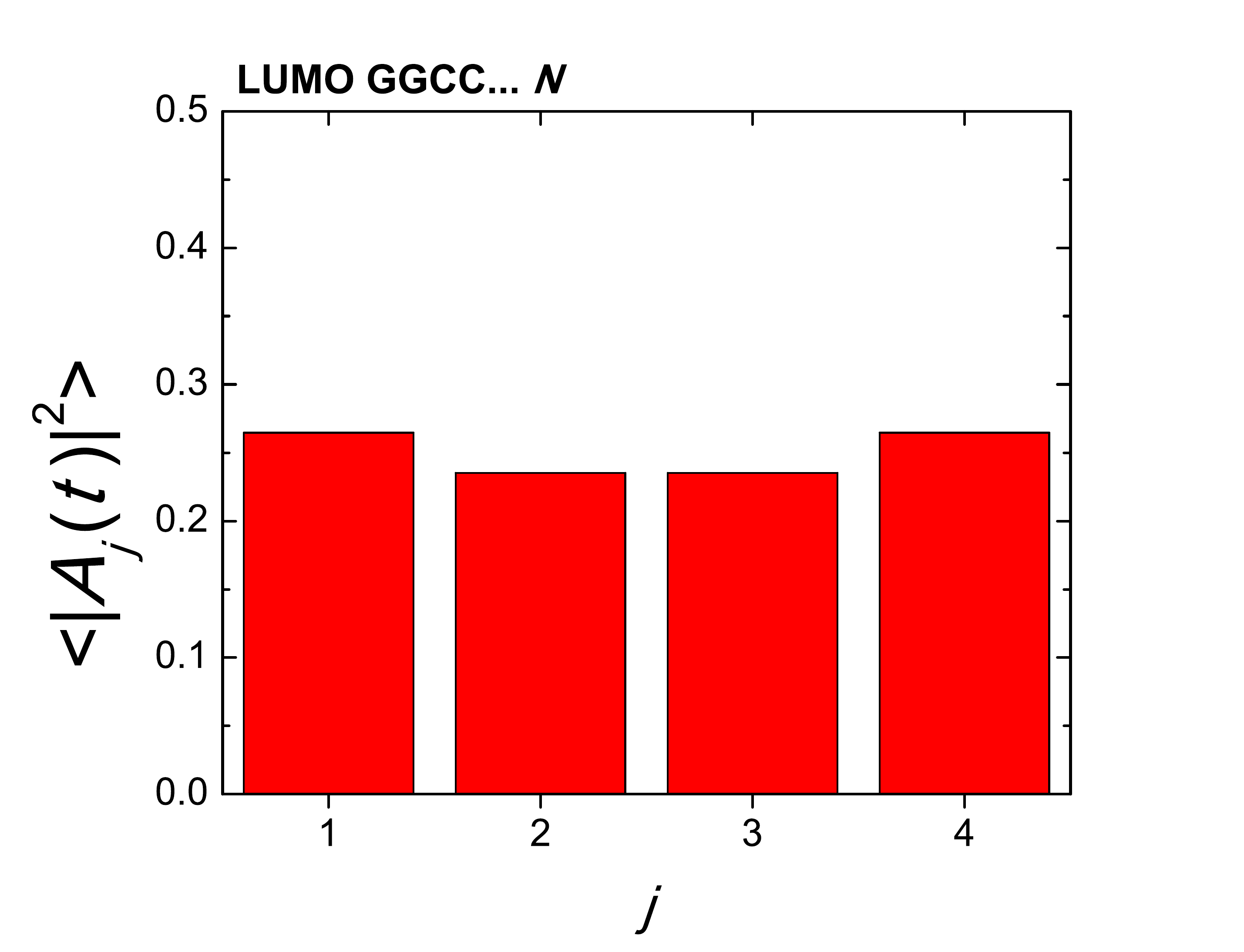}
\includegraphics[width=0.4\textwidth]{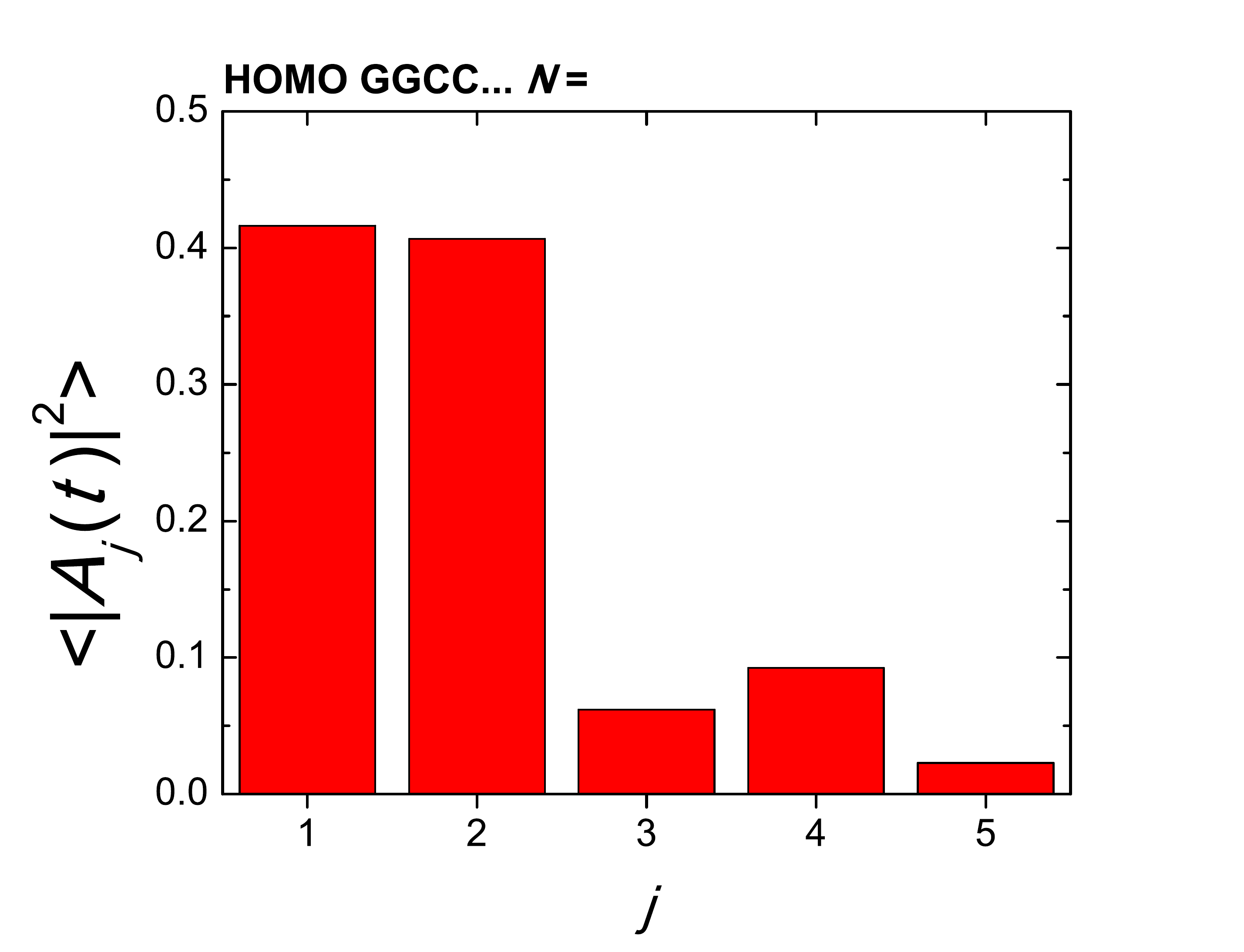}
\includegraphics[width=0.4\textwidth]{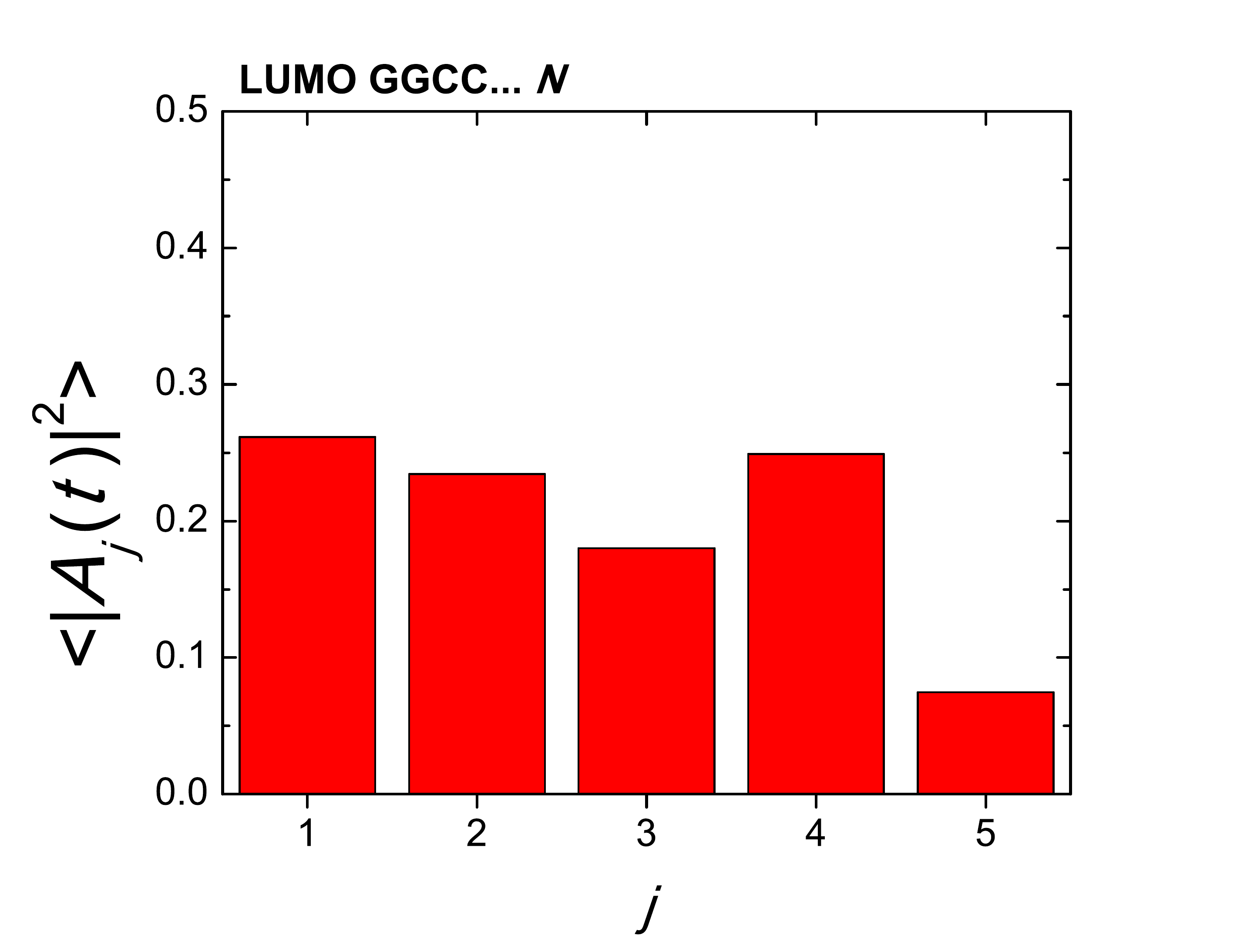}
\includegraphics[width=0.4\textwidth]{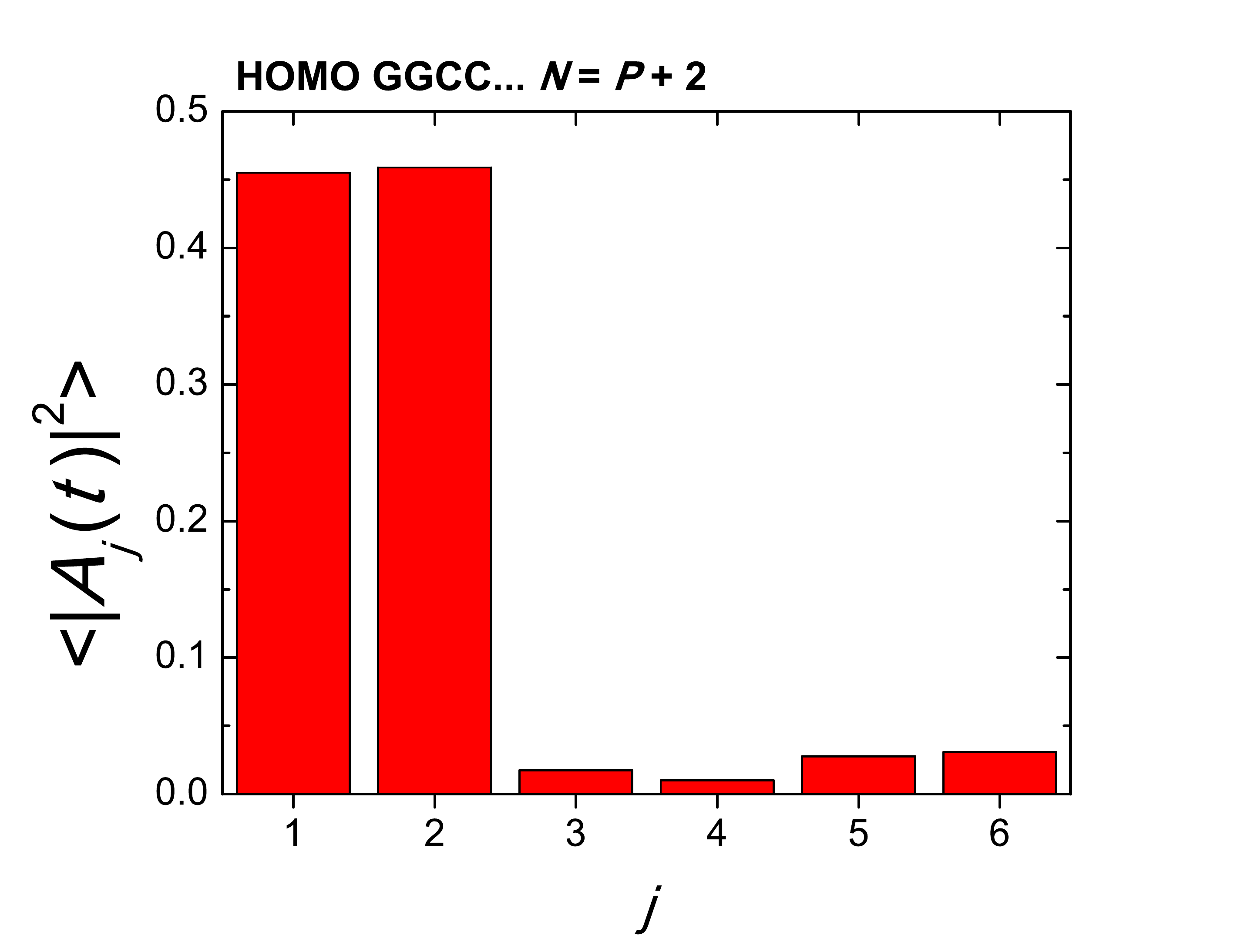}
\includegraphics[width=0.4\textwidth]{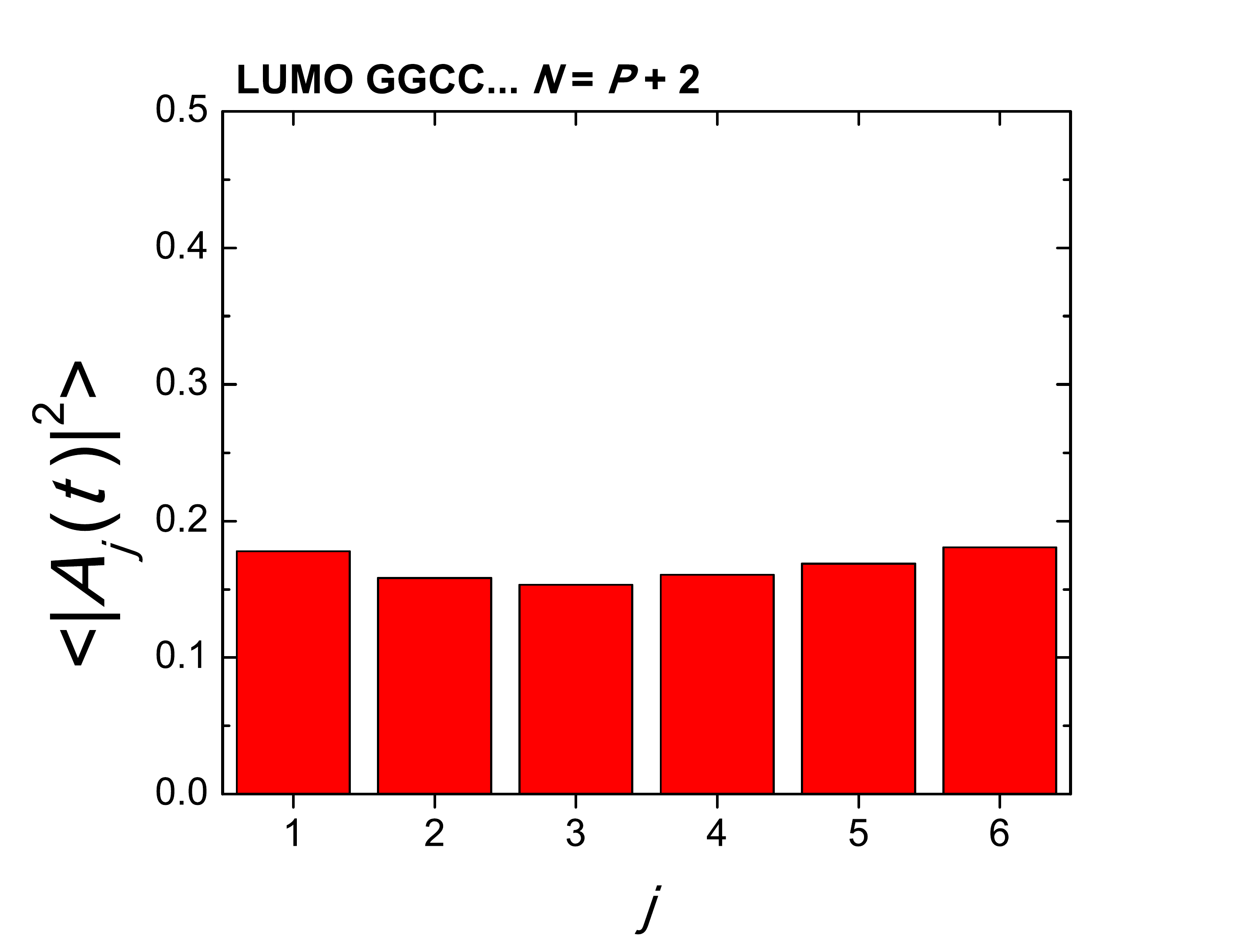}
\includegraphics[width=0.4\textwidth]{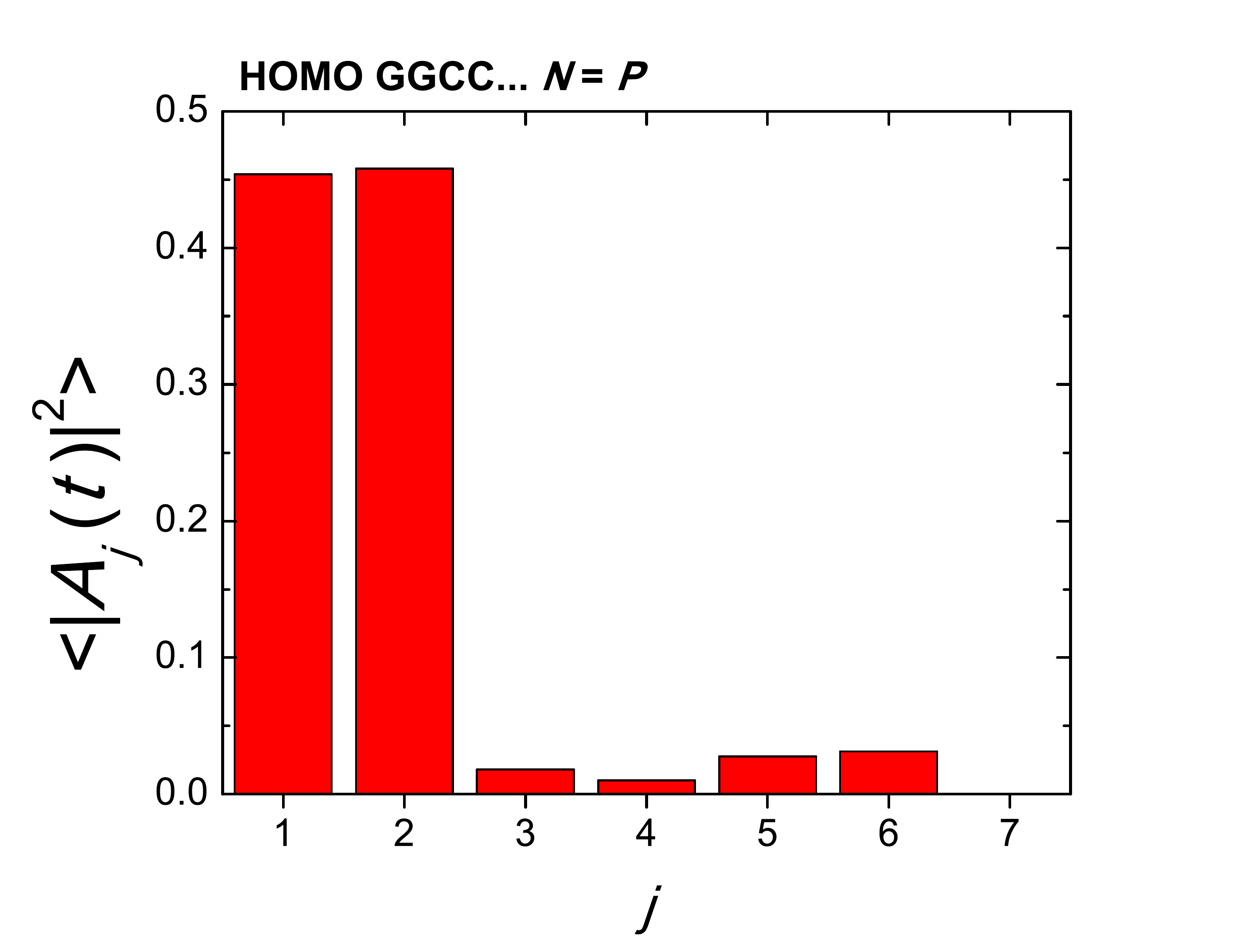}
\includegraphics[width=0.4\textwidth]{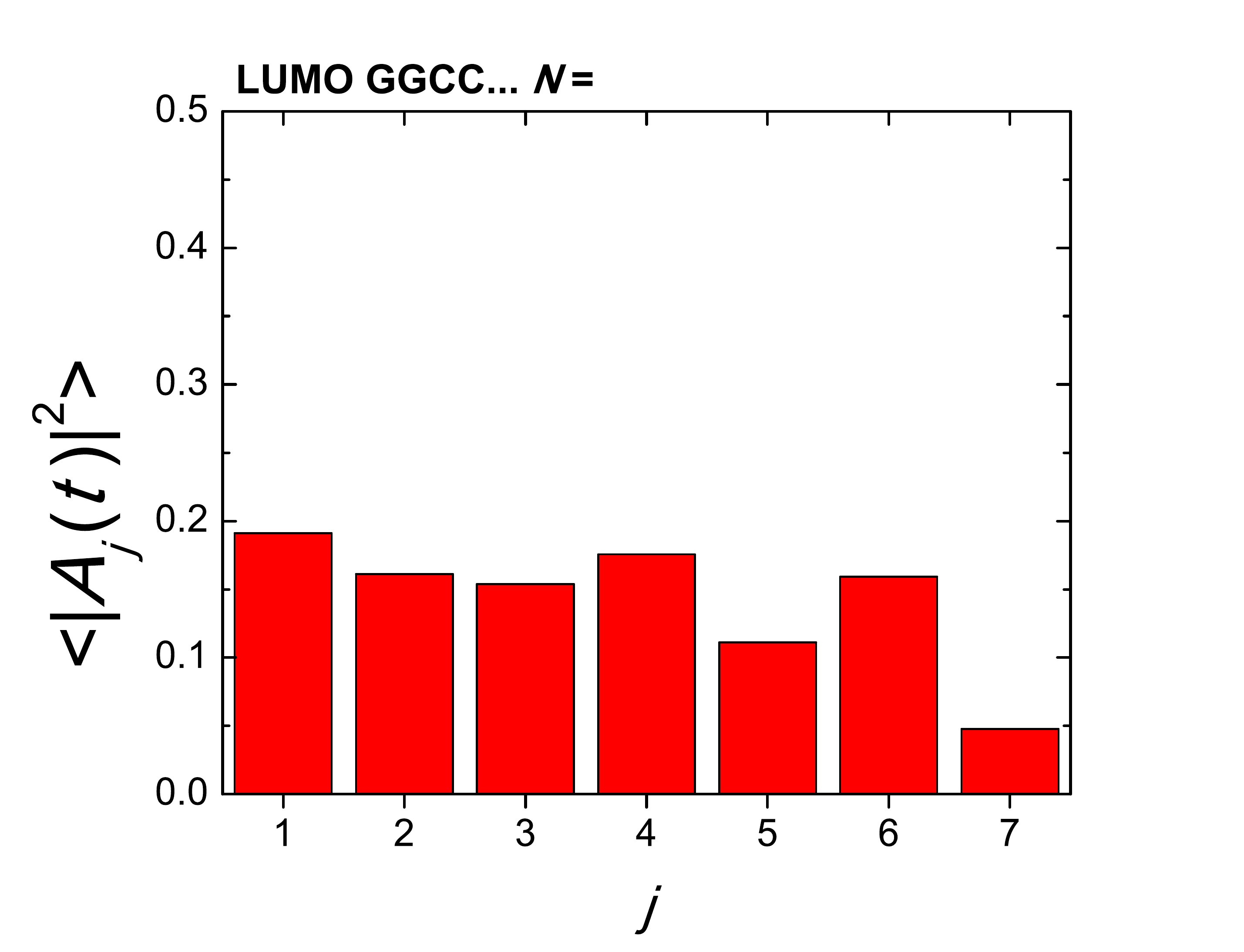}
\caption{Mean (over time) probabilities to find the extra carrier at each monomer $j$, having placed it initially at the first monomer, for I4 (GGCC...) polymers, for the HOMO (left) and the LUMO (right). $N = P + \tau$, $\tau = 0, 1, \dots, P-1$.}
\label{fig:ProbabilitiesHL-I4}
\end{figure*}

\begin{figure*}[!h]
\includegraphics[width=0.4\textwidth]{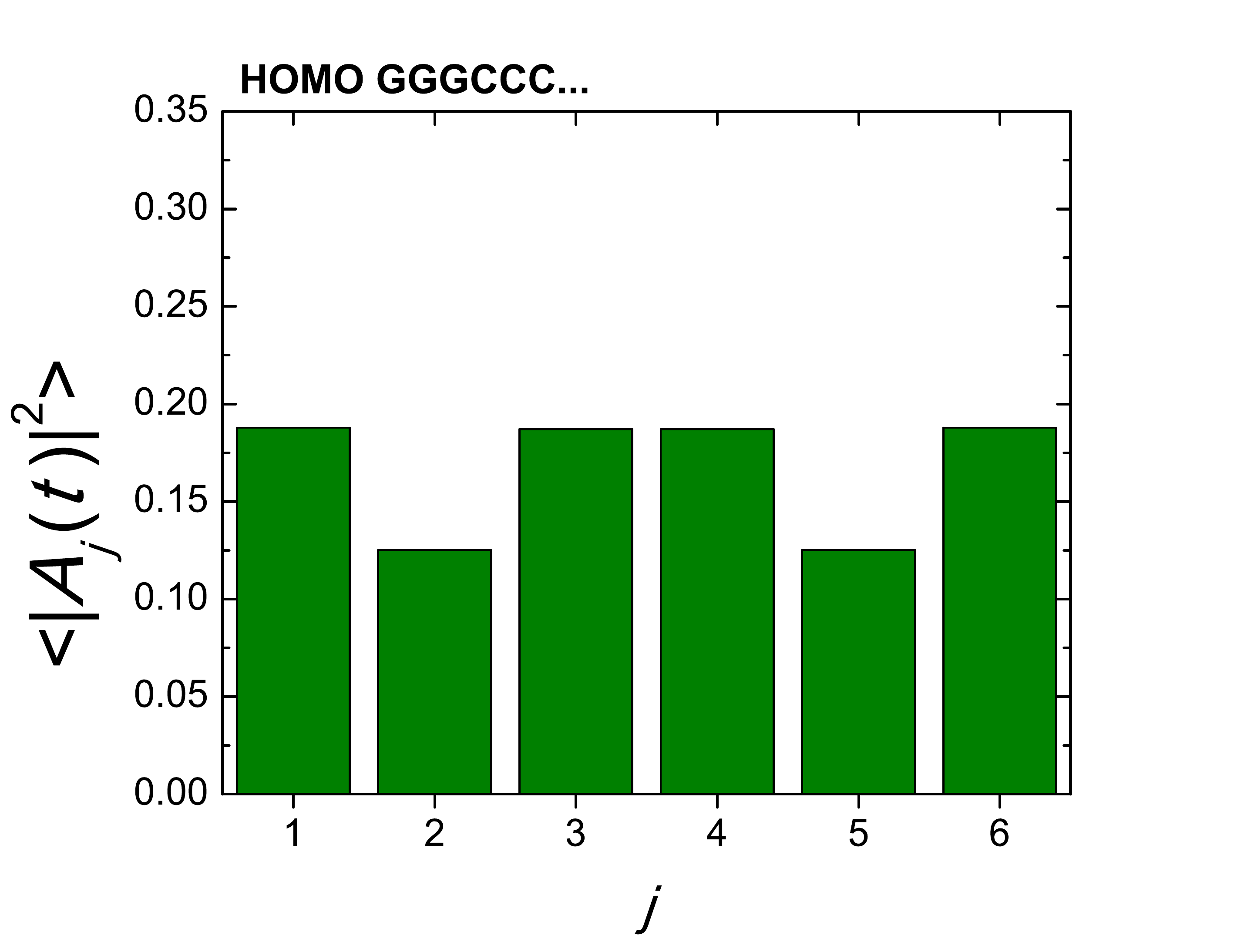}
\includegraphics[width=0.4\textwidth]{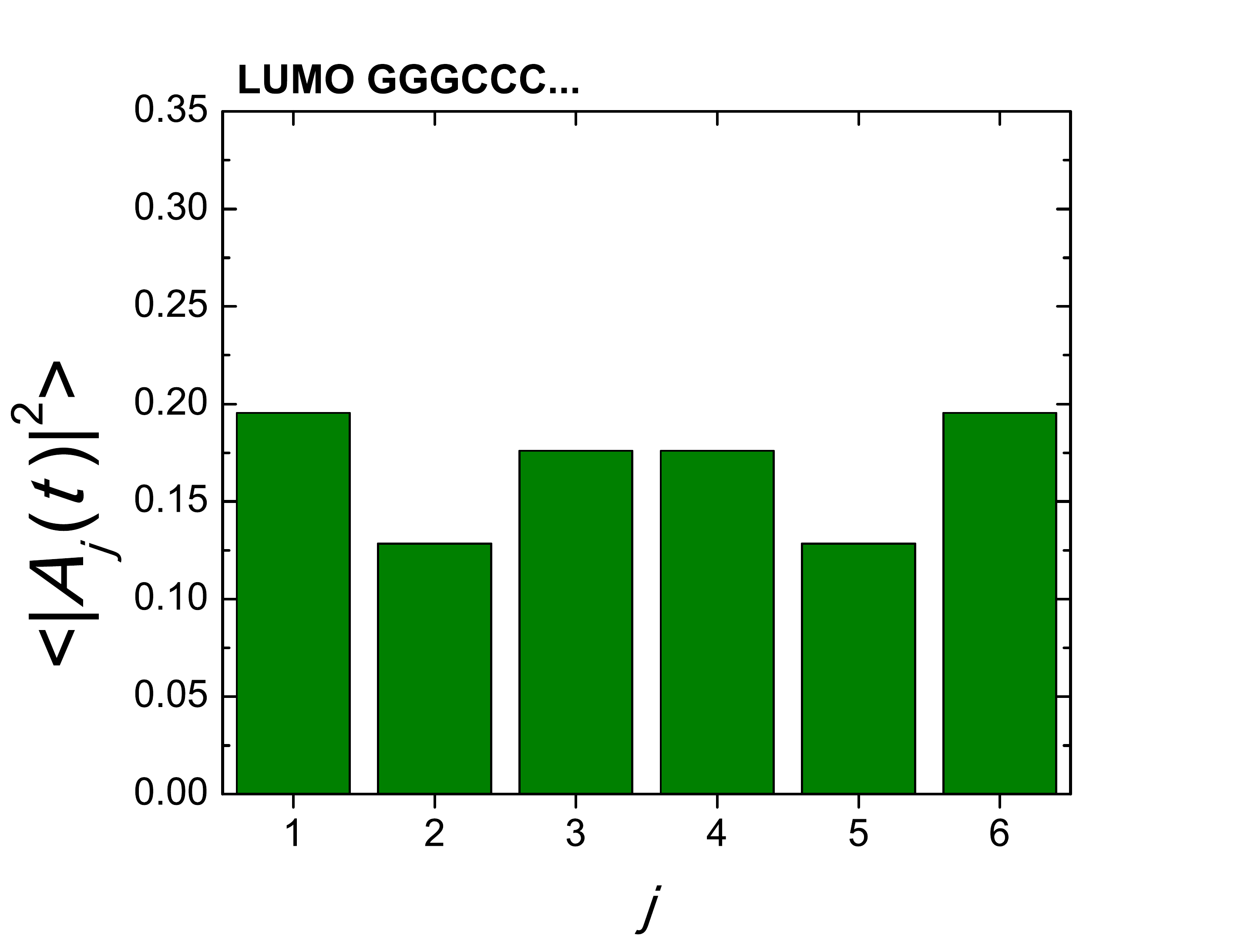}
\includegraphics[width=0.4\textwidth]{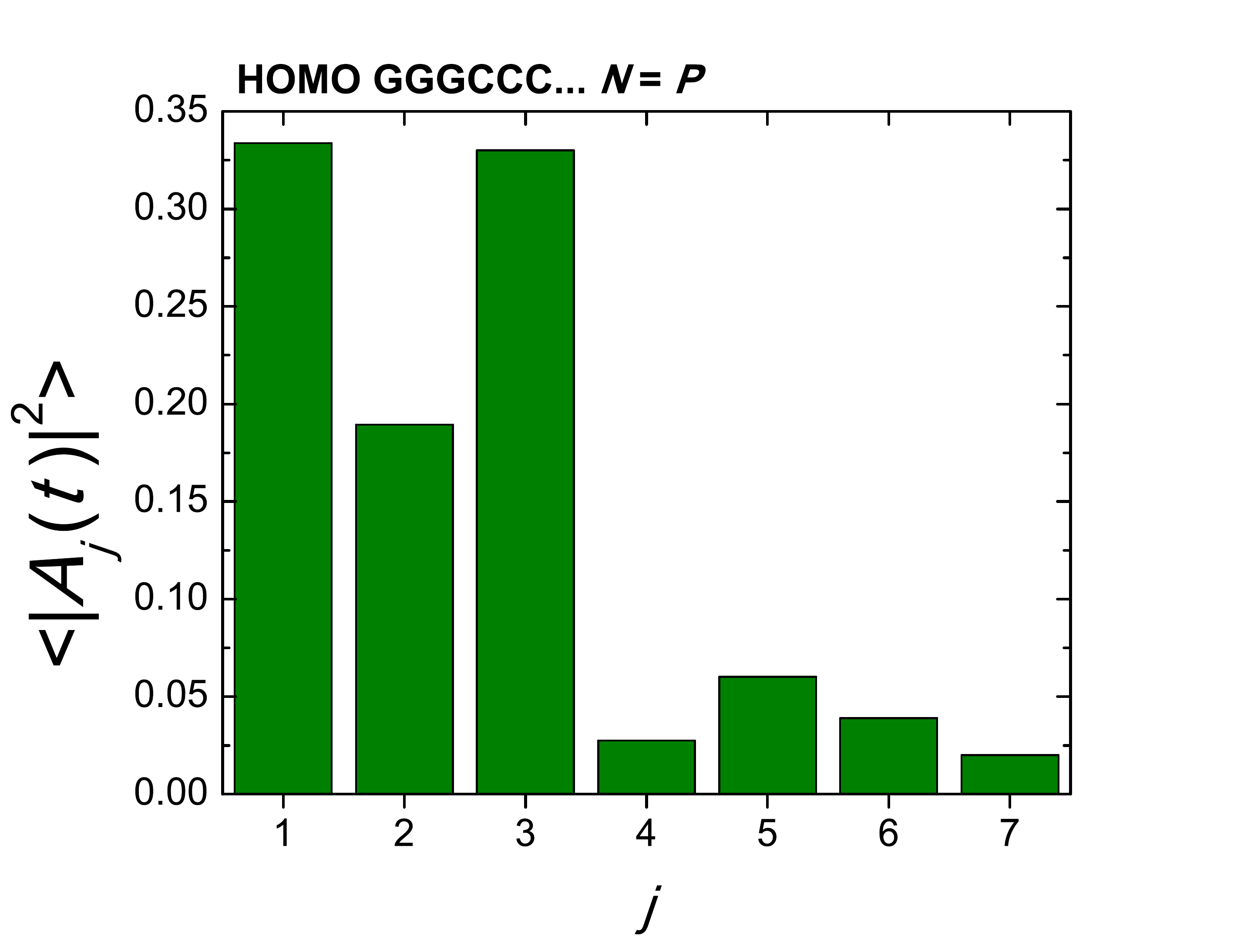}
\includegraphics[width=0.4\textwidth]{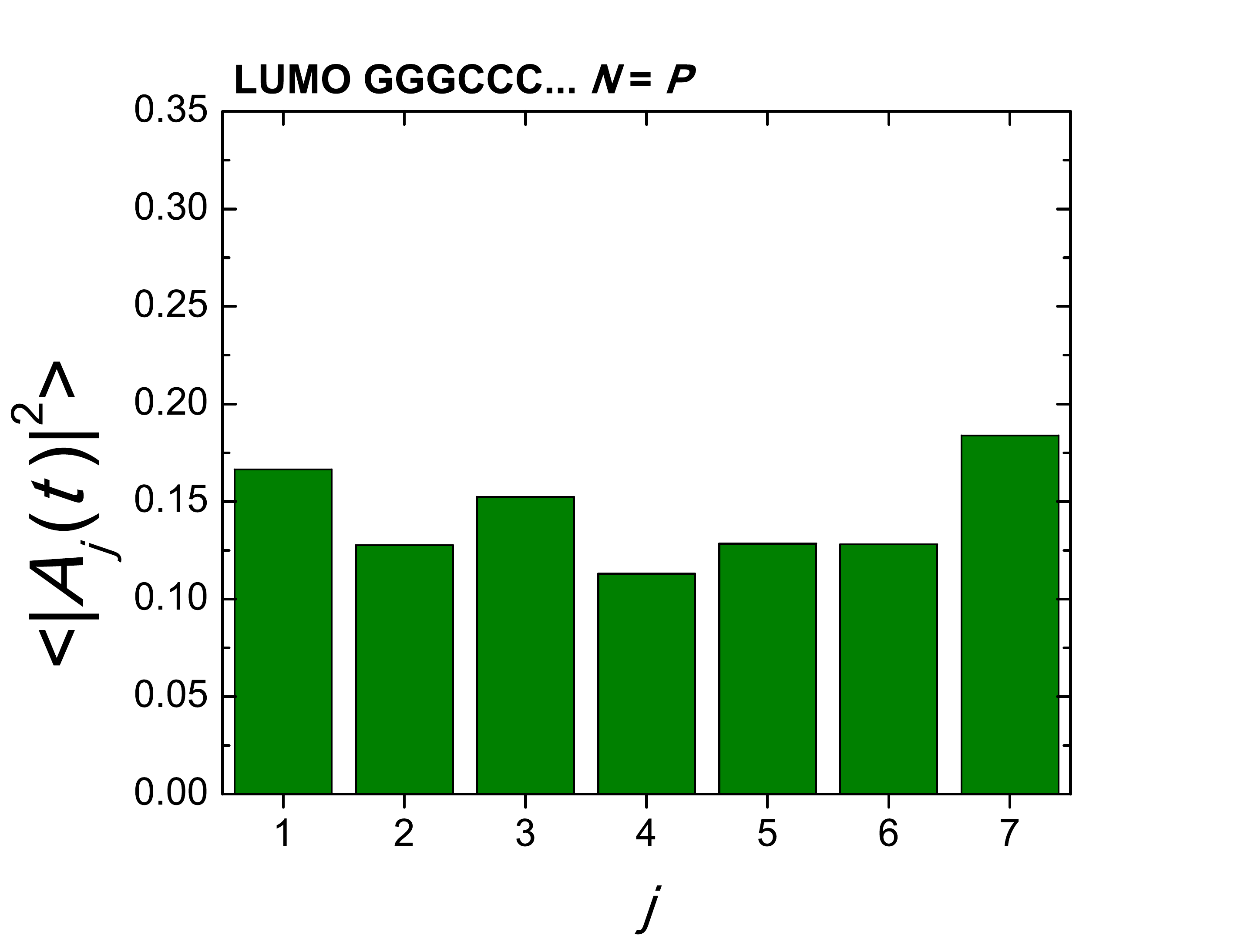}
\includegraphics[width=0.4\textwidth]{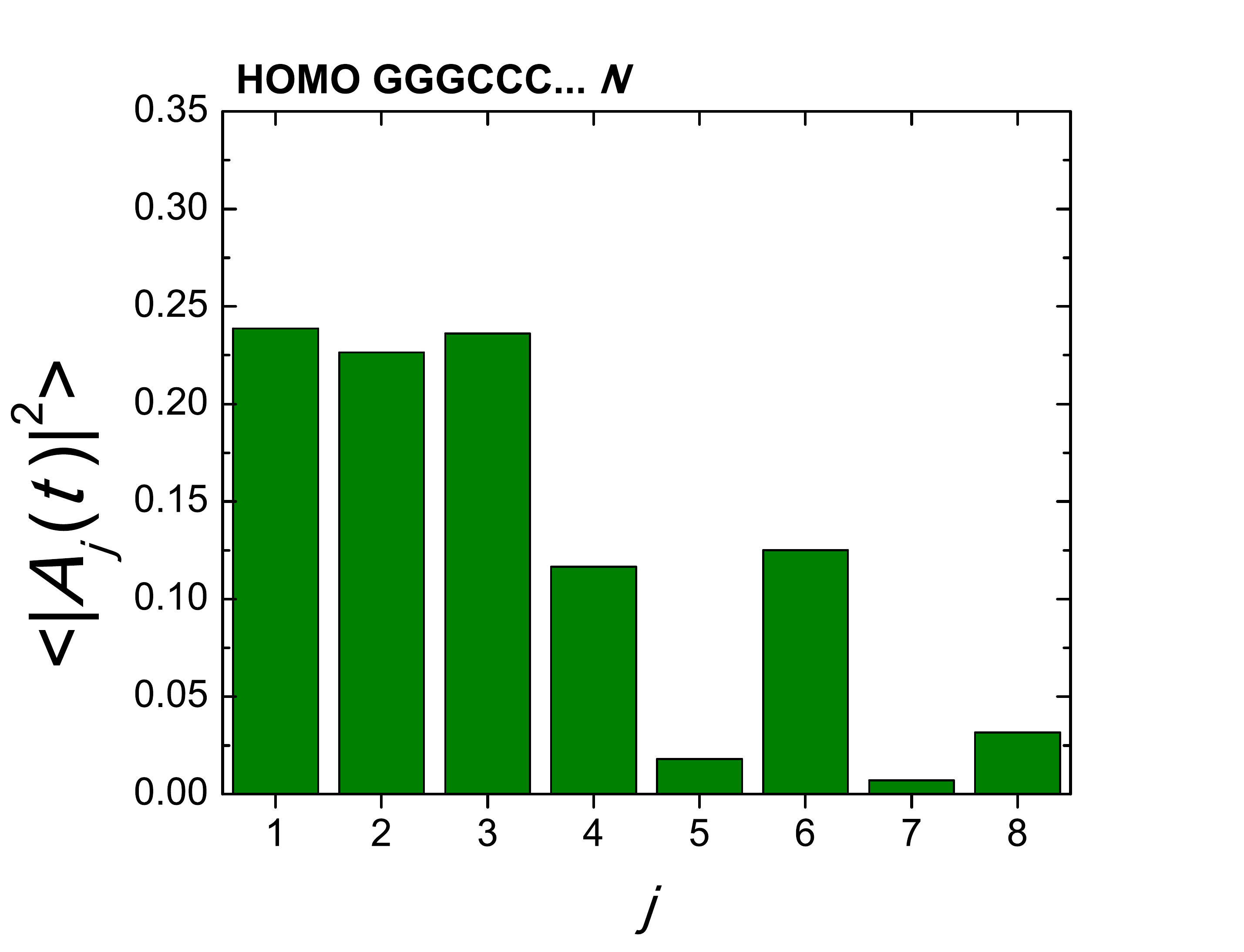}
\includegraphics[width=0.4\textwidth]{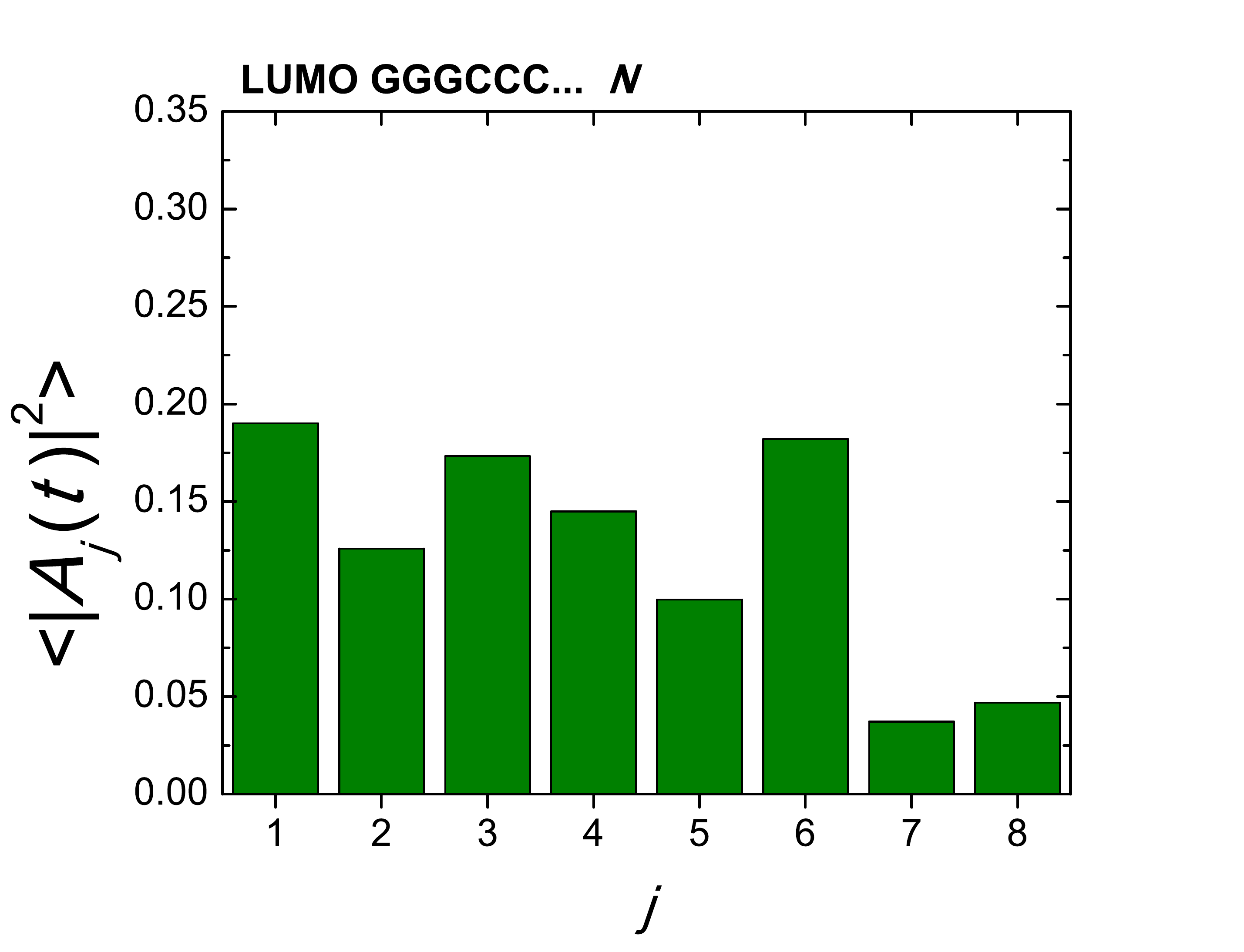}
\caption{Mean (over time) probabilities to find the extra carrier at each monomer $j$, having placed it initially at the first monomer, for I6 (GGGCCC...) polymers, for the HOMO (left) and the LUMO (right). $N = P + \tau$, $\tau = 0, 1, \dots, P-1$. \emph{Continued at the next page...}}
\label{fig:ProbabilitiesHL-I6}
\end{figure*}
\begin{figure*}[!h]
\addtocounter{figure}{-1}
\includegraphics[width=0.4\textwidth]{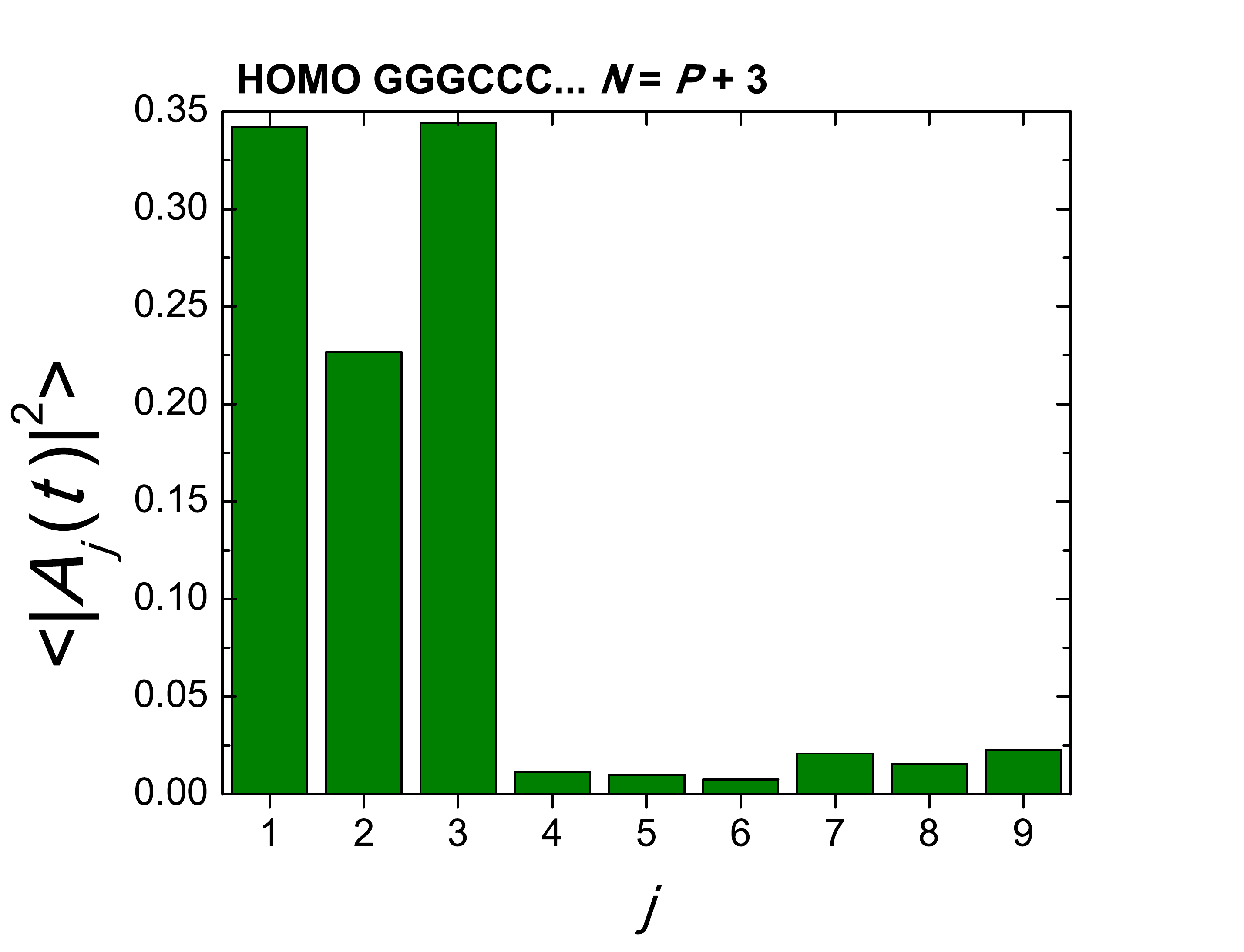}
\includegraphics[width=0.4\textwidth]{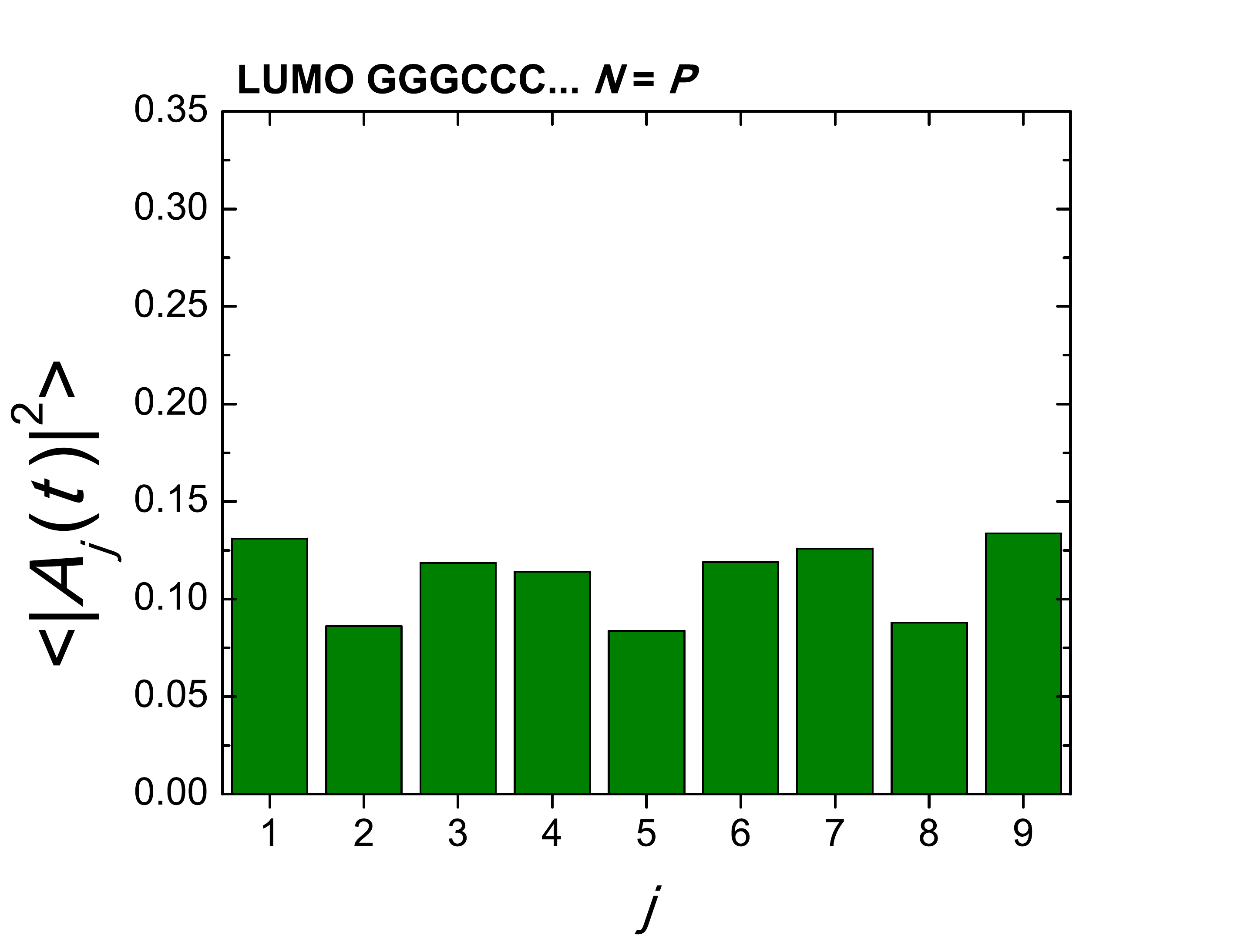}
\includegraphics[width=0.4\textwidth]{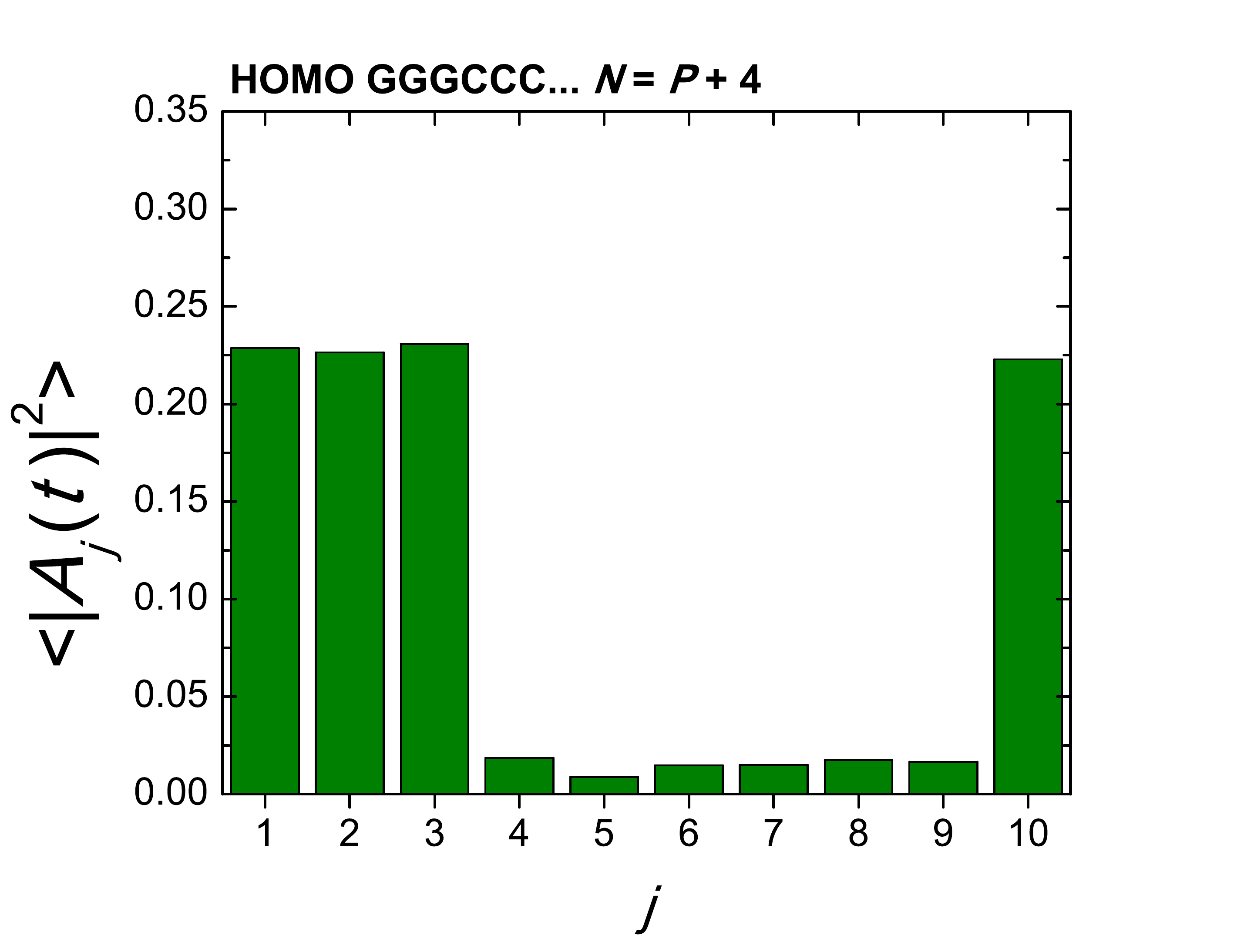}
\includegraphics[width=0.4\textwidth]{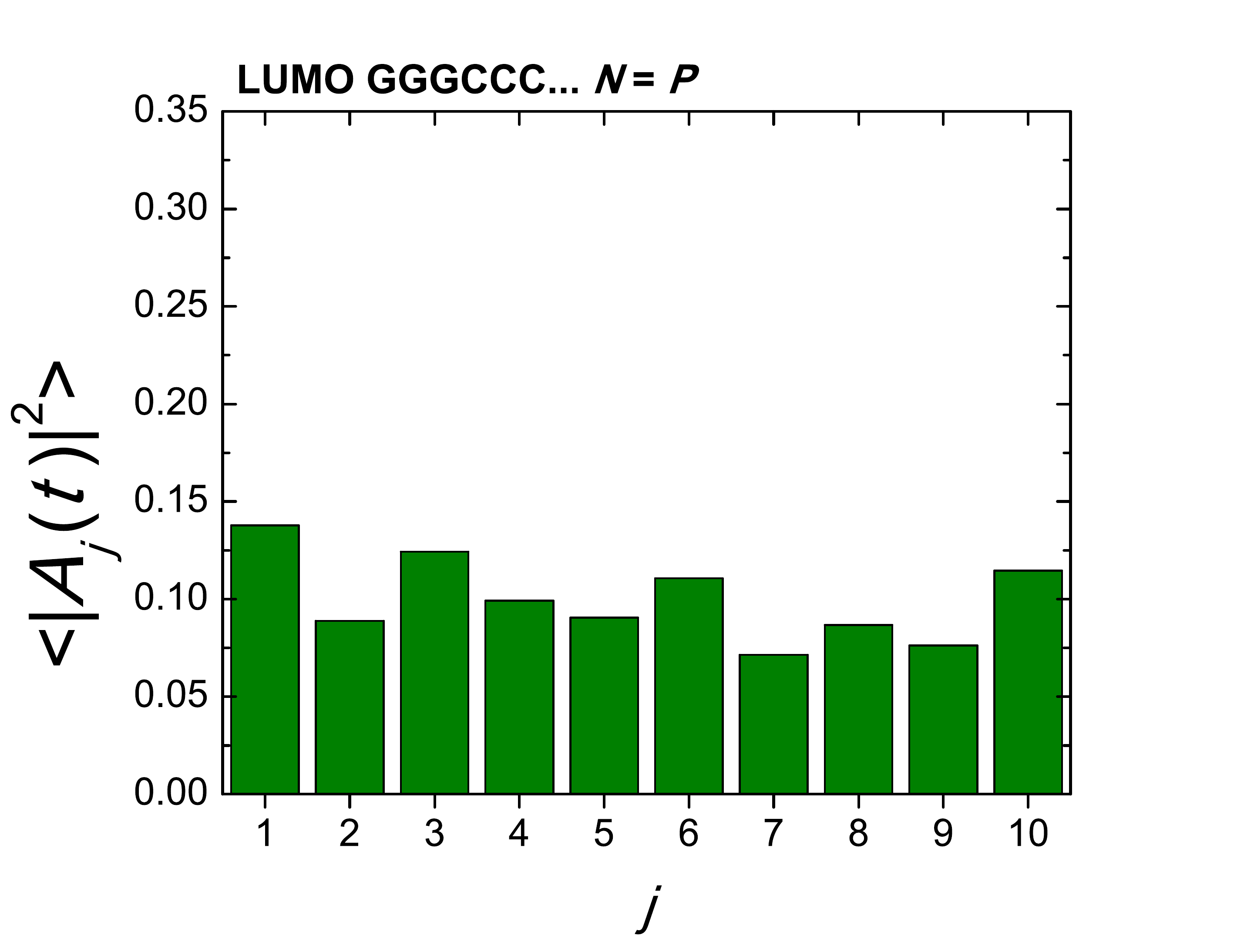}
\includegraphics[width=0.4\textwidth]{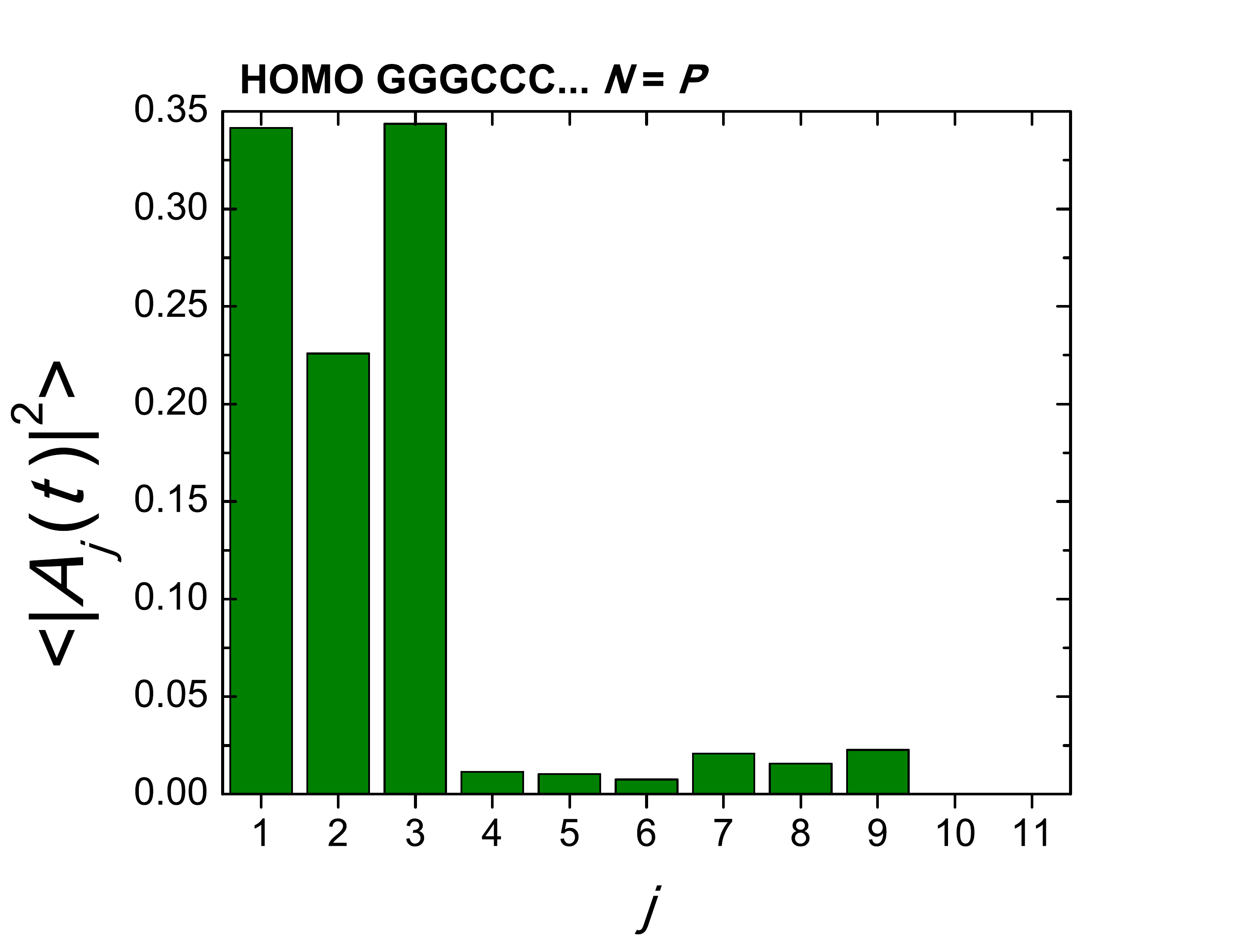}
\includegraphics[width=0.4\textwidth]{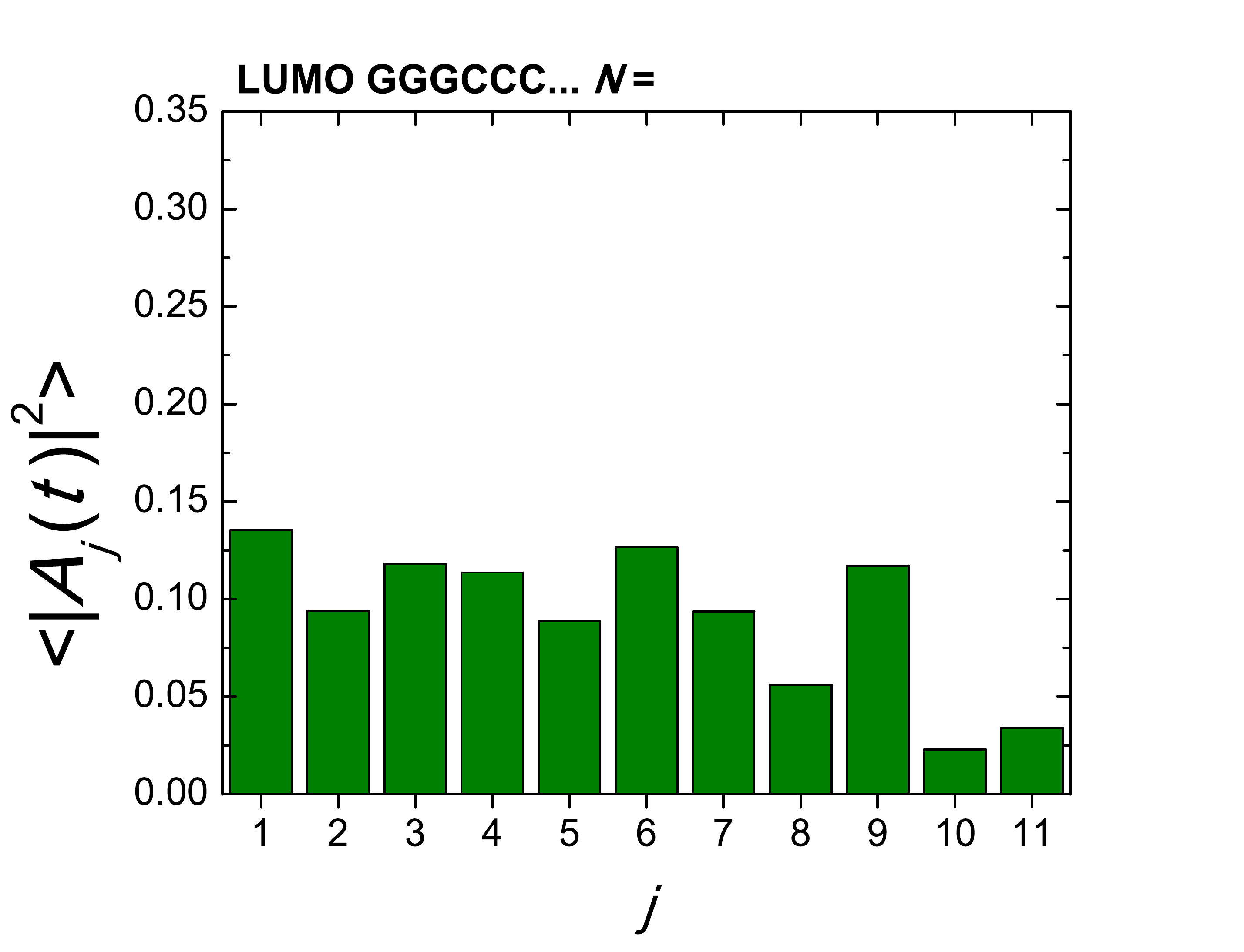}
\caption{\emph{...Continued from the previous page.} Mean (over time) probabilities to find the extra carrier at each monomer $j$, having placed it initially at the first monomer, for I6 (GGGCCC...) polymers, for the HOMO (left) and the LUMO (right). $N = P + \tau$, $\tau = 0, 1, \dots, P-1$.}
\end{figure*}

\begin{figure*}[!h]
\includegraphics[width=0.4\textwidth]{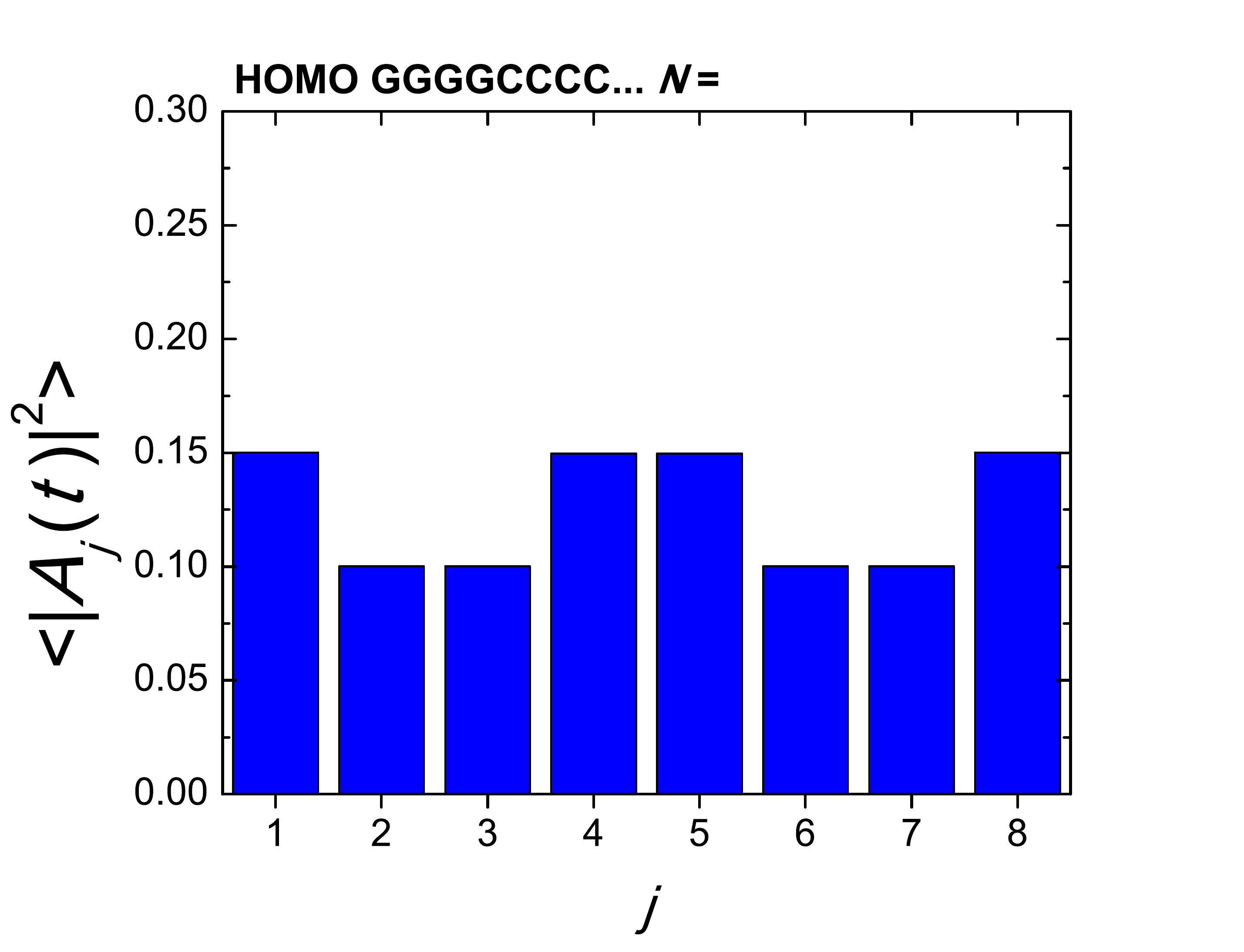}
\includegraphics[width=0.4\textwidth]{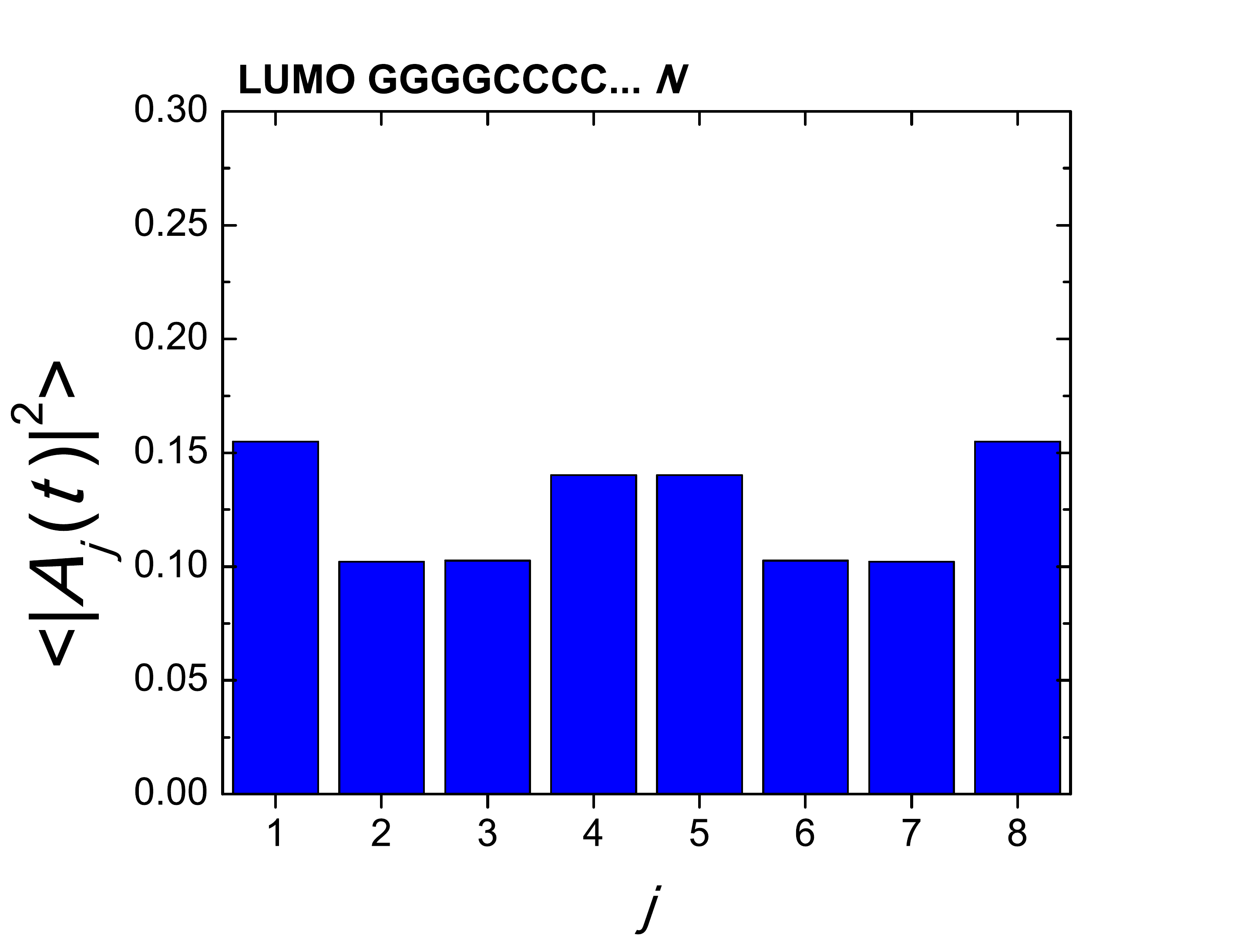}
\includegraphics[width=0.4\textwidth]{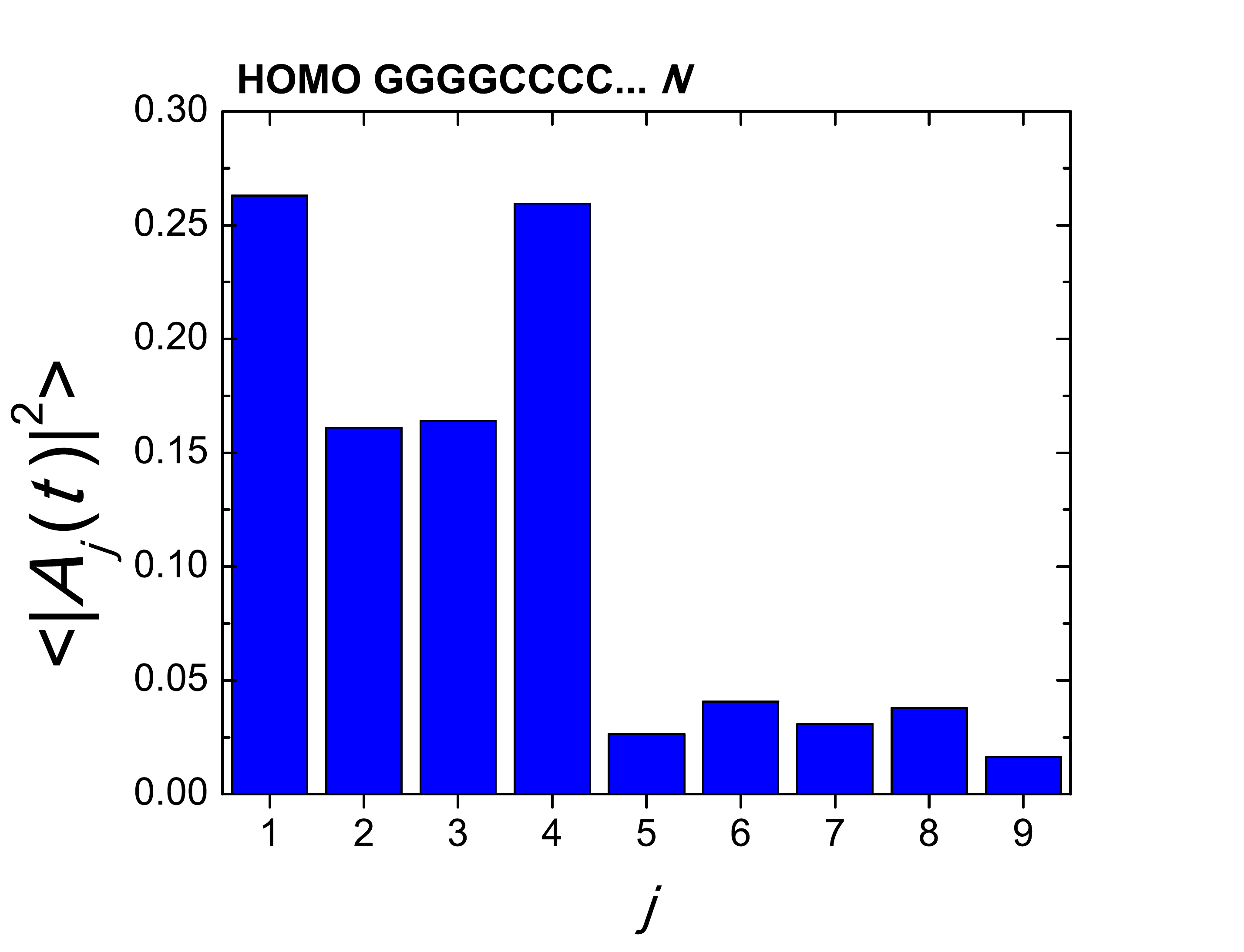}
\includegraphics[width=0.4\textwidth]{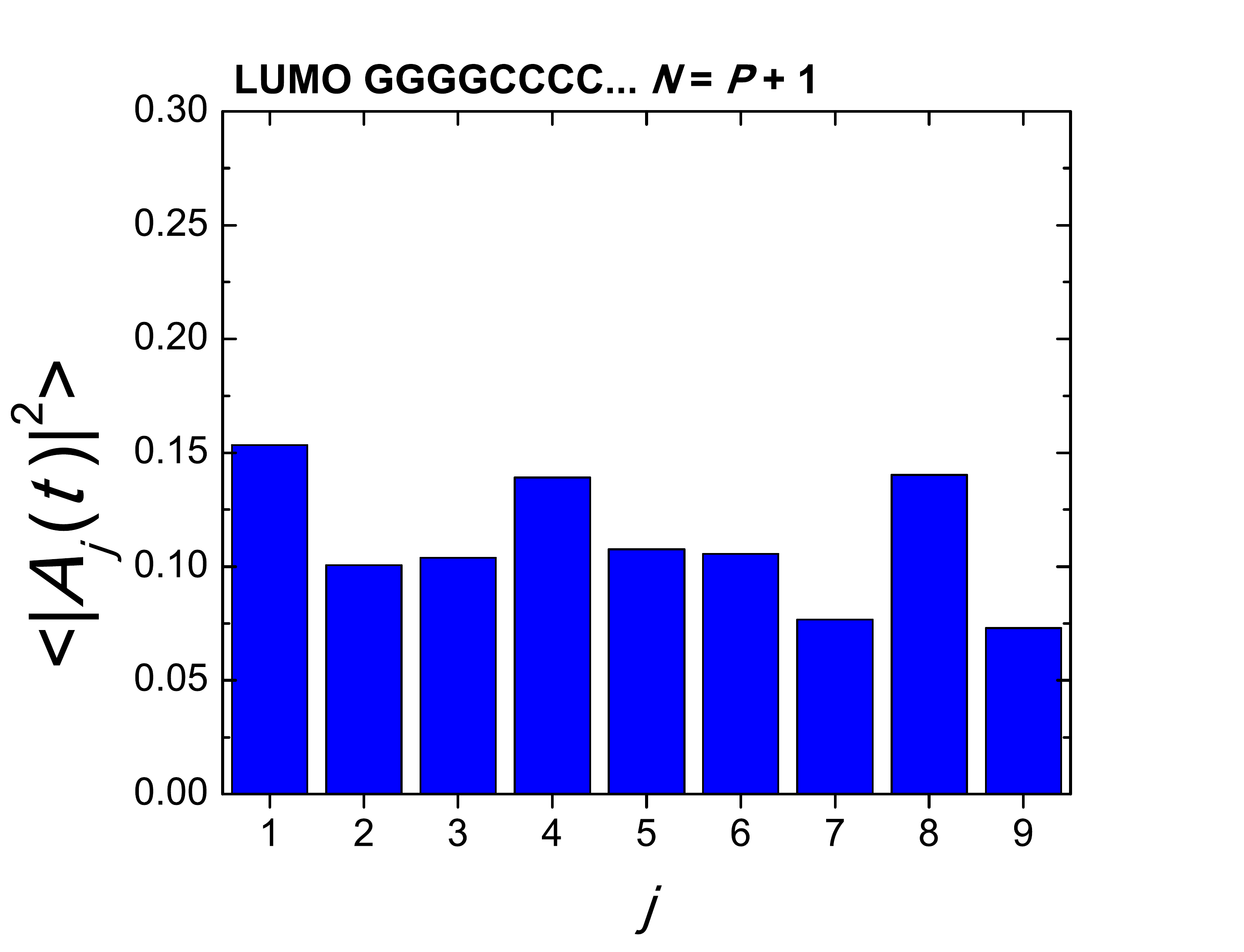}
\includegraphics[width=0.4\textwidth]{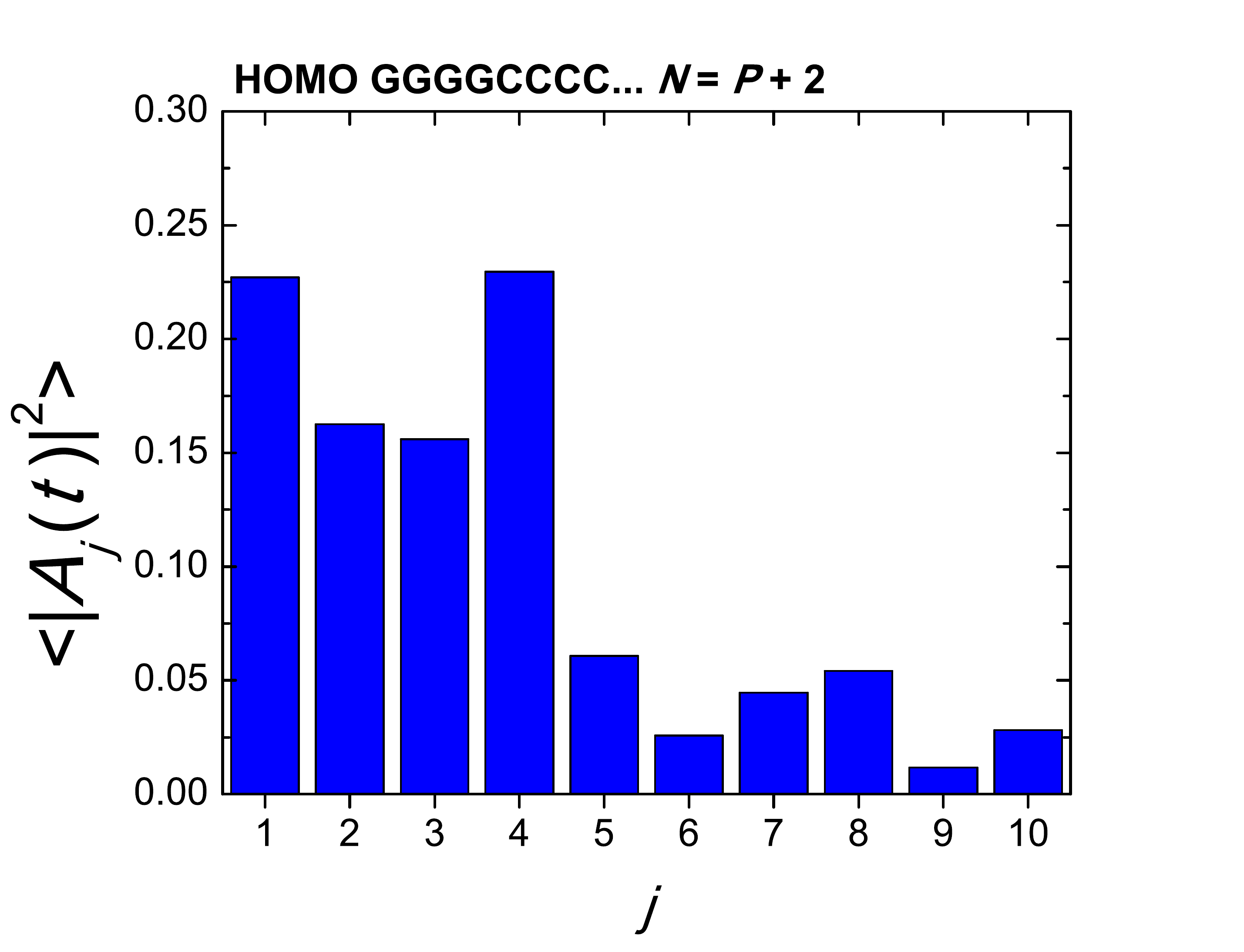}
\includegraphics[width=0.4\textwidth]{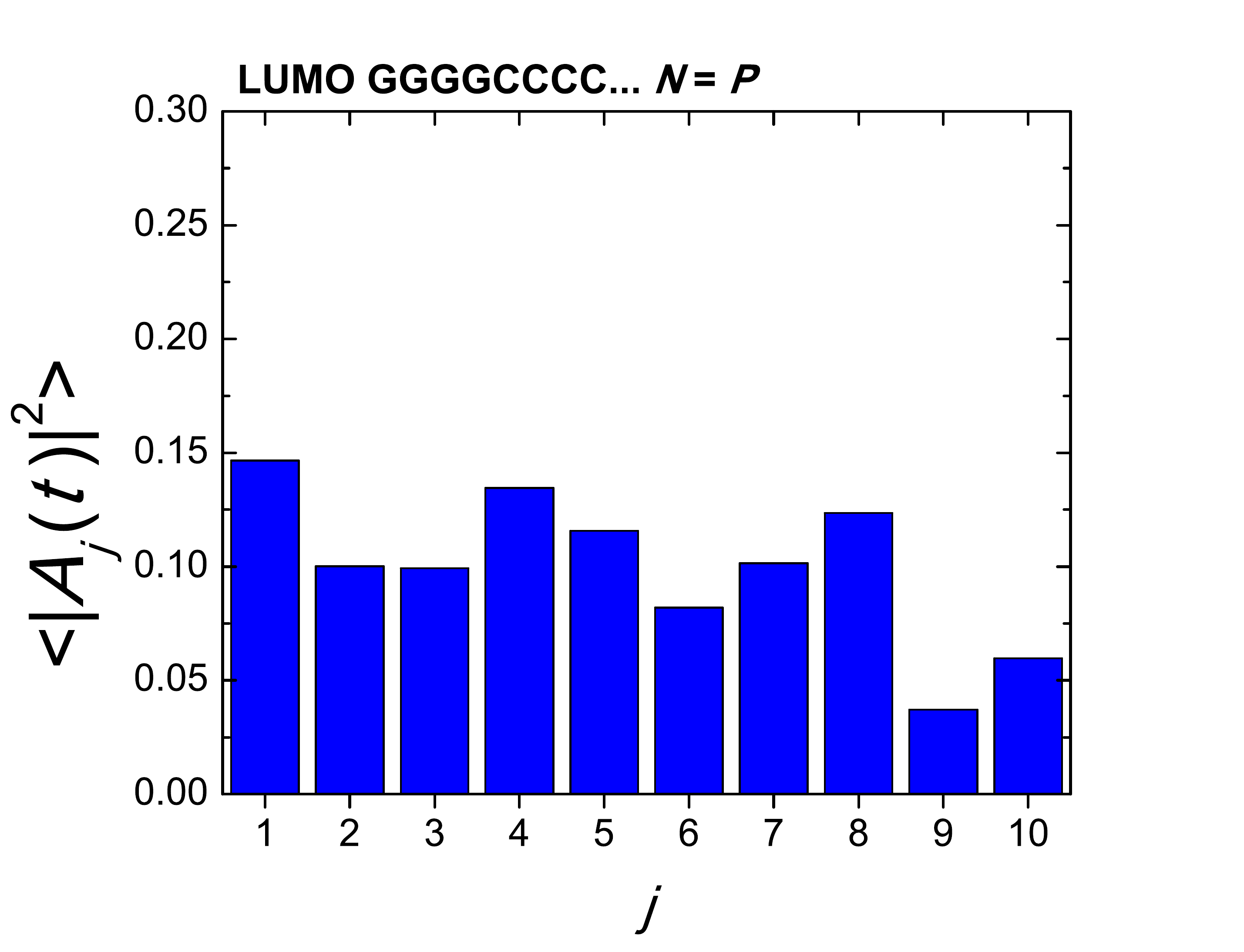}
\includegraphics[width=0.4\textwidth]{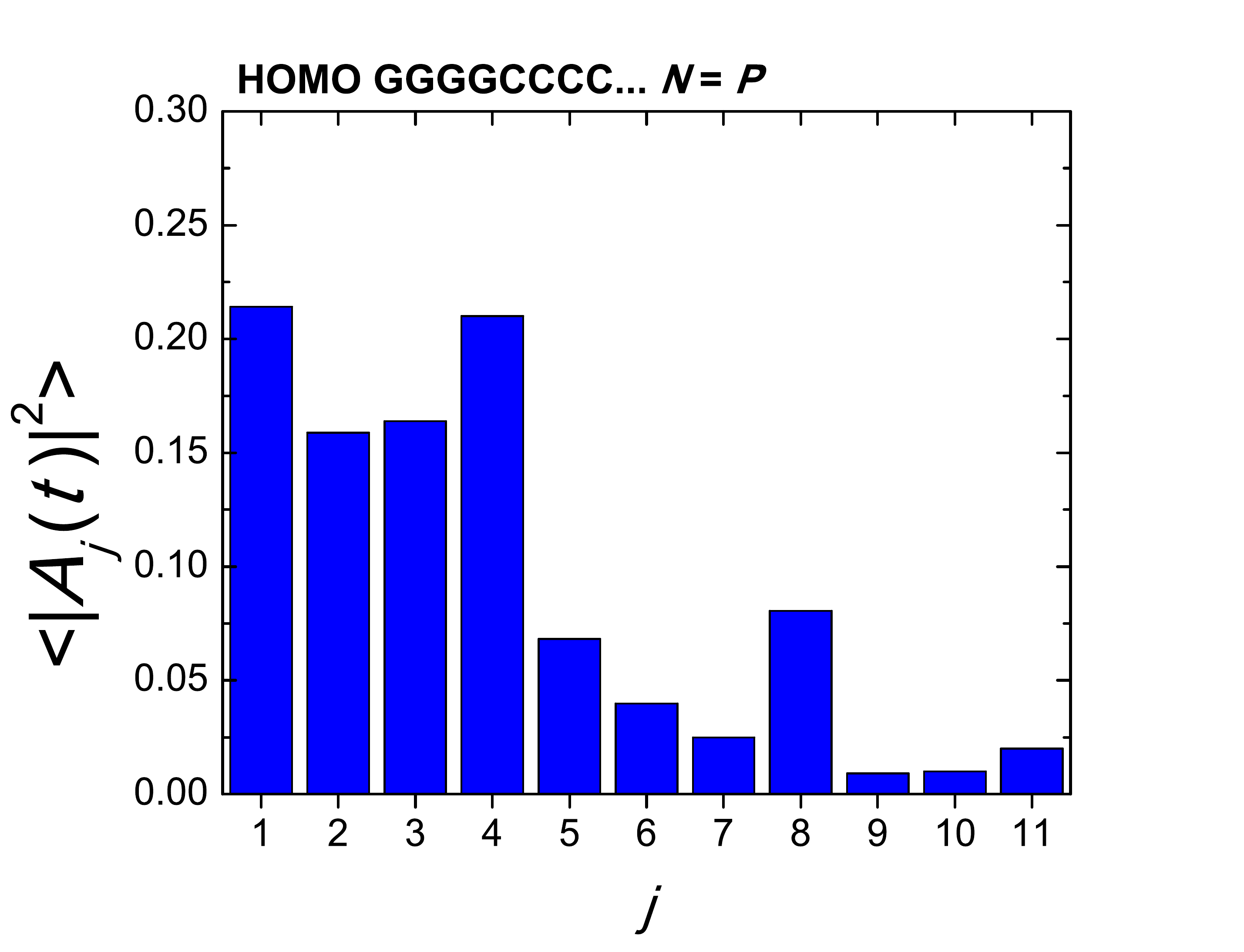}
\includegraphics[width=0.4\textwidth]{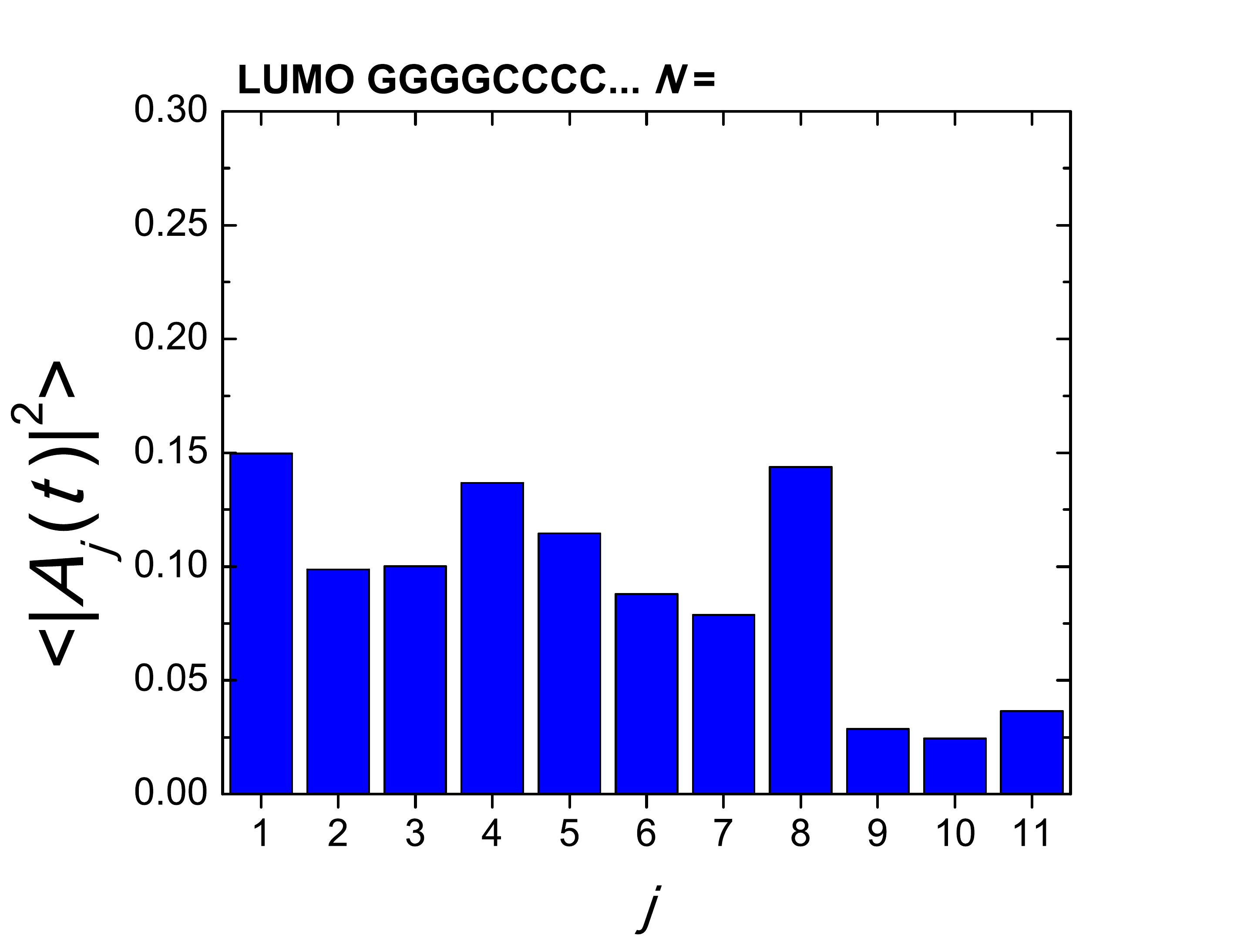}
\caption{Mean (over time) probabilities to find the extra carrier at each monomer $j$, having placed it initially at the first monomer, for I8 (GGGGCCCC...) polymers, for the HOMO (left) and the LUMO (right). $N = P + \tau$, $\tau = 0, 1, \dots, P-1$. \emph{Continued at the next page...}}
\label{fig:ProbabilitiesHL-I8}
\end{figure*}
\begin{figure*}[!h]
\addtocounter{figure}{-1}
\includegraphics[width=0.4\textwidth]{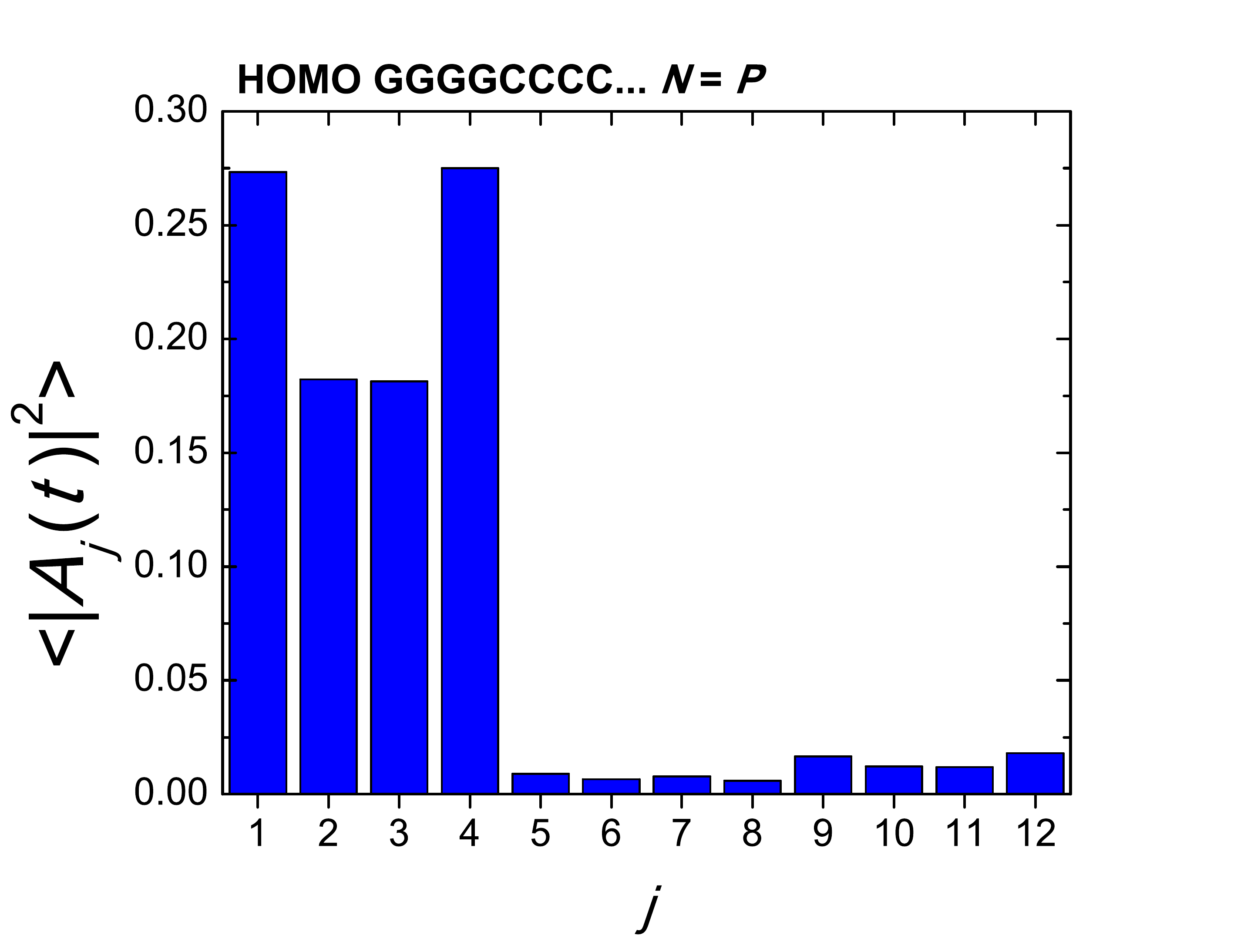}
\includegraphics[width=0.4\textwidth]{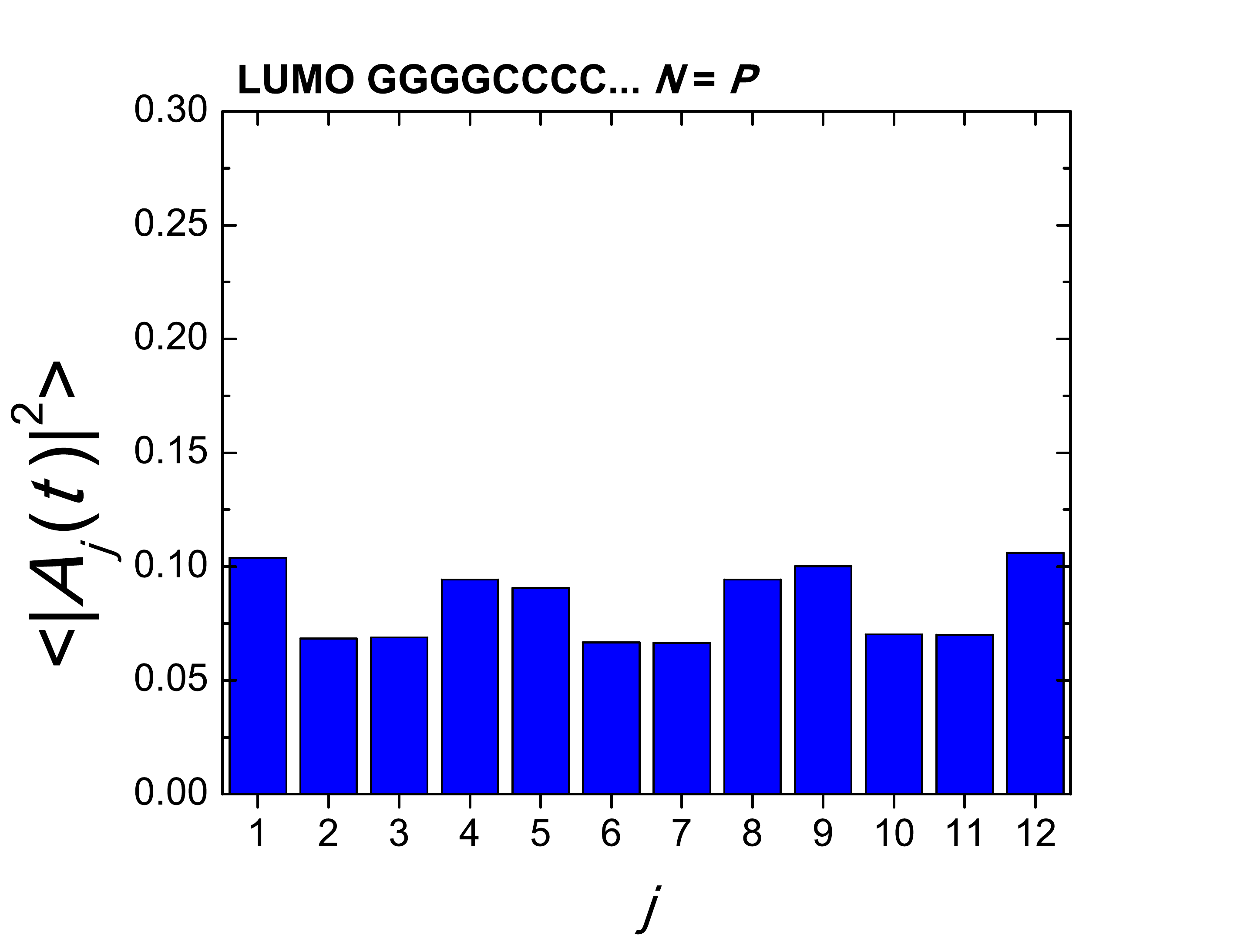}
\includegraphics[width=0.4\textwidth]{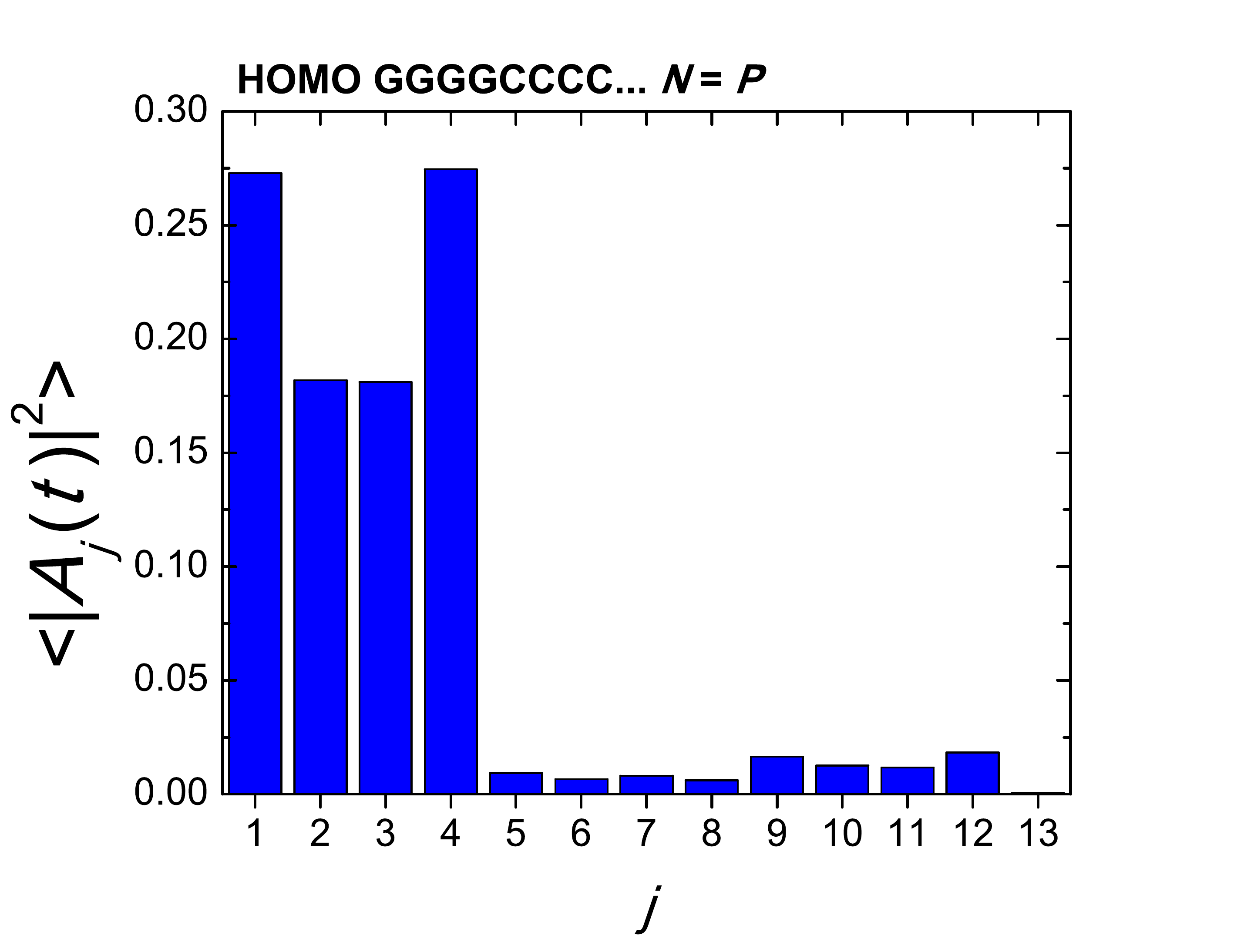}
\includegraphics[width=0.4\textwidth]{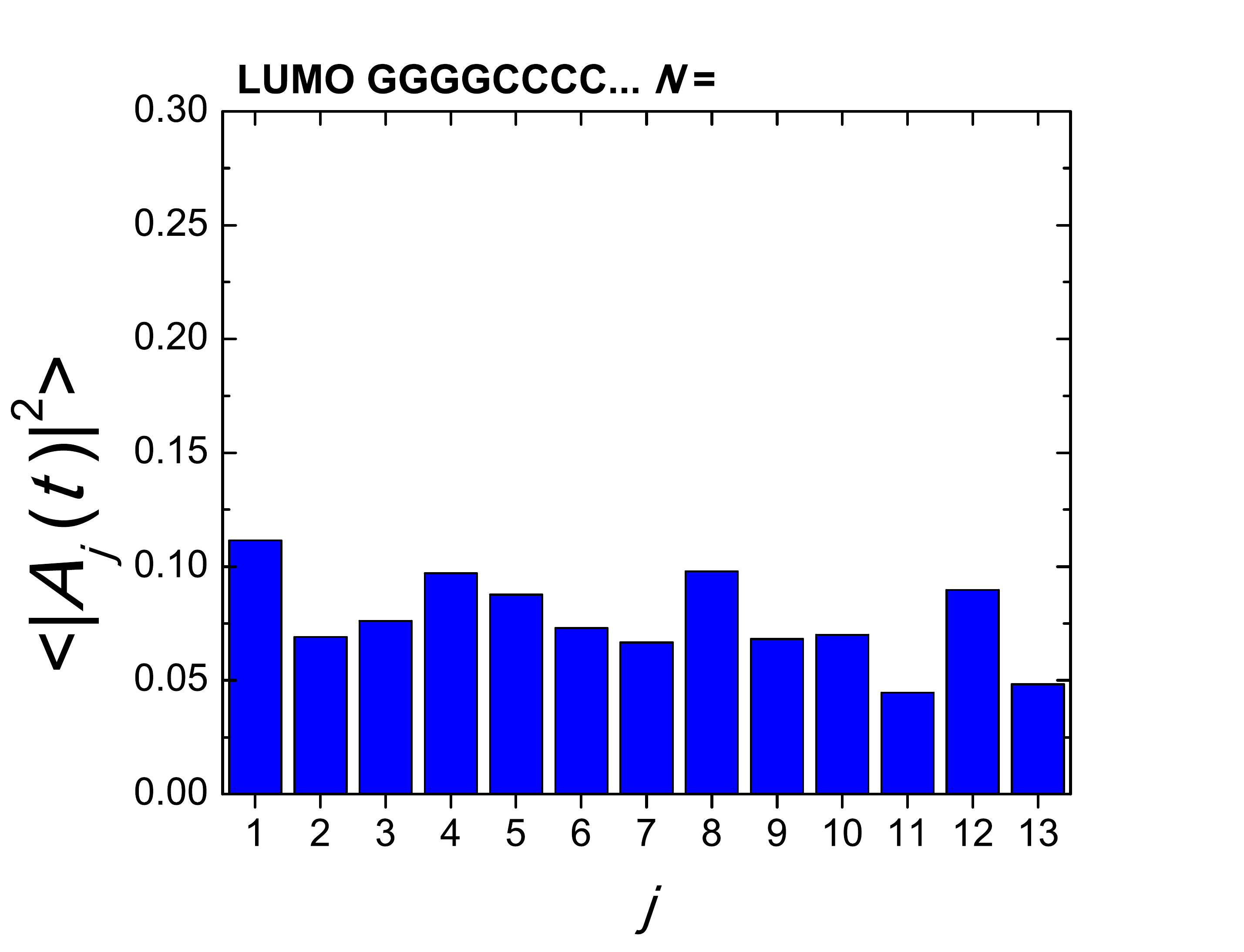}
\includegraphics[width=0.4\textwidth]{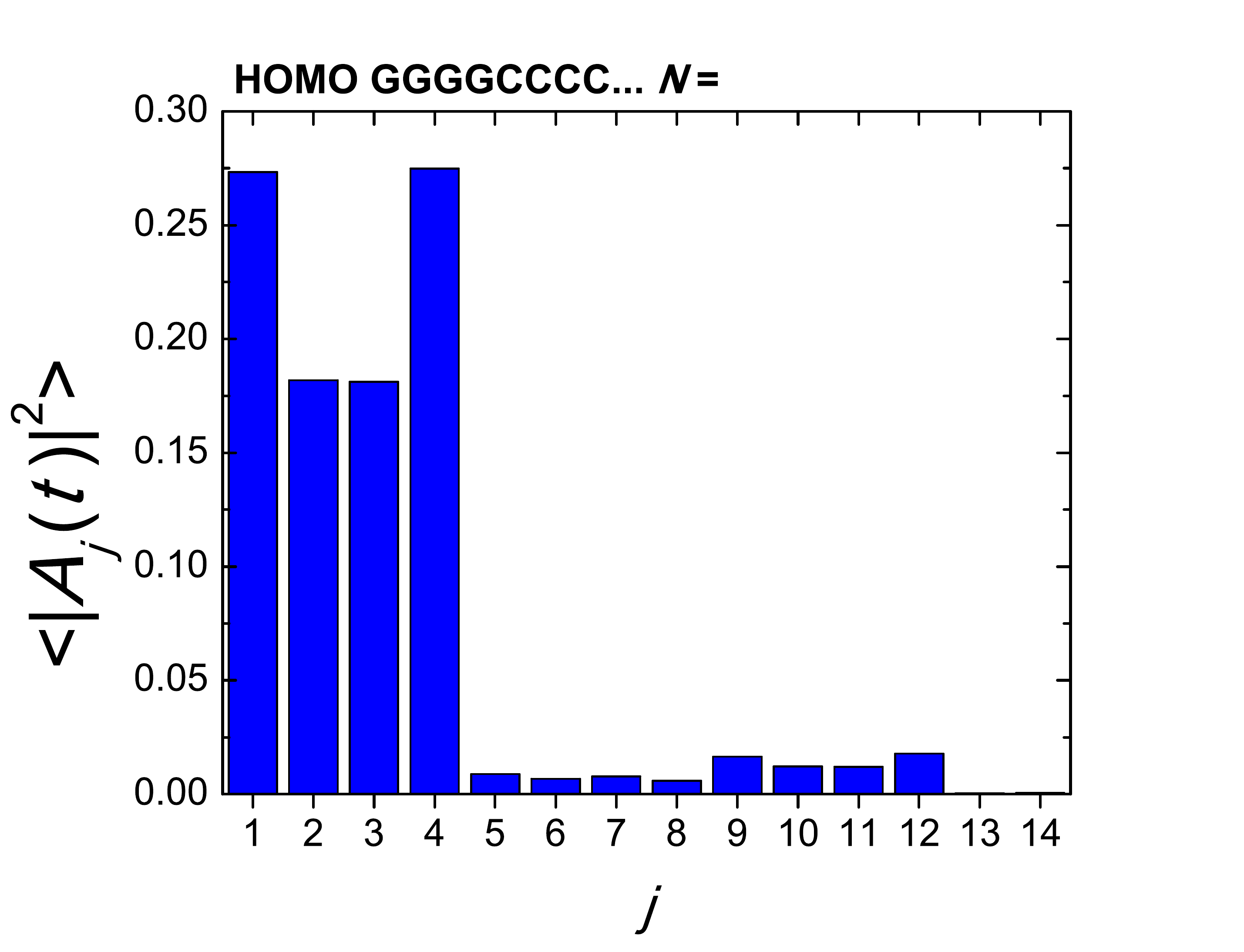}
\includegraphics[width=0.4\textwidth]{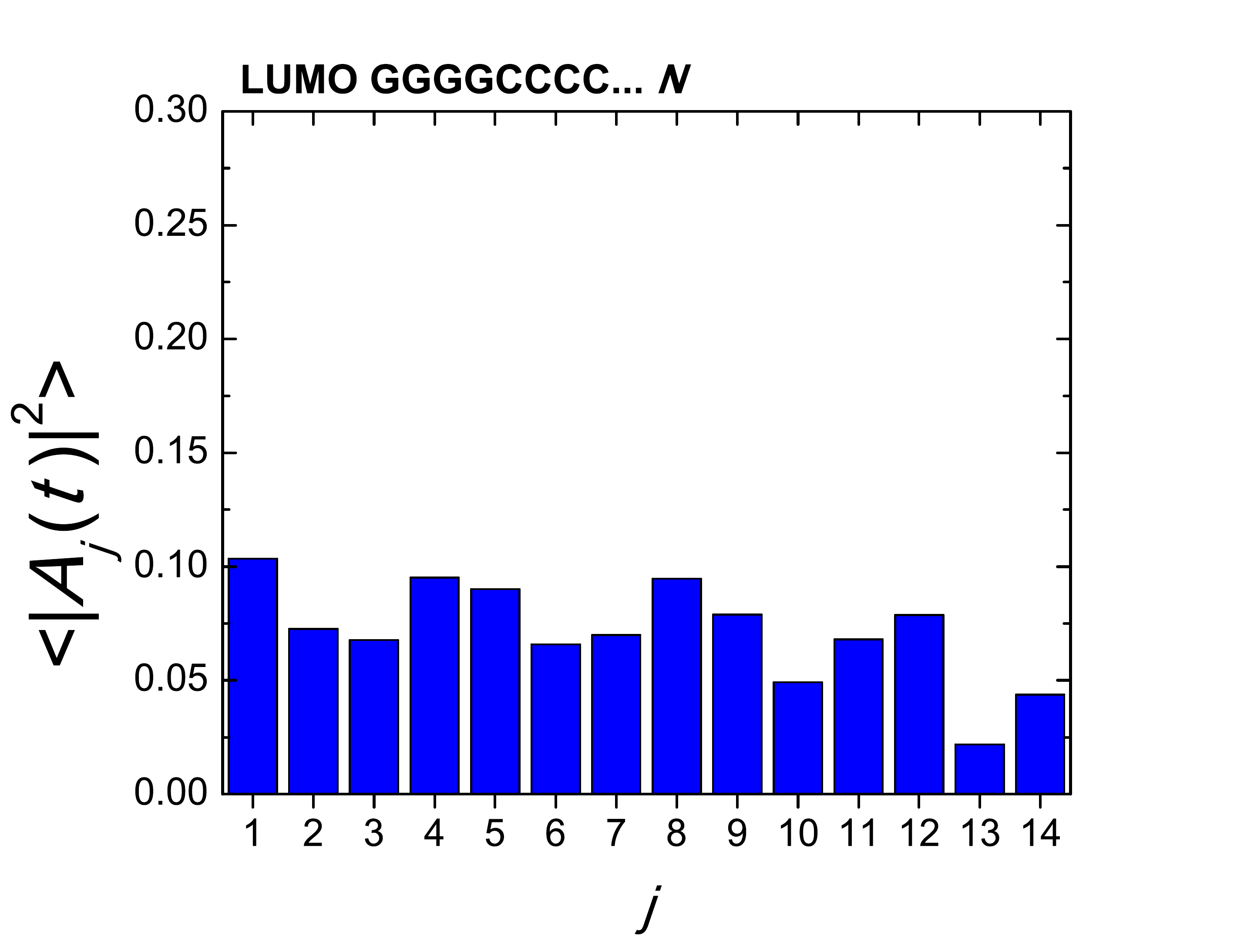}
\includegraphics[width=0.4\textwidth]{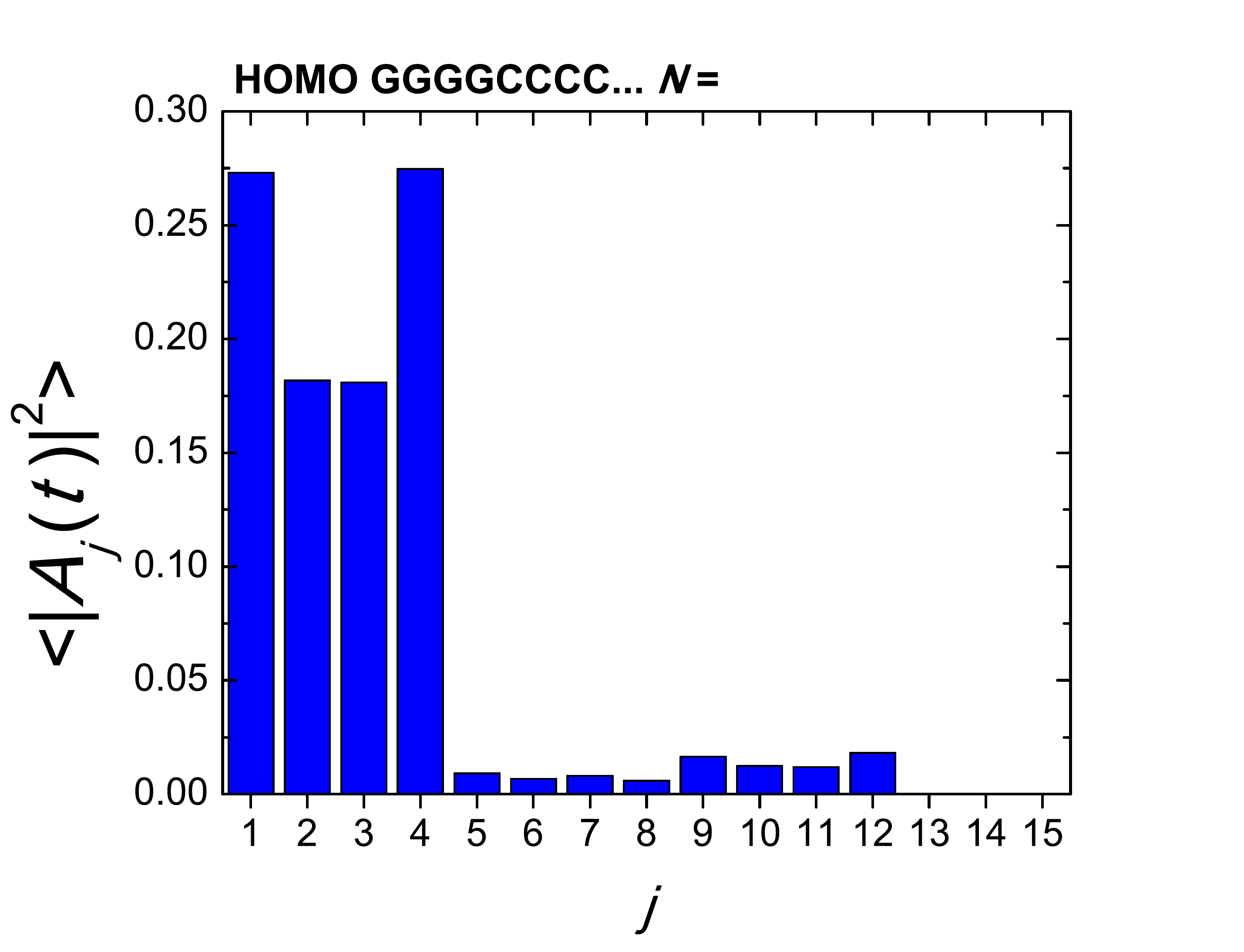}
\includegraphics[width=0.4\textwidth]{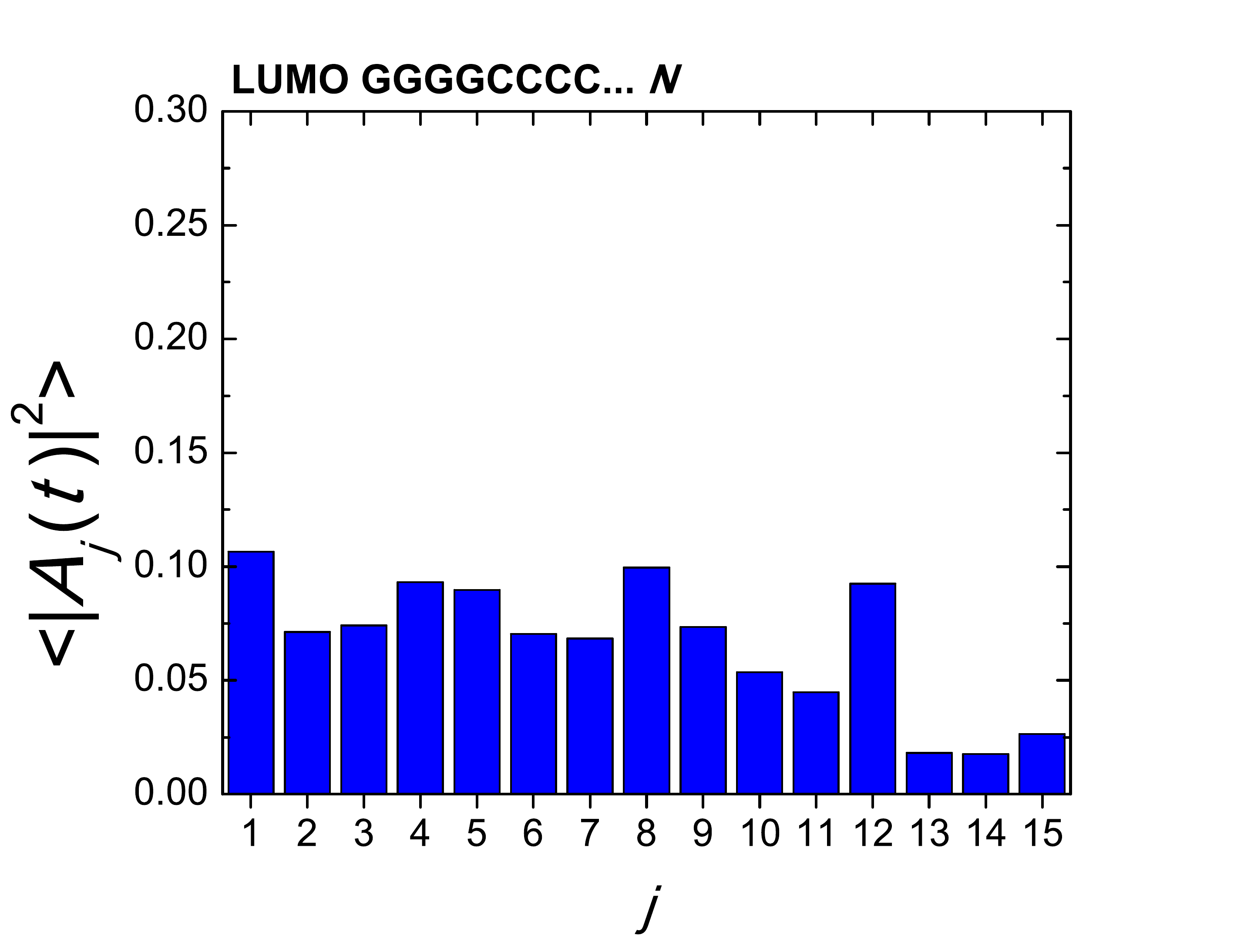}
\caption{\emph{...Continued from the previous page.} Mean (over time) probabilities to find the extra carrier at each monomer $j$, having placed it initially at the first monomer, for I8 (GGGGCCCC...) polymers, for the HOMO (left) and the LUMO (right). $N = P + \tau$, $\tau = 0, 1, \dots, P-1$.}
\end{figure*}

\begin{figure*}[!h]
\includegraphics[width=0.4\textwidth]{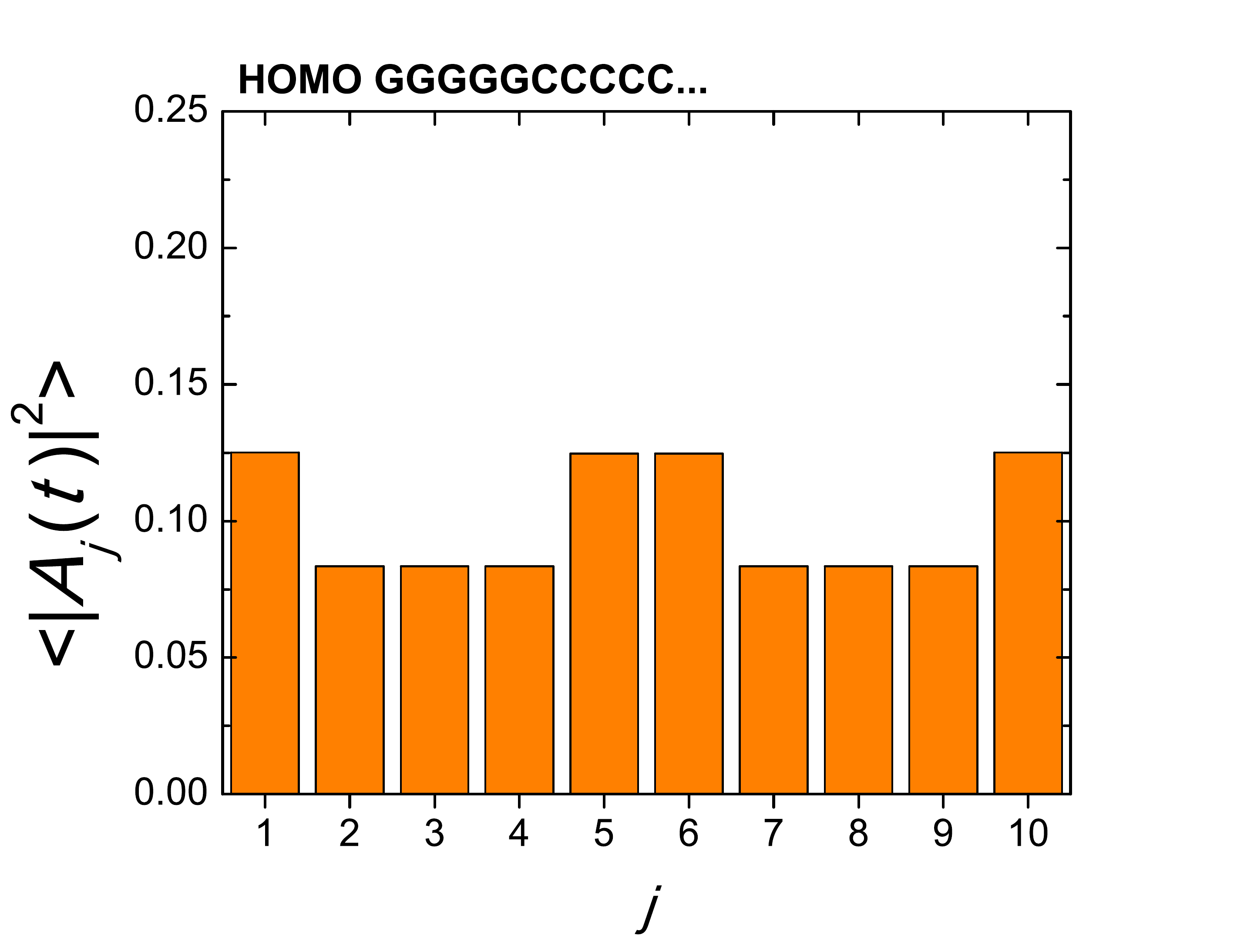}
\includegraphics[width=0.4\textwidth]{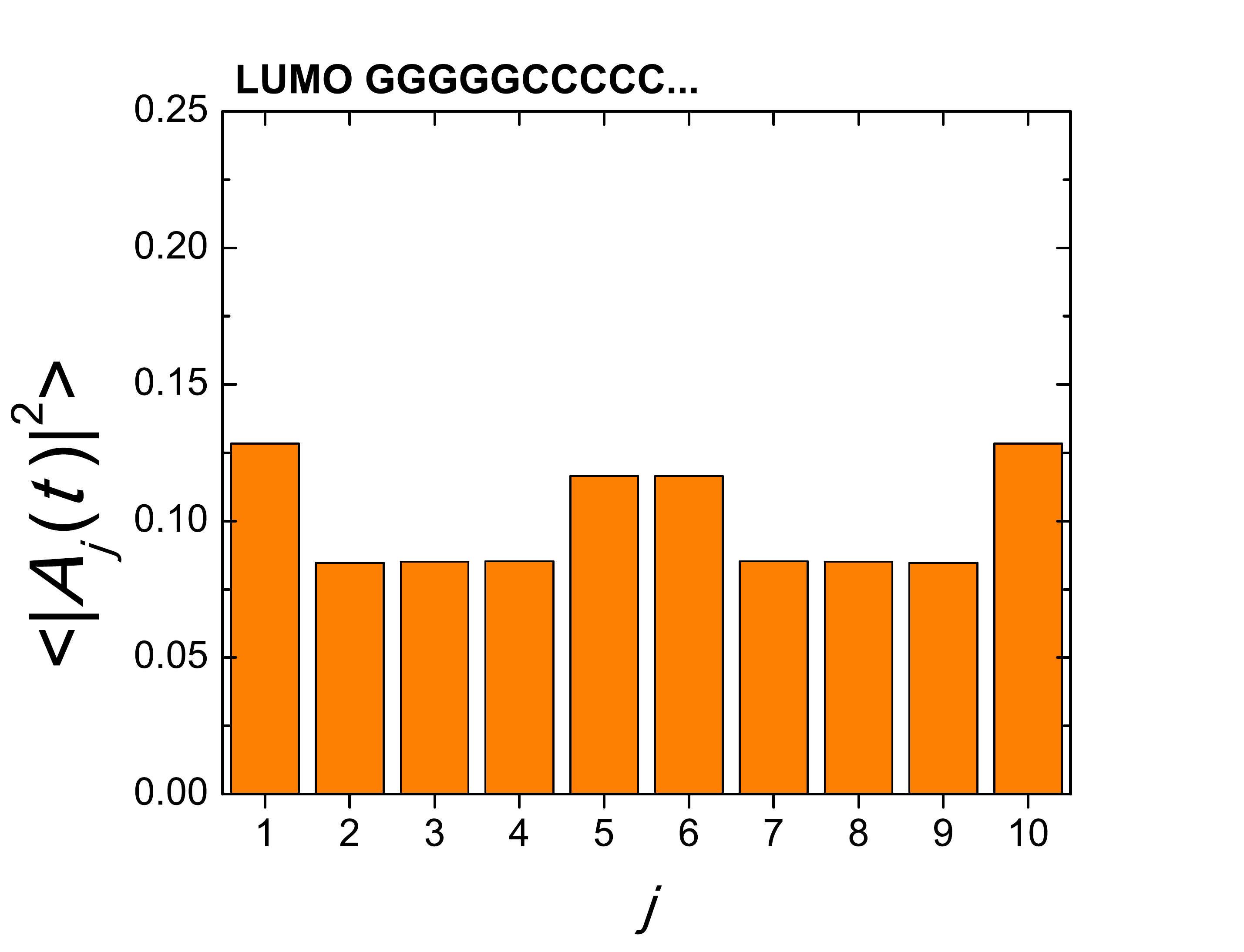}
\includegraphics[width=0.4\textwidth]{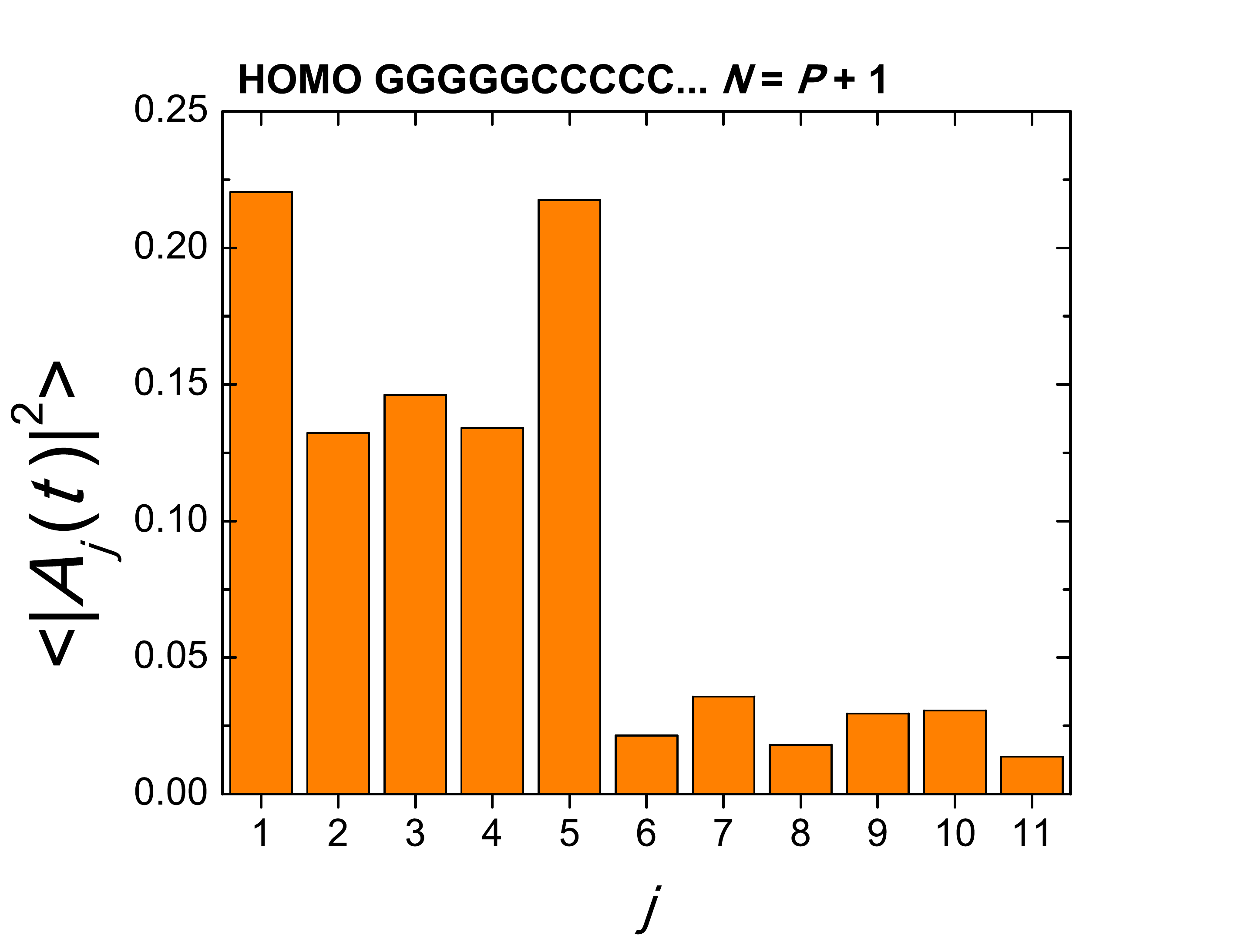}
\includegraphics[width=0.4\textwidth]{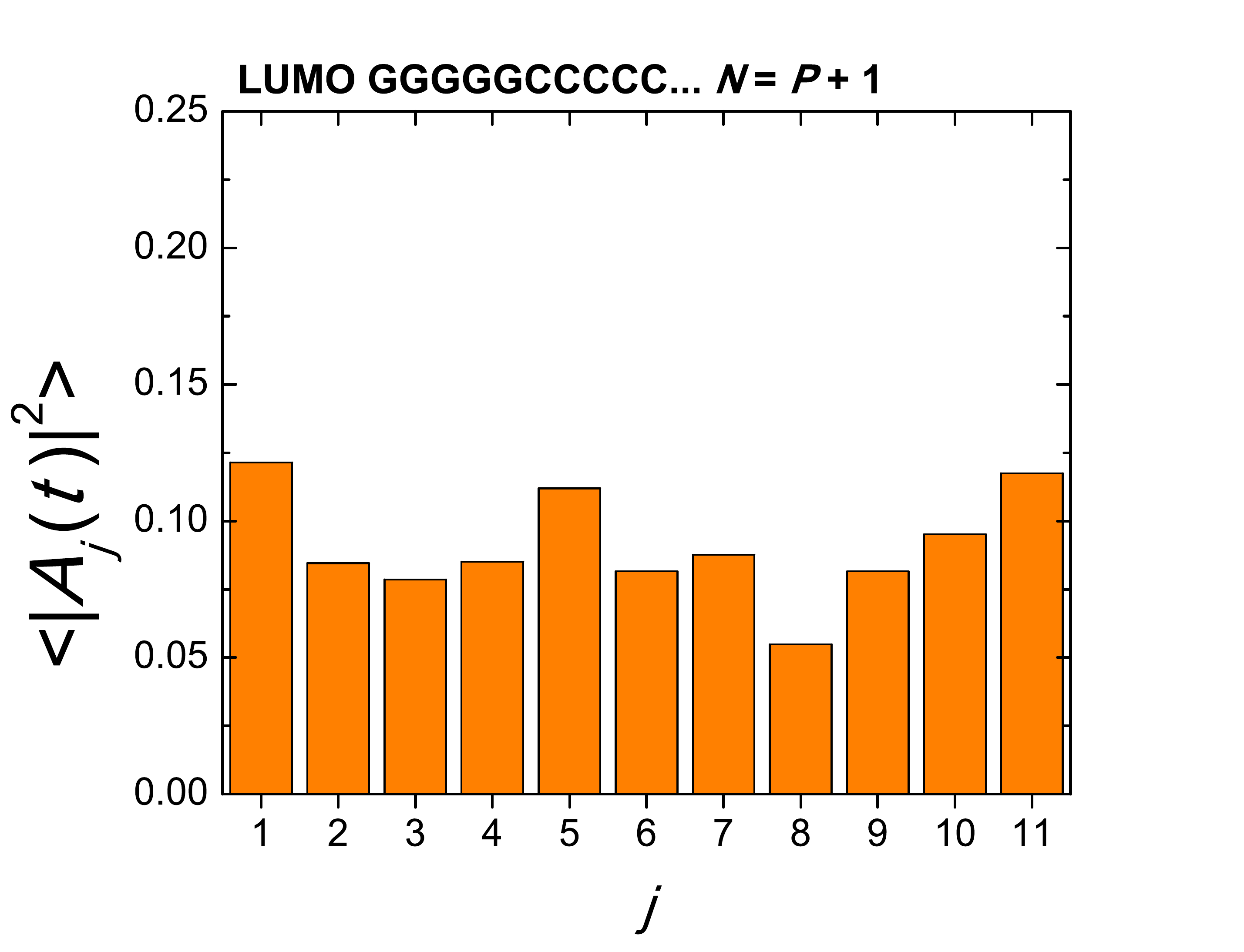}
\includegraphics[width=0.4\textwidth]{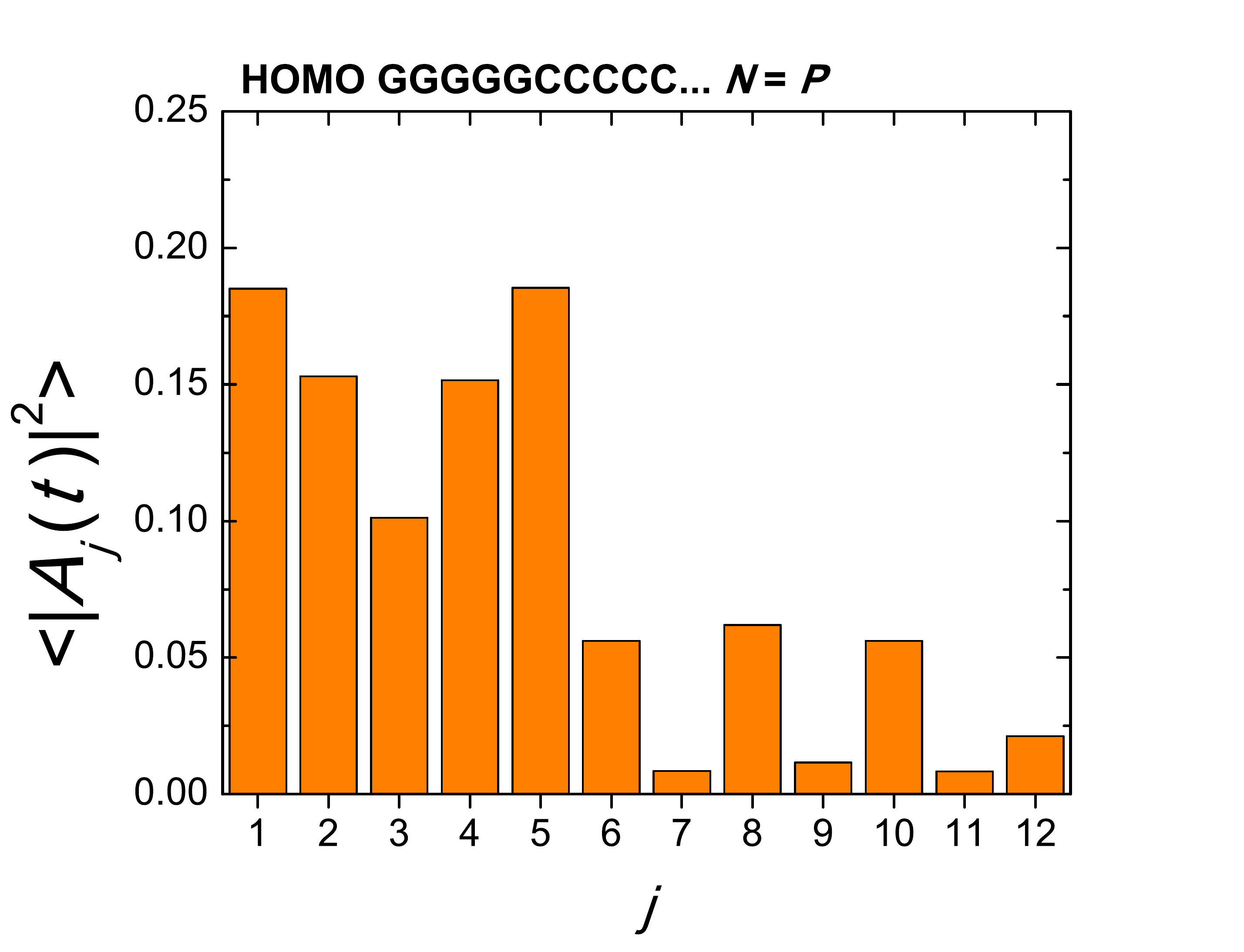}
\includegraphics[width=0.4\textwidth]{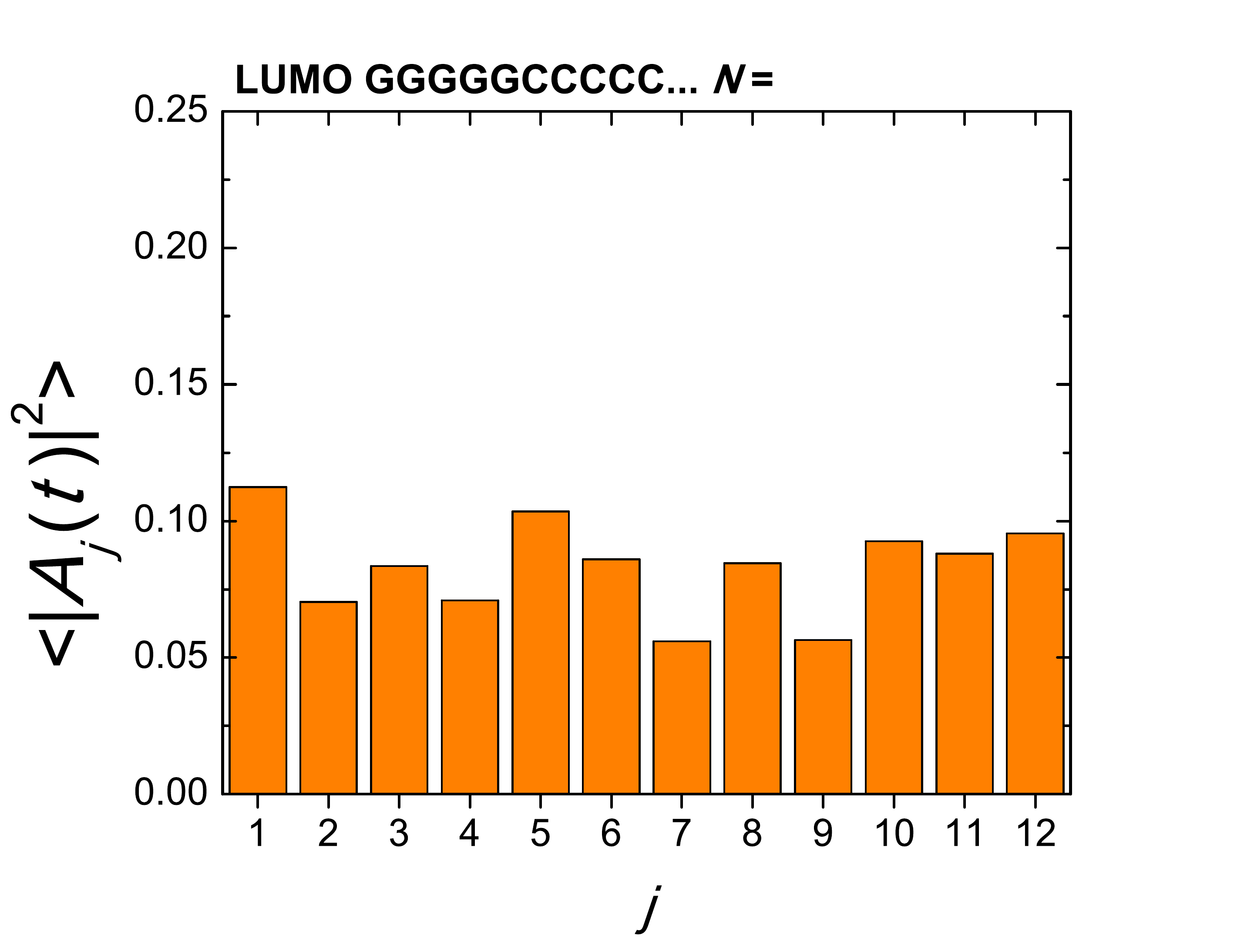}
\includegraphics[width=0.4\textwidth]{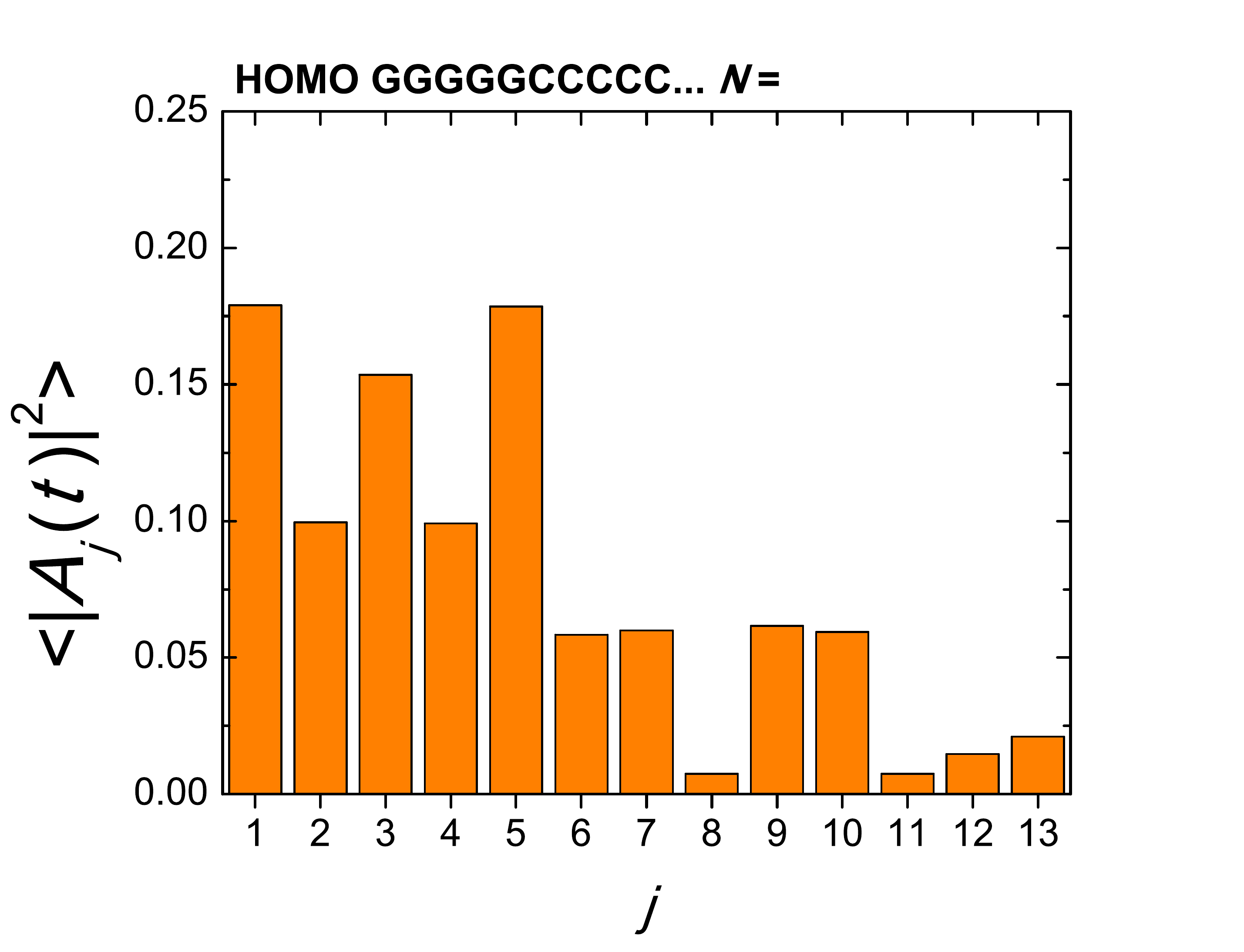}
\includegraphics[width=0.4\textwidth]{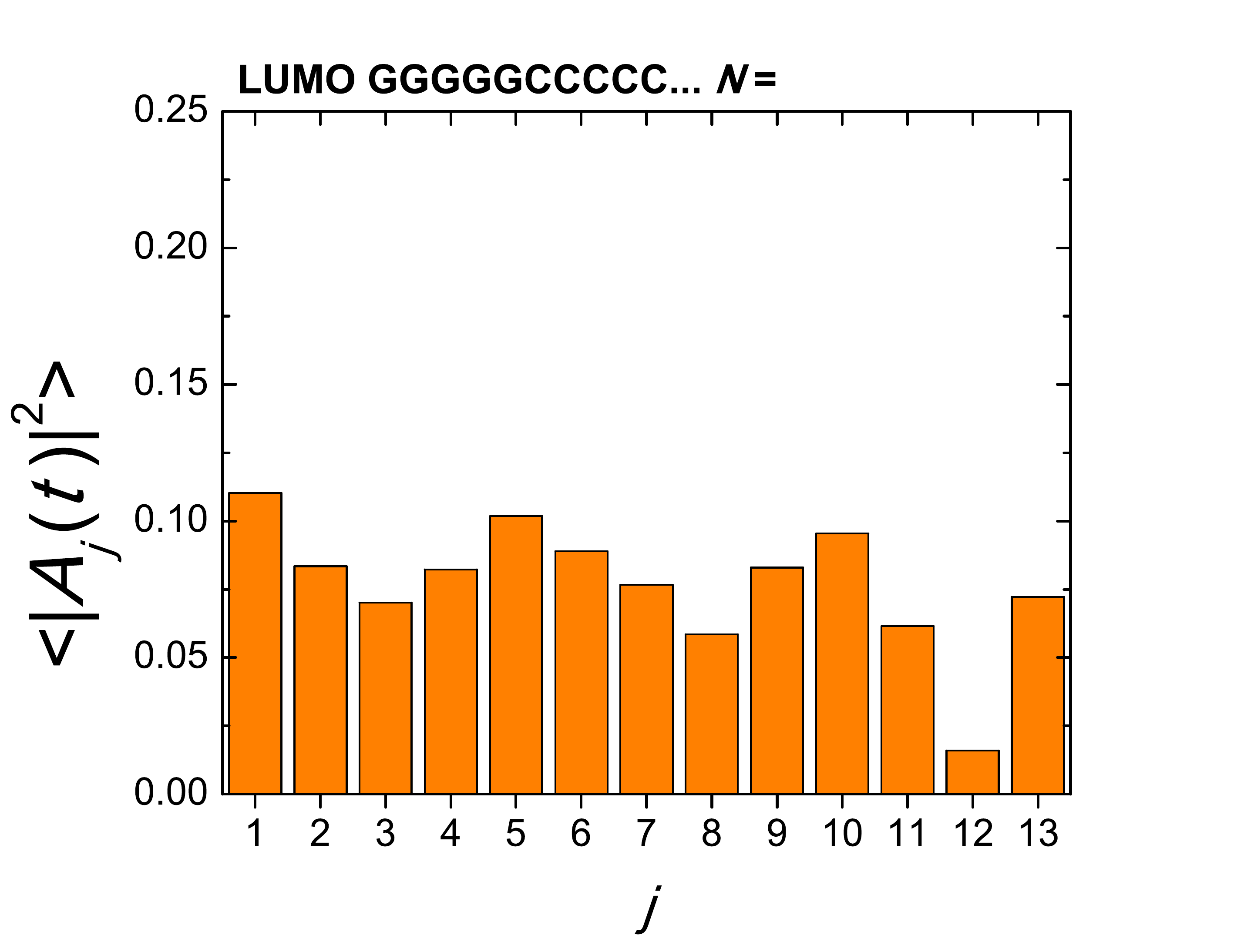}
\caption{Mean (over time) probabilities to find the extra carrier at each monomer $j$, having placed it initially at the first monomer, for I10 (GGGGGCCCCC...) polymers, for the HOMO (left) and the LUMO (right). $N = P + \tau$, $\tau = 0, 1, \dots, P-1$. \emph{Continued at the next page...}}
\label{fig:ProbabilitiesHL-I10}
\end{figure*}
\begin{figure*}[!h]
\addtocounter{figure}{-1}
\includegraphics[width=0.4\textwidth]{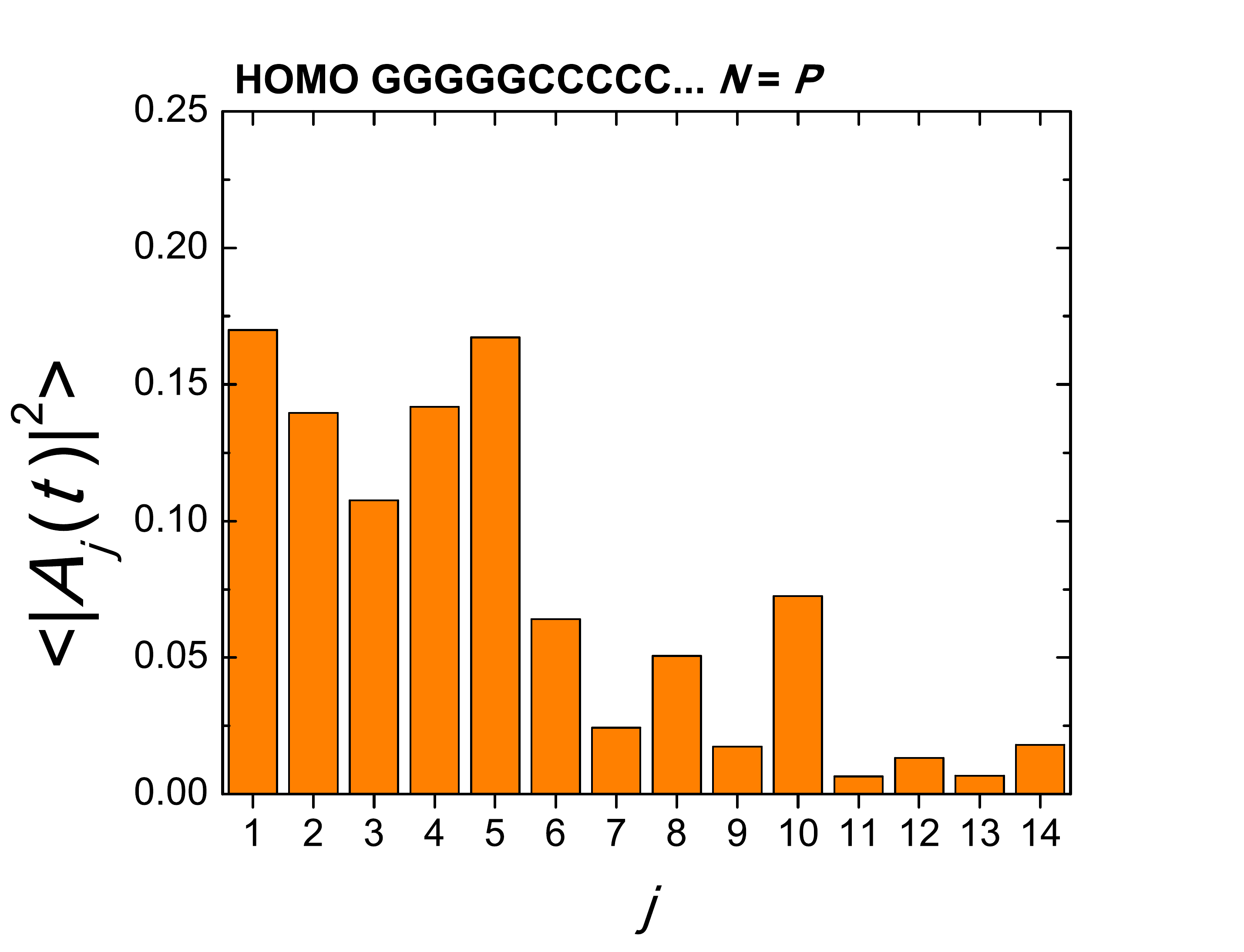}
\includegraphics[width=0.4\textwidth]{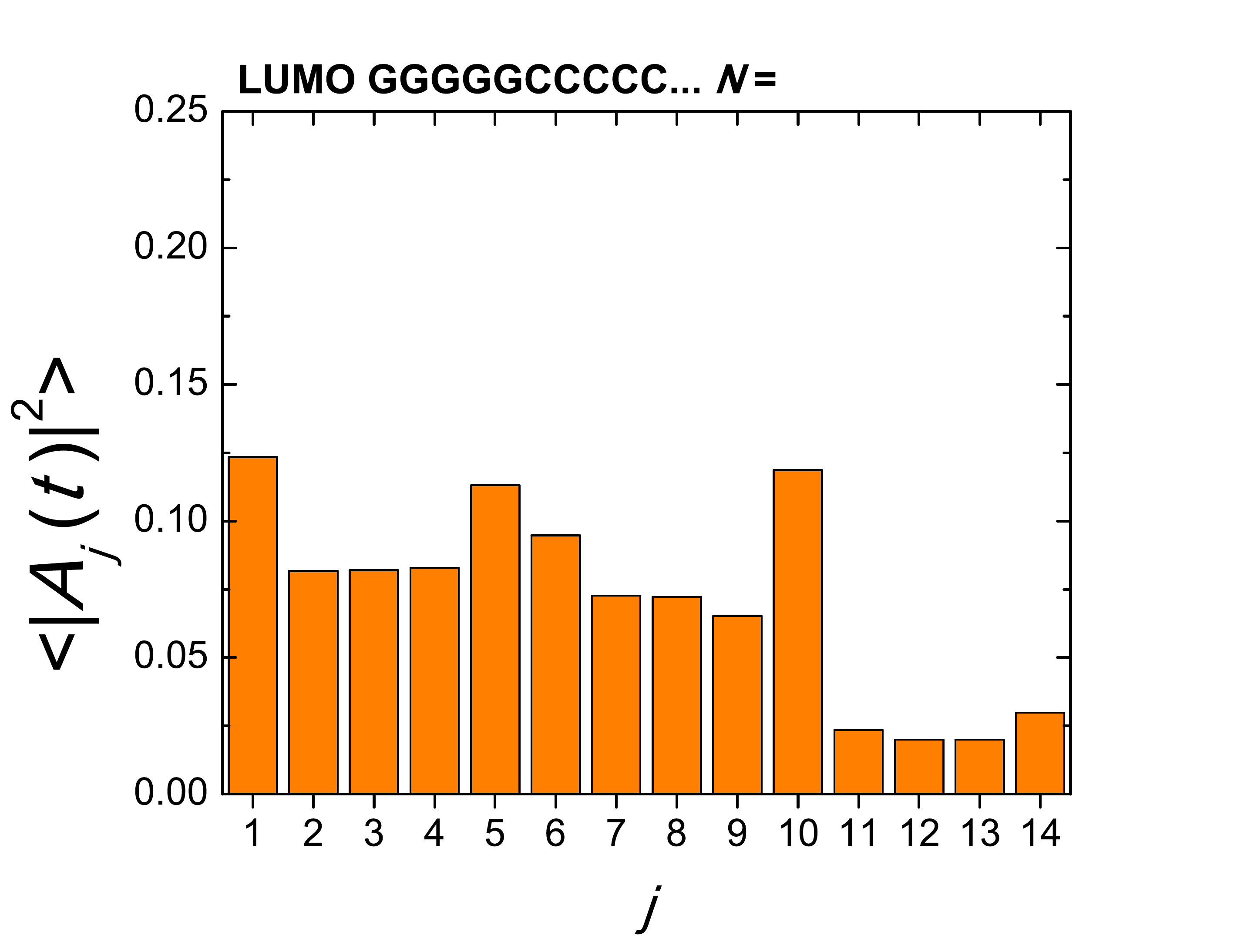}
\includegraphics[width=0.4\textwidth]{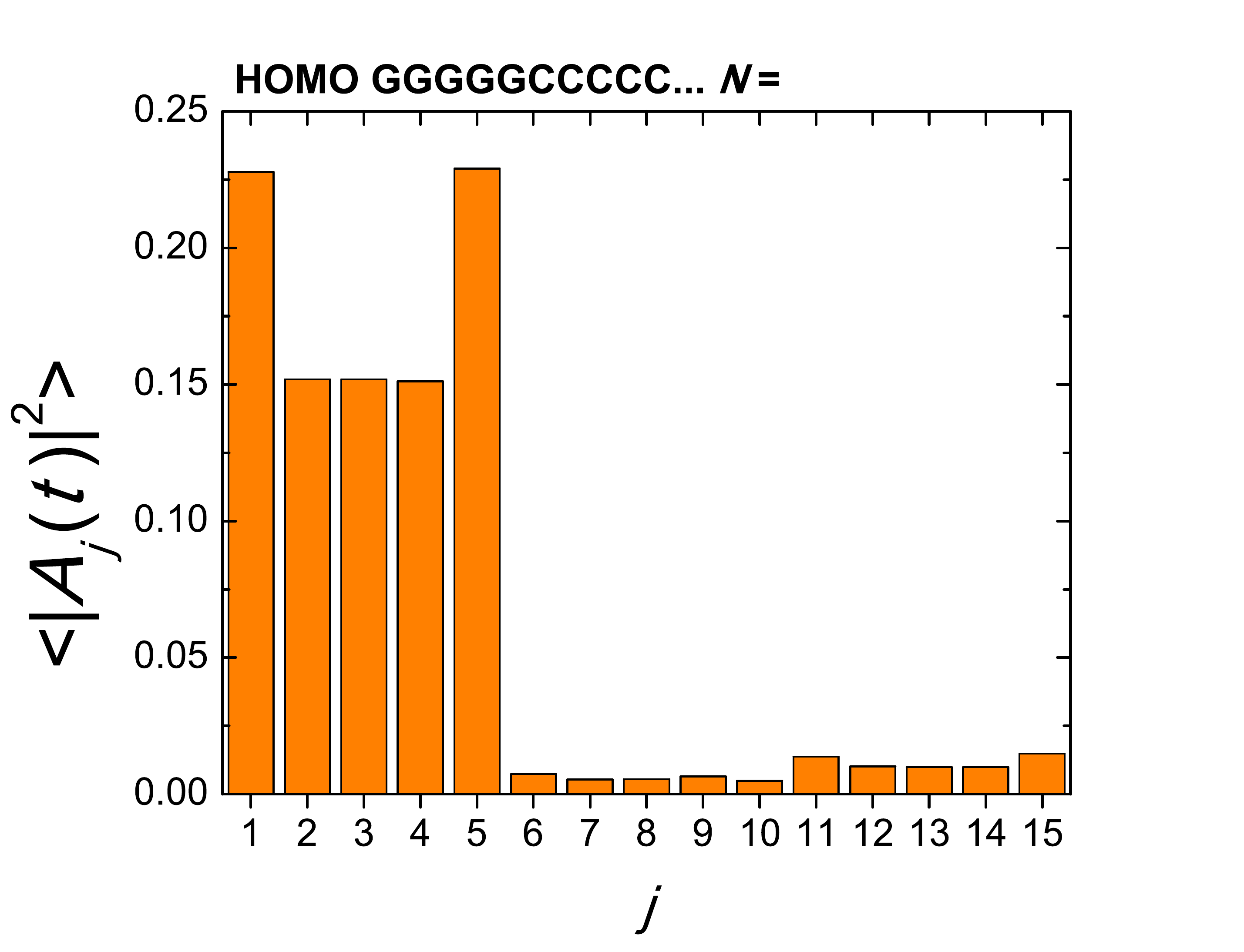}
\includegraphics[width=0.4\textwidth]{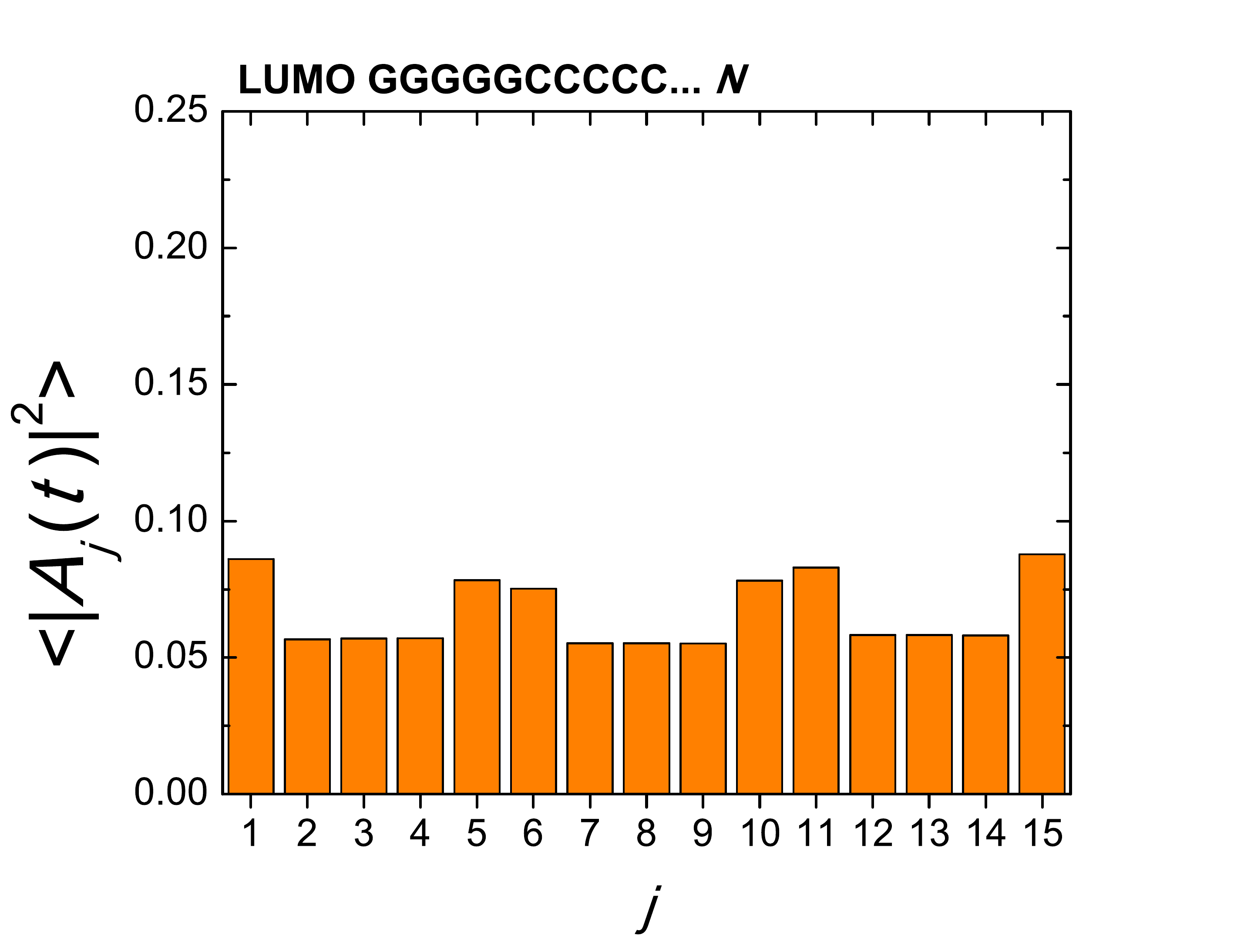}
\includegraphics[width=0.4\textwidth]{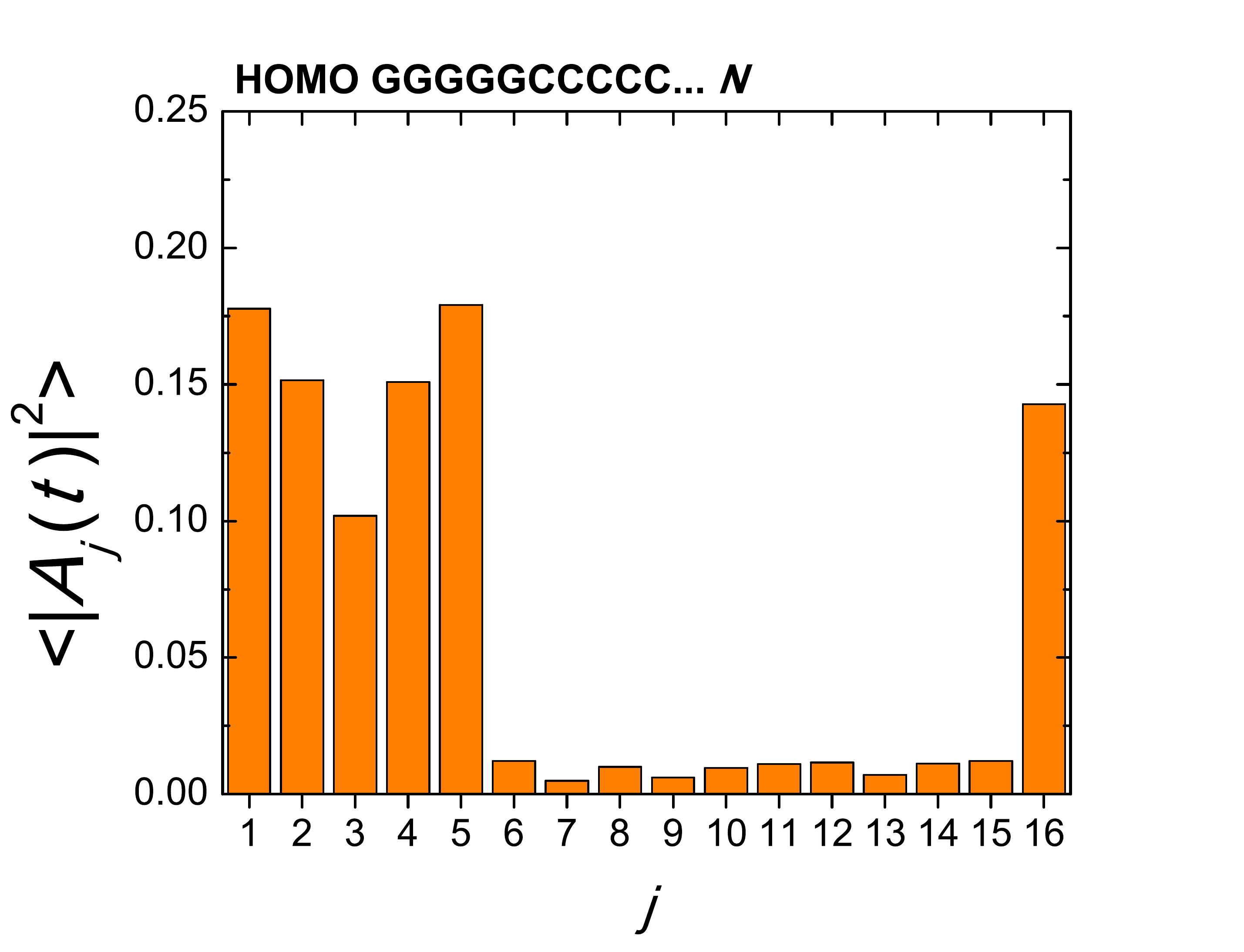}
\includegraphics[width=0.4\textwidth]{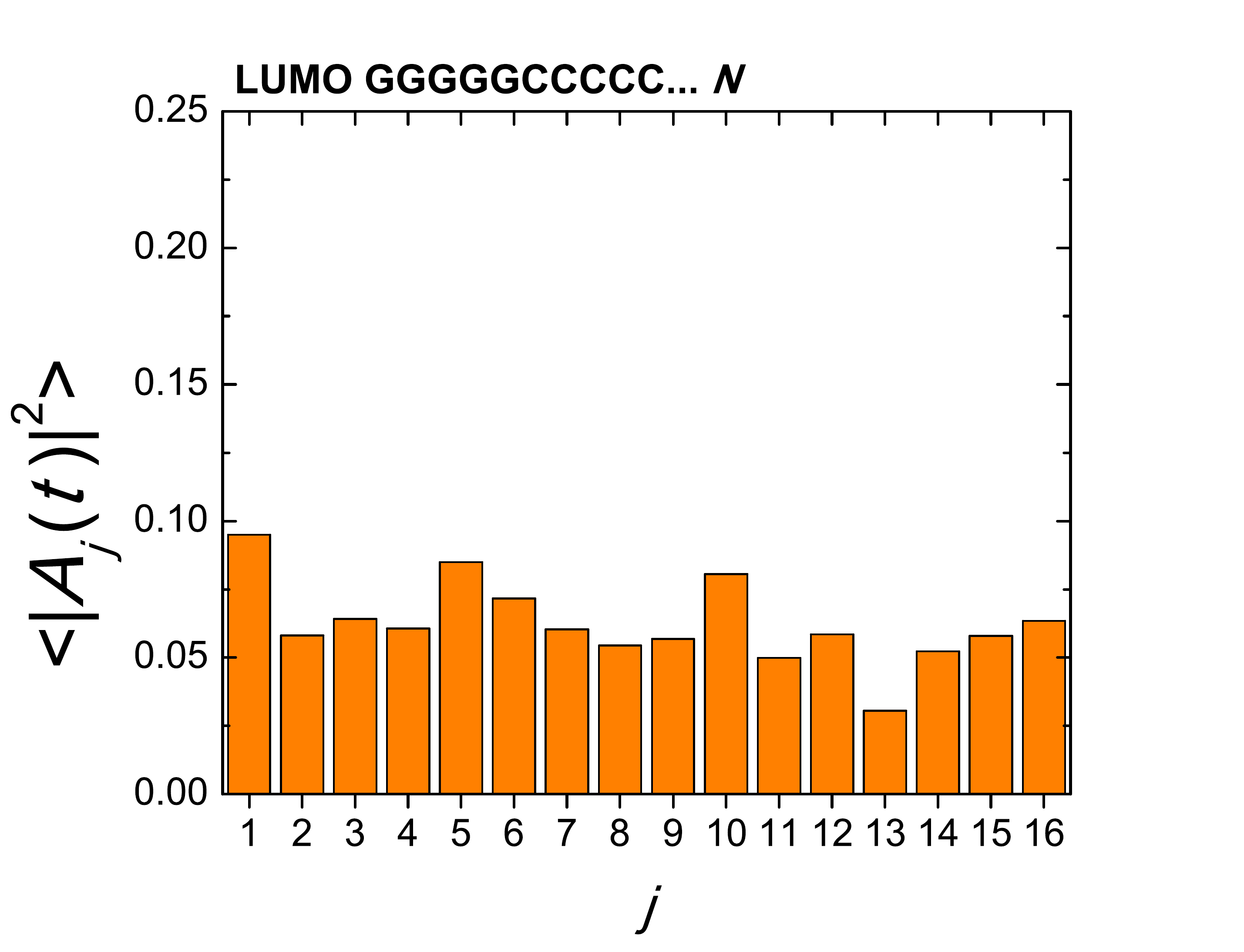}
\includegraphics[width=0.4\textwidth]{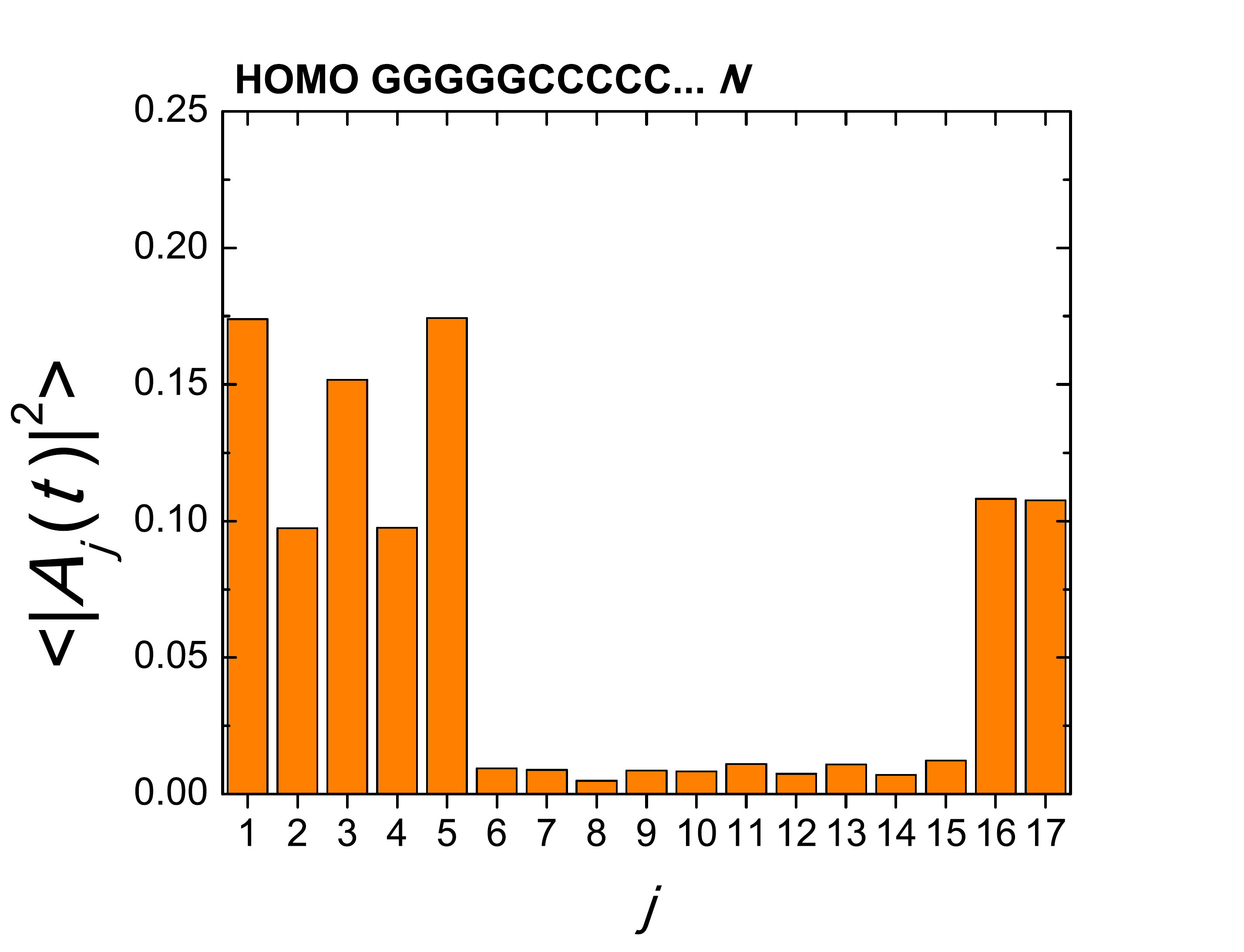}
\includegraphics[width=0.4\textwidth]{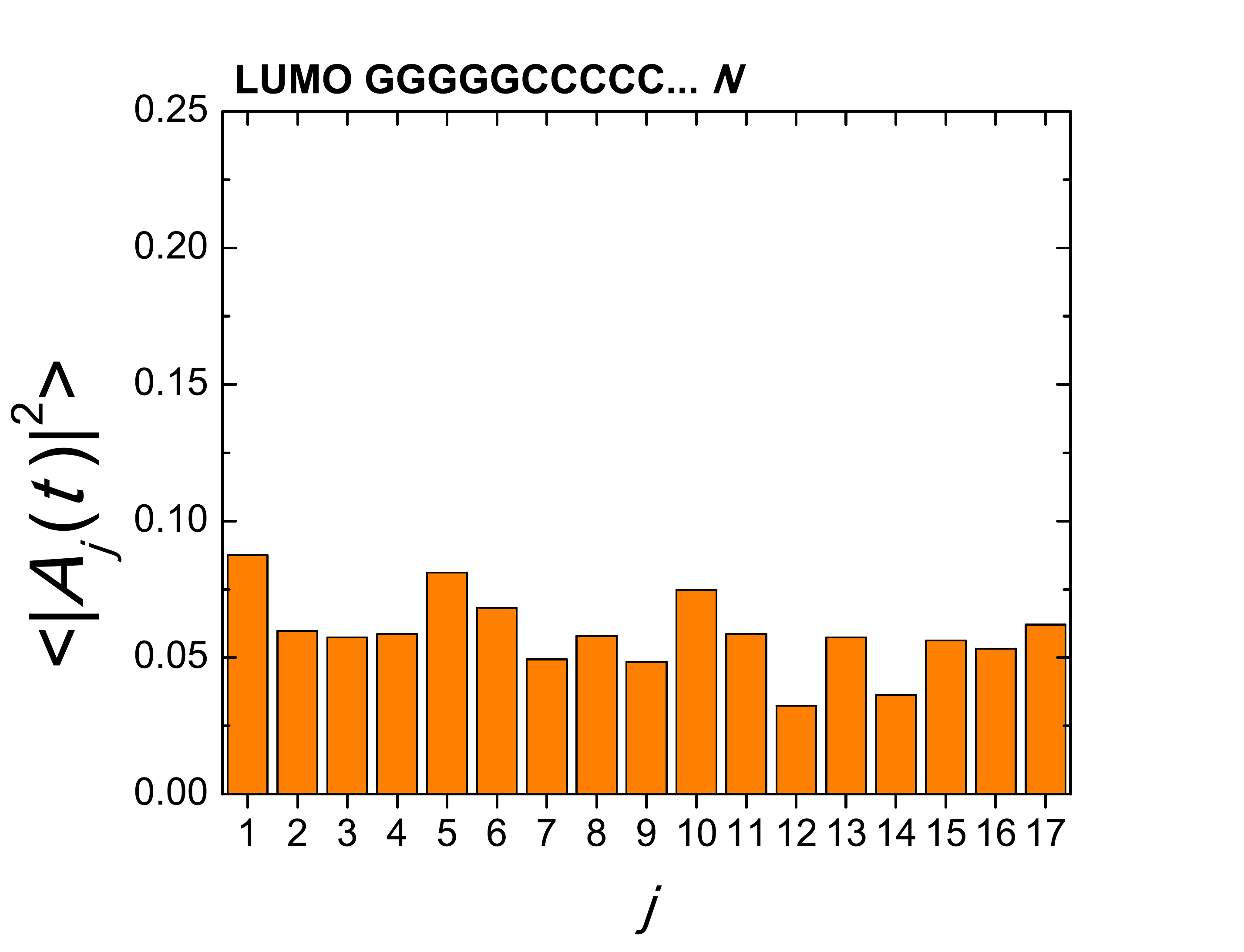}
\caption{\emph{...Continued from the previous page.} Mean (over time) probabilities to find the extra carrier at each monomer $j$, having placed it initially at the first monomer, for I10  (GGGGGCCCCC...) polymers, for the HOMO (left) and the LUMO (right). $N = P + \tau$, $\tau = 0, 1, \dots, P-1$. \emph{Continued at the next page...}}
\end{figure*}
\begin{figure*}[!h]
\addtocounter{figure}{-1}
\includegraphics[width=0.4\textwidth]{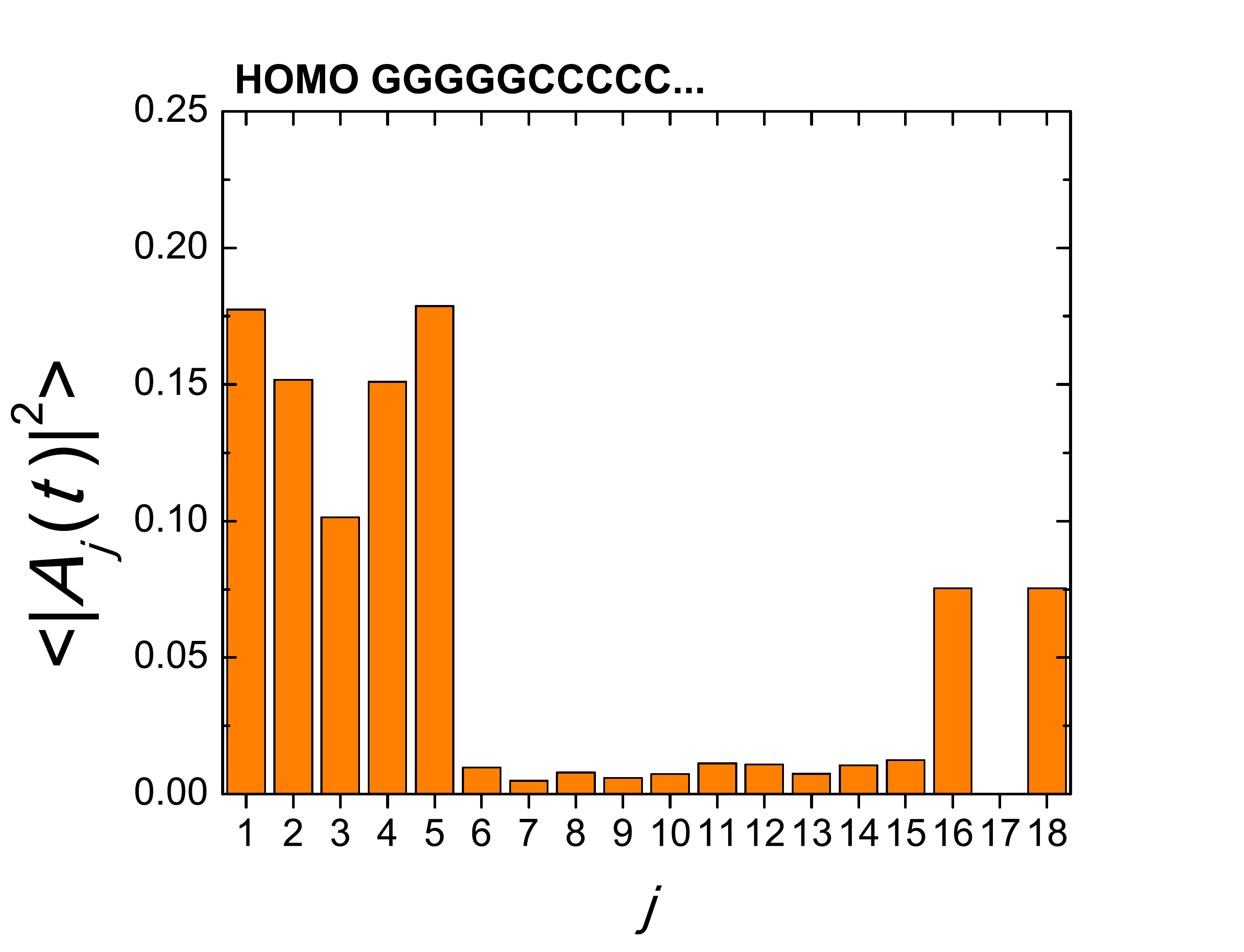}
\includegraphics[width=0.4\textwidth]{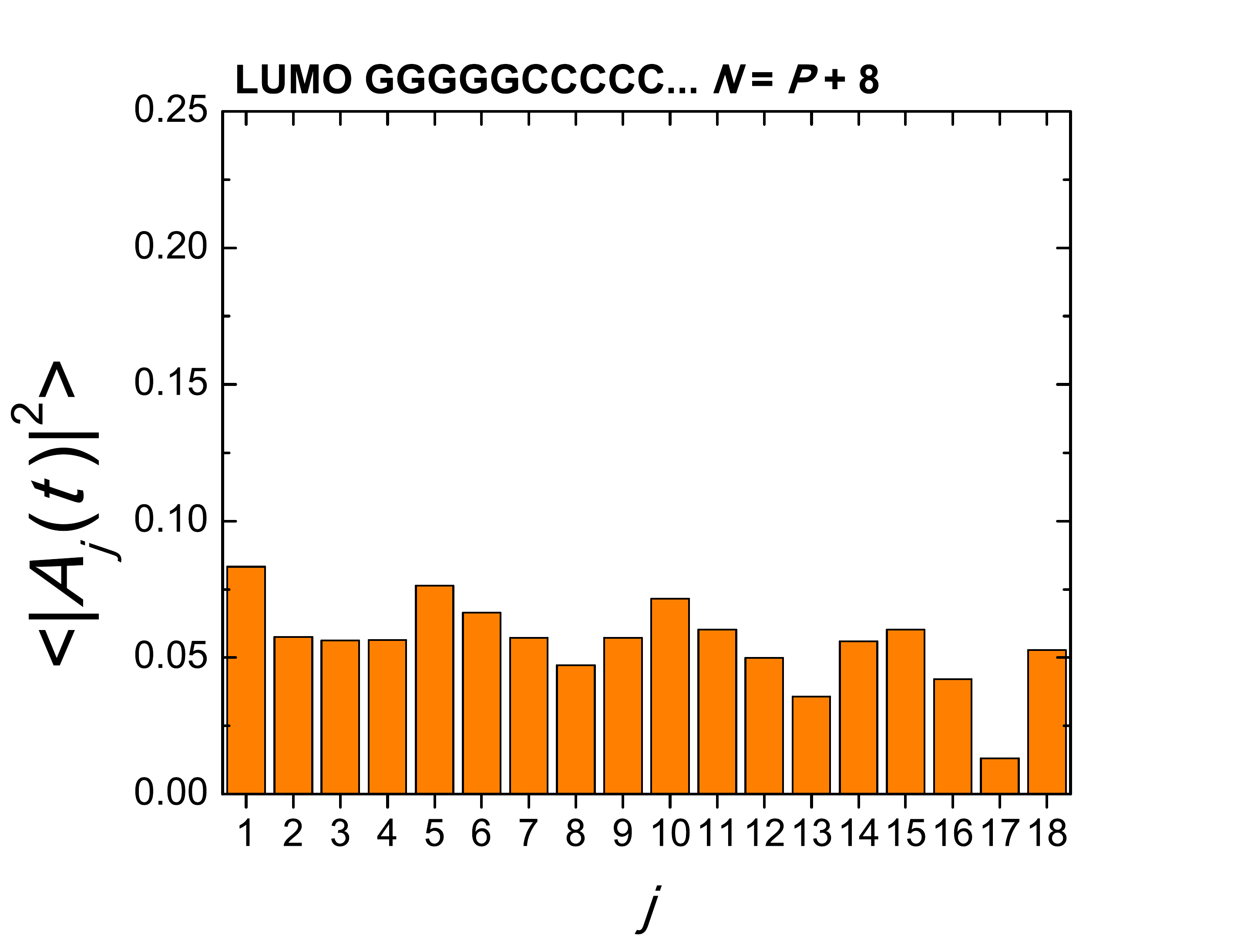}
\includegraphics[width=0.4\textwidth]{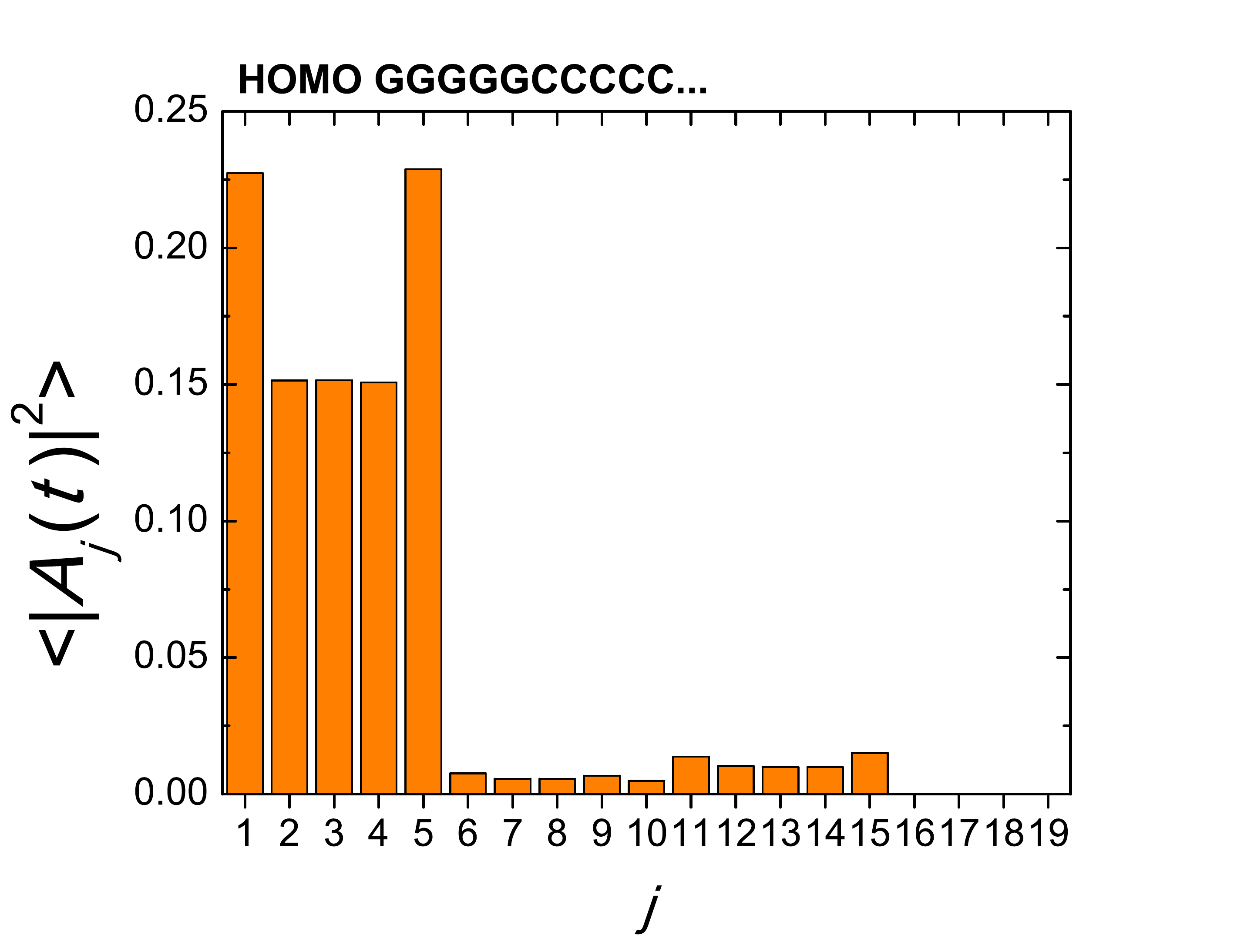}
\includegraphics[width=0.4\textwidth]{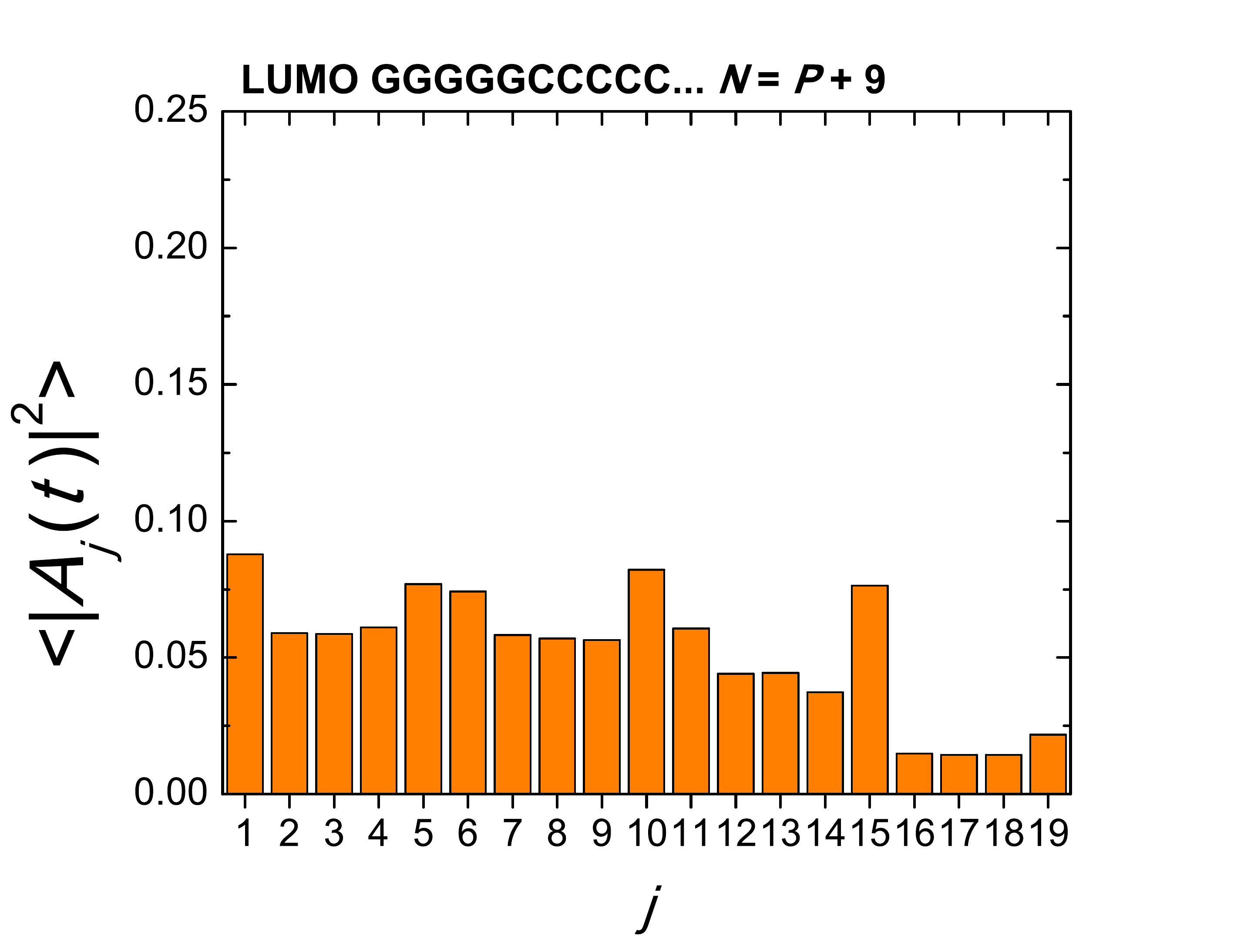}
\caption{\emph{...Continued from the previous page.} Mean (over time) probabilities to find the extra carrier at each monomer $j$, having placed it initially at the first monomer, for I10 (GGGGGCCCCC...) polymers, for the HOMO (left) and the LUMO (right). $N = P + \tau$, $\tau = 0, 1, \dots, P-1$.}
\end{figure*}

\begin{figure*}[!h]
\includegraphics[width=0.4\textwidth]{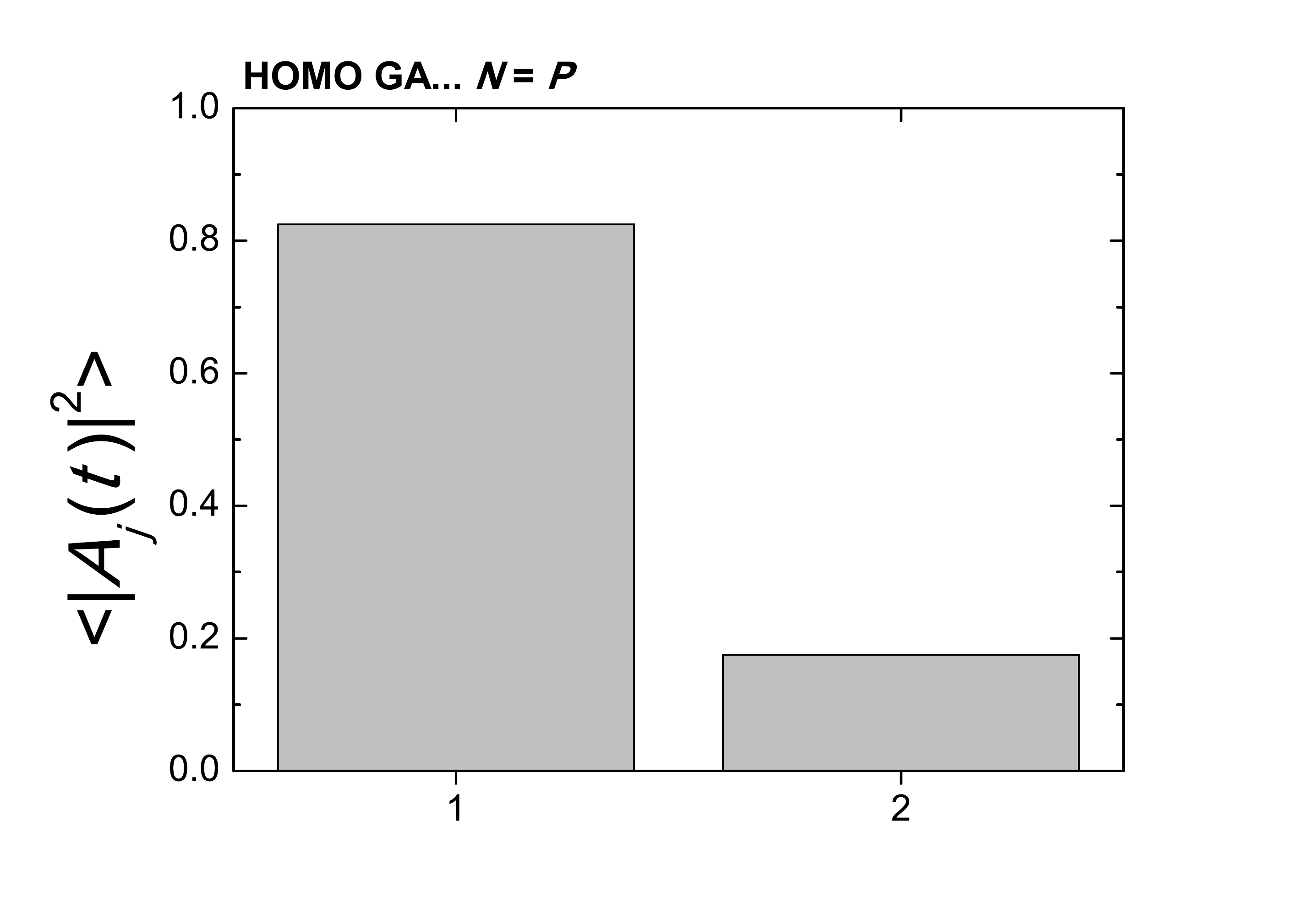}
\includegraphics[width=0.4\textwidth]{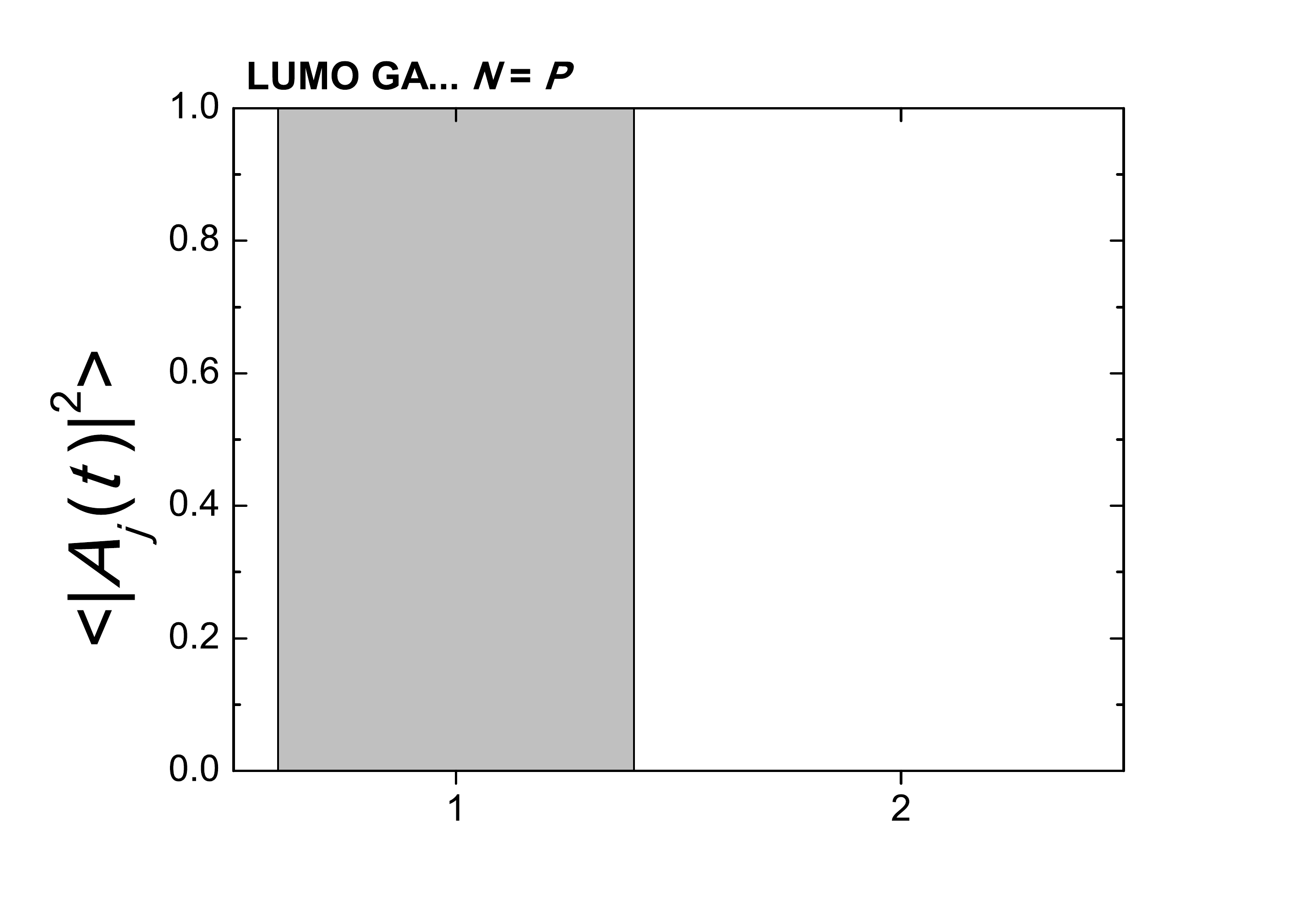} 
\includegraphics[width=0.4\textwidth]{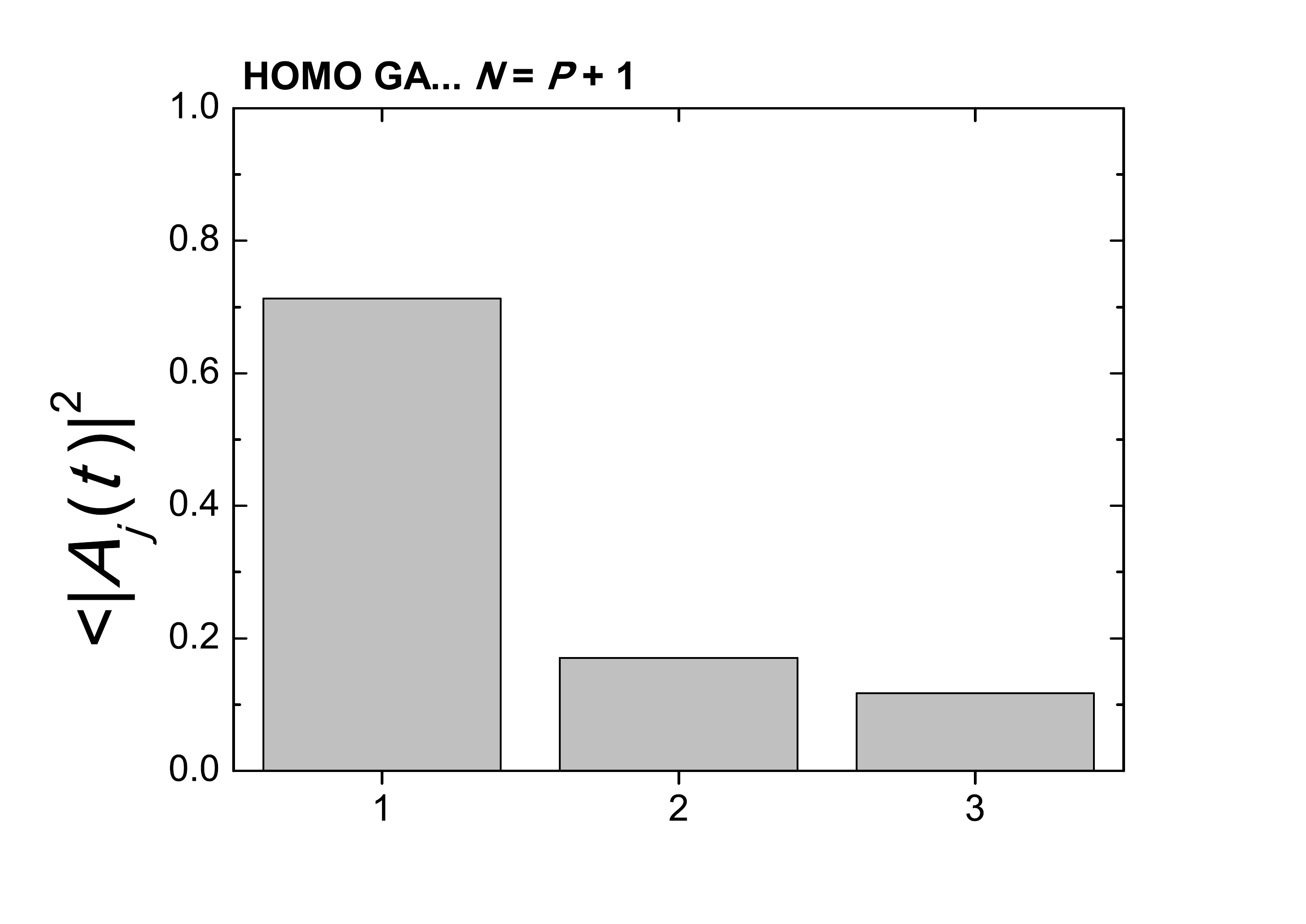}
\includegraphics[width=0.4\textwidth]{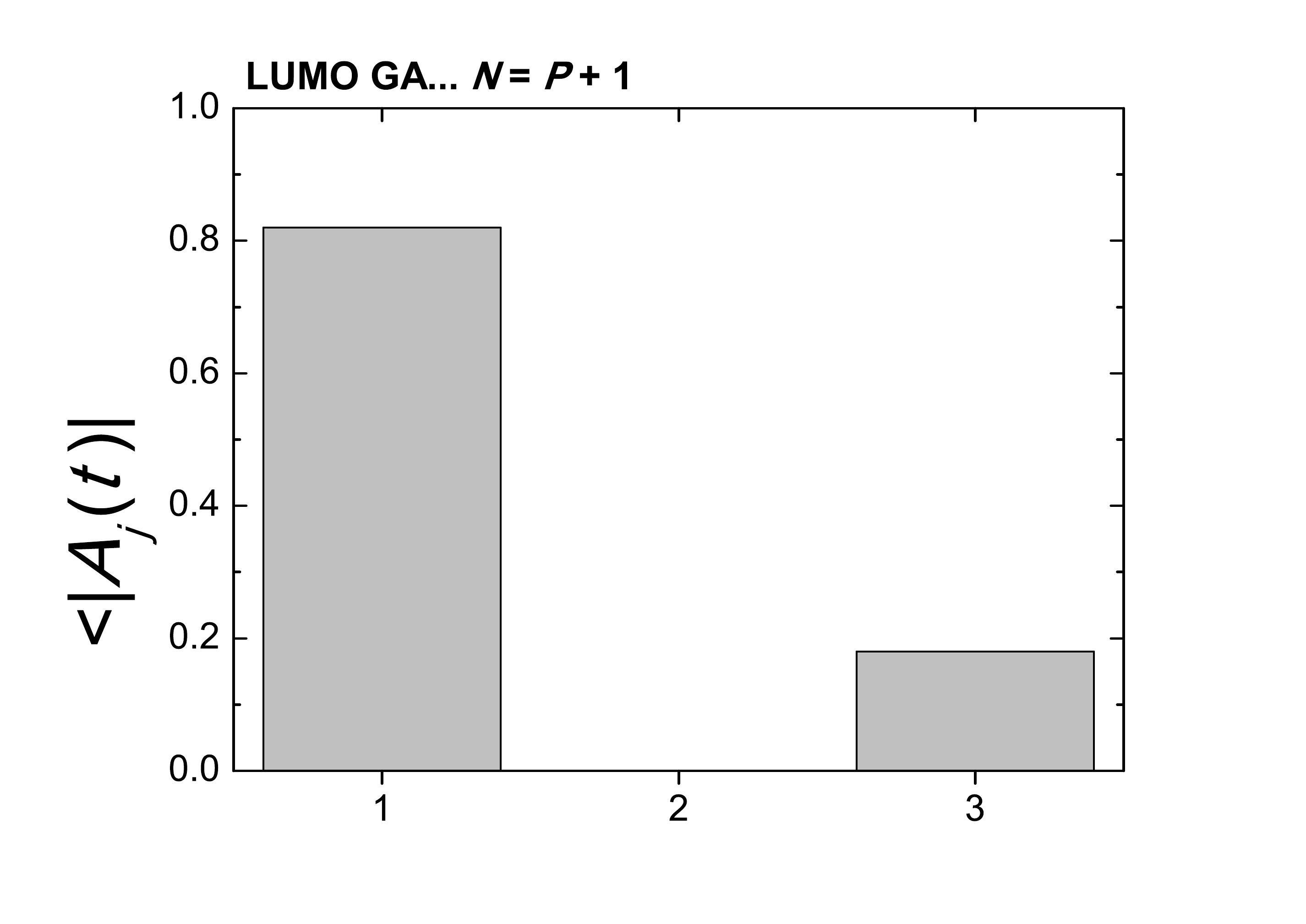}
\caption{Mean (over time) probabilities to find the extra carrier at each monomer $j$, having placed it initially at the first monomer, for D2 (GA...) polymers, for the HOMO (left) and the LUMO (right). $N = P + \tau$, $\tau = 0, 1, \dots, P-1$.}
\label{fig:ProbabilitiesHL-D2}
\end{figure*}

\begin{figure*}[!h]
\includegraphics[width=0.4\textwidth]{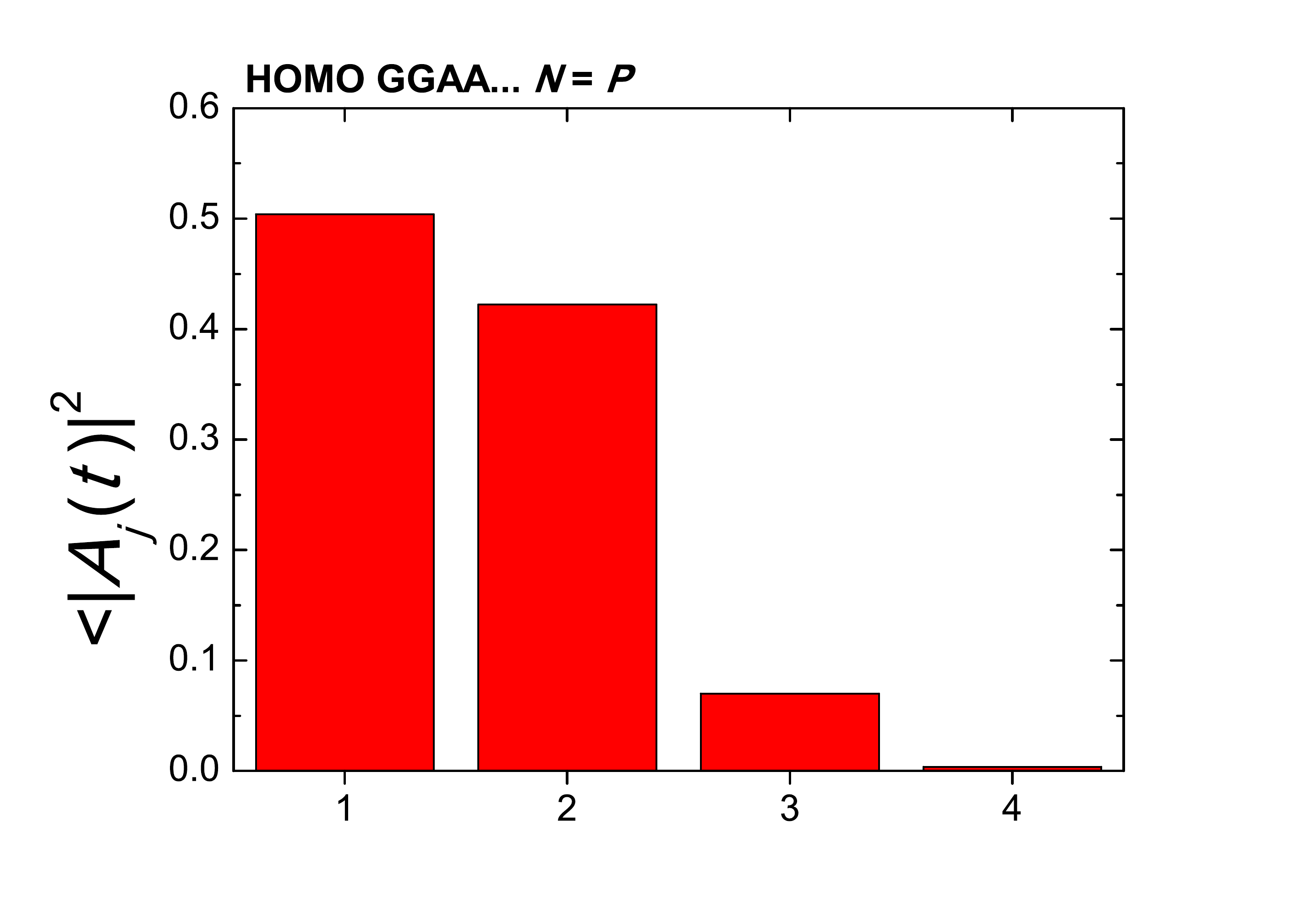}
\includegraphics[width=0.4\textwidth]{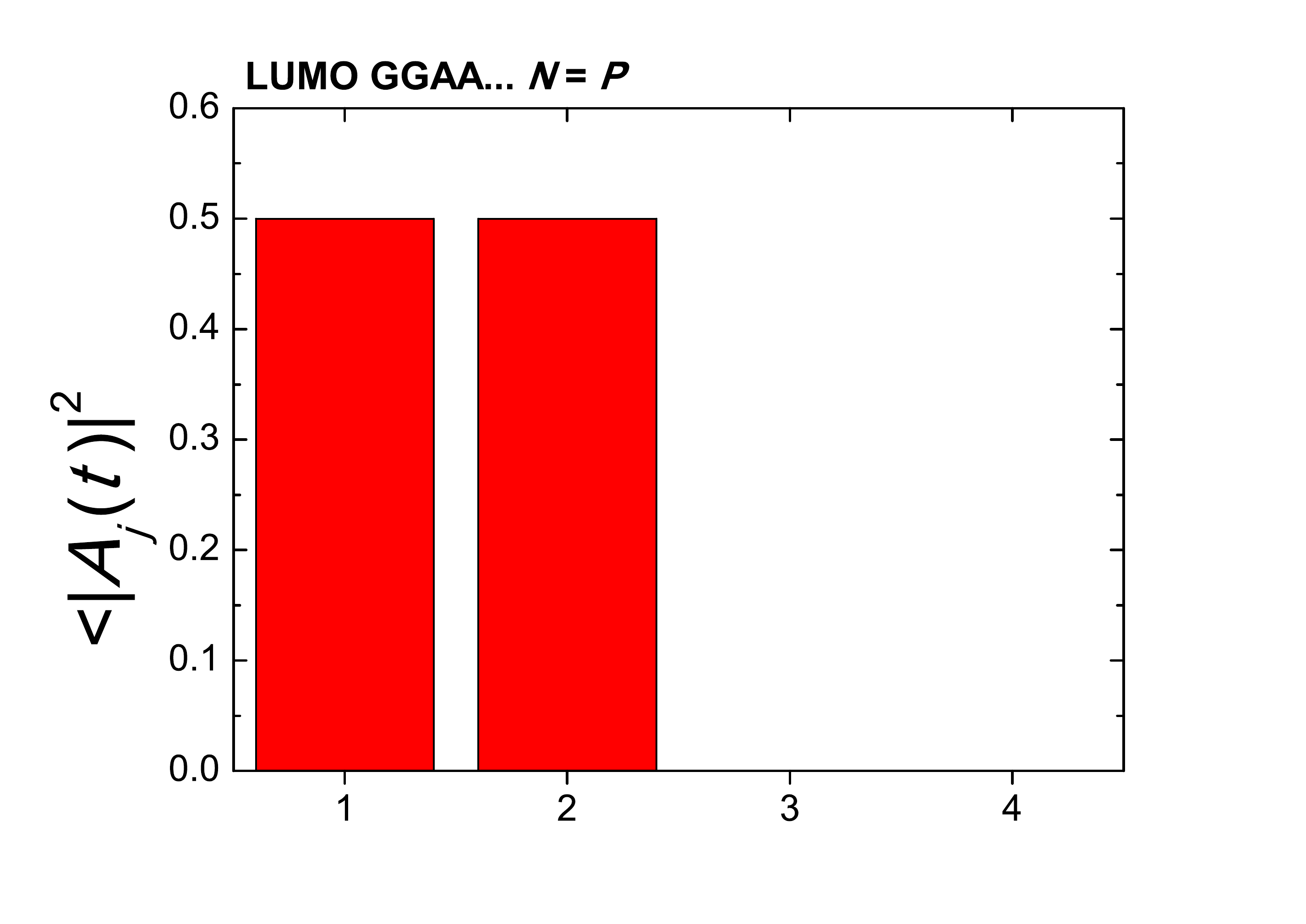}
\includegraphics[width=0.4\textwidth]{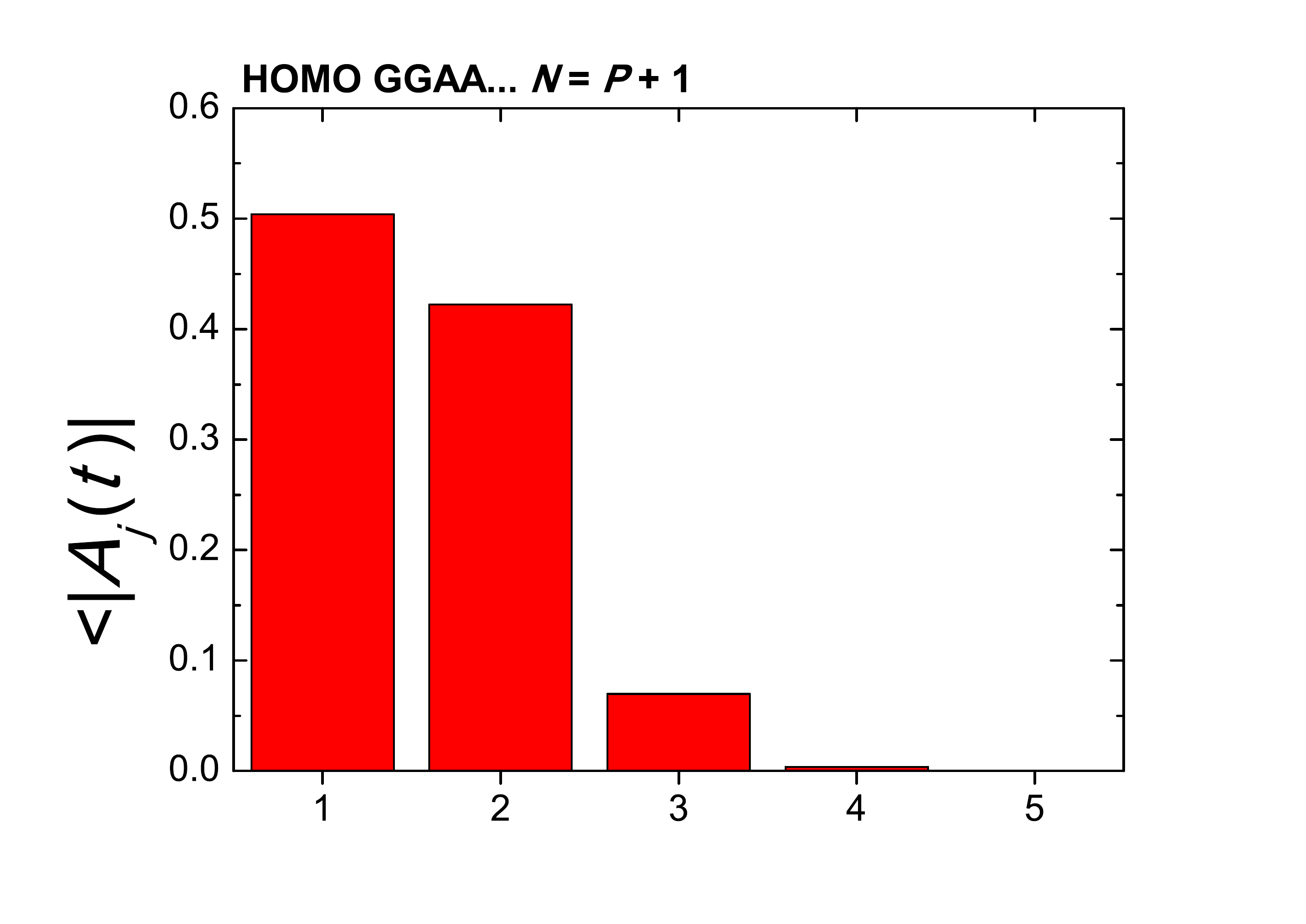}
\includegraphics[width=0.4\textwidth]{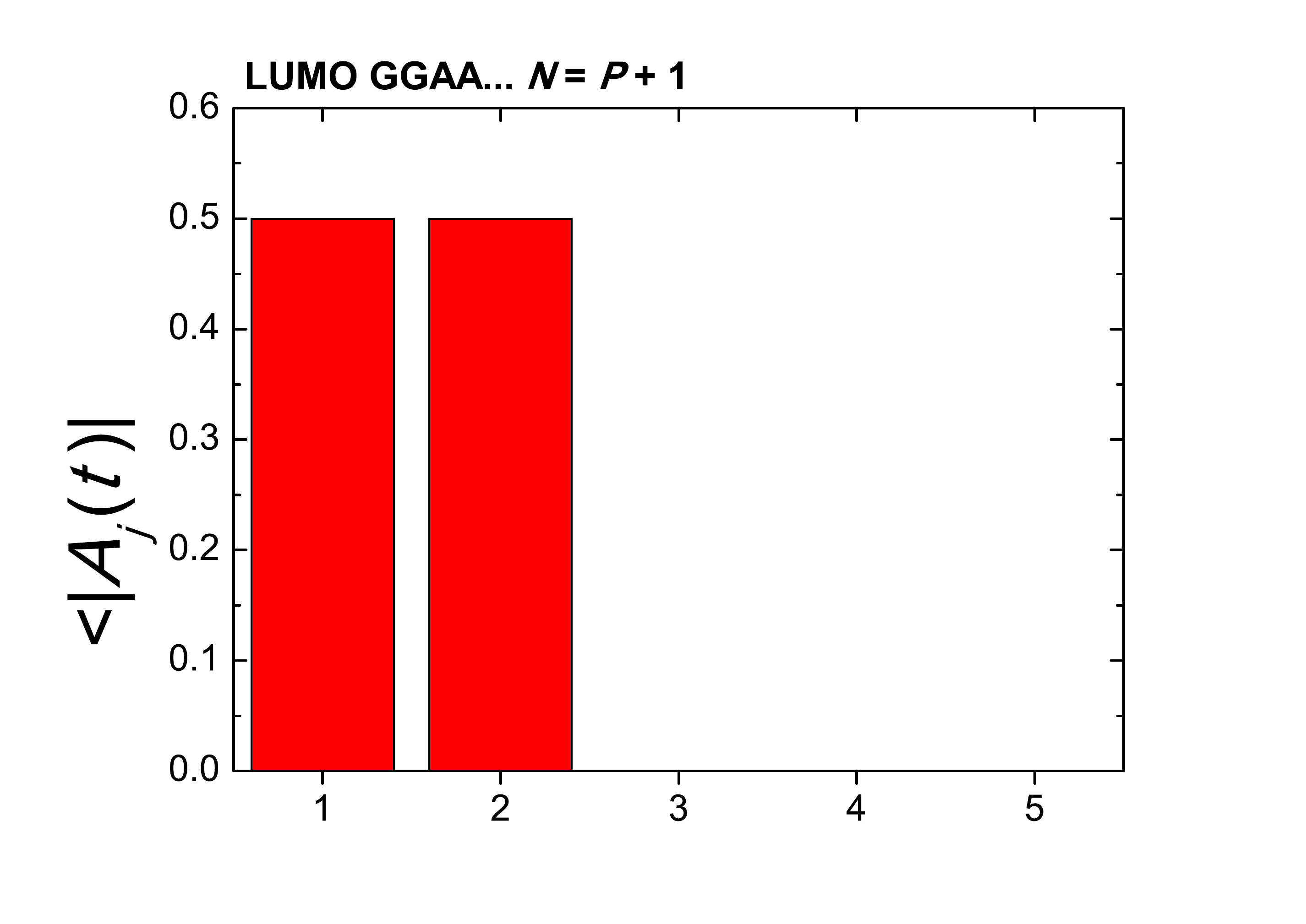}
\includegraphics[width=0.4\textwidth]{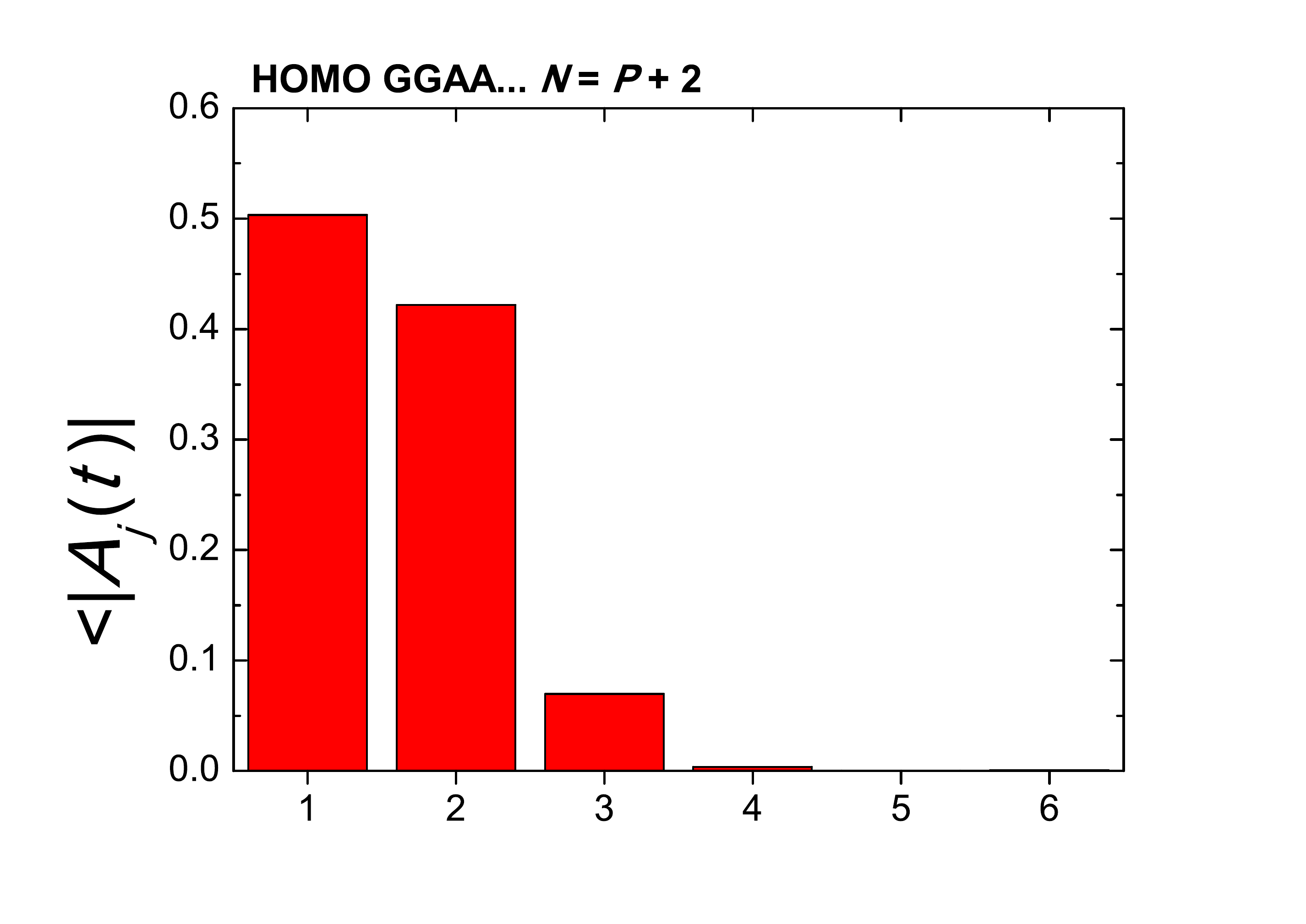}
\includegraphics[width=0.4\textwidth]{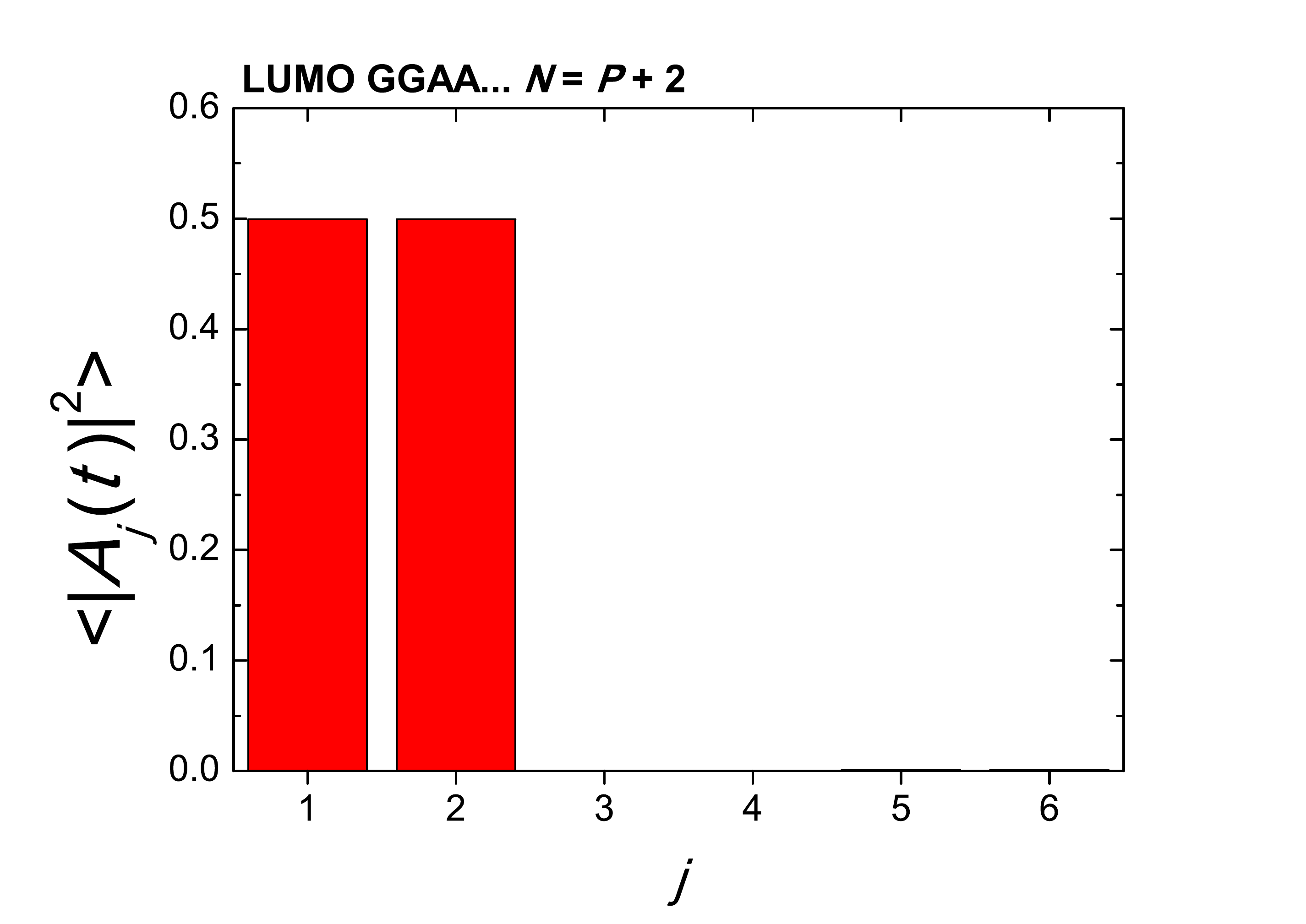}
\includegraphics[width=0.4\textwidth]{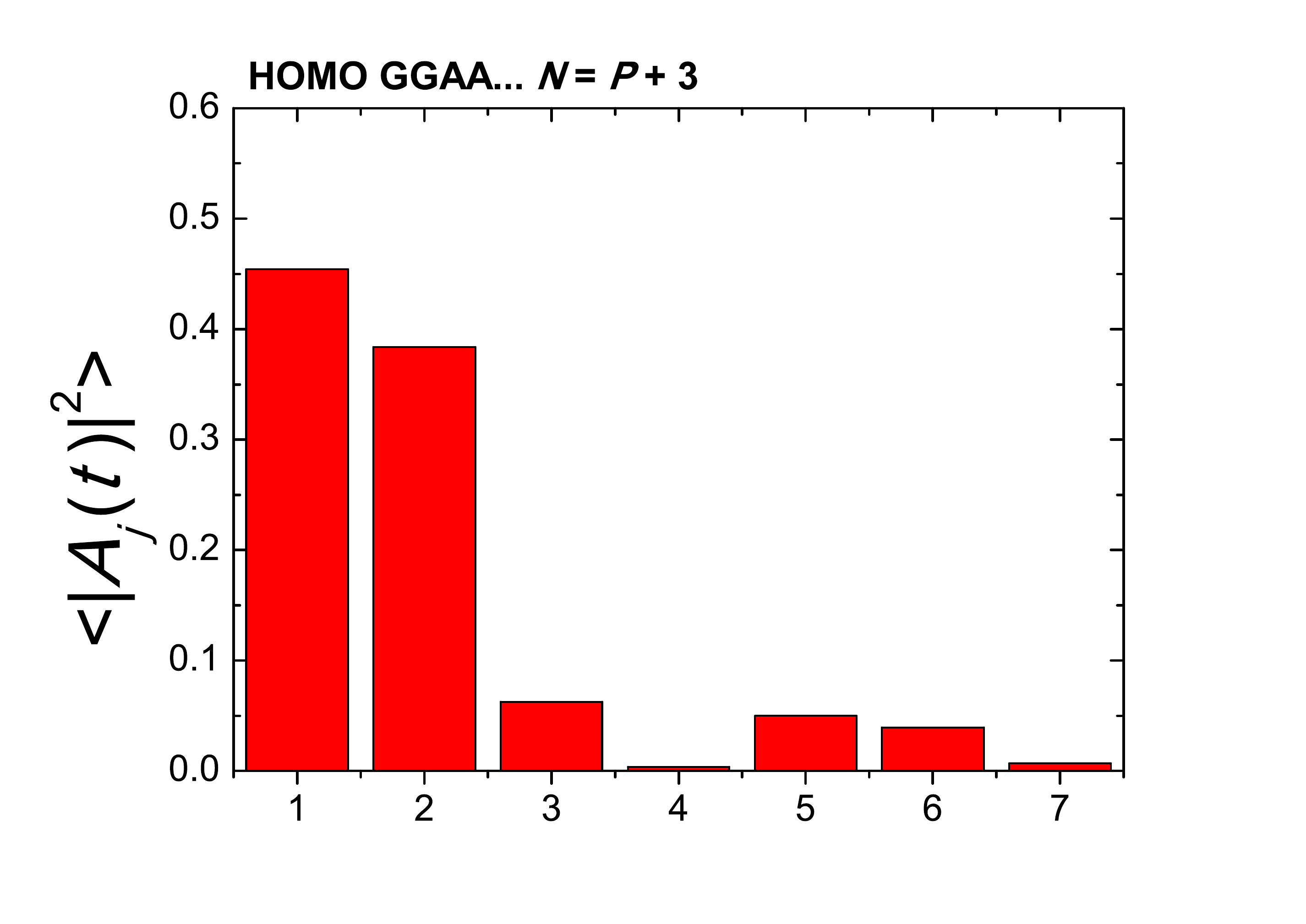}
\includegraphics[width=0.4\textwidth]{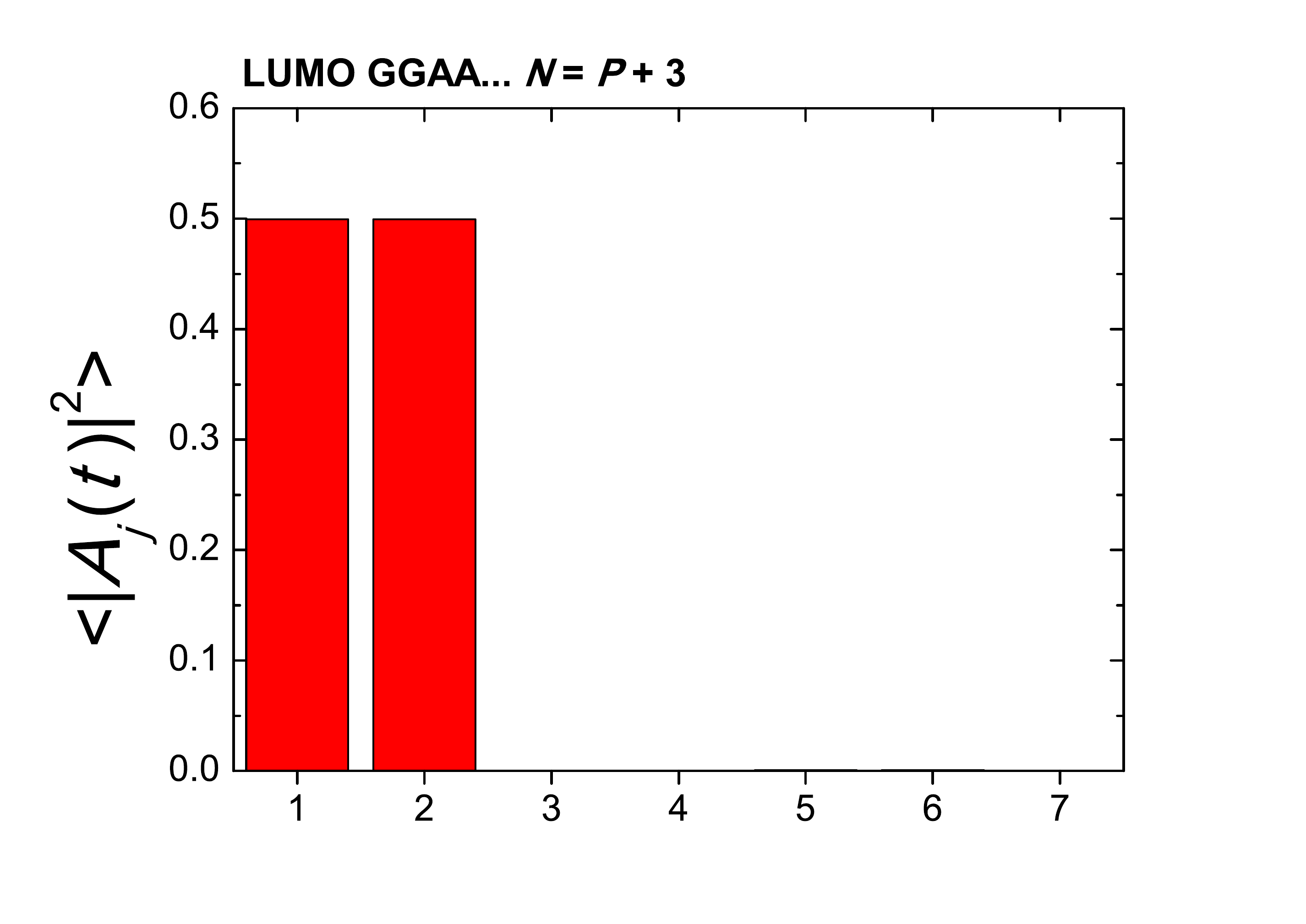}
\caption{Mean (over time) probabilities to find the extra carrier at each monomer $j$, having placed it initially at the first monomer, for D4 (GGAA...) polymers, for the HOMO (left) and the LUMO (right). $N = P + \tau$, $\tau = 0, 1, \dots, P-1$.}
\label{fig:ProbabilitiesHL-D4}
\end{figure*}

\begin{figure*}[!h]
\includegraphics[width=0.4\textwidth]{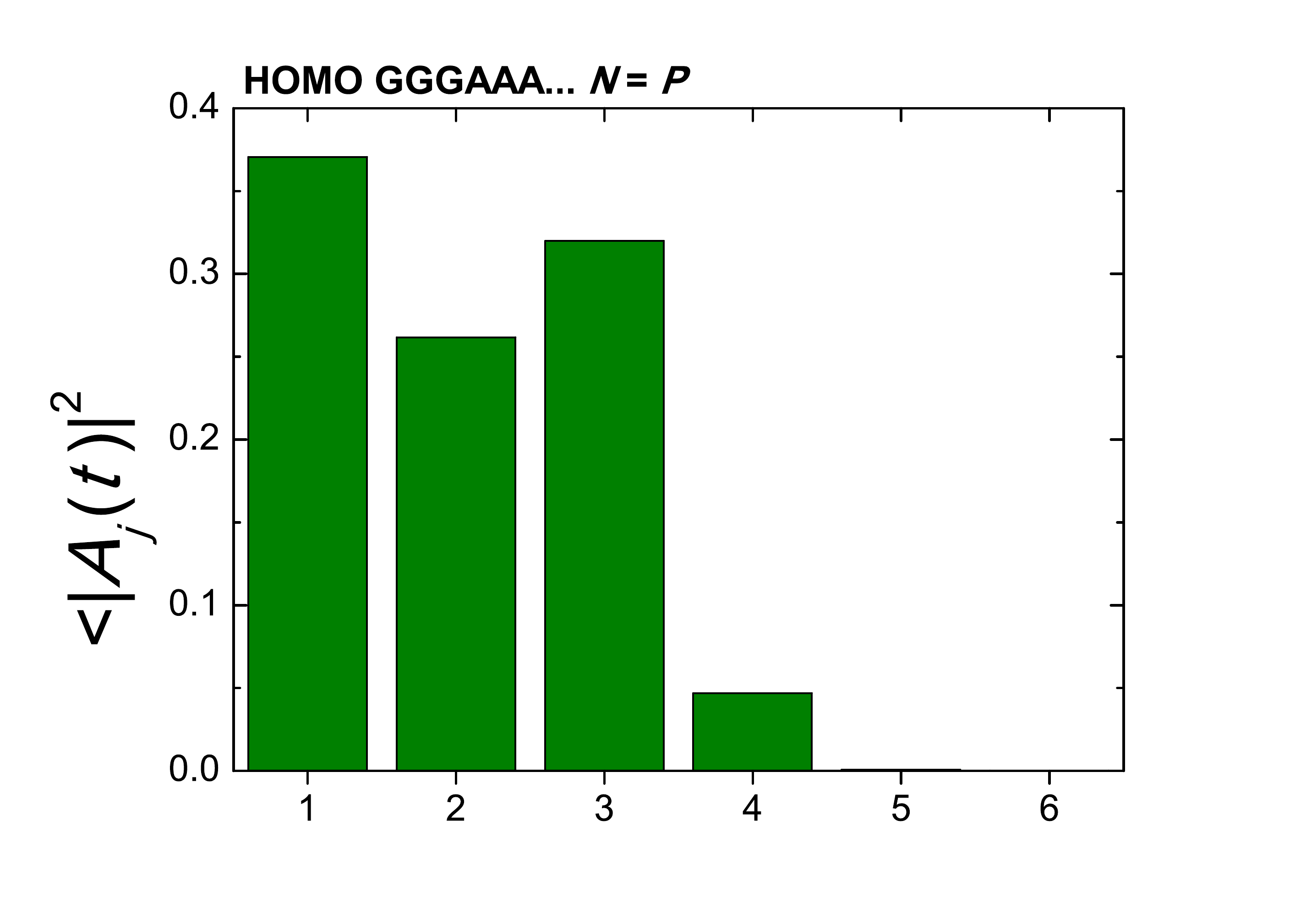}
\includegraphics[width=0.4\textwidth]{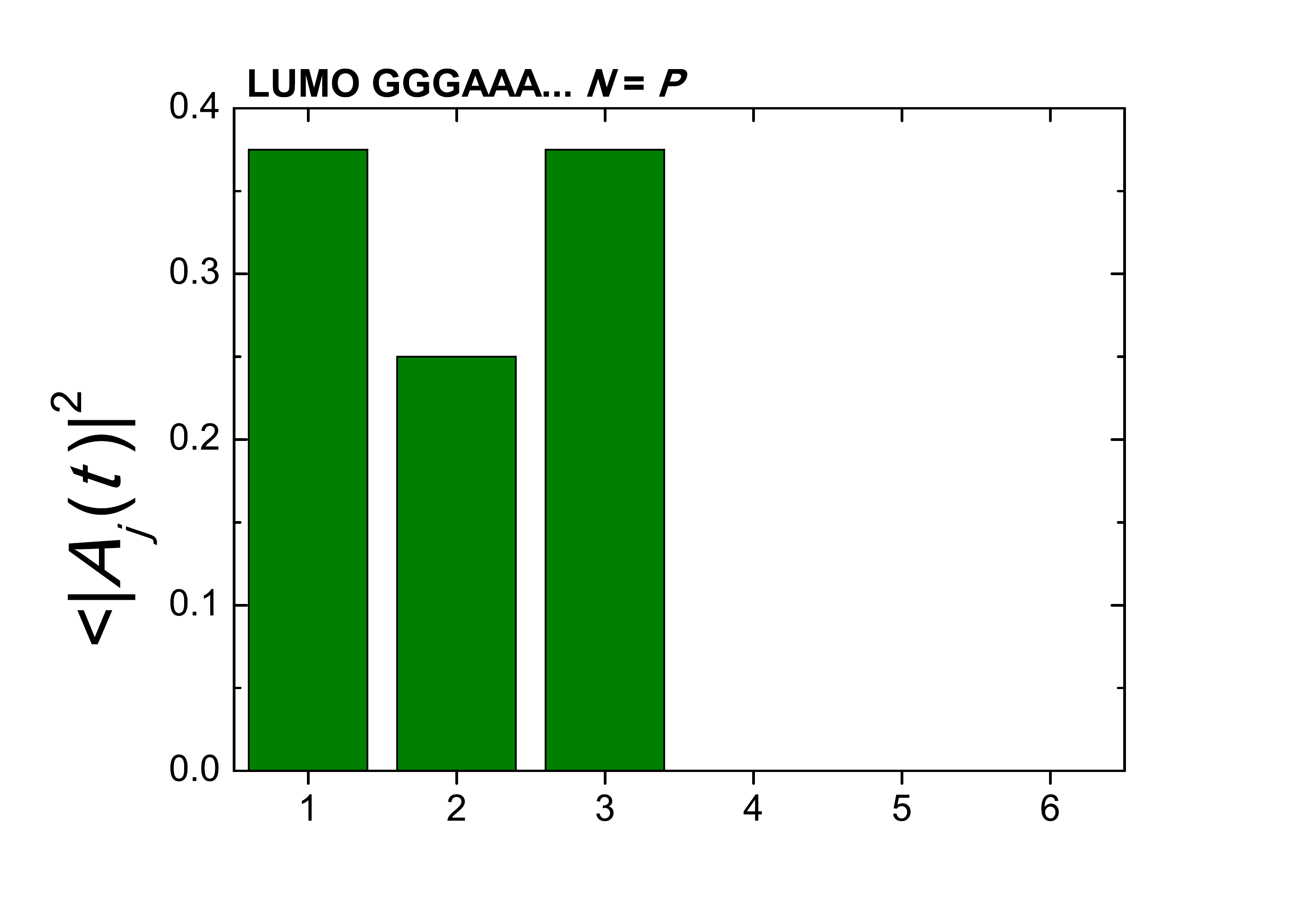}
\includegraphics[width=0.4\textwidth]{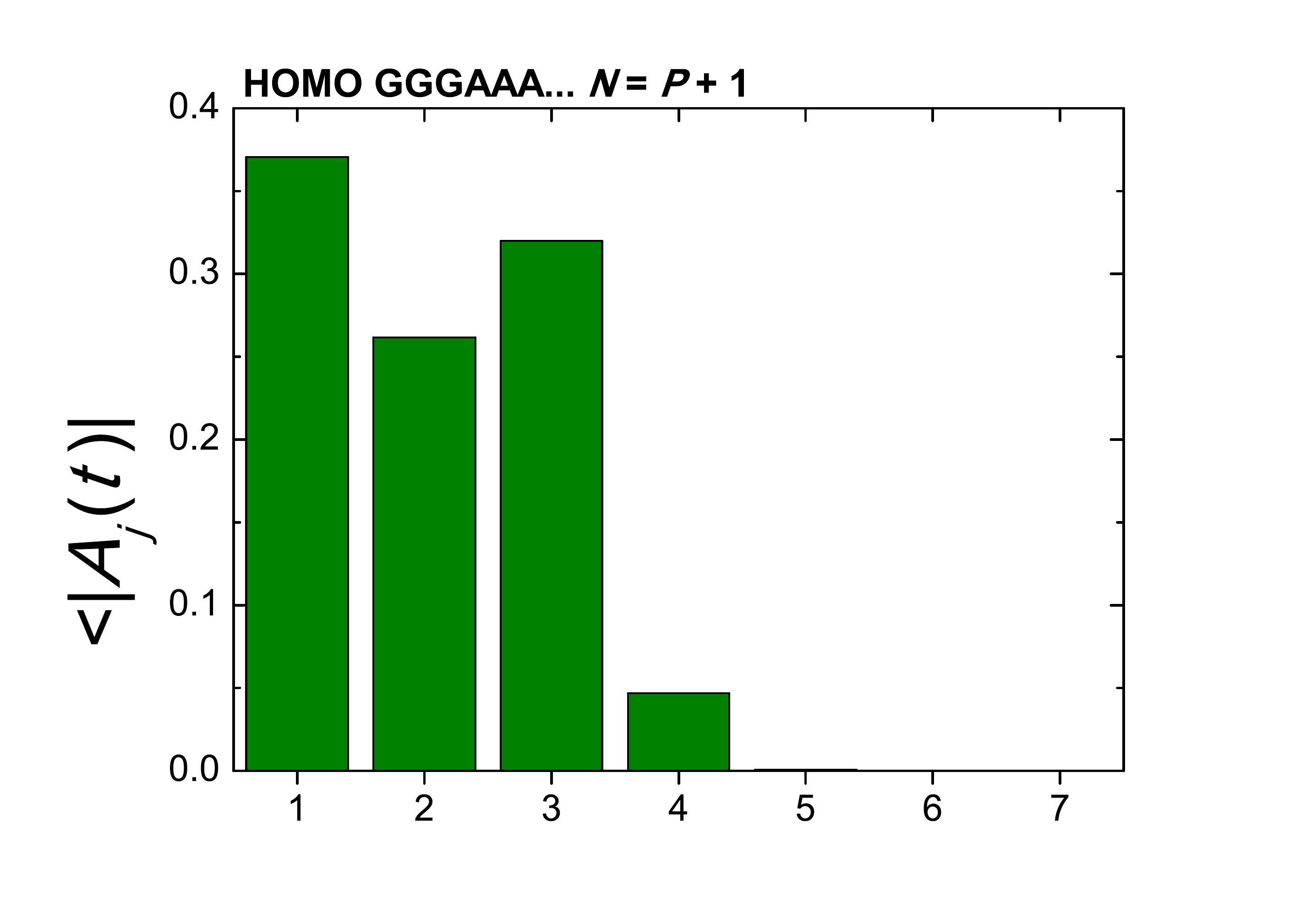}
\includegraphics[width=0.4\textwidth]{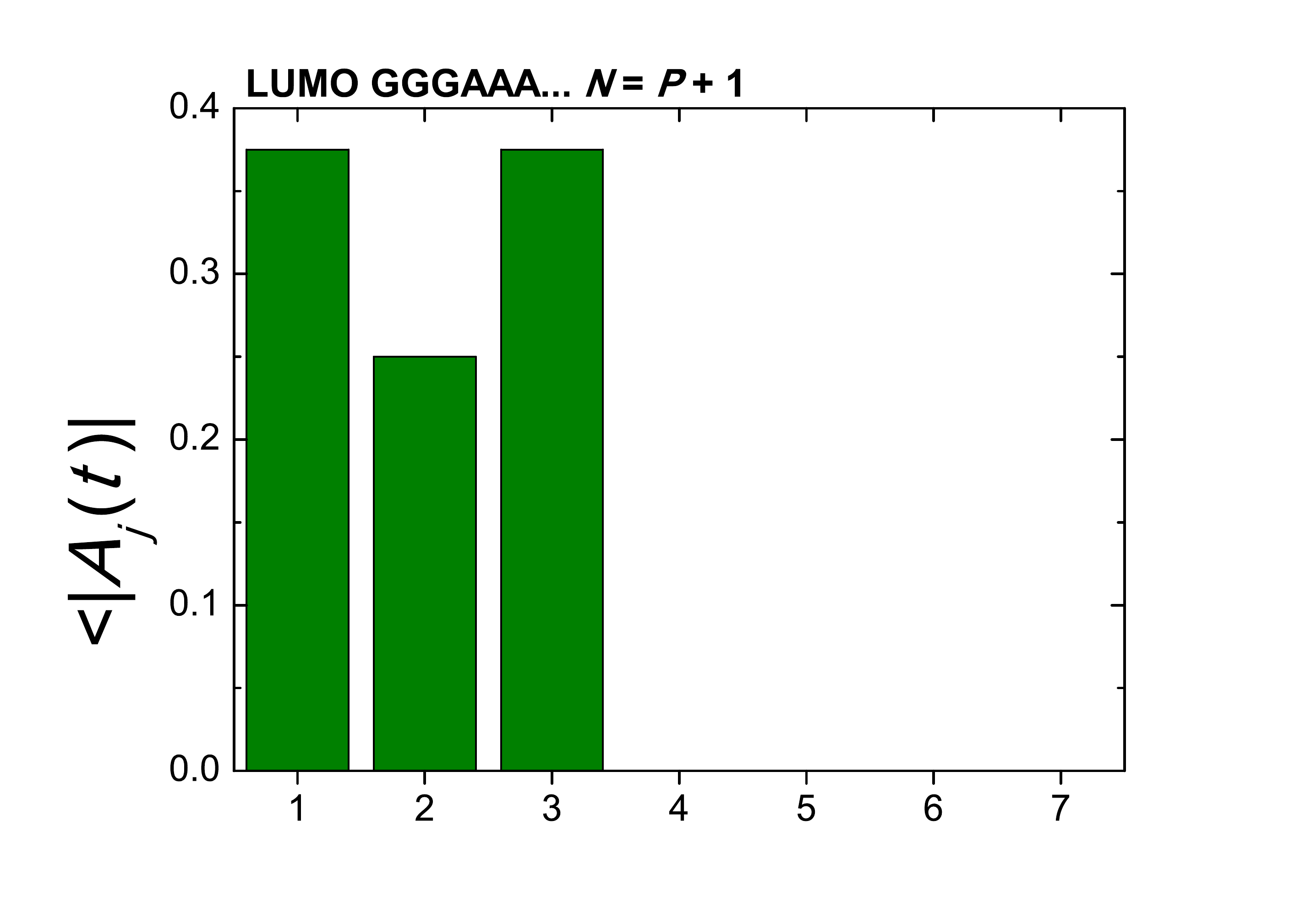}
\includegraphics[width=0.4\textwidth]{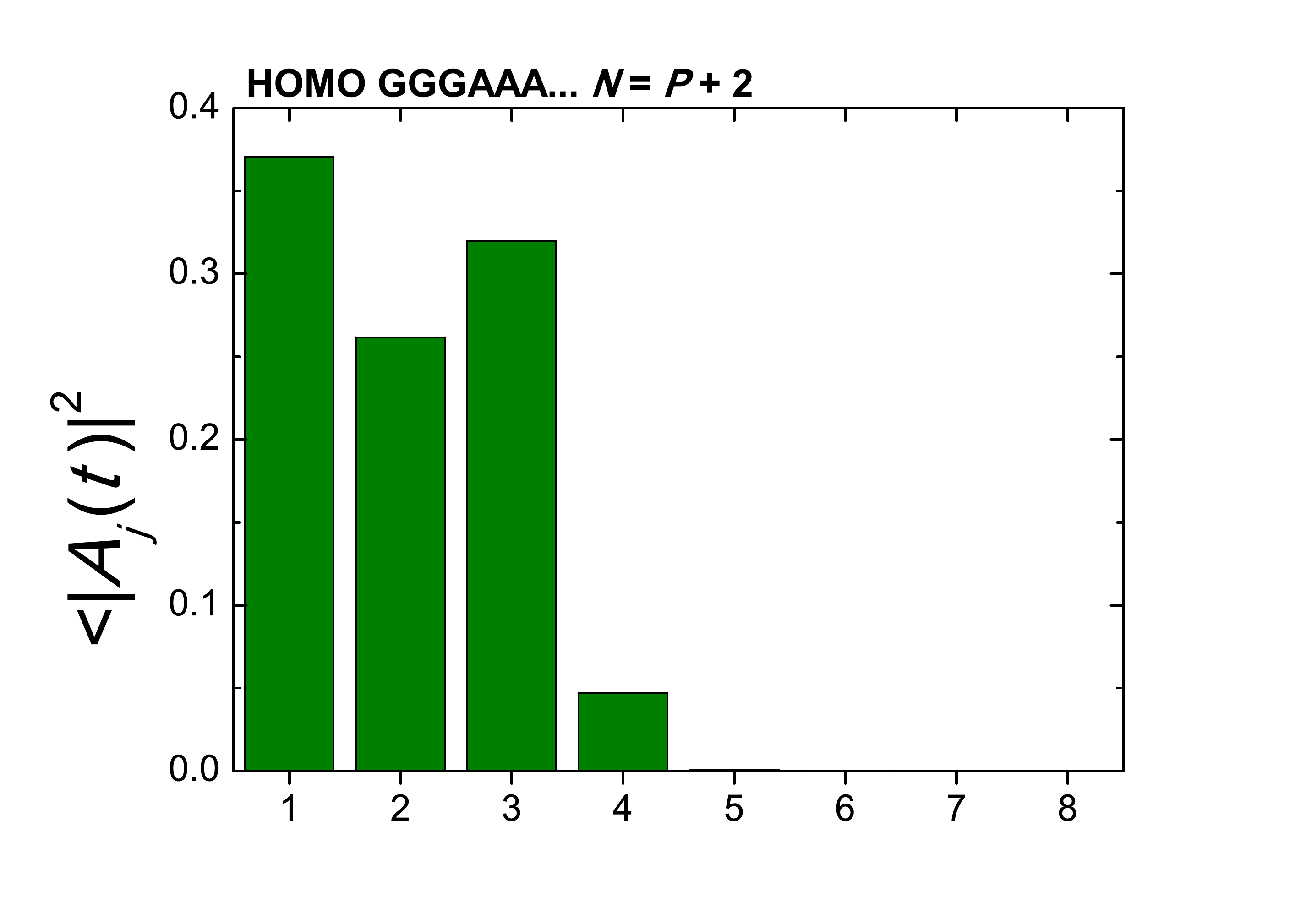}
\includegraphics[width=0.4\textwidth]{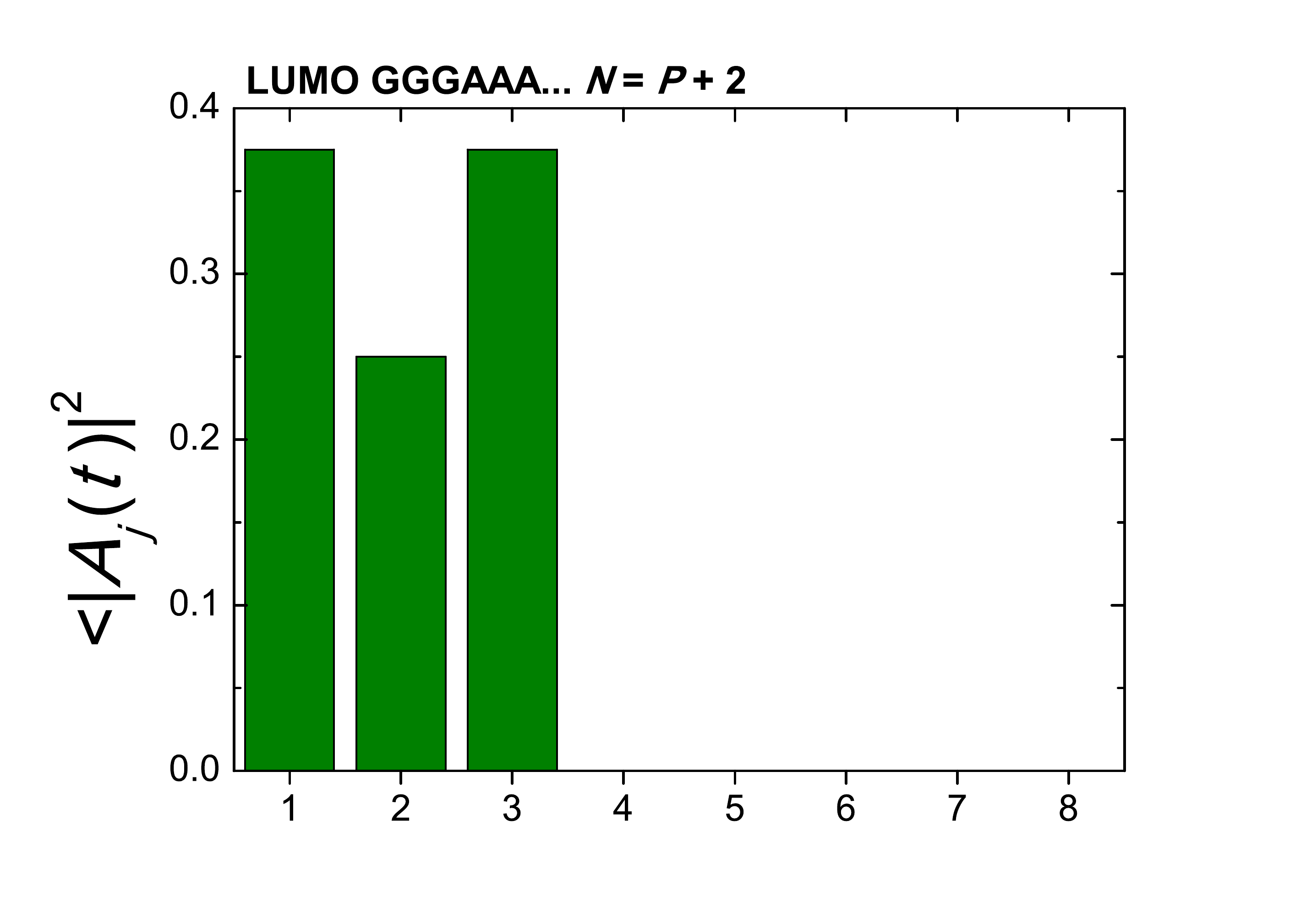}
\caption{Mean (over time) probabilities to find the extra carrier at each monomer $j$, having placed it initially at the first monomer, for D6 (GGGAAA...) polymers, for the HOMO (left) and the LUMO (right). $N = P + \tau$, $\tau = 0, 1, \dots, P-1$. \emph{Continued at the next page...}}
\label{fig:ProbabilitiesHL-D6}
\end{figure*}
\begin{figure*}[!h]
\addtocounter{figure}{-1}
\includegraphics[width=0.4\textwidth]{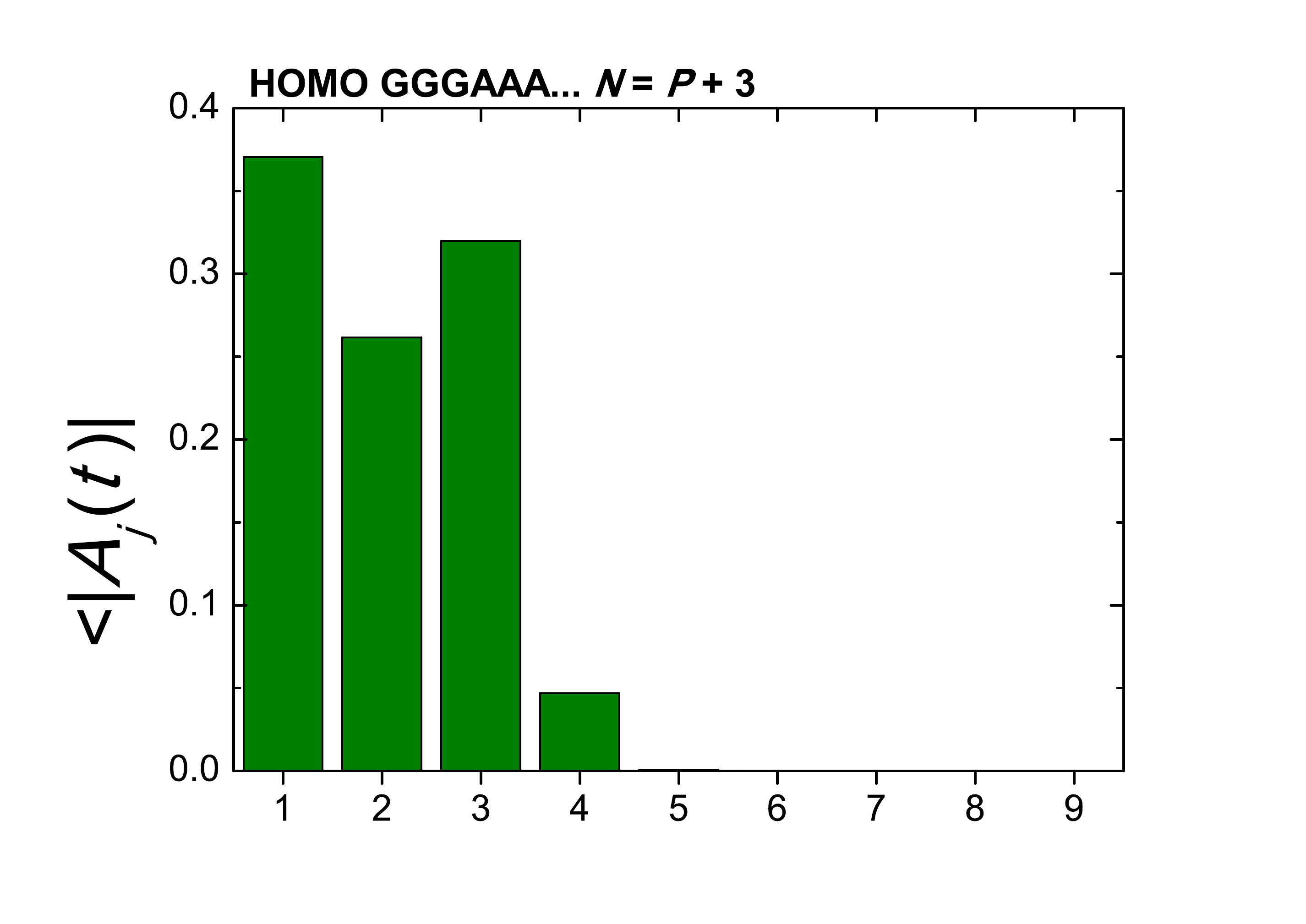}
\includegraphics[width=0.4\textwidth]{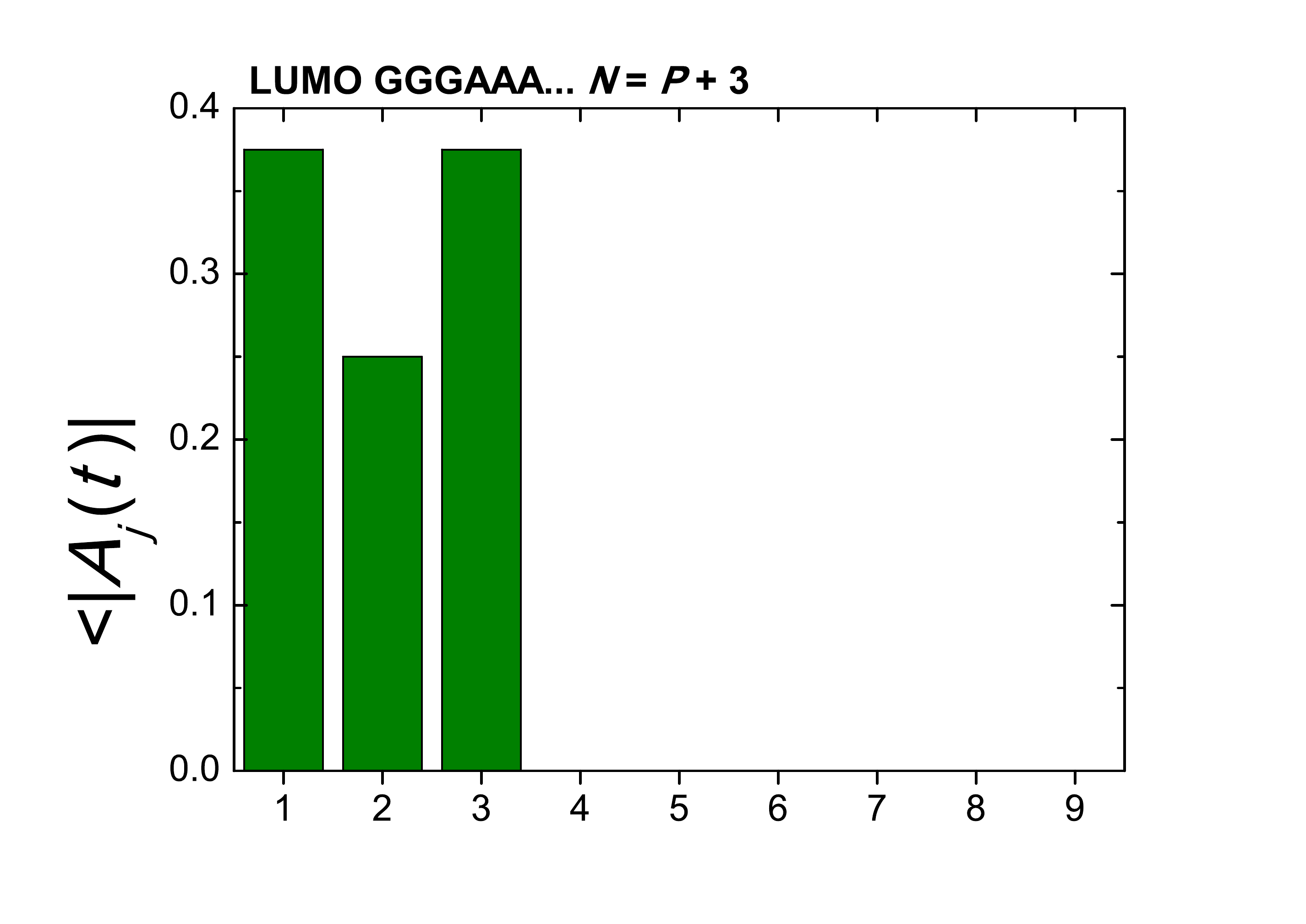}
\includegraphics[width=0.4\textwidth]{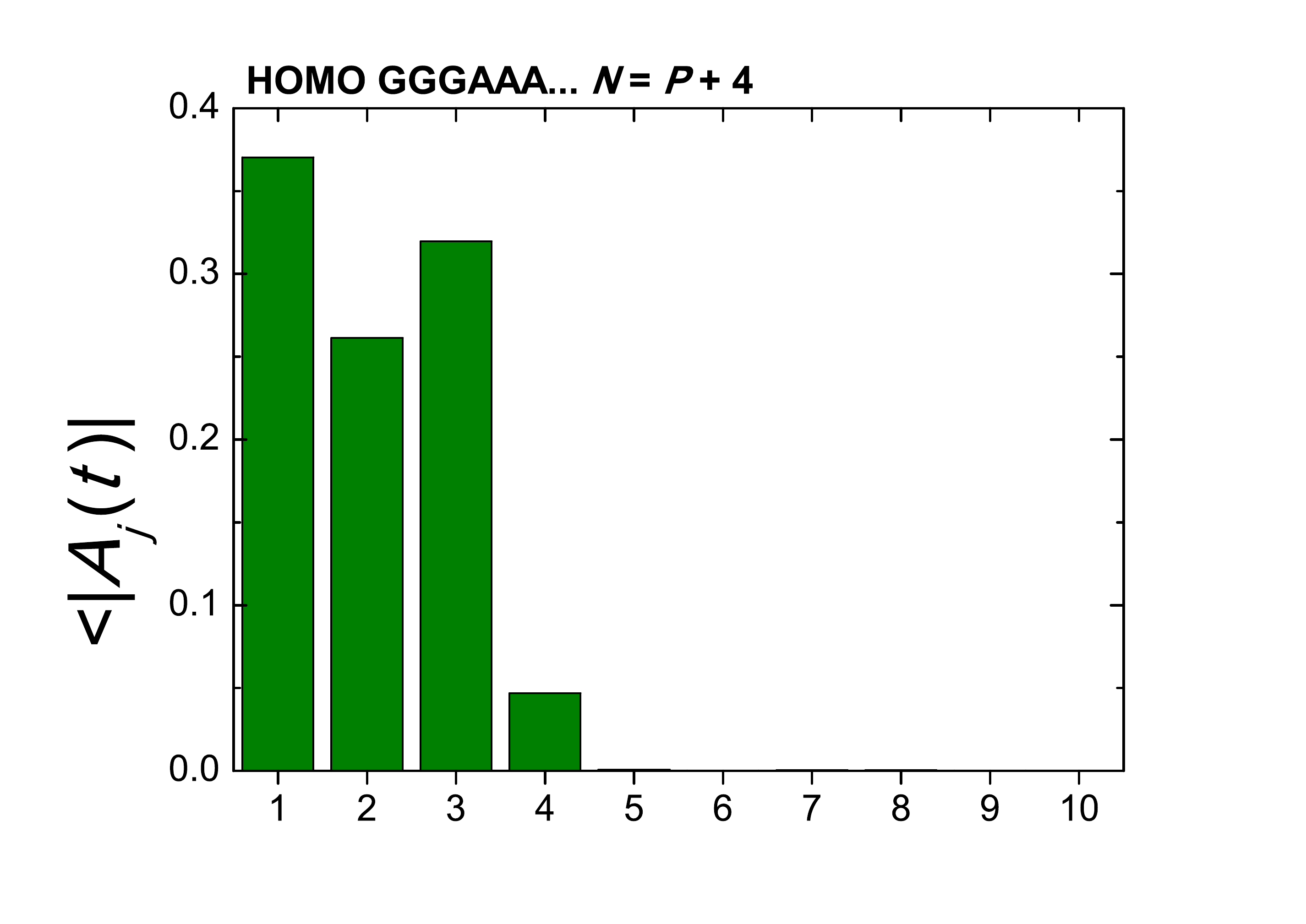}
\includegraphics[width=0.4\textwidth]{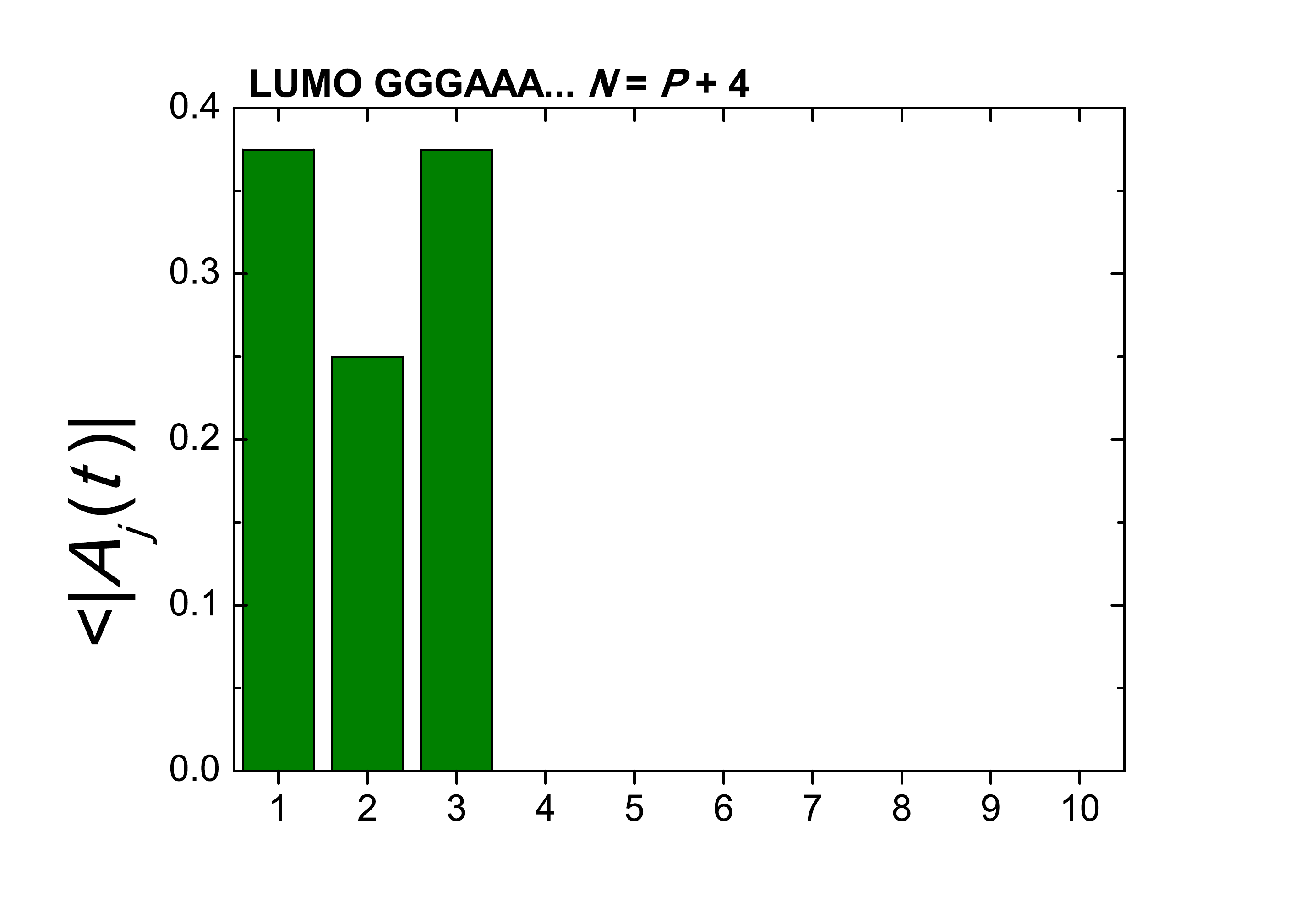}
\includegraphics[width=0.4\textwidth]{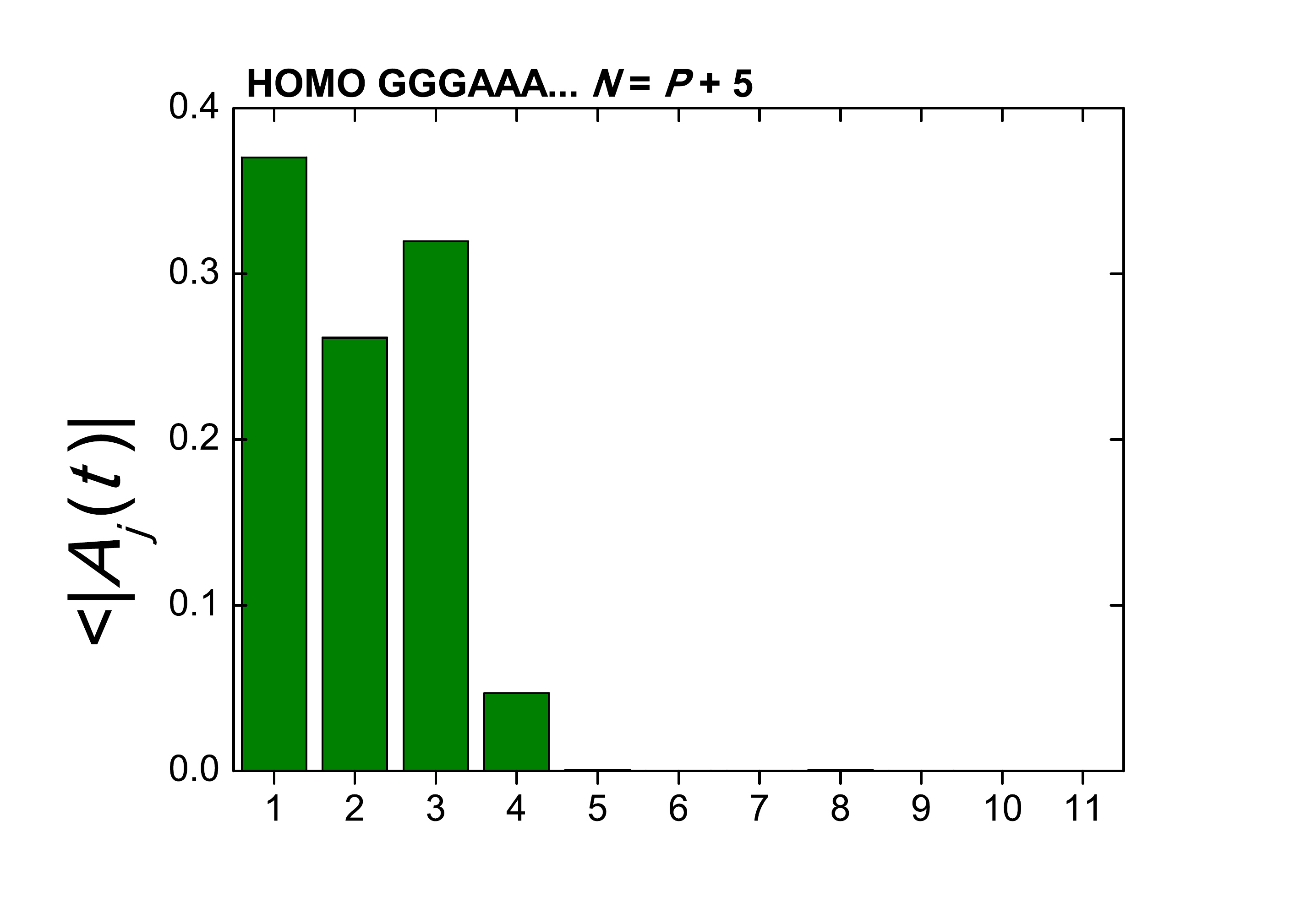}
\includegraphics[width=0.4\textwidth]{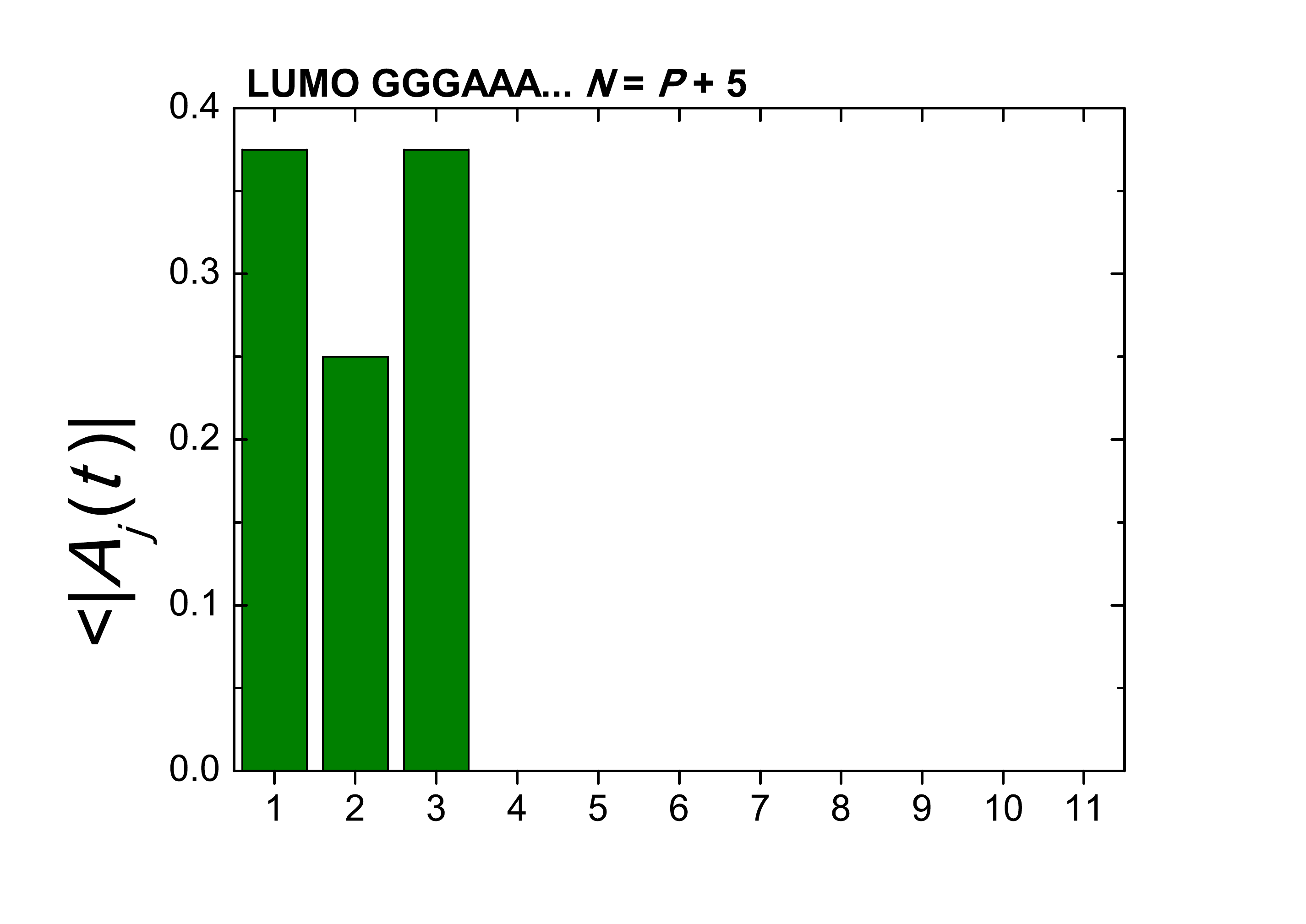}
\caption{\emph{...Continued from the previous page.} Mean (over time) probabilities to find the extra carrier at each monomer $j$, having placed it initially at the first monomer, for D6 (GGGAAA...) polymers, for the HOMO (left) and the LUMO (right). $N = P + \tau$, $\tau = 0, 1, \dots, P-1$.}
\end{figure*}

\begin{figure*}[!h]
\includegraphics[width=0.4\textwidth]{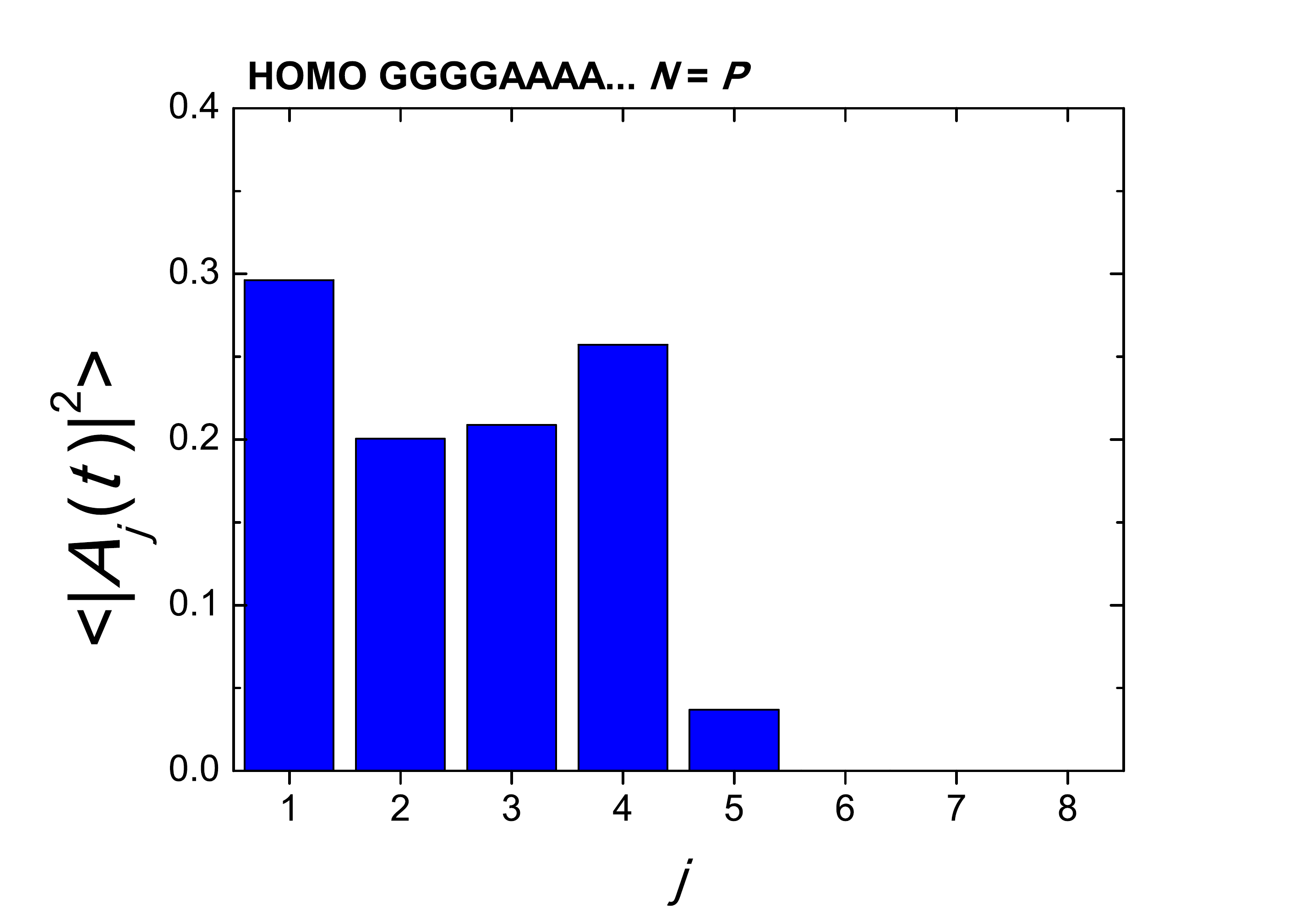}
\includegraphics[width=0.4\textwidth]{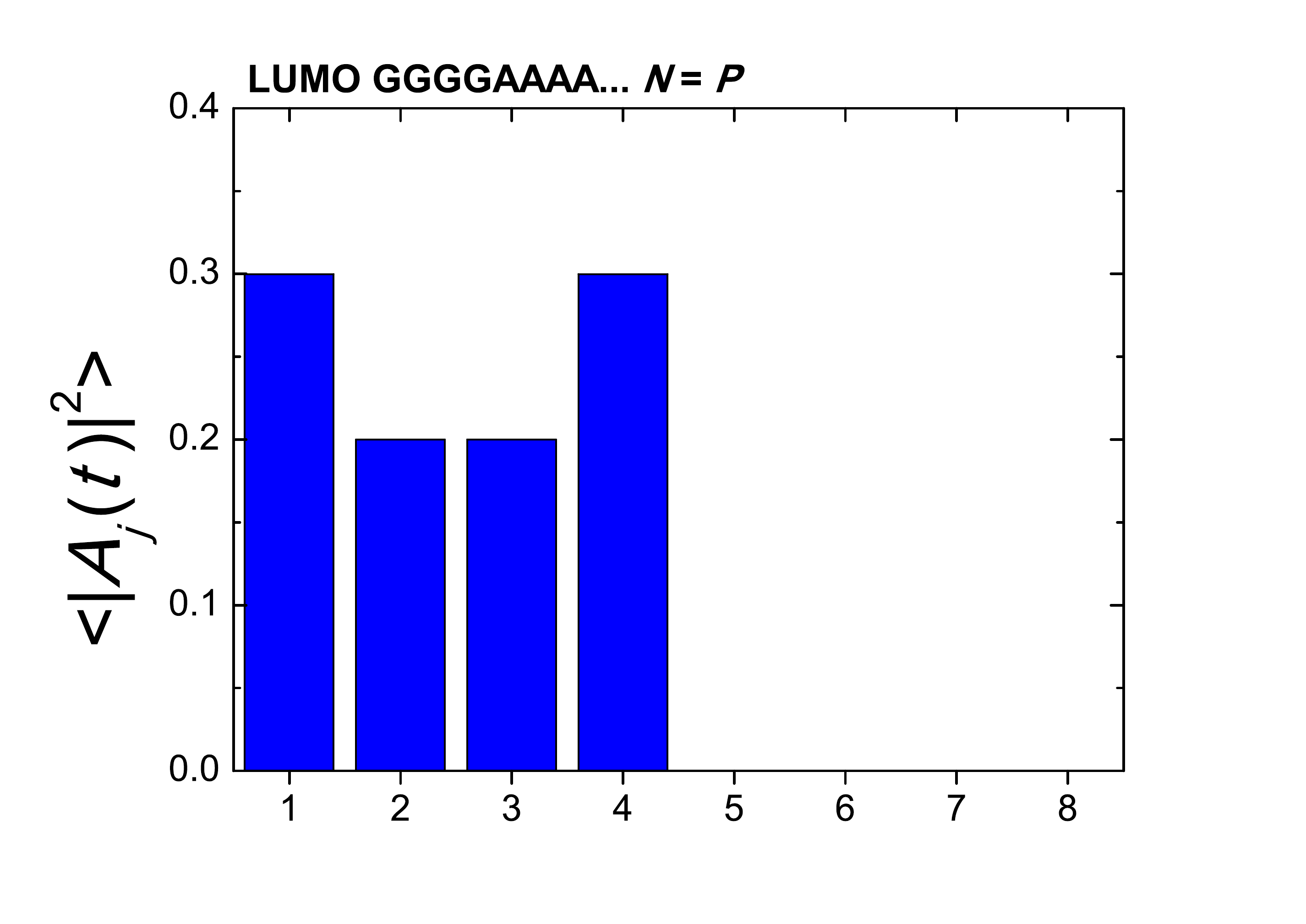}
\includegraphics[width=0.4\textwidth]{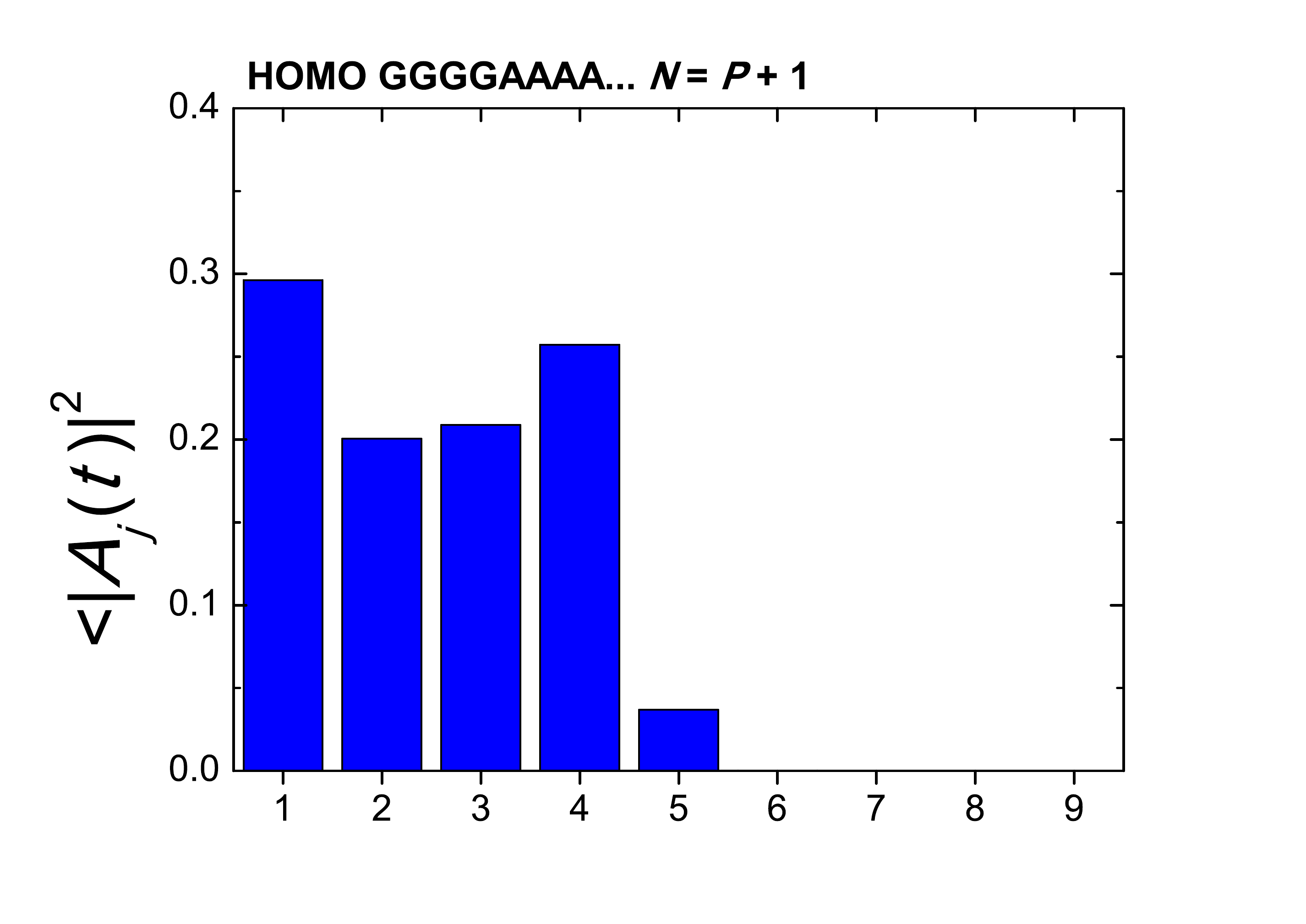}
\includegraphics[width=0.4\textwidth]{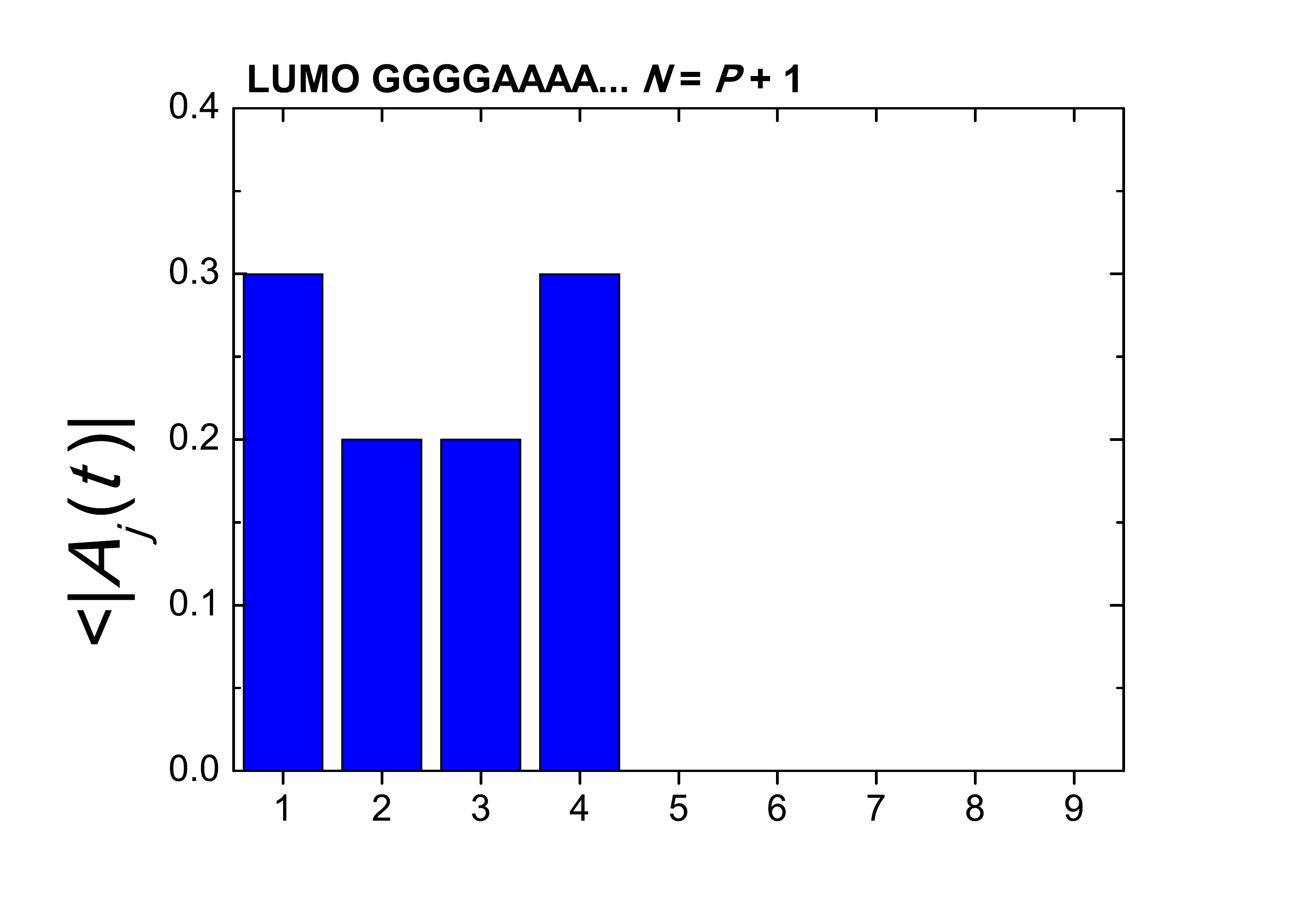}
\includegraphics[width=0.4\textwidth]{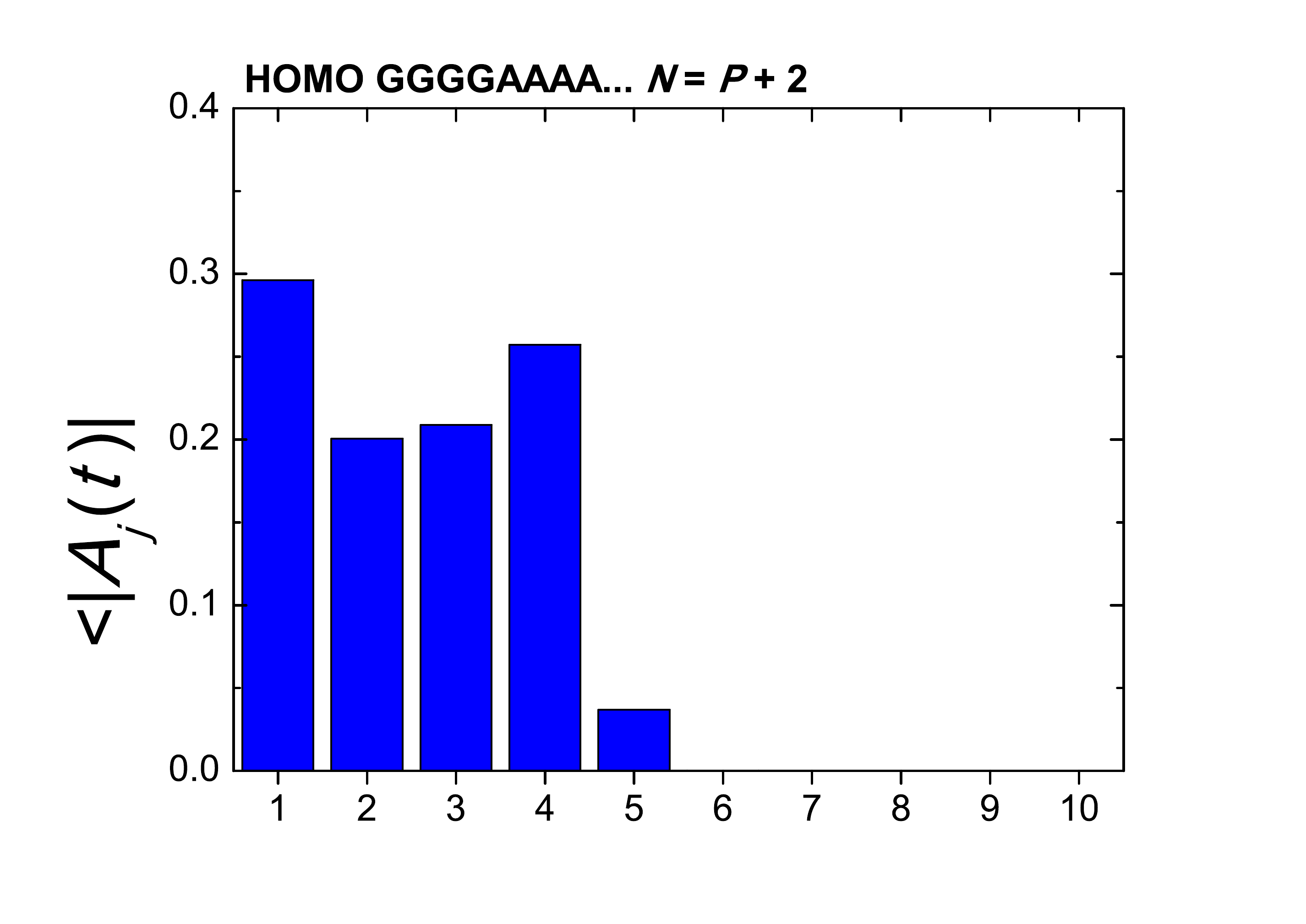}
\includegraphics[width=0.4\textwidth]{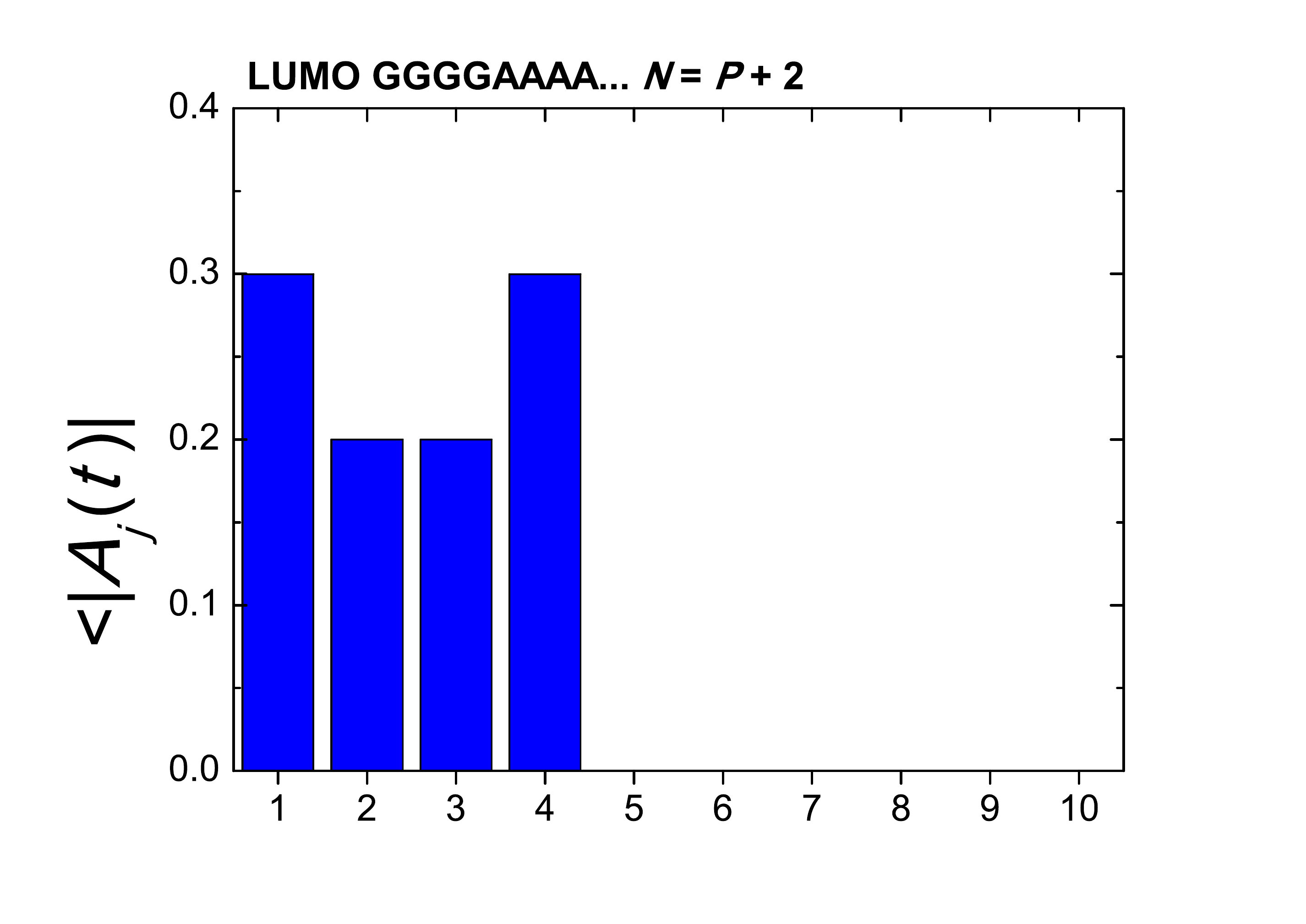}
\includegraphics[width=0.4\textwidth]{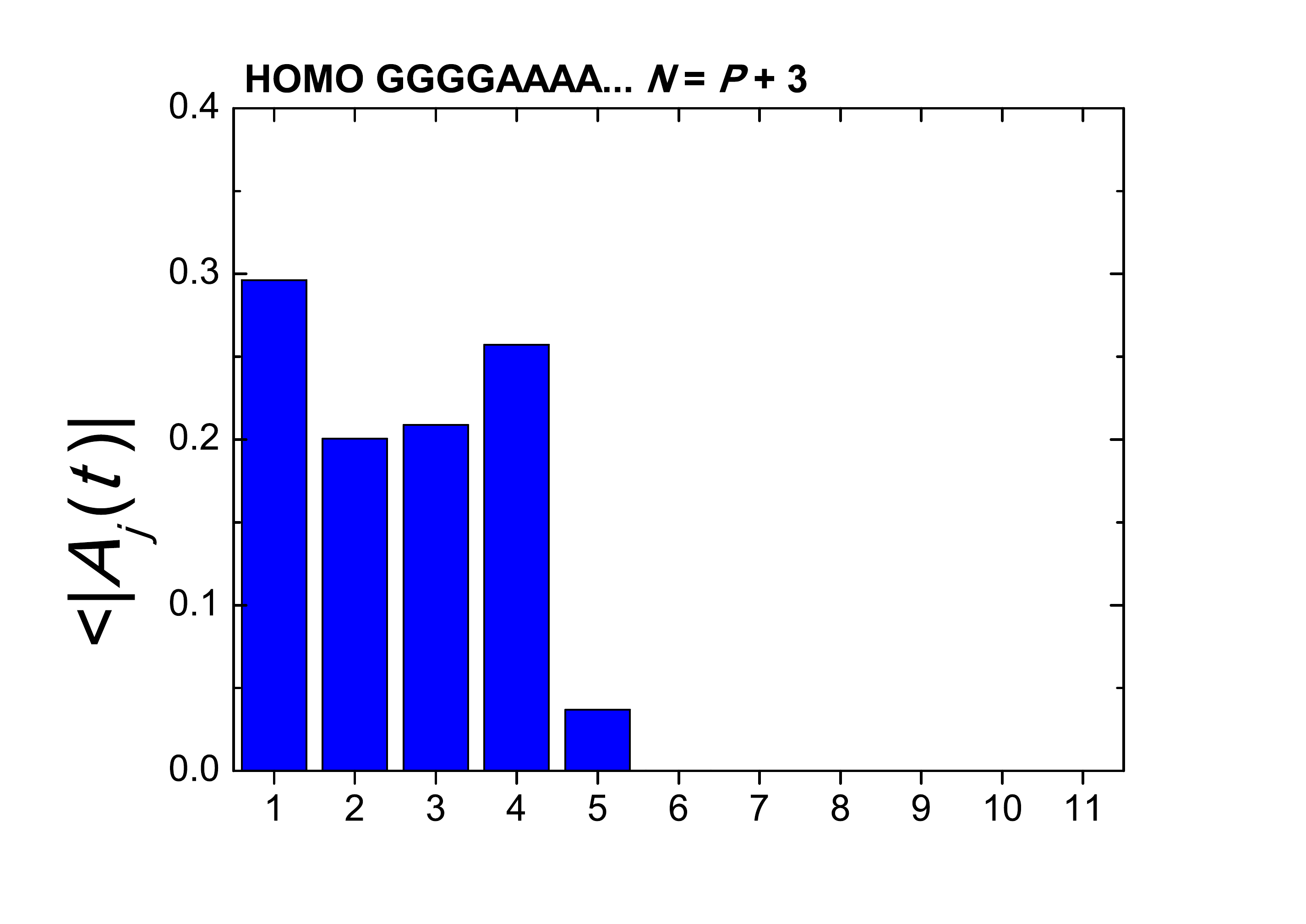}
\includegraphics[width=0.4\textwidth]{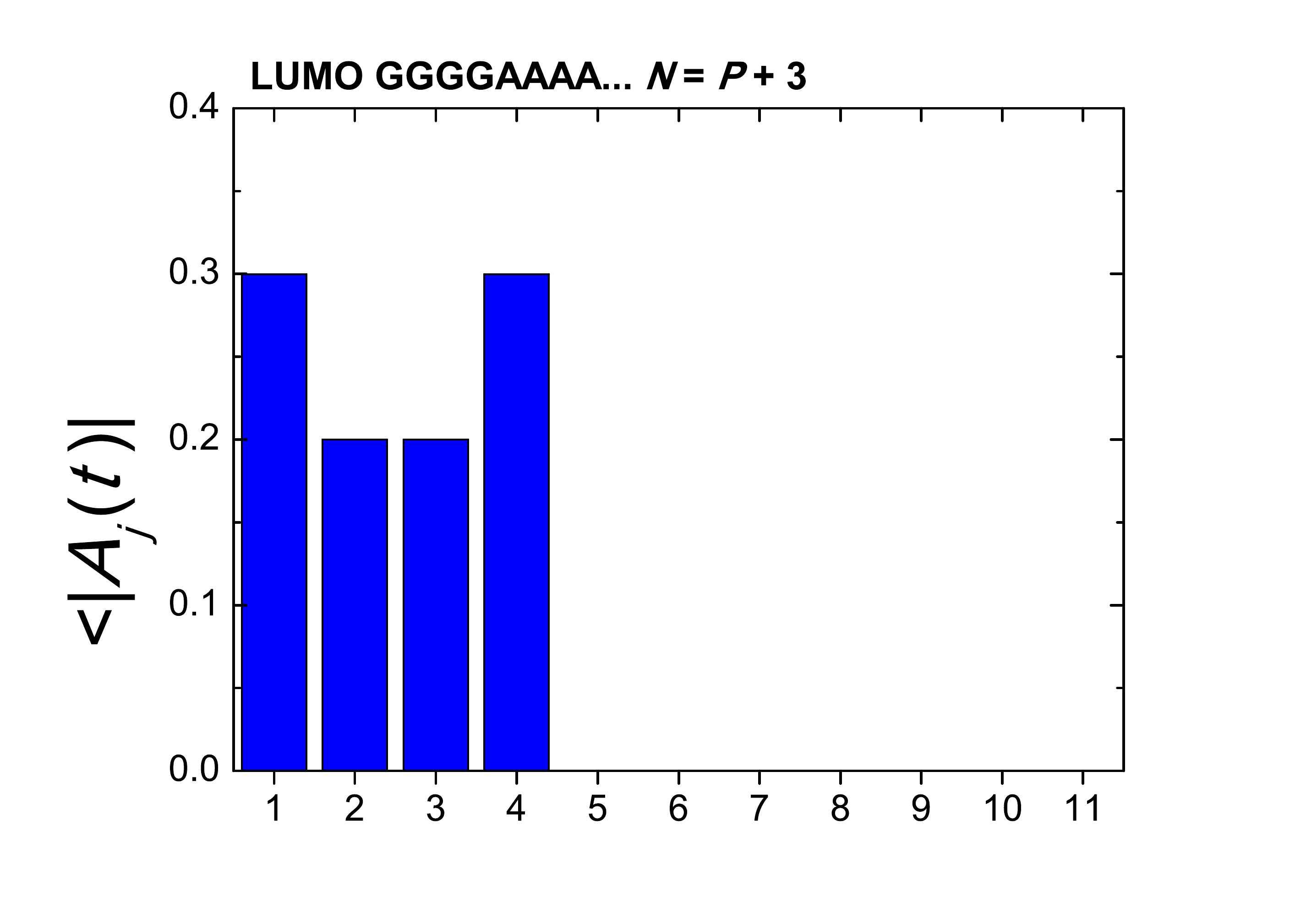}
\caption{Mean (over time) probabilities to find the extra carrier at each monomer $j$, having placed it initially at the first monomer, for D8 (GGGGAAAA...) polymers, for the HOMO (left) and the LUMO (right). $N = P + \tau$, $\tau = 0, 1, \dots, P-1$. \emph{Continued at the next page...}}
\label{fig:ProbabilitiesHL-D8}
\end{figure*}
\begin{figure*}[!h]
\addtocounter{figure}{-1}
\includegraphics[width=0.4\textwidth]{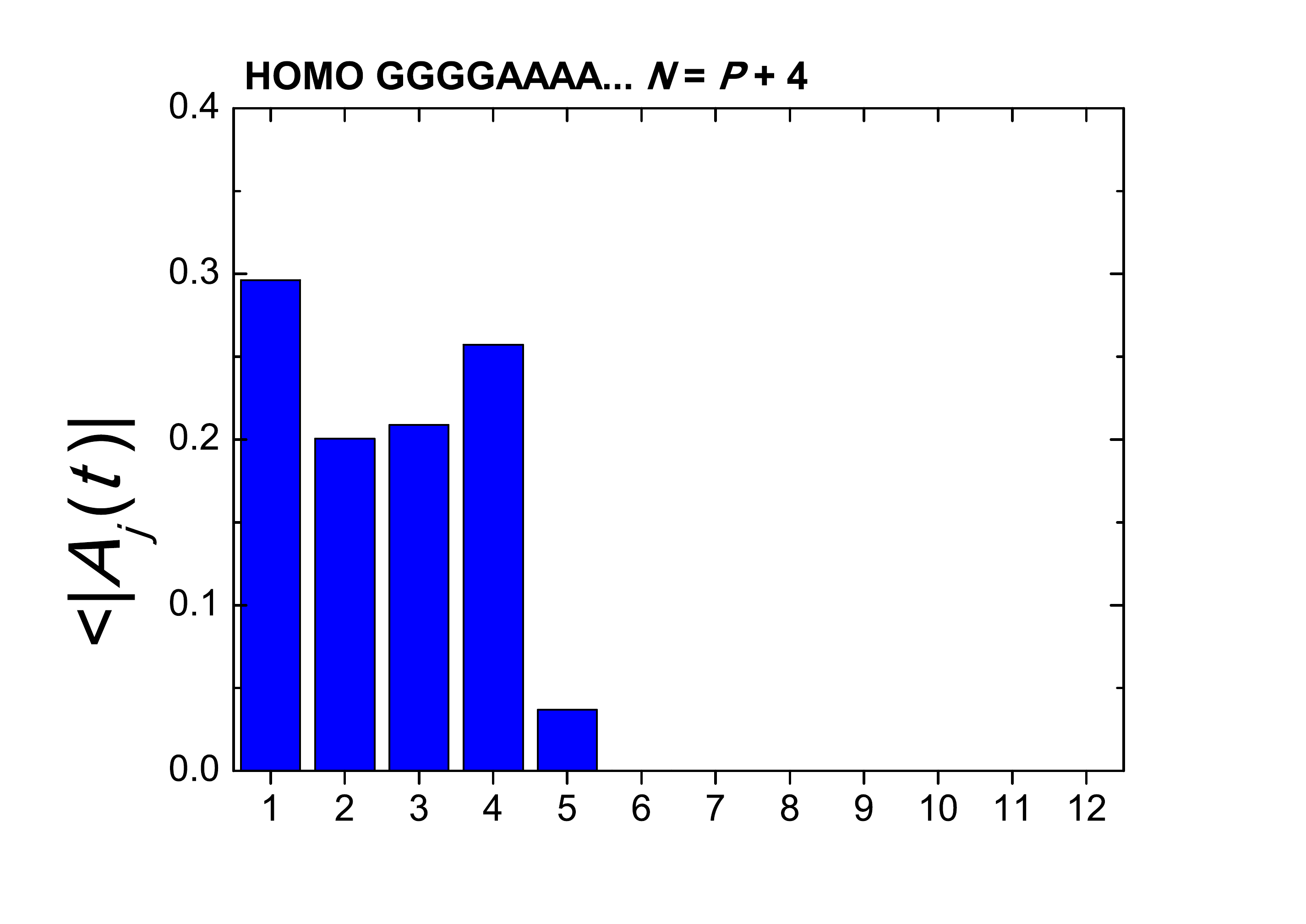}
\includegraphics[width=0.4\textwidth]{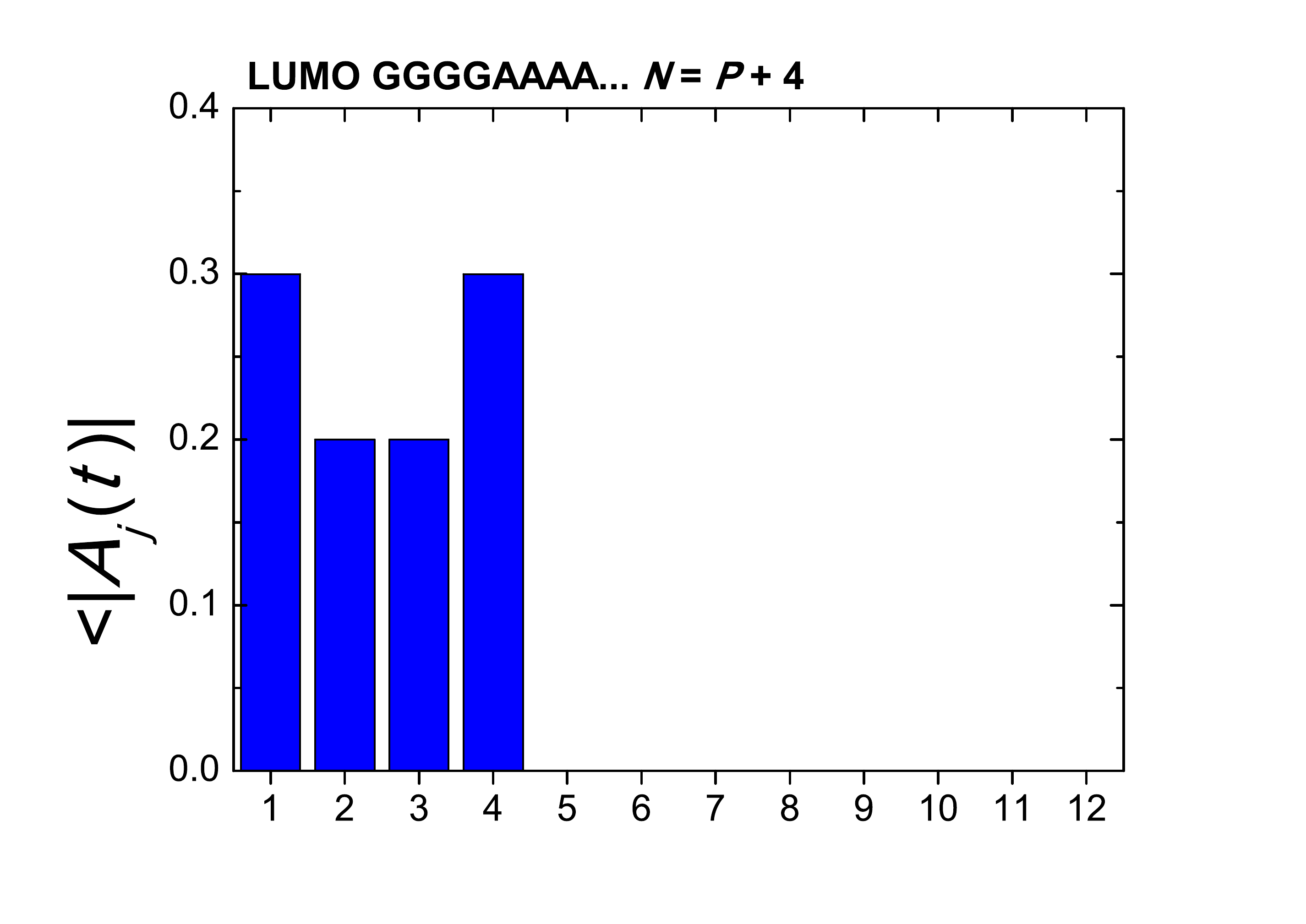}
\includegraphics[width=0.4\textwidth]{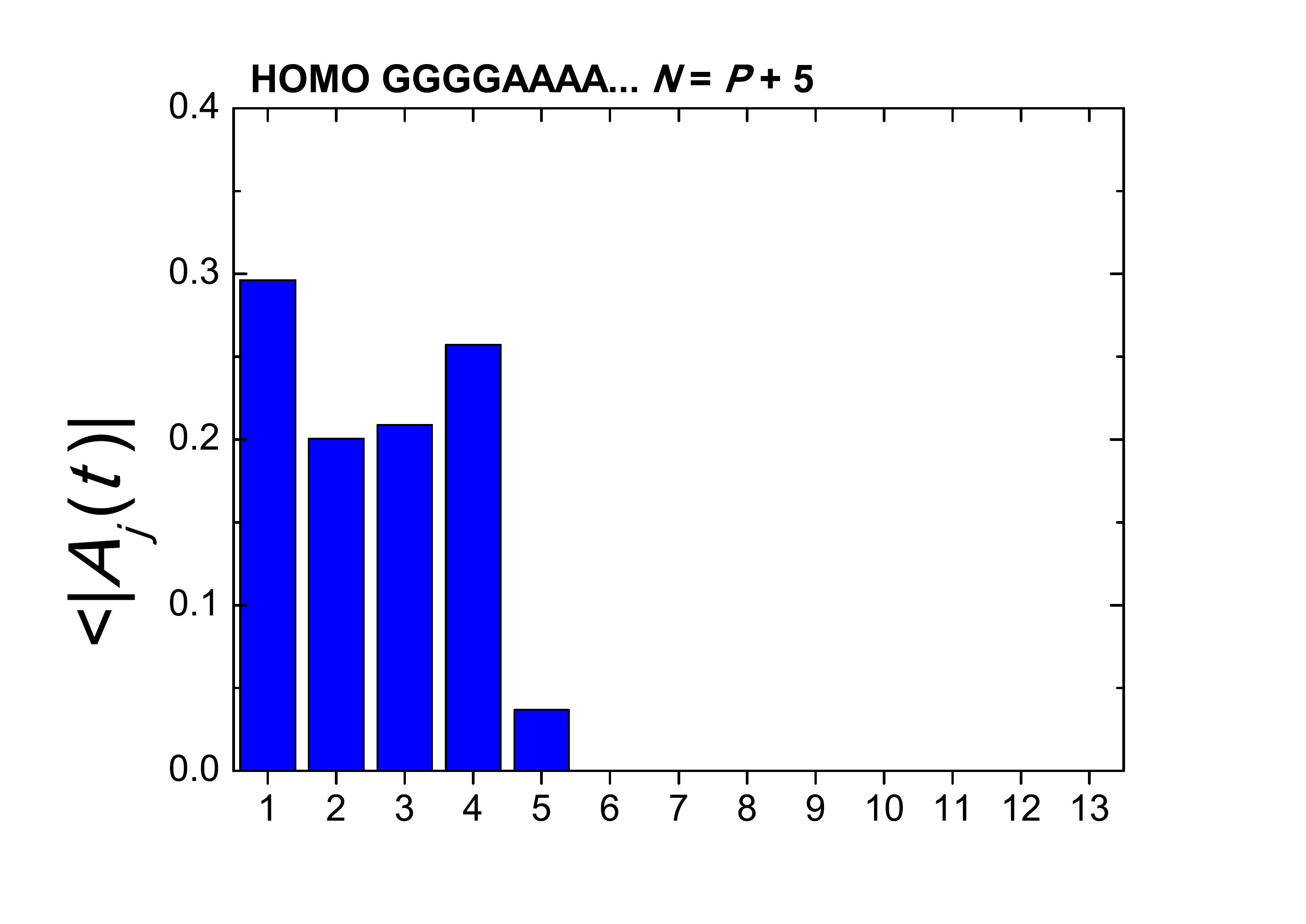}
\includegraphics[width=0.4\textwidth]{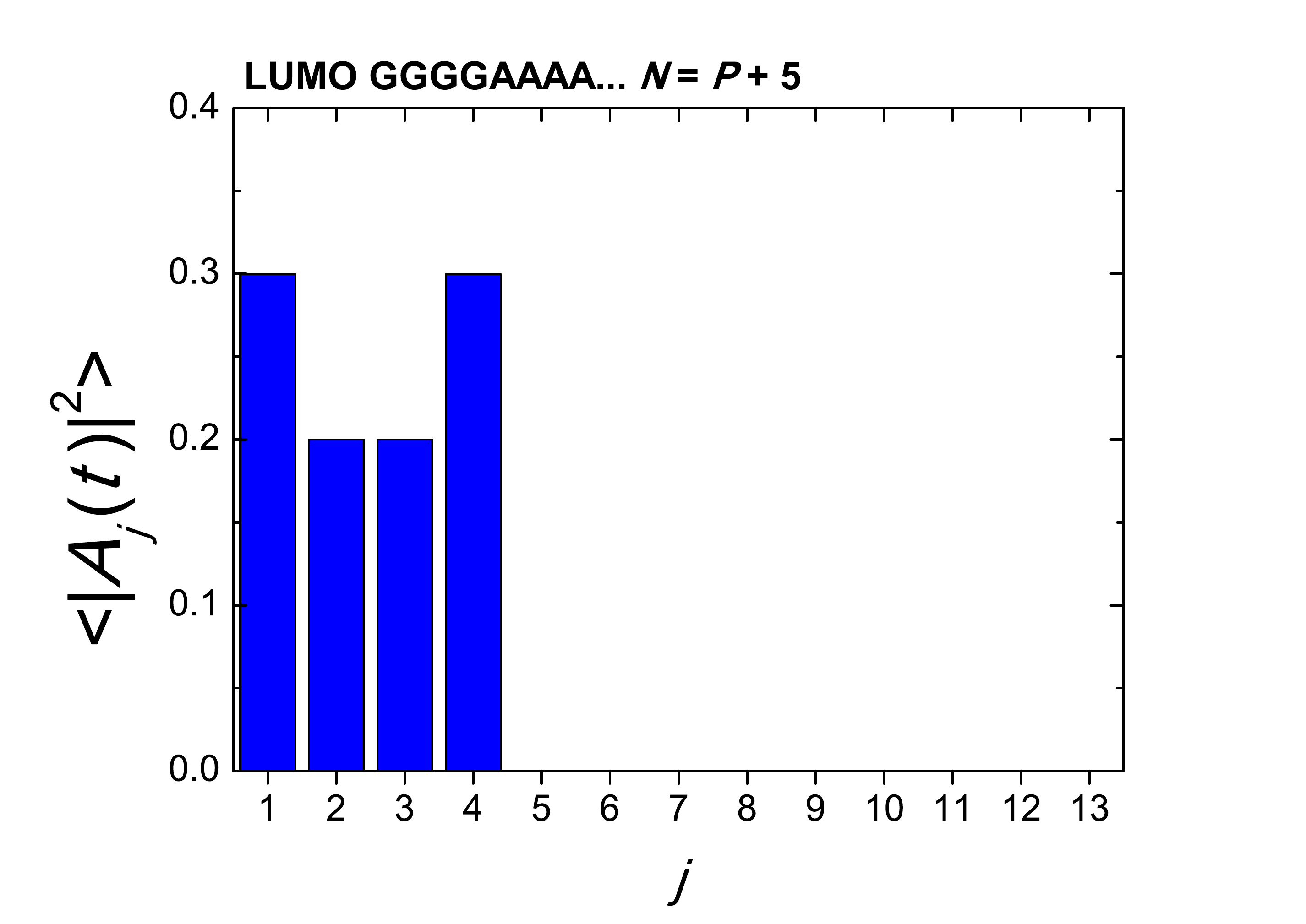}
\includegraphics[width=0.4\textwidth]{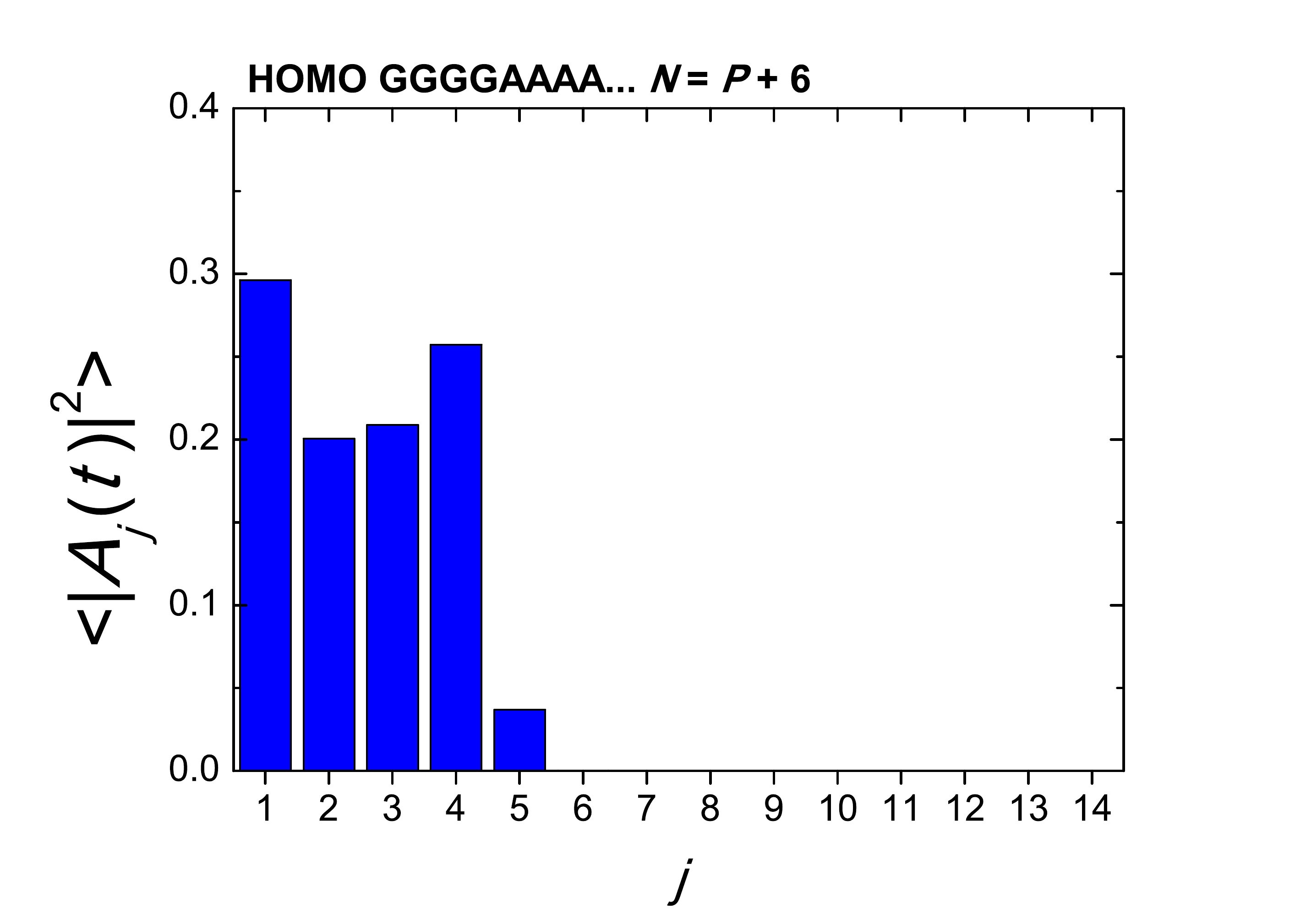}
\includegraphics[width=0.4\textwidth]{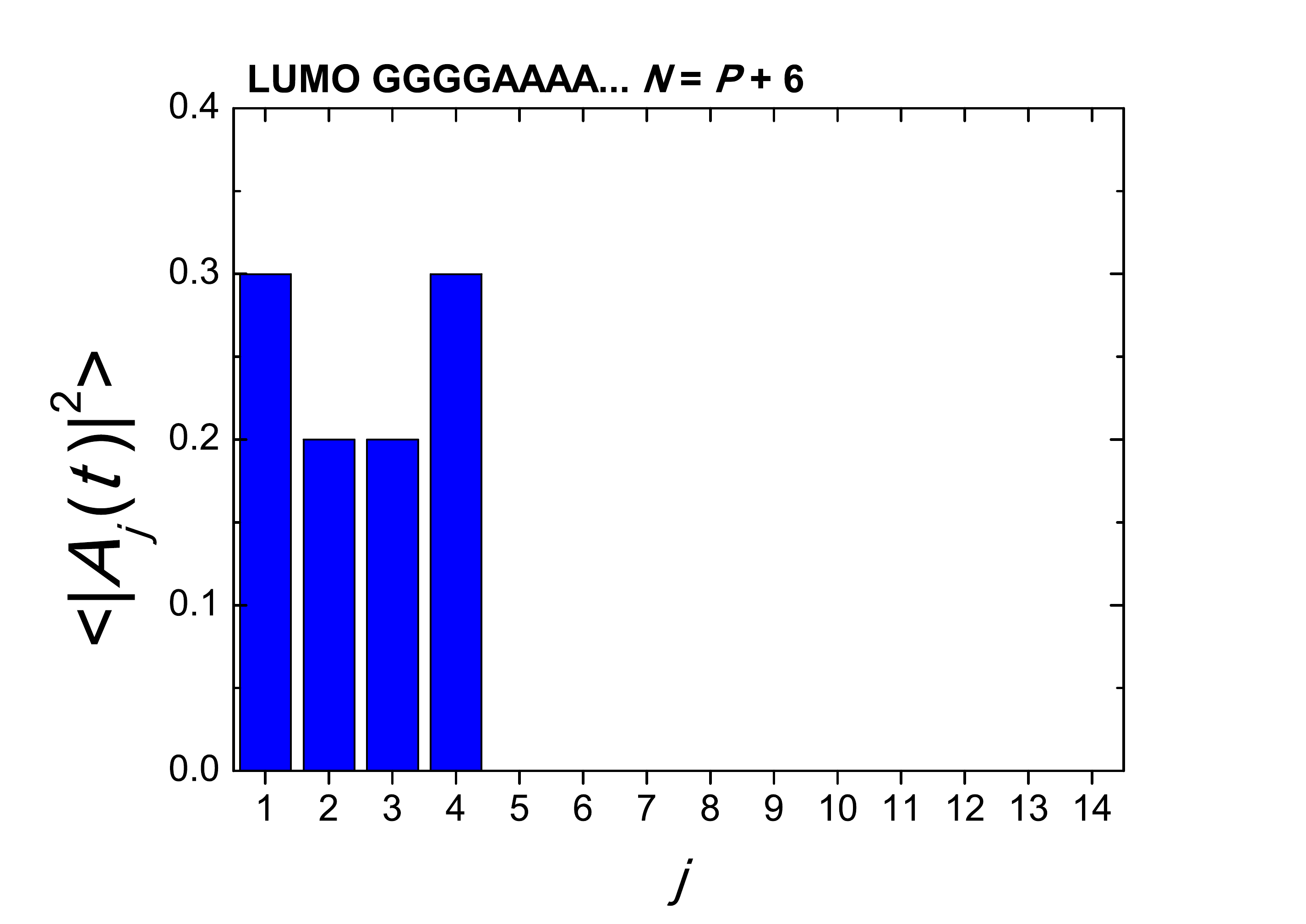}
\includegraphics[width=0.4\textwidth]{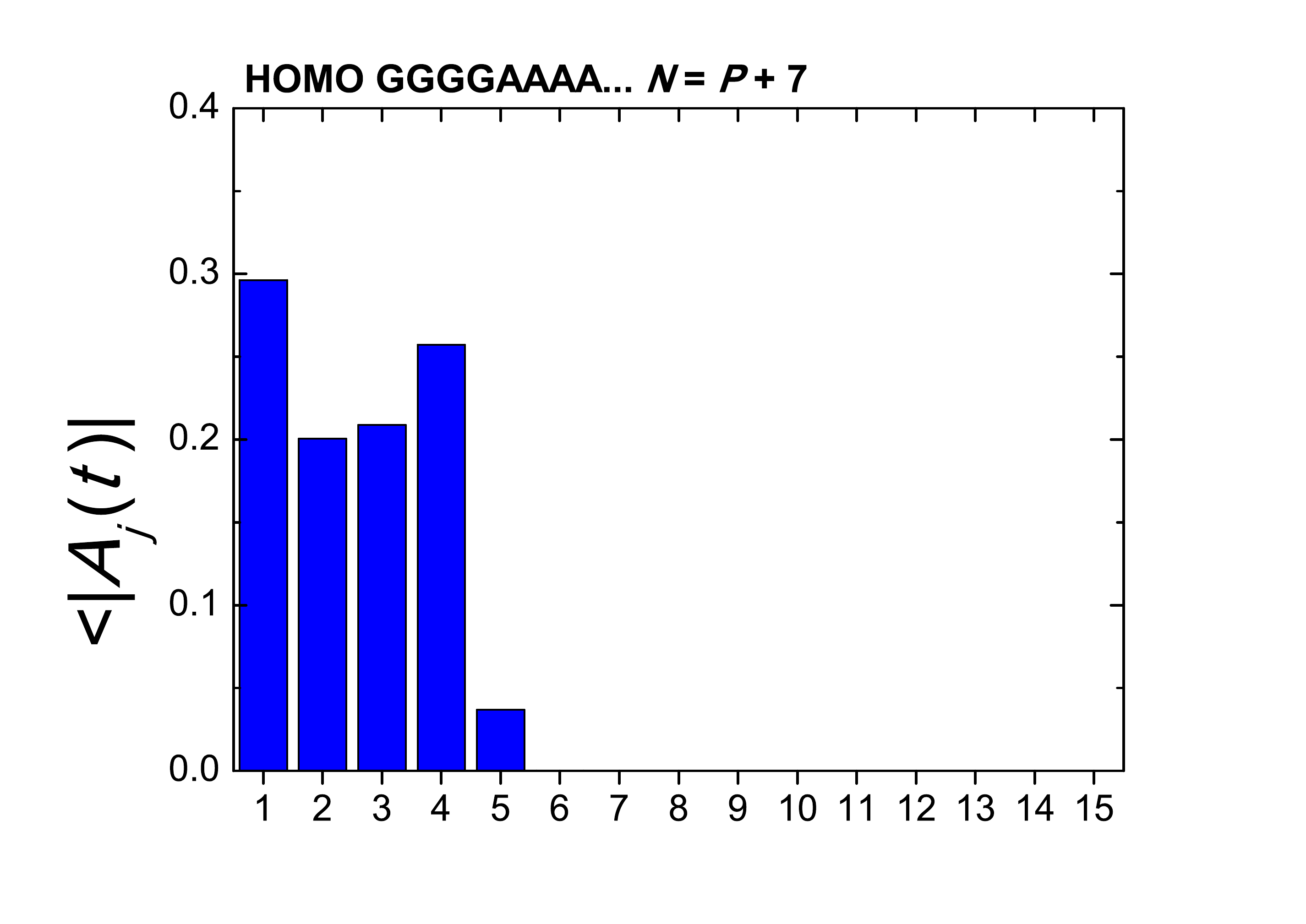}
\includegraphics[width=0.4\textwidth]{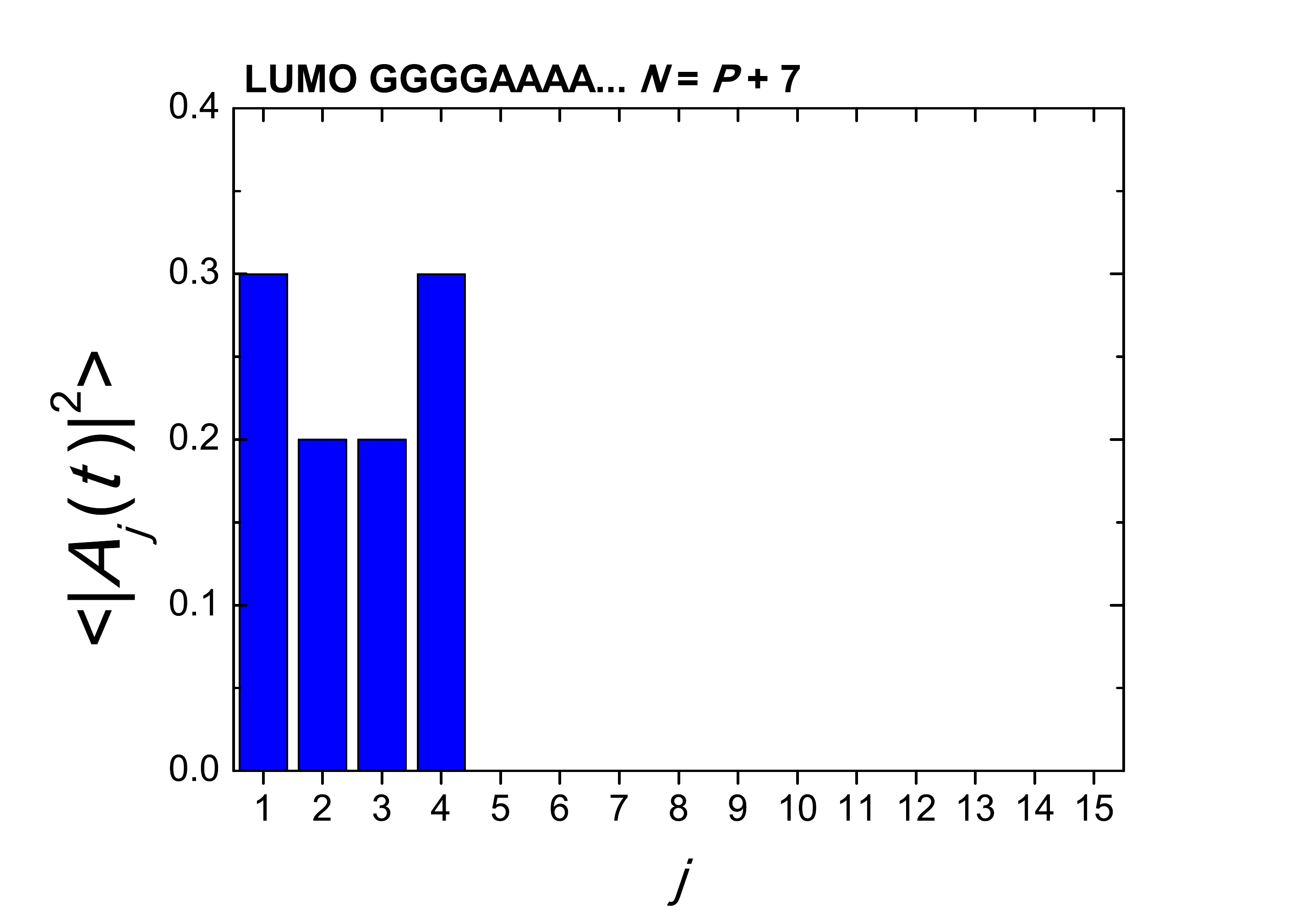}
\caption{\emph{...Continued from the previous page.} Mean (over time) probabilities to find the extra carrier at each monomer $j$, having placed it initially at the first monomer, for D8 (GGGGAAAA...) polymers, for the HOMO (left) and the LUMO (right). $N = P + \tau$, $\tau = 0, 1, \dots, P-1$.}
\end{figure*}

\begin{figure*}[!h]
\includegraphics[width=0.4\textwidth]{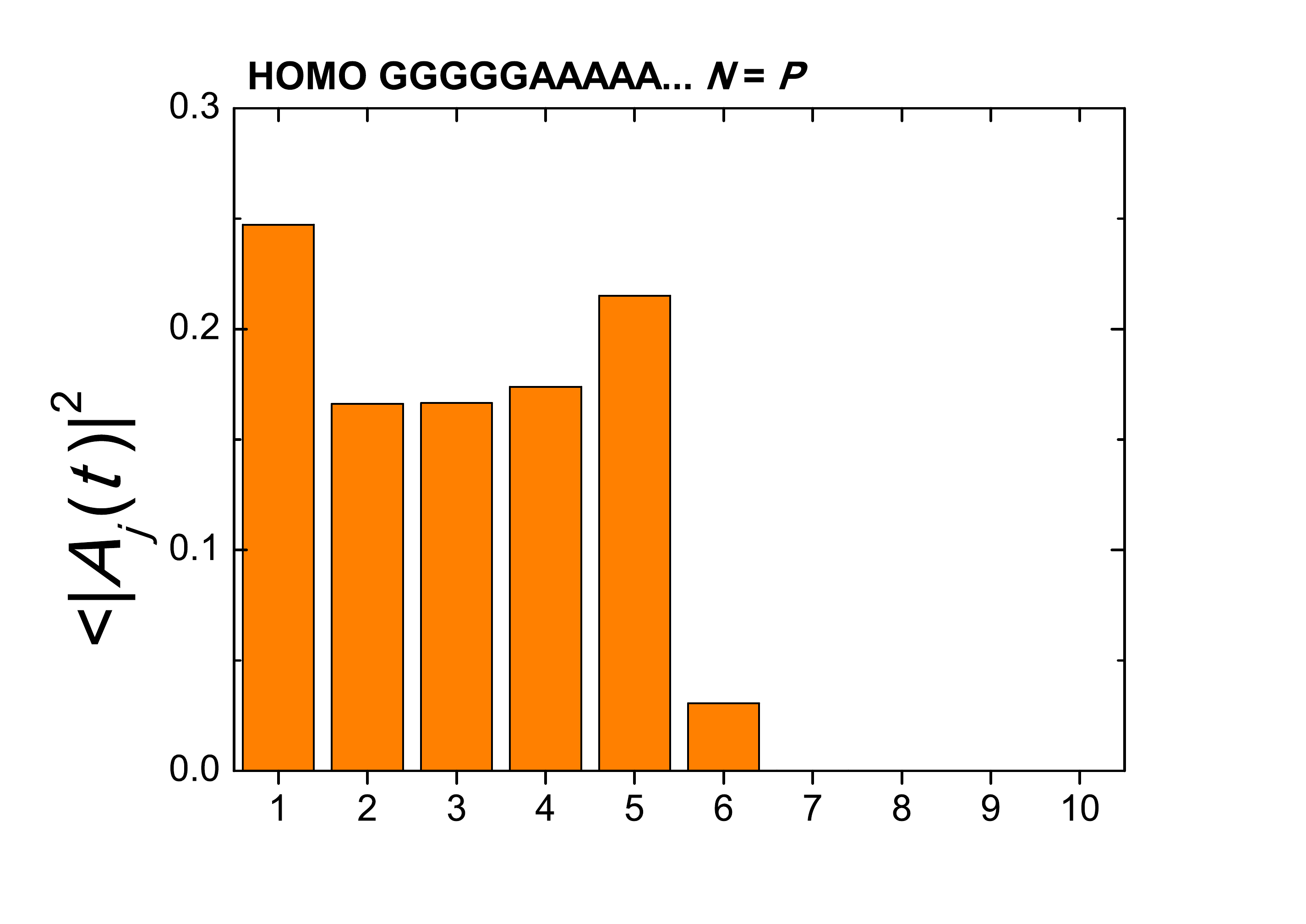}
\includegraphics[width=0.4\textwidth]{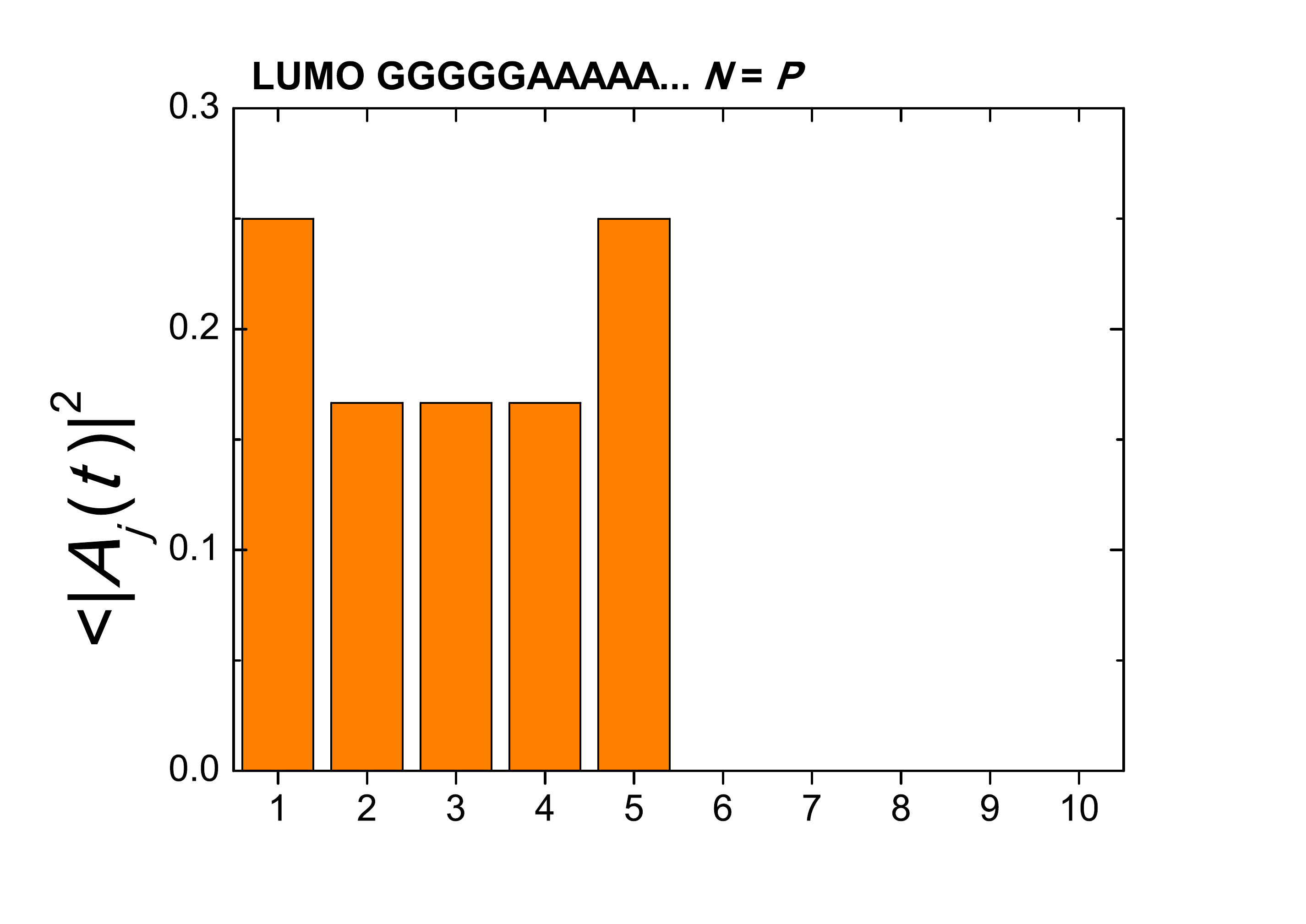}
\includegraphics[width=0.4\textwidth]{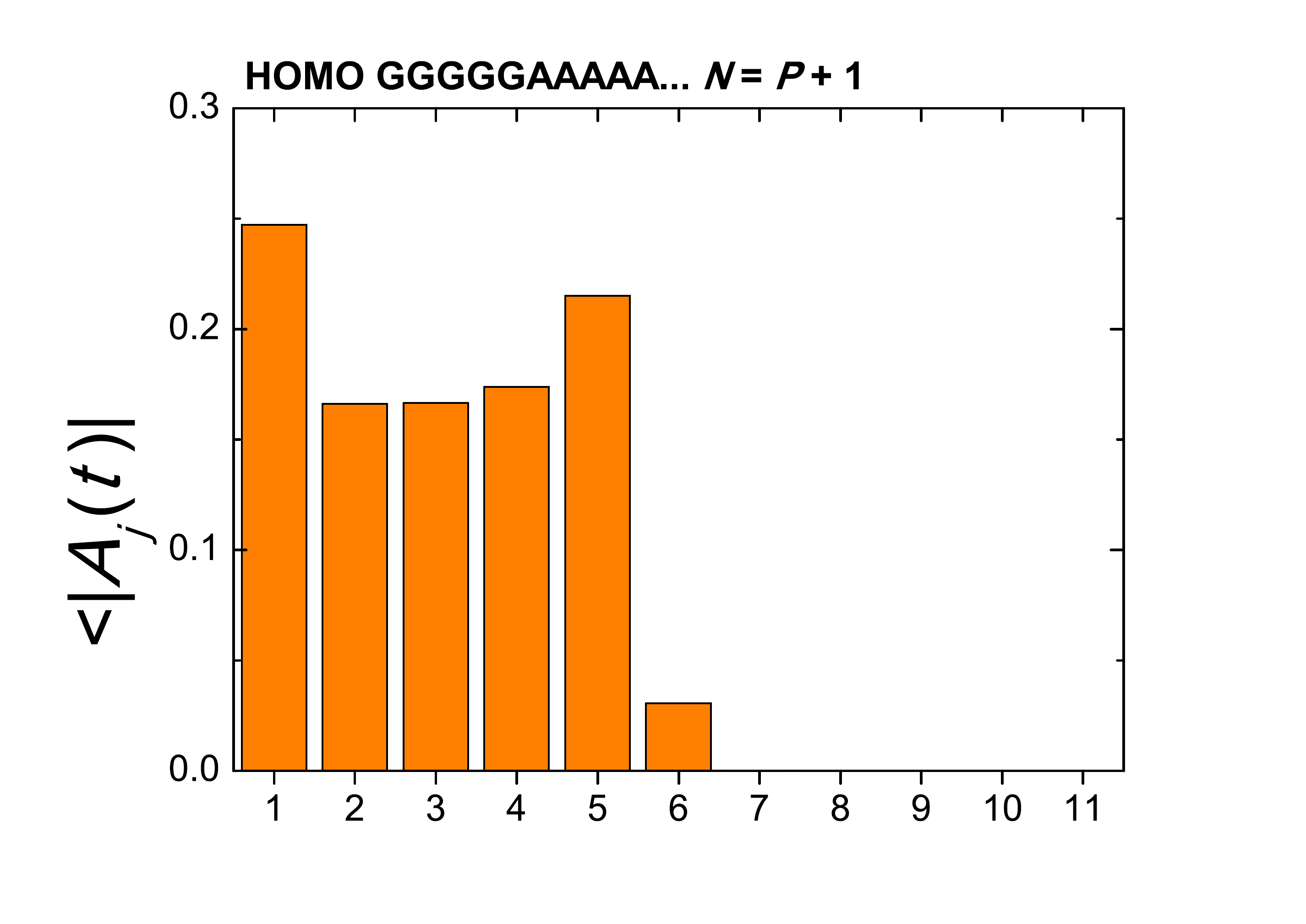}
\includegraphics[width=0.4\textwidth]{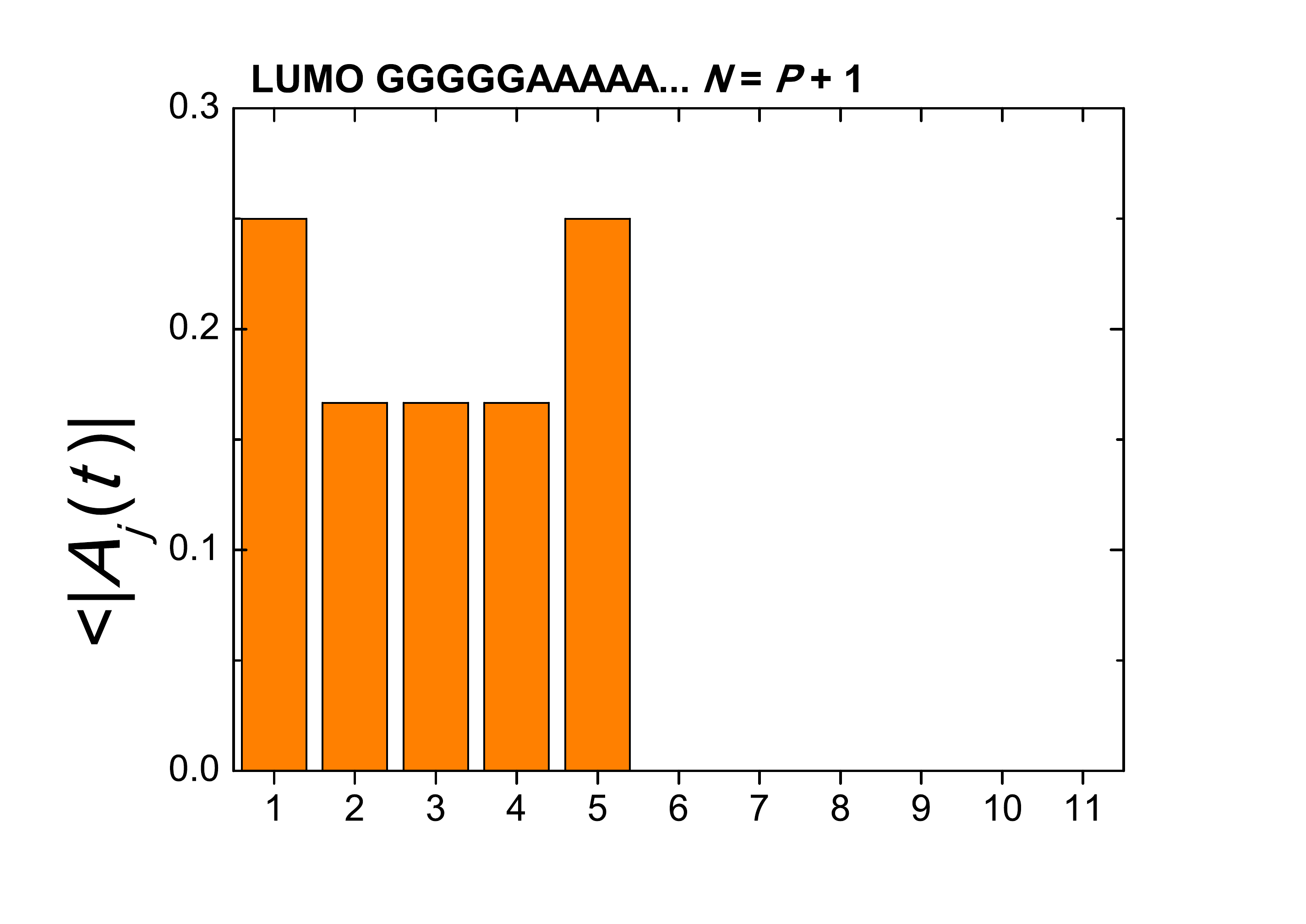}
\includegraphics[width=0.4\textwidth]{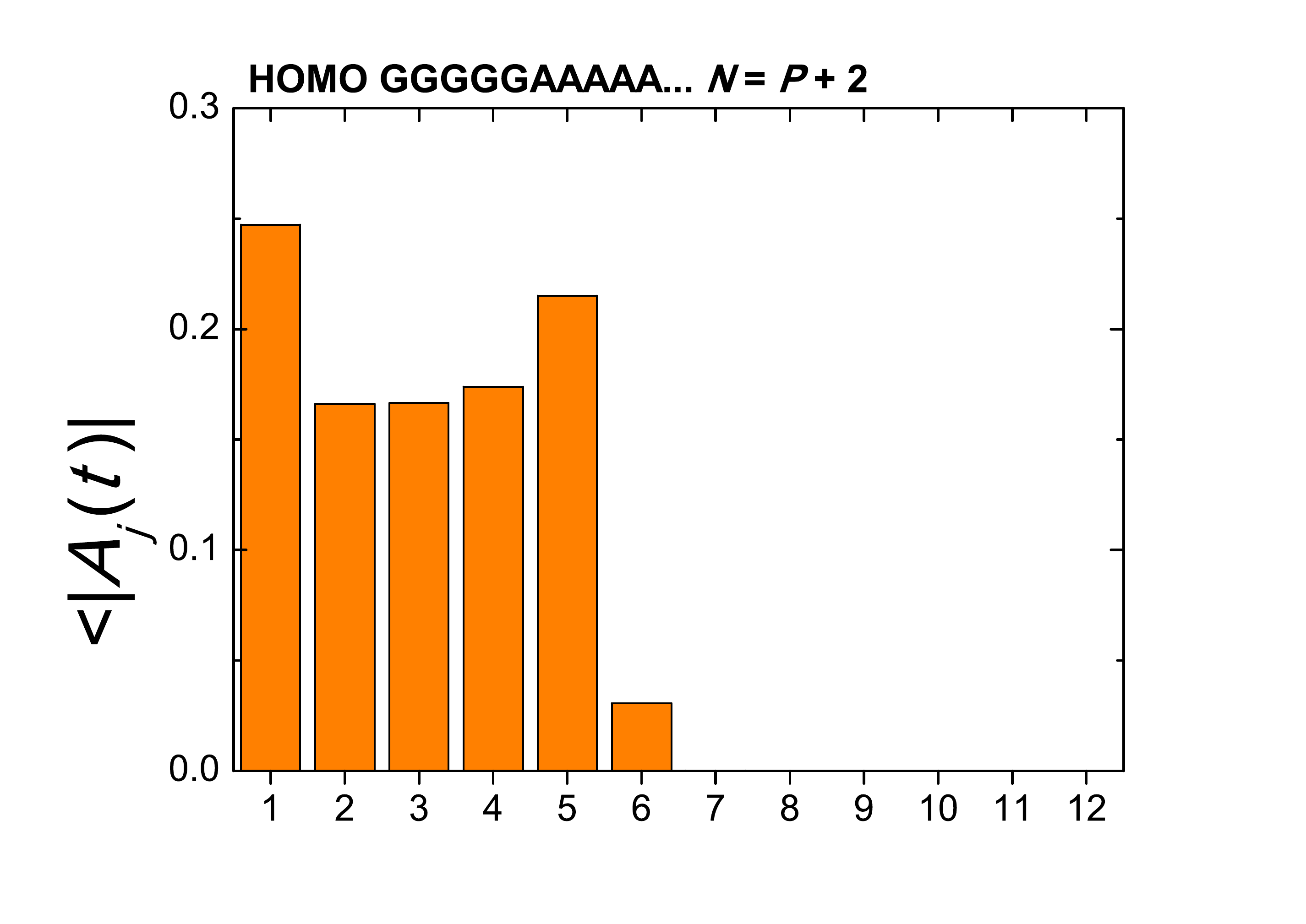}
\includegraphics[width=0.4\textwidth]{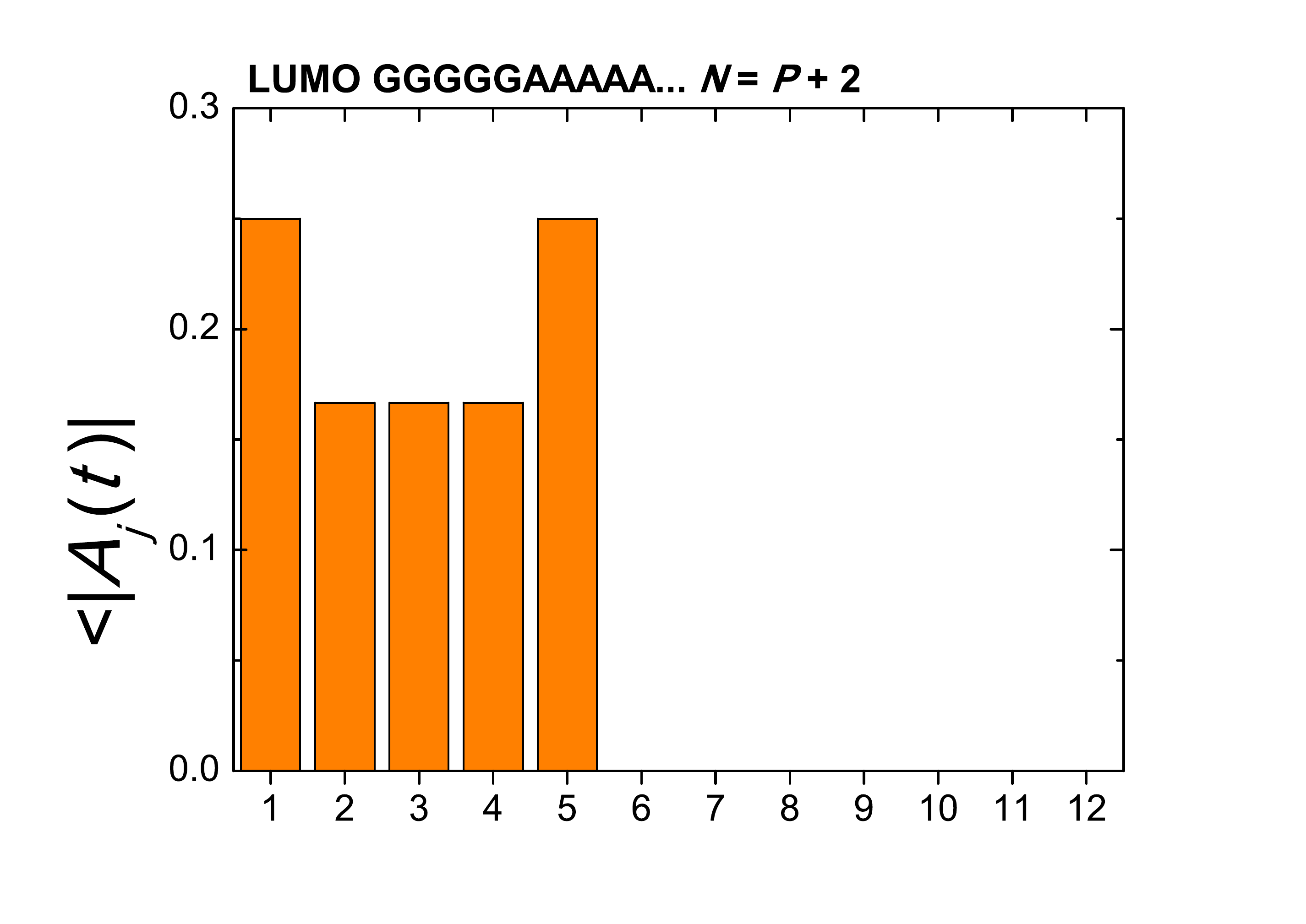}
\includegraphics[width=0.4\textwidth]{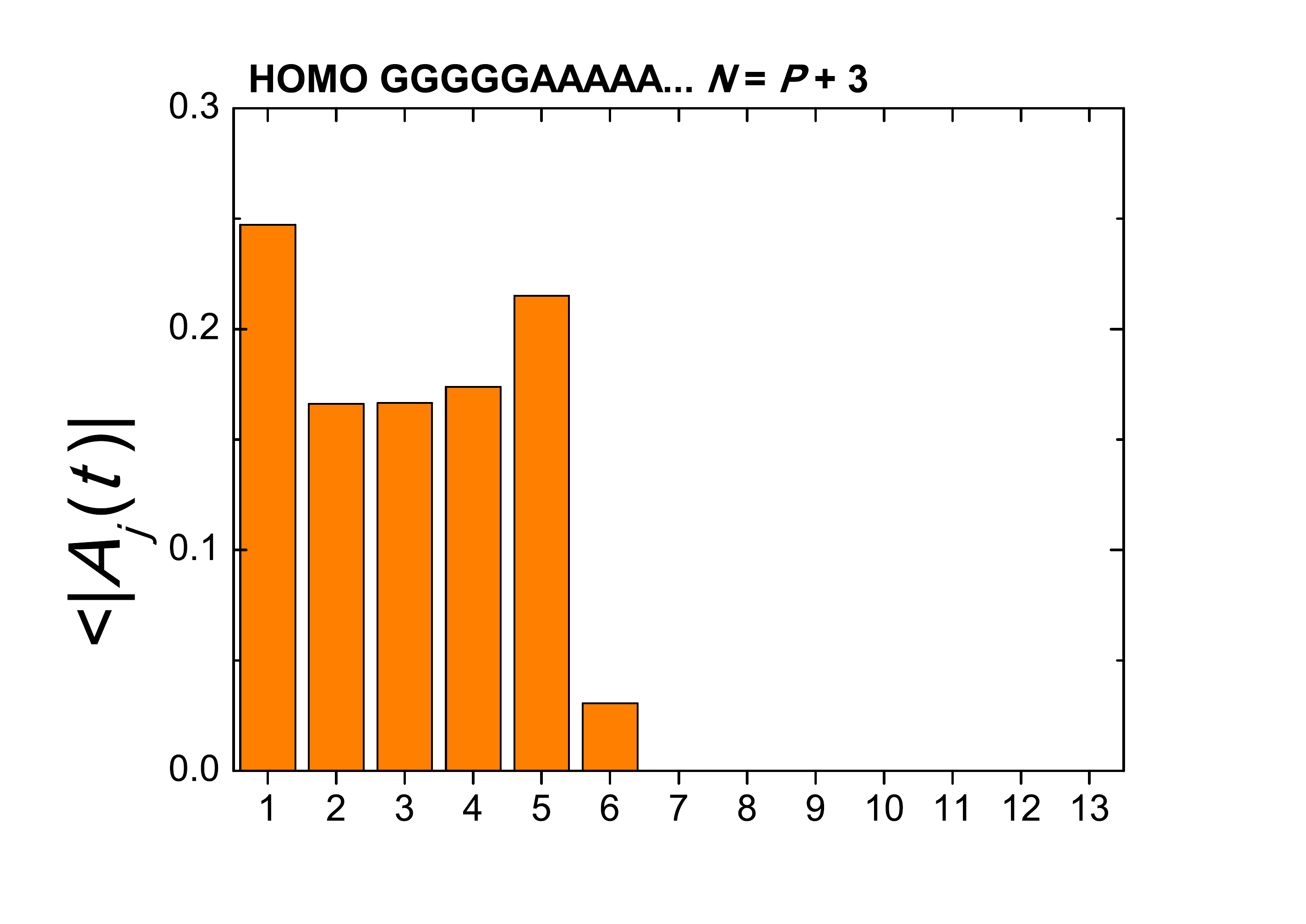}
\includegraphics[width=0.4\textwidth]{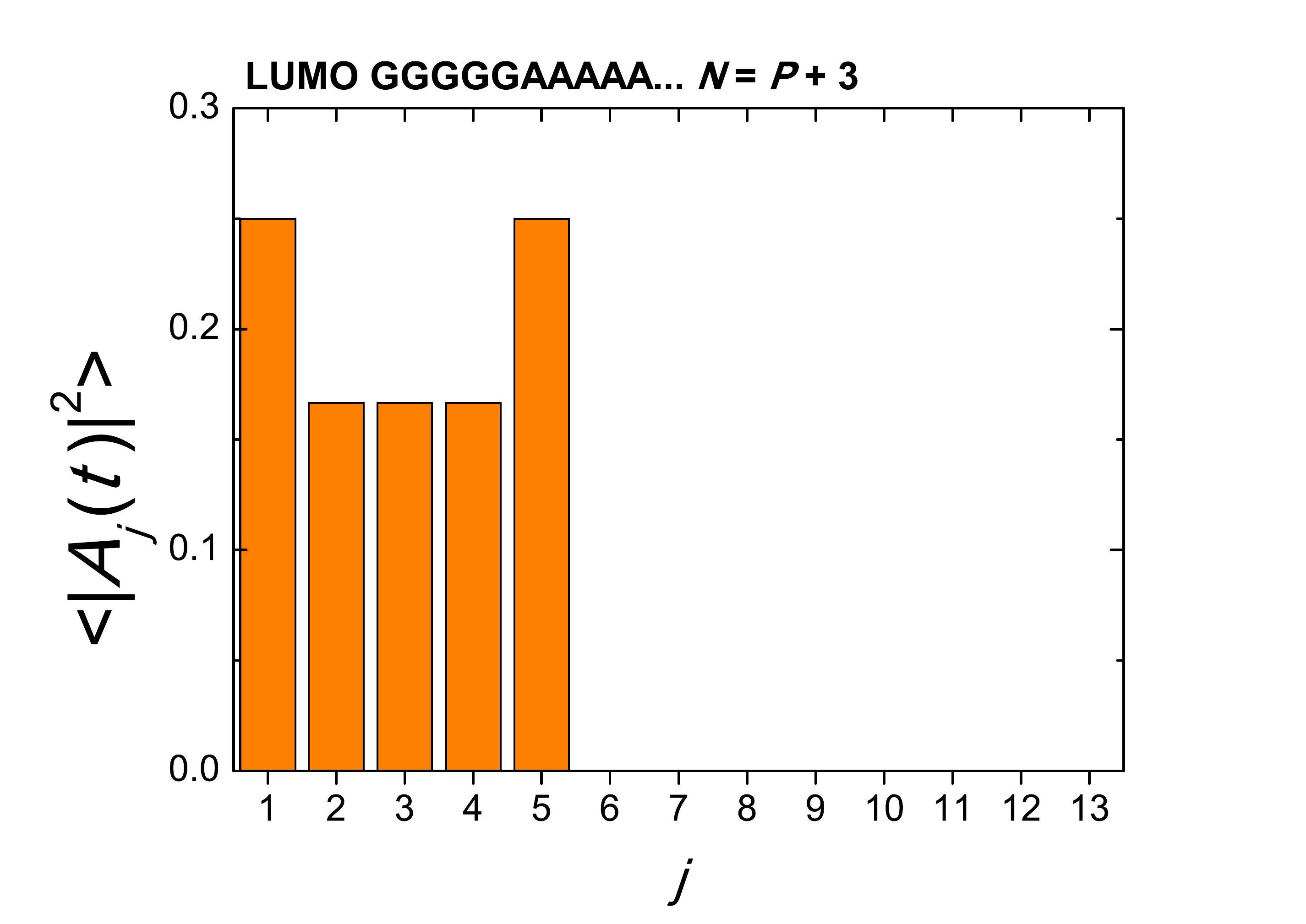}
\caption{Mean (over time) probabilities to find the extra carrier at each monomer $j$, having placed it initially at the first monomer, for D10 (GGGGGAAAAA...) polymers, for the HOMO (left) and the LUMO (right). $N = P + \tau$, $\tau = 0, 1, \dots, P-1$. \emph{Continued at the next page...}}
\label{fig:ProbabilitiesHL-D10}
\end{figure*}
\begin{figure*}[!h]
\addtocounter{figure}{-1}
\includegraphics[width=0.4\textwidth]{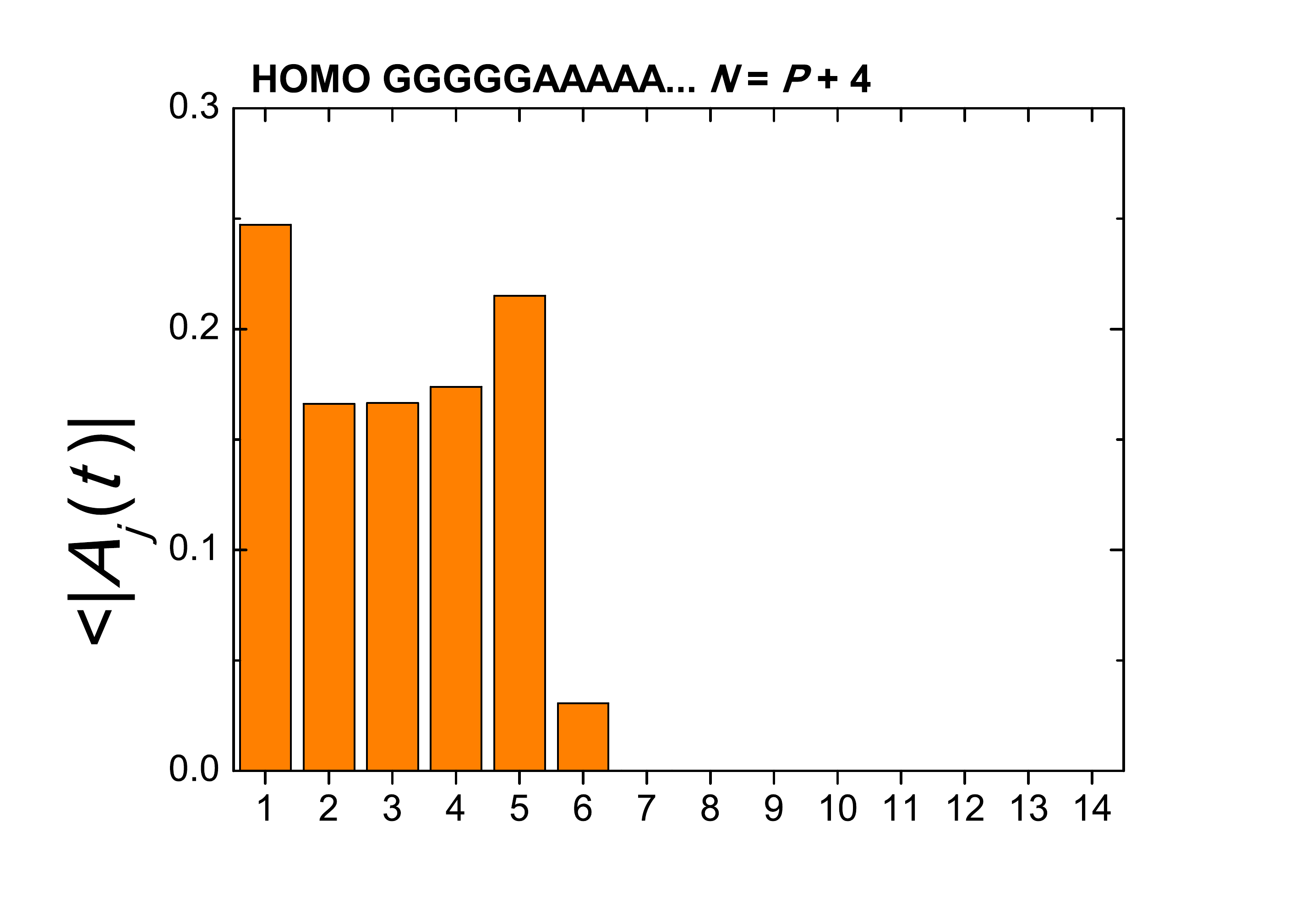}
\includegraphics[width=0.4\textwidth]{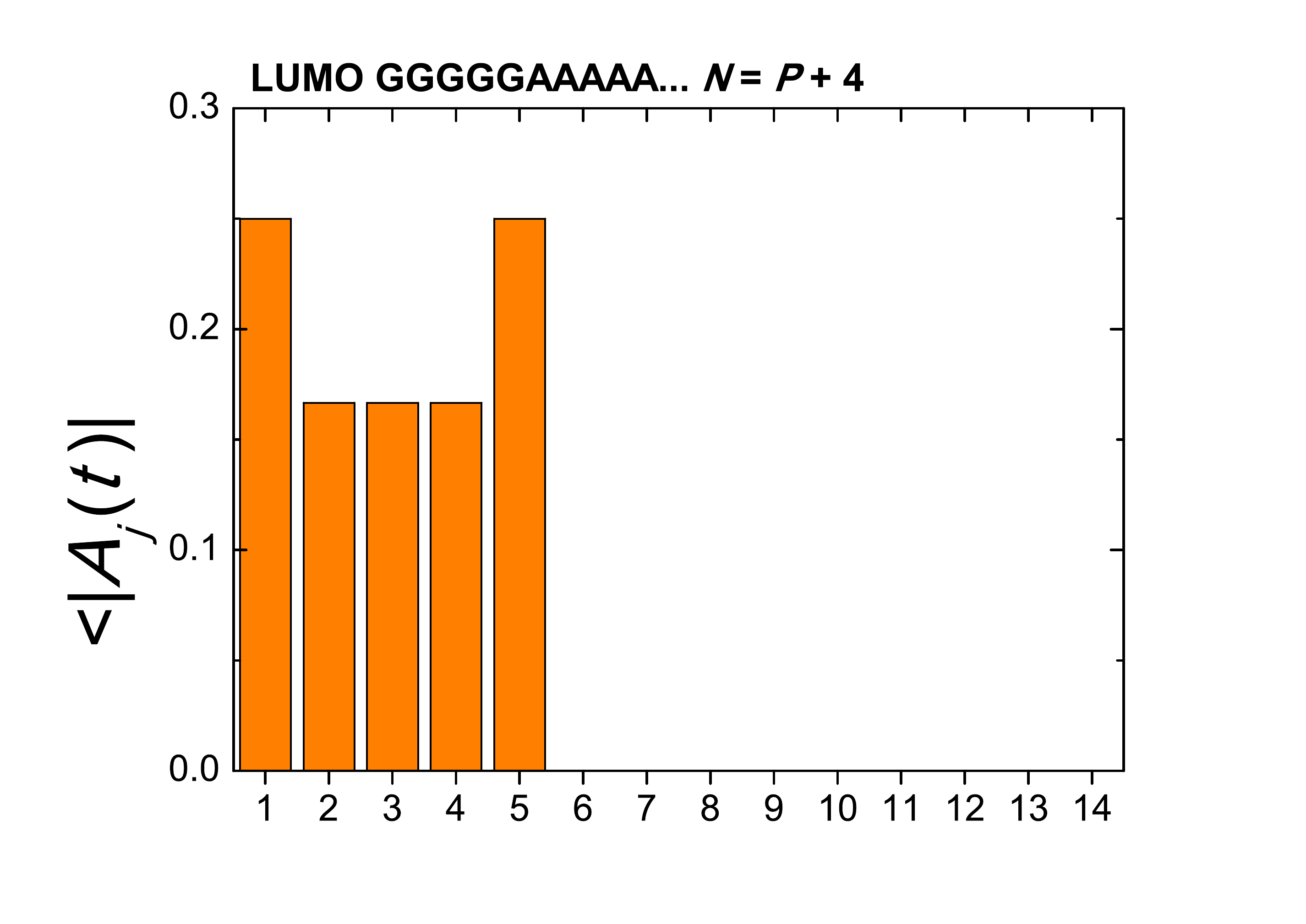}
\includegraphics[width=0.4\textwidth]{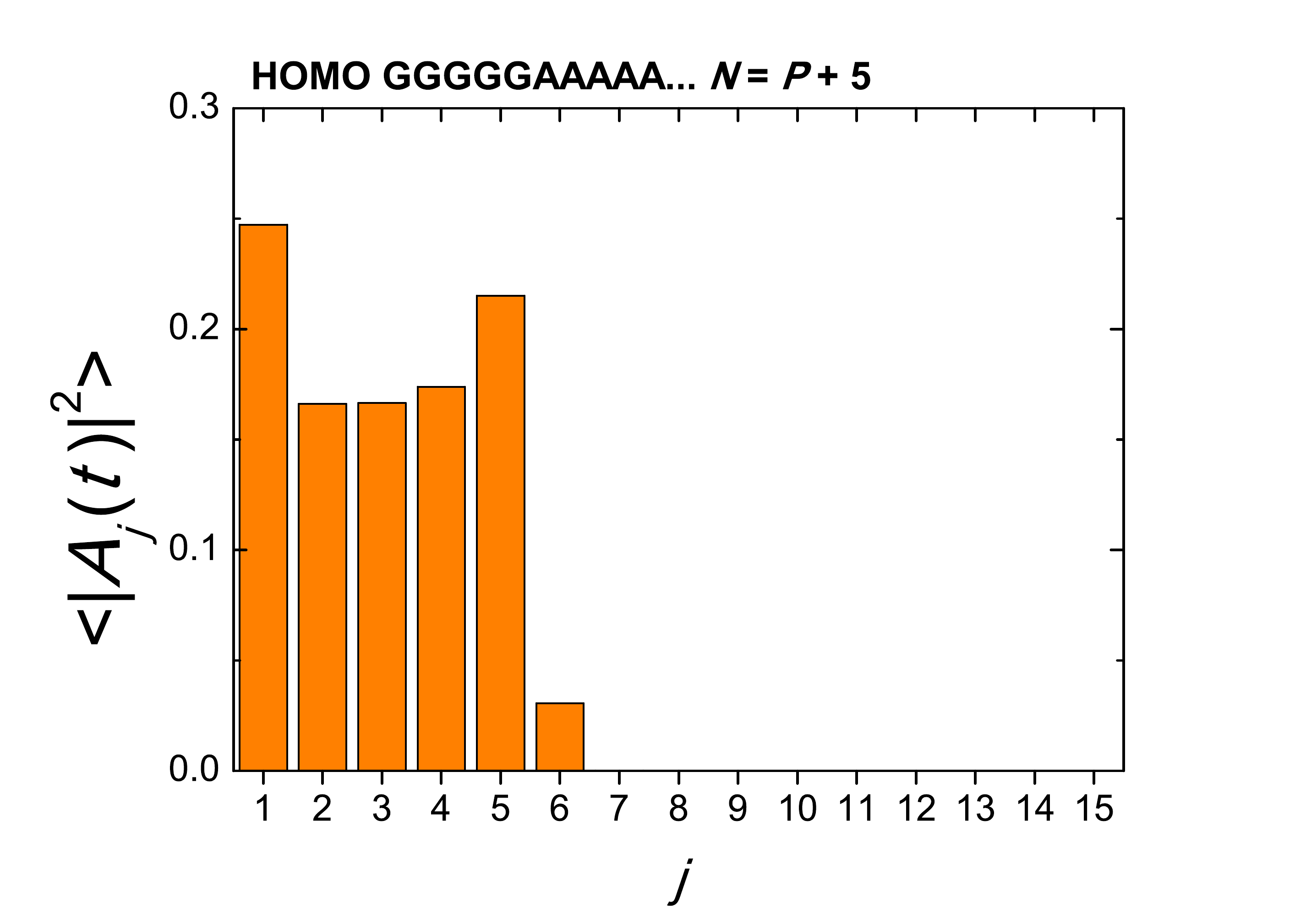}
\includegraphics[width=0.4\textwidth]{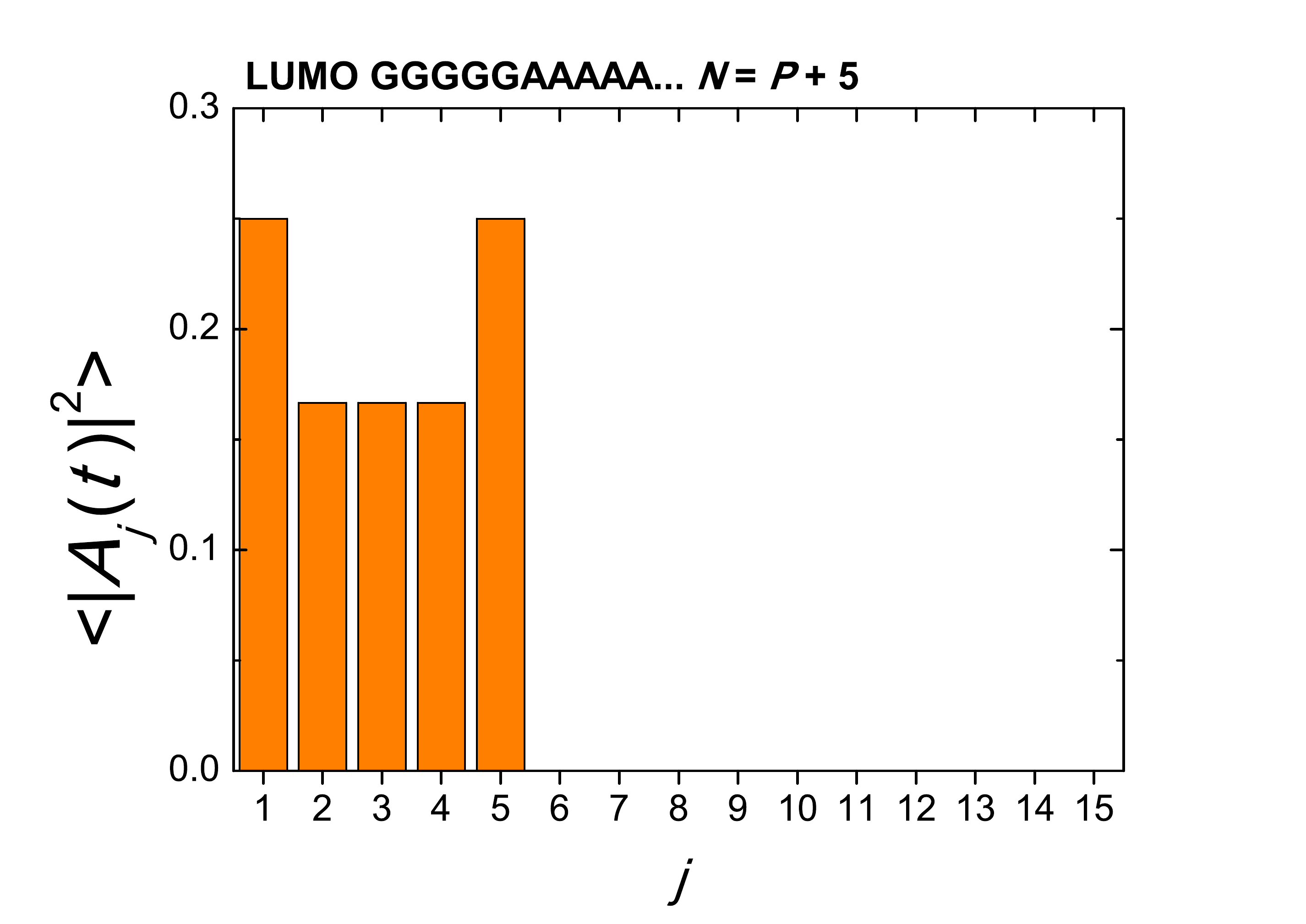}
\includegraphics[width=0.4\textwidth]{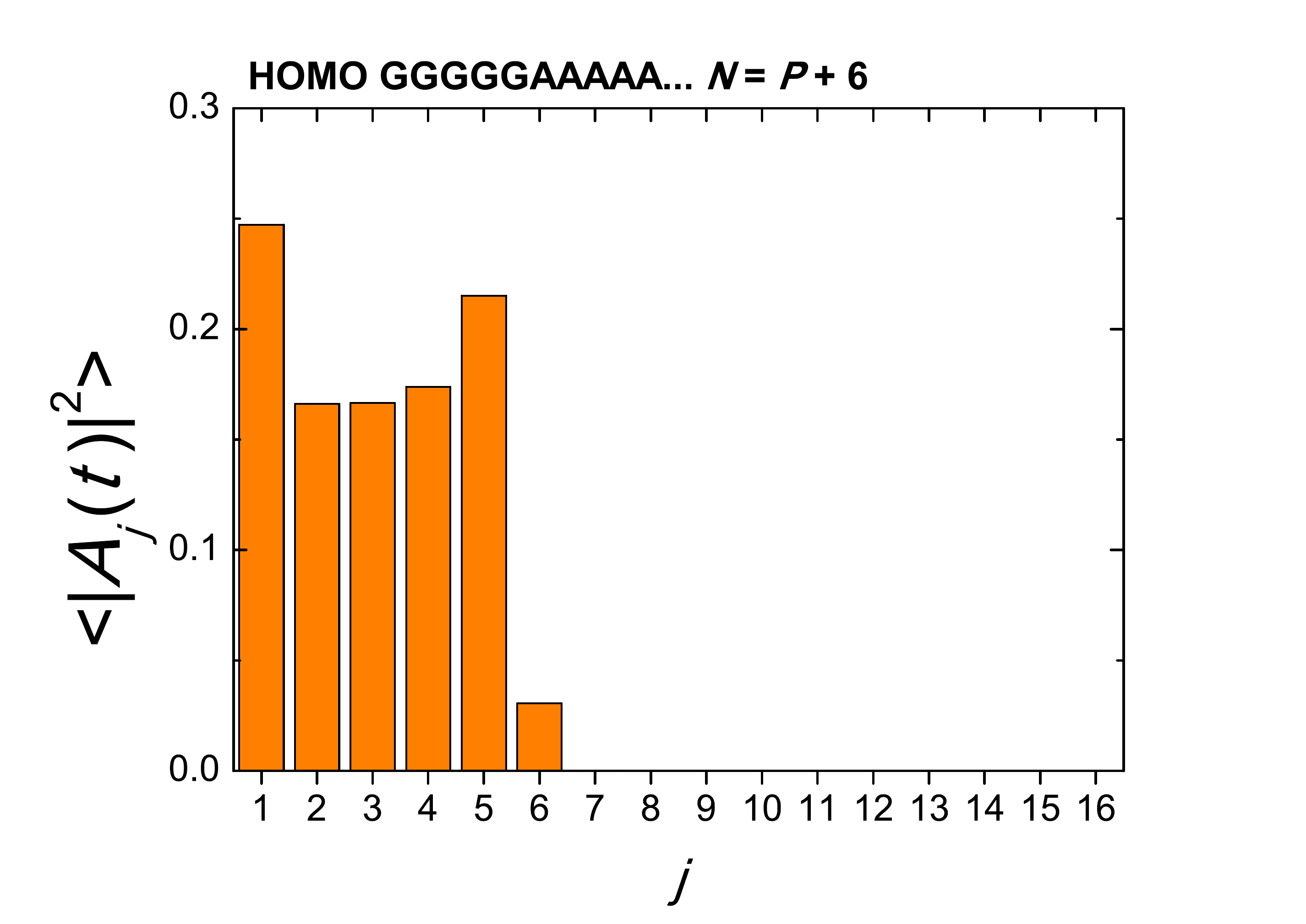}
\includegraphics[width=0.4\textwidth]{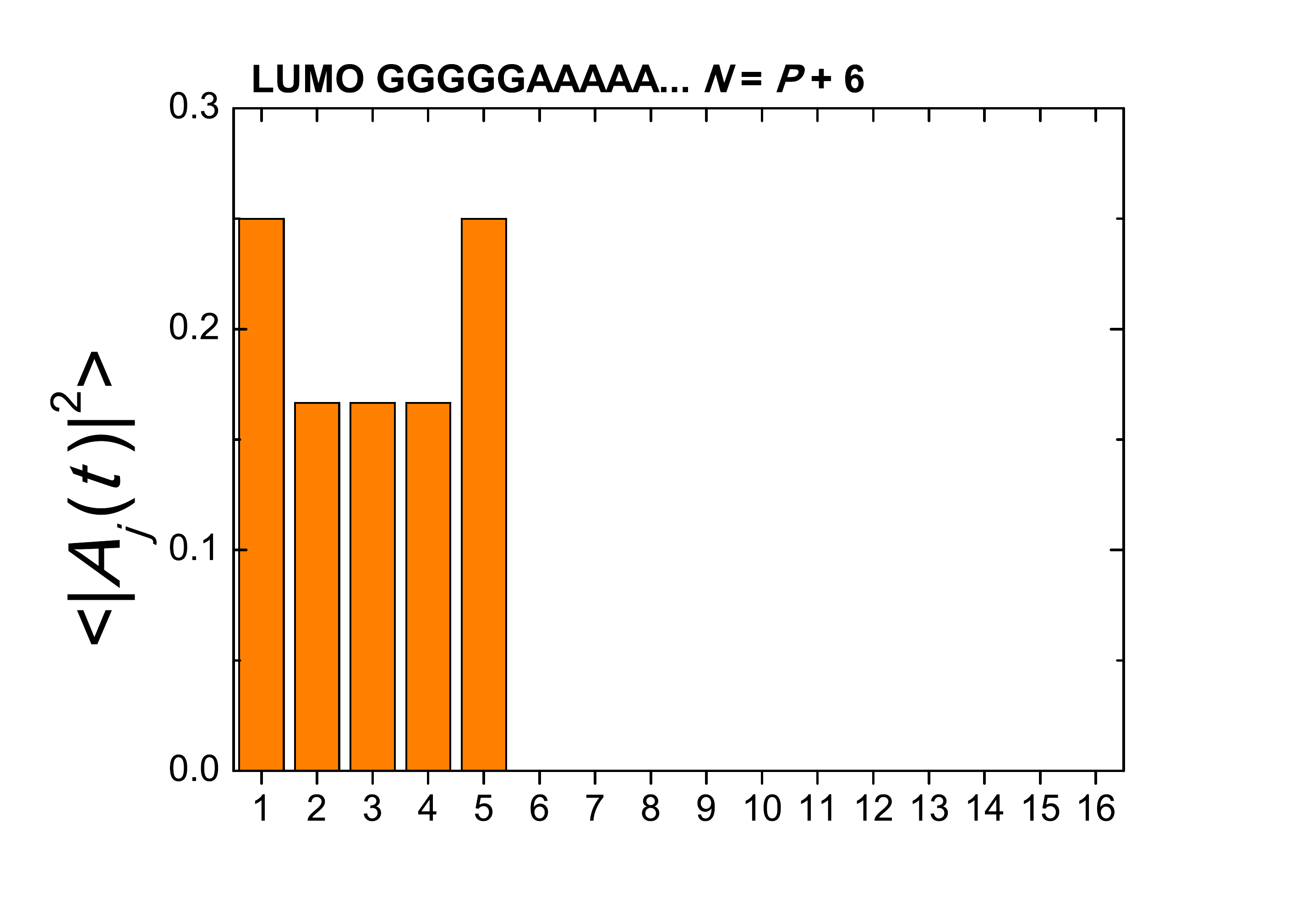}
\includegraphics[width=0.4\textwidth]{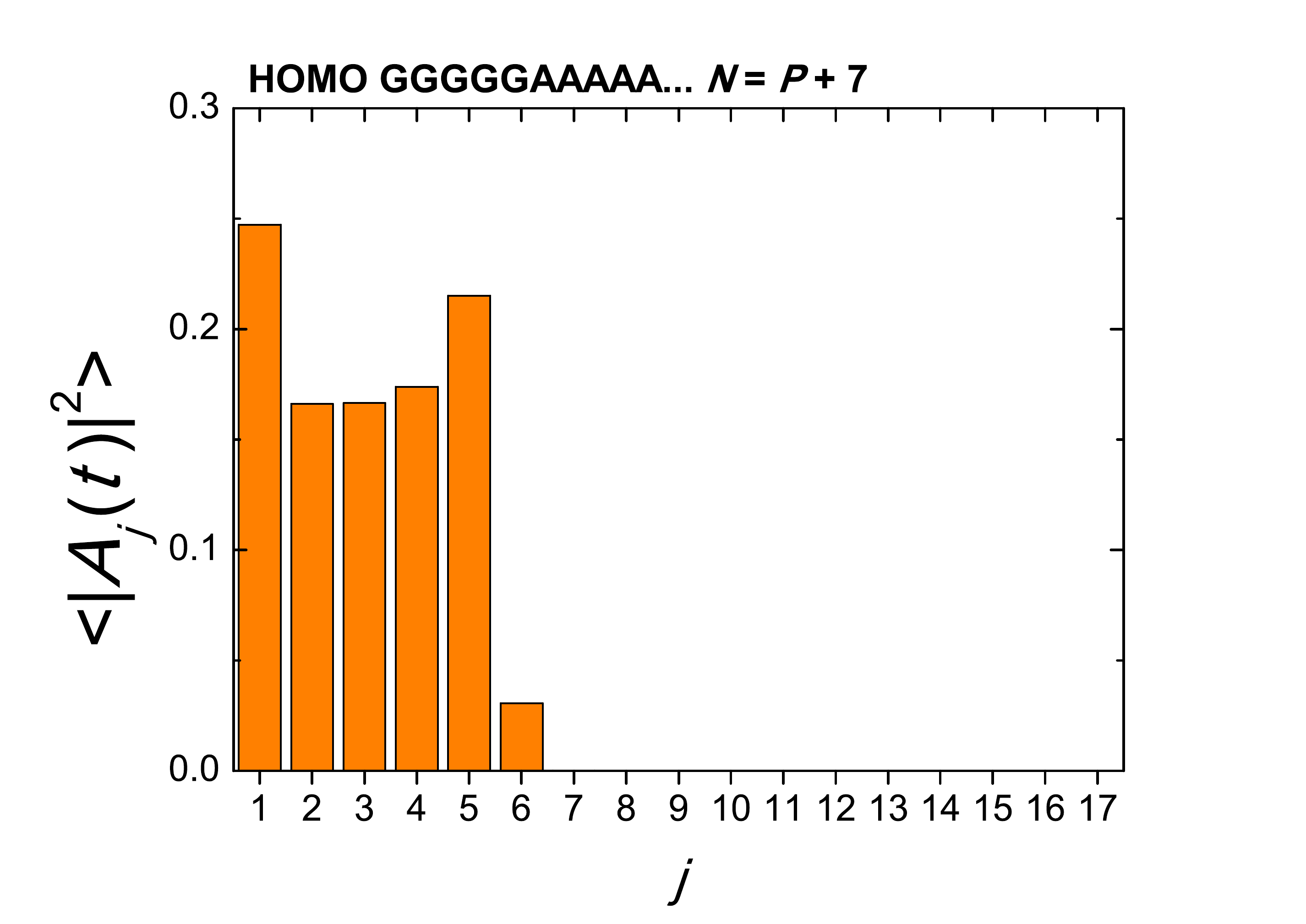}
\includegraphics[width=0.4\textwidth]{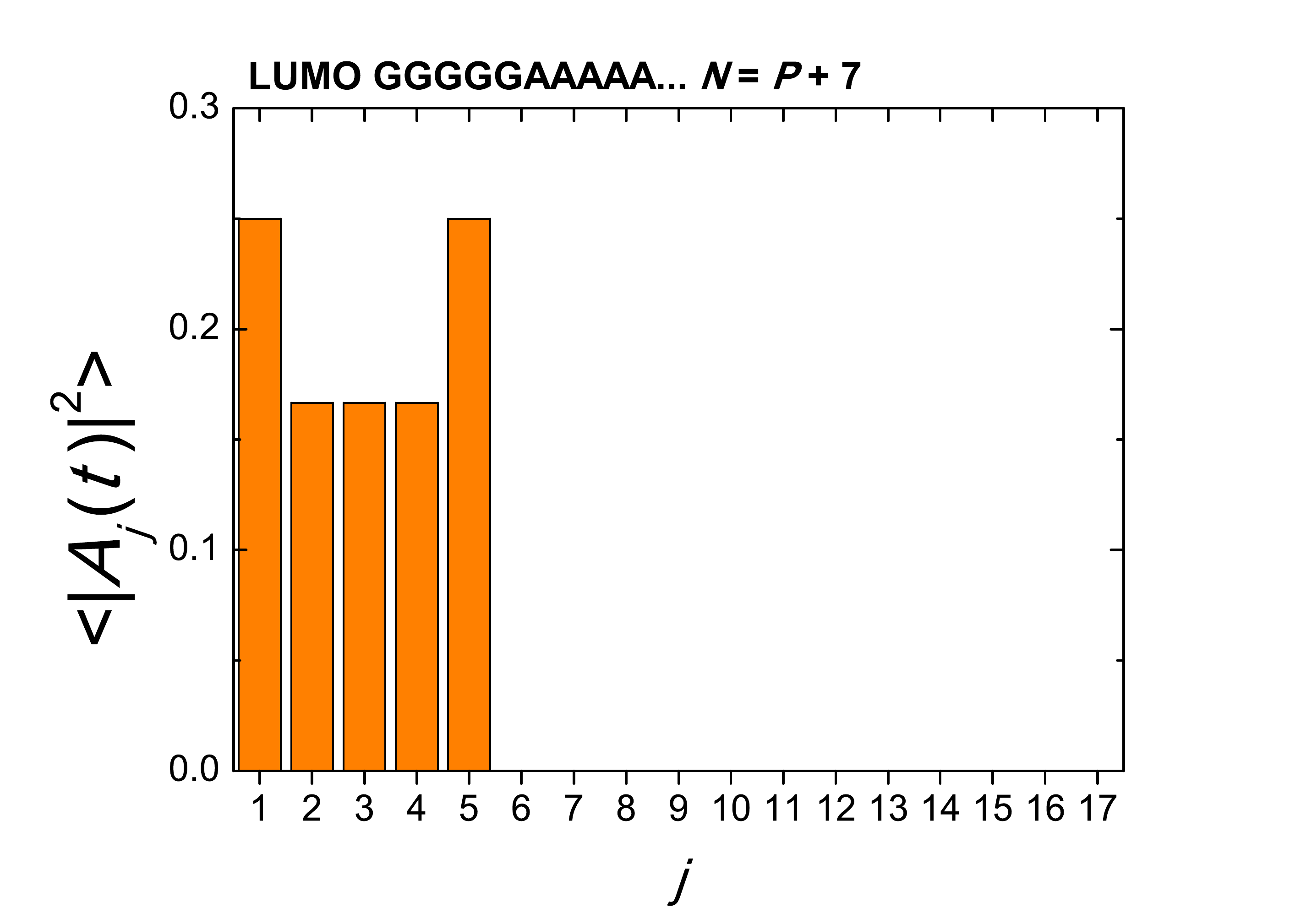}
\caption{\emph{...Continued from the previous page.} Mean (over time) probabilities to find the extra carrier at each monomer $j$, having placed it initially at the first monomer, for D10 (GGGGGAAAAA...) polymers, for the HOMO (left) and the LUMO (right). $N = P + \tau$, $\tau = 0, 1, \dots, P-1$. \emph{Continued at the next page...}}
\end{figure*}
\begin{figure*}[!h]
\addtocounter{figure}{-1}
\includegraphics[width=0.4\textwidth]{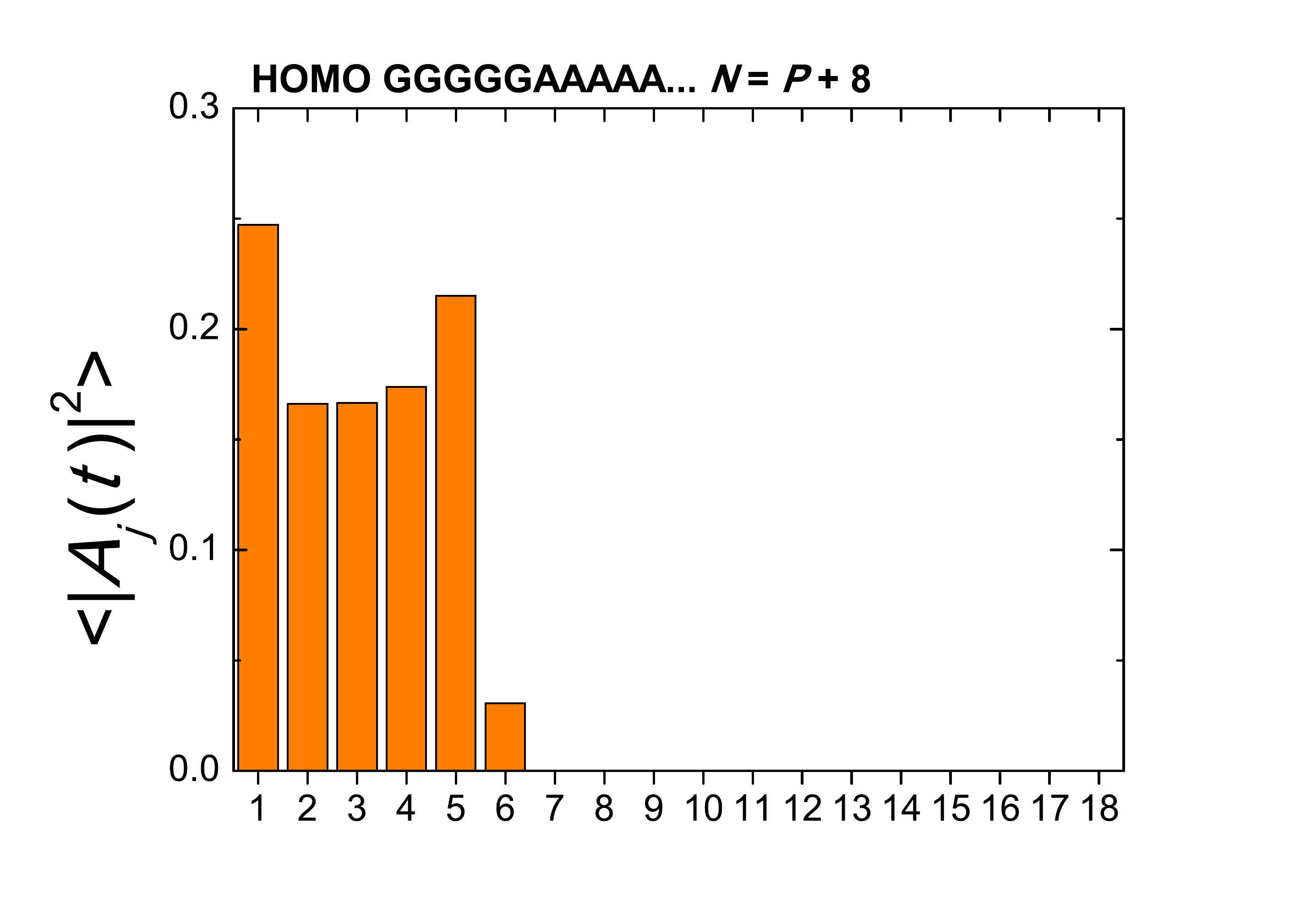}
\includegraphics[width=0.4\textwidth]{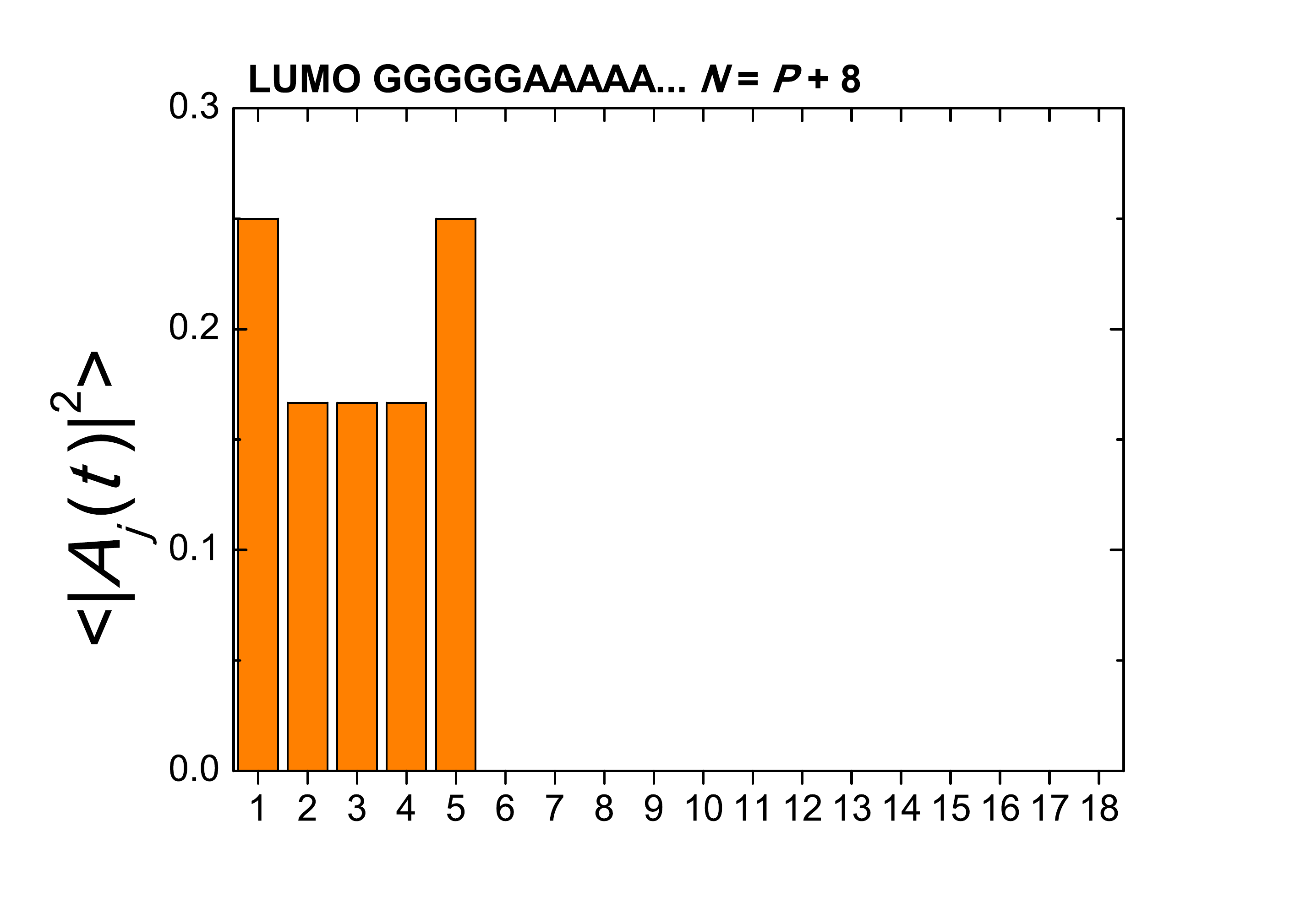}
\includegraphics[width=0.4\textwidth]{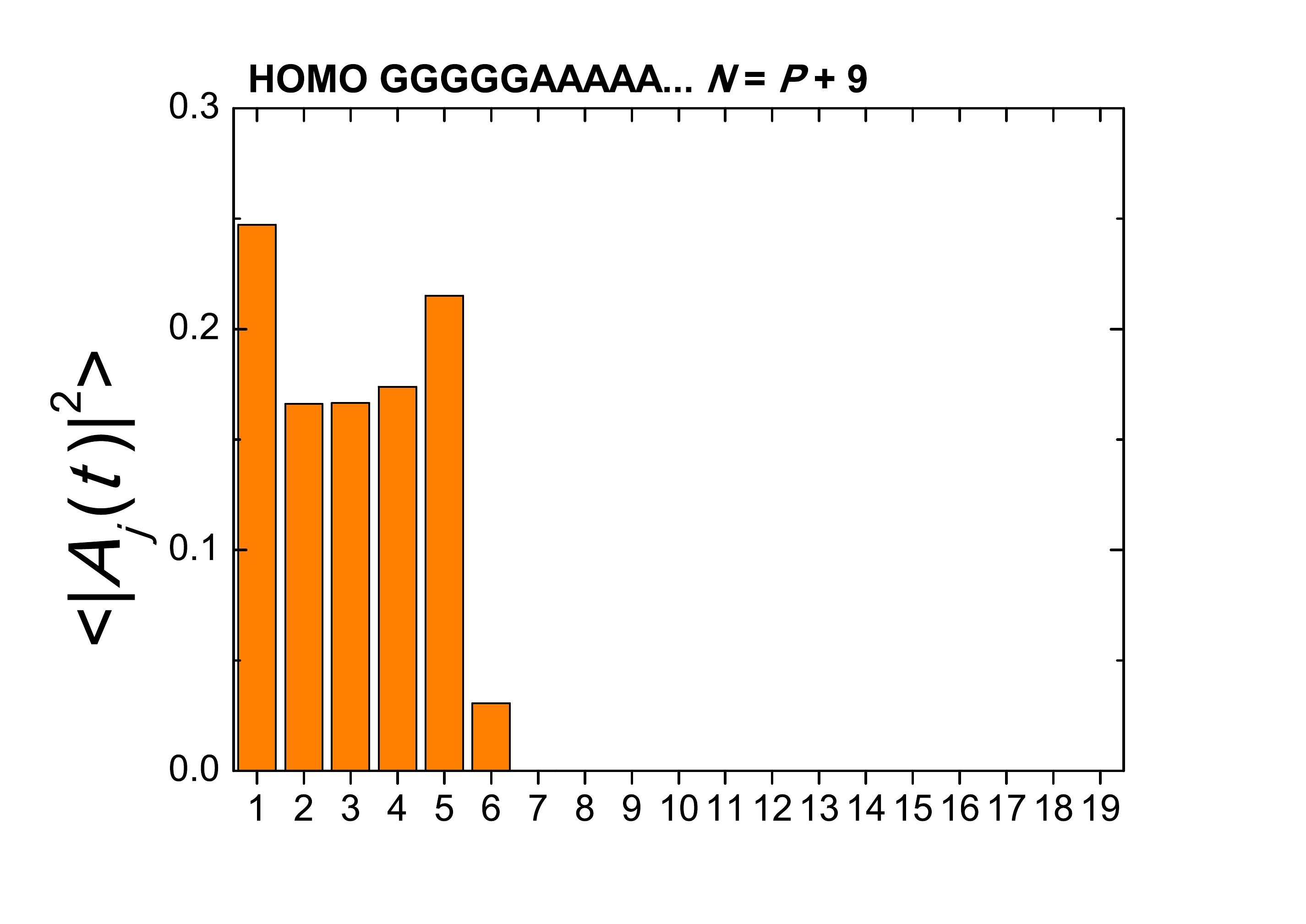}
\includegraphics[width=0.4\textwidth]{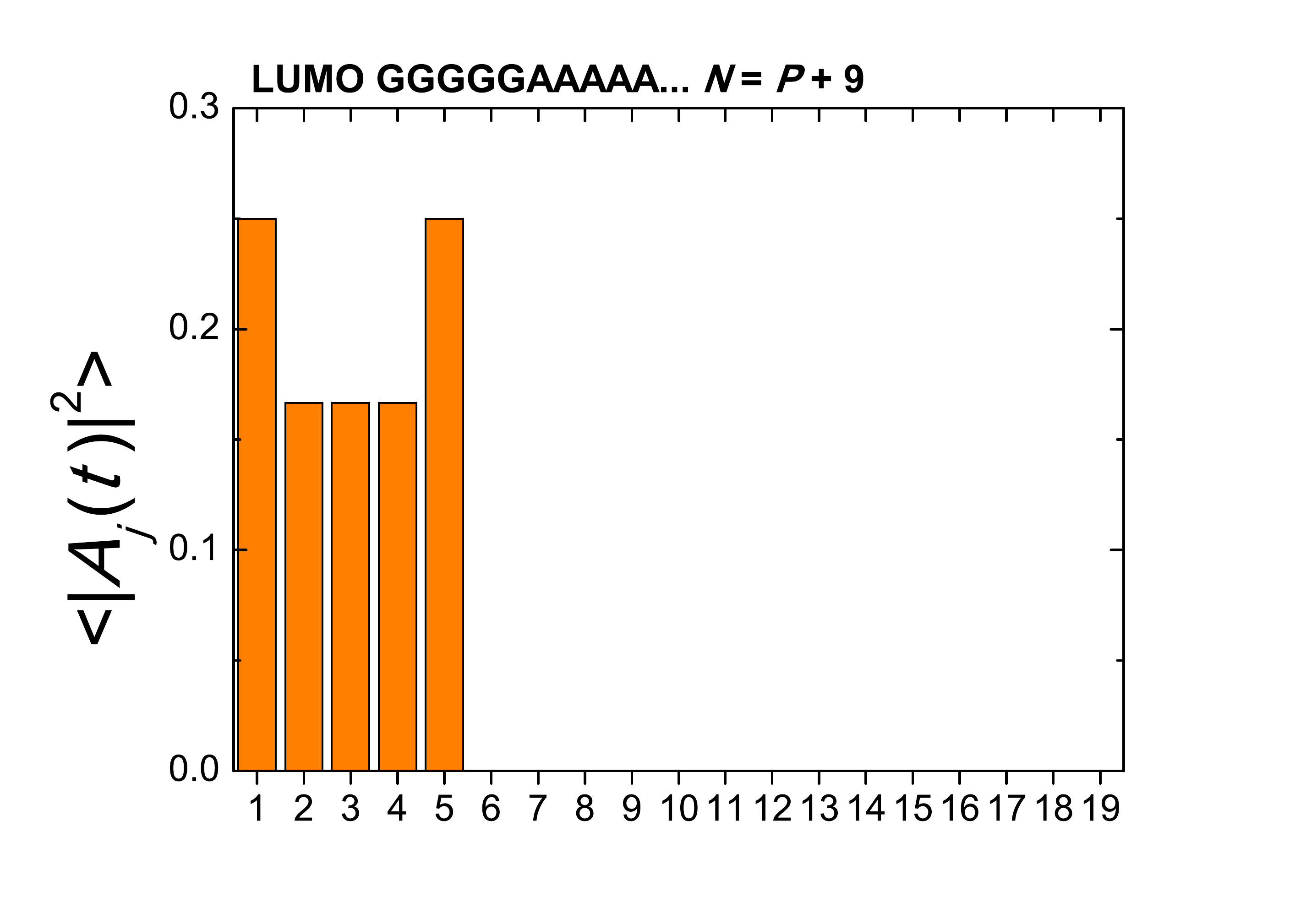}
\caption{\emph{...Continued from the previous page.} Mean (over time) probabilities to find the extra carrier at each monomer $j$, having placed it initially at the first monomer, for D10 (GGGGGAAAAA...) polymers, for the HOMO (left) and the LUMO (right). $N = P + \tau$, $\tau = 0, 1, \dots, P-1$.}
\end{figure*}

\begin{figure}[!h]
	\includegraphics[width=0.4\textwidth]{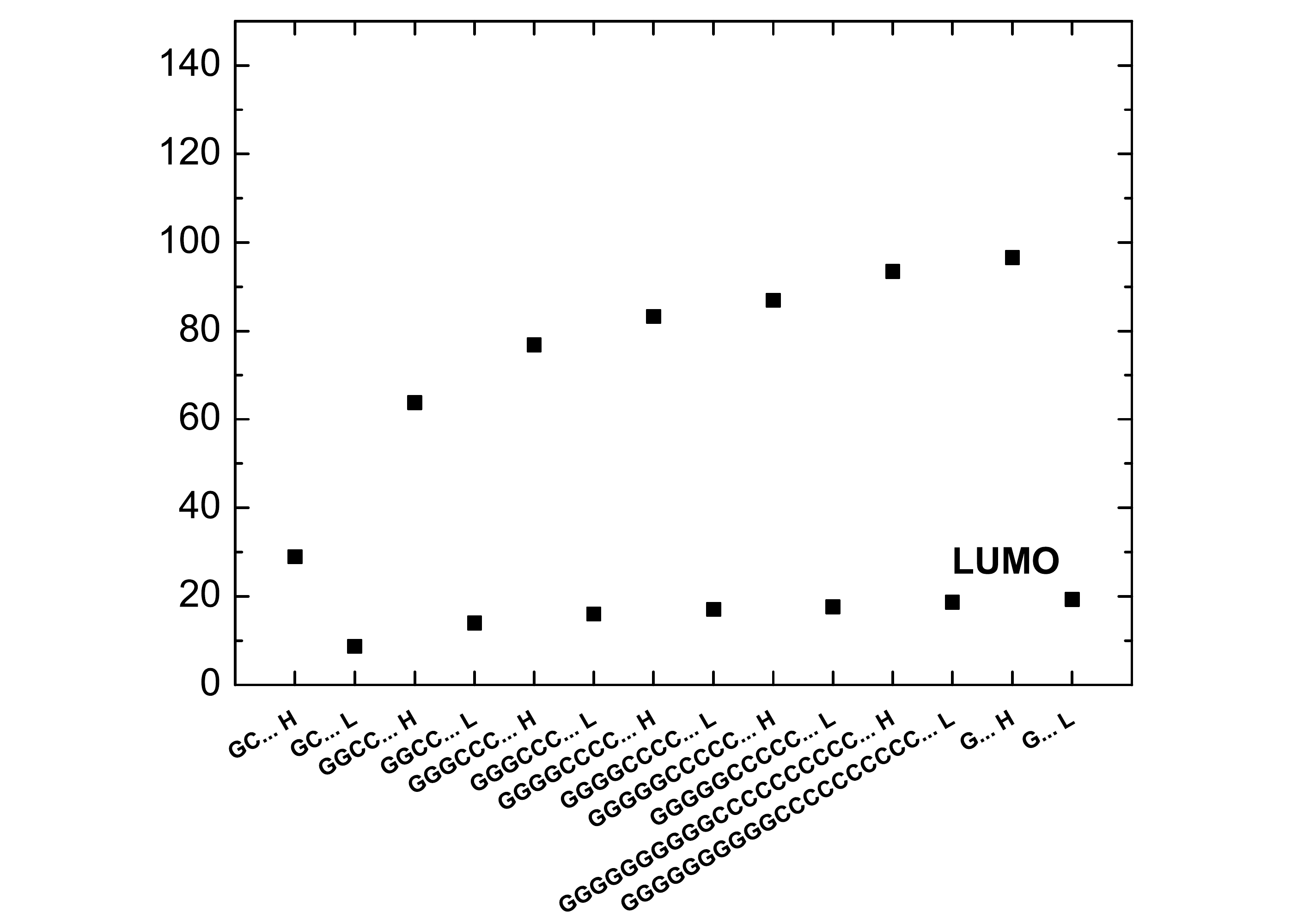}
	\includegraphics[width=0.4\textwidth]{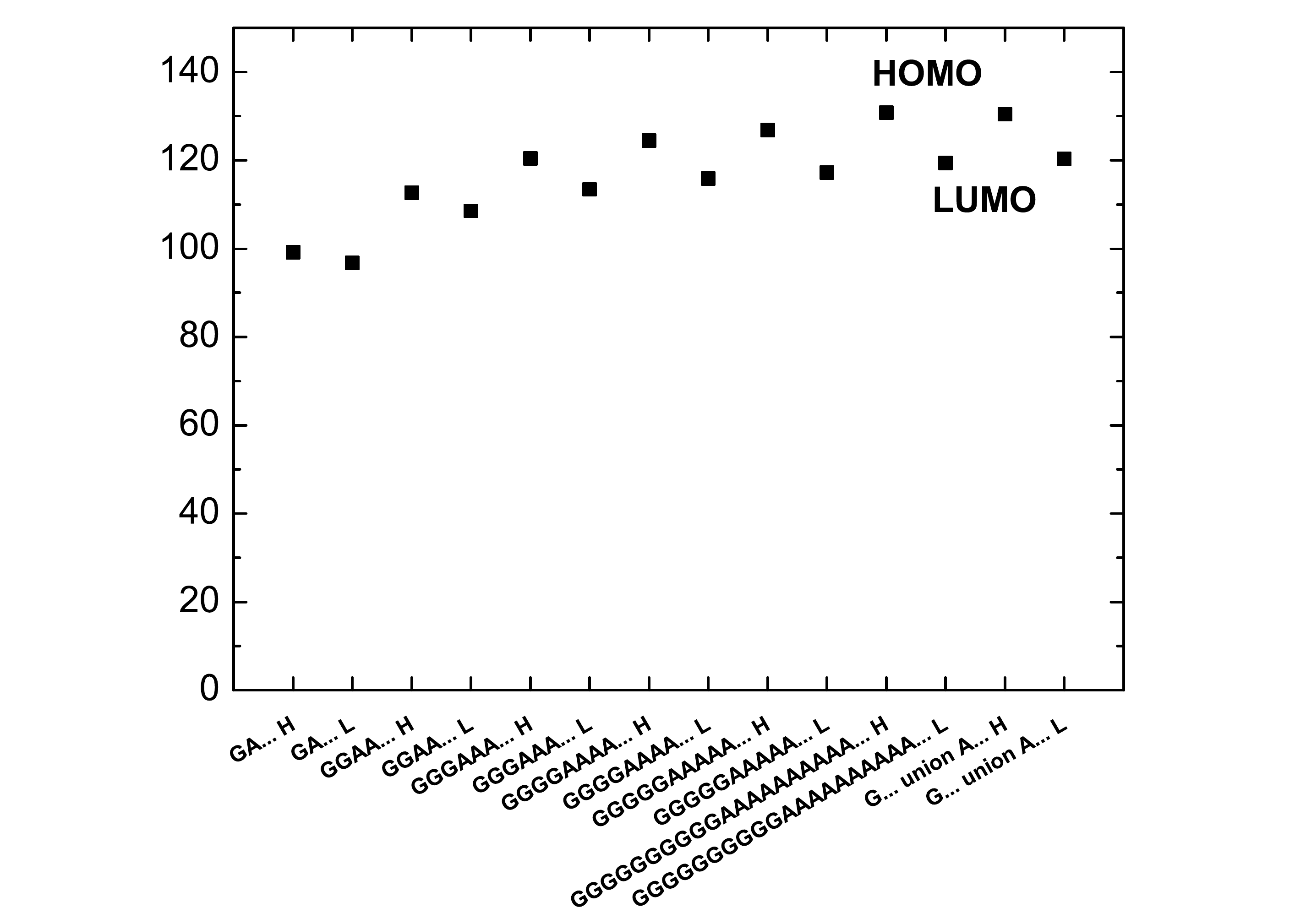}
	\caption{The maximum frequency of the Fourier spectrum, for the HOMO and the LUMO regime. 
		[Left panel]
		I1 (G...), 
		I2 (GC...), 
		I4 (GGCC...), 
		I6 (GGGCCC...), 
		I8 (GGGGCCCC...),
		I10 (GGGGGCCCCC...), and
		I20 (GGGGGGGGGGCCCCCCCCCC...) polymers.
		[Right panel]
		D2 (GA...), 
		D4 (GGAA...), 
		D6 (GGGAAA...), 
		D8 (GGGGAAAA...),
		D10 (GGGGGAAAAA...), and
		D20 (GGGGGGGGGGAAAAAAAAAA...) polymers.
	}
	\label{fig:maxfN60HL}
\end{figure}

\clearpage

\begin{figure*} [h!]
	\includegraphics[width=0.45\textwidth]{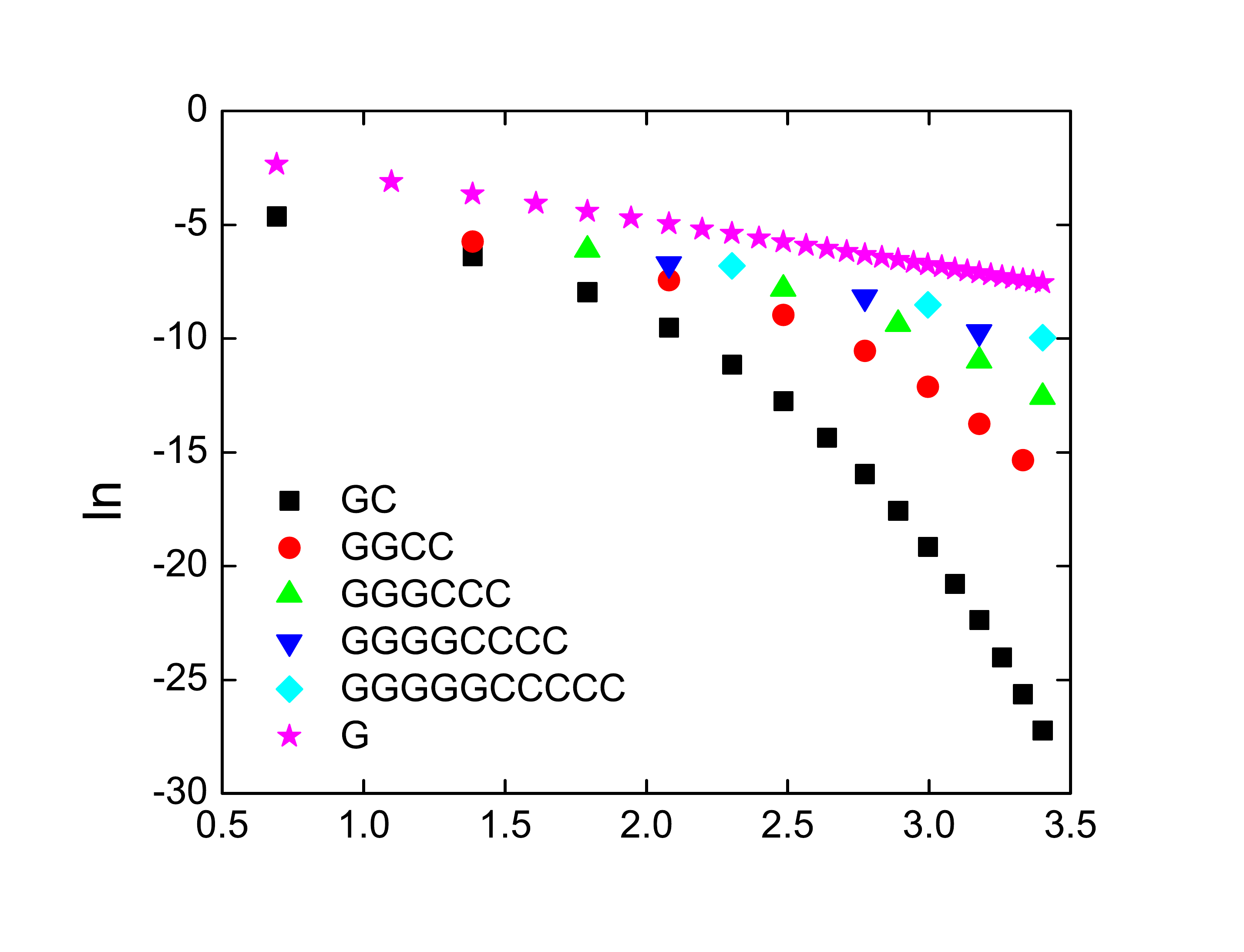}
	\includegraphics[width=0.45\textwidth]{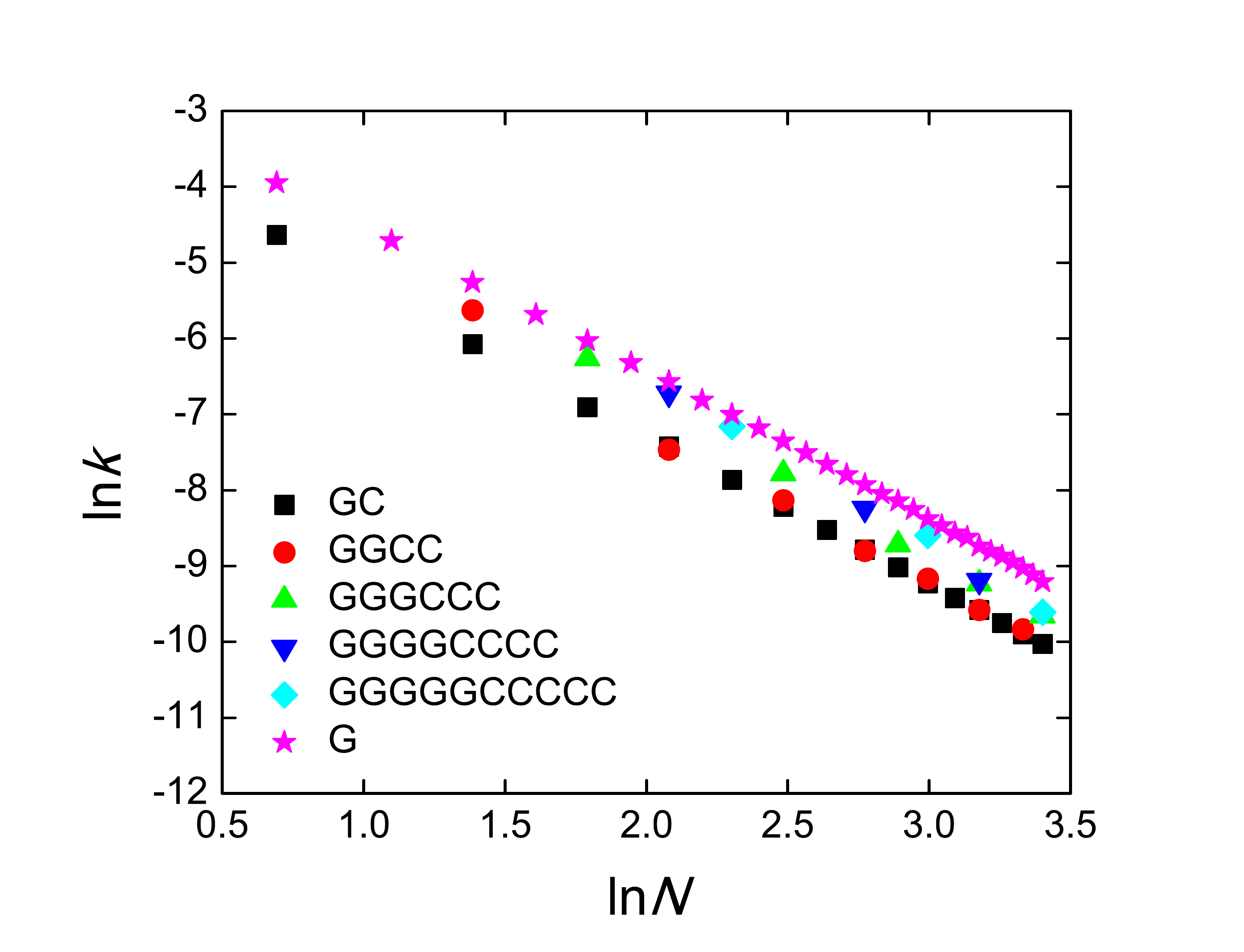}
	\includegraphics[width=0.45\textwidth]{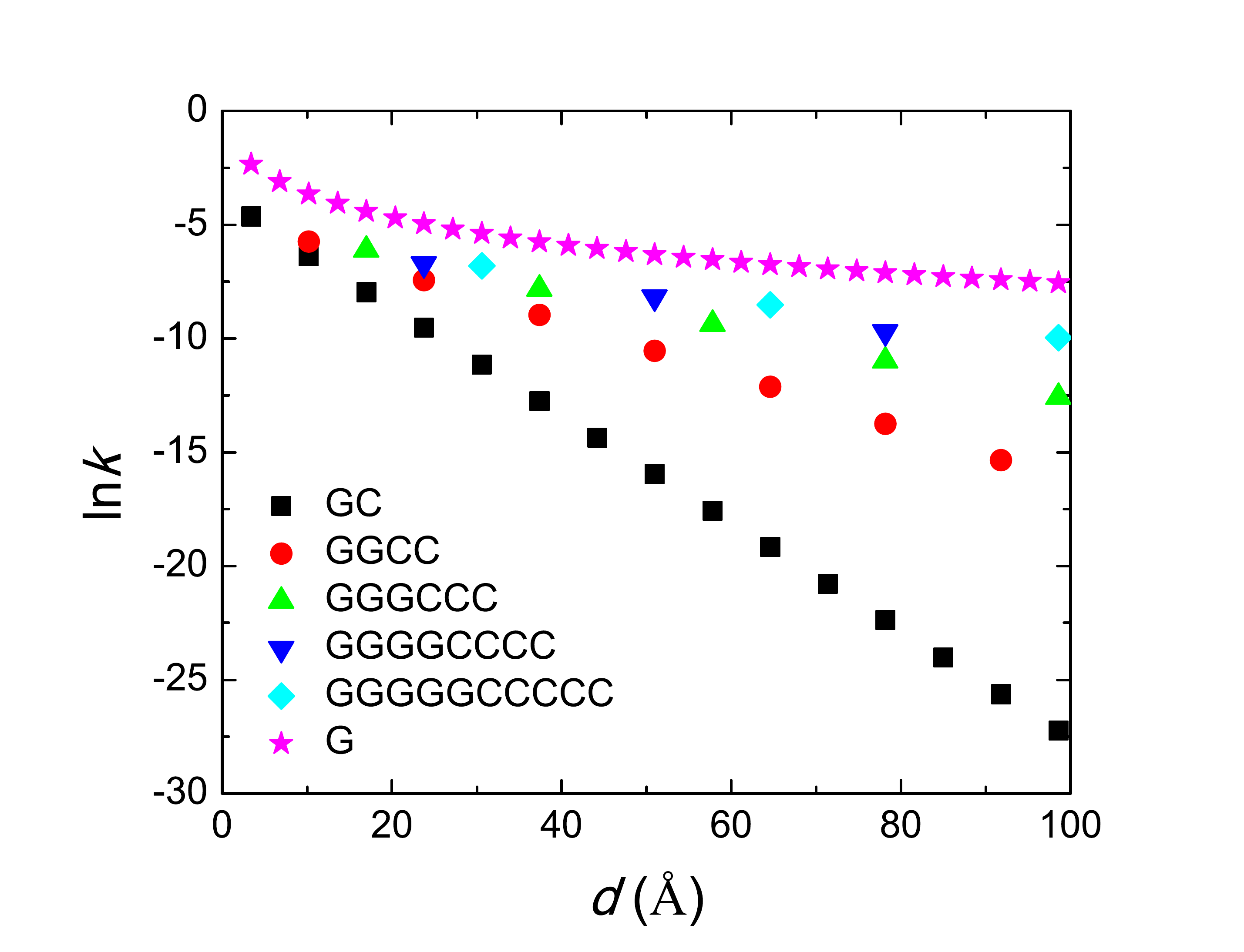}
	\includegraphics[width=0.45\textwidth]{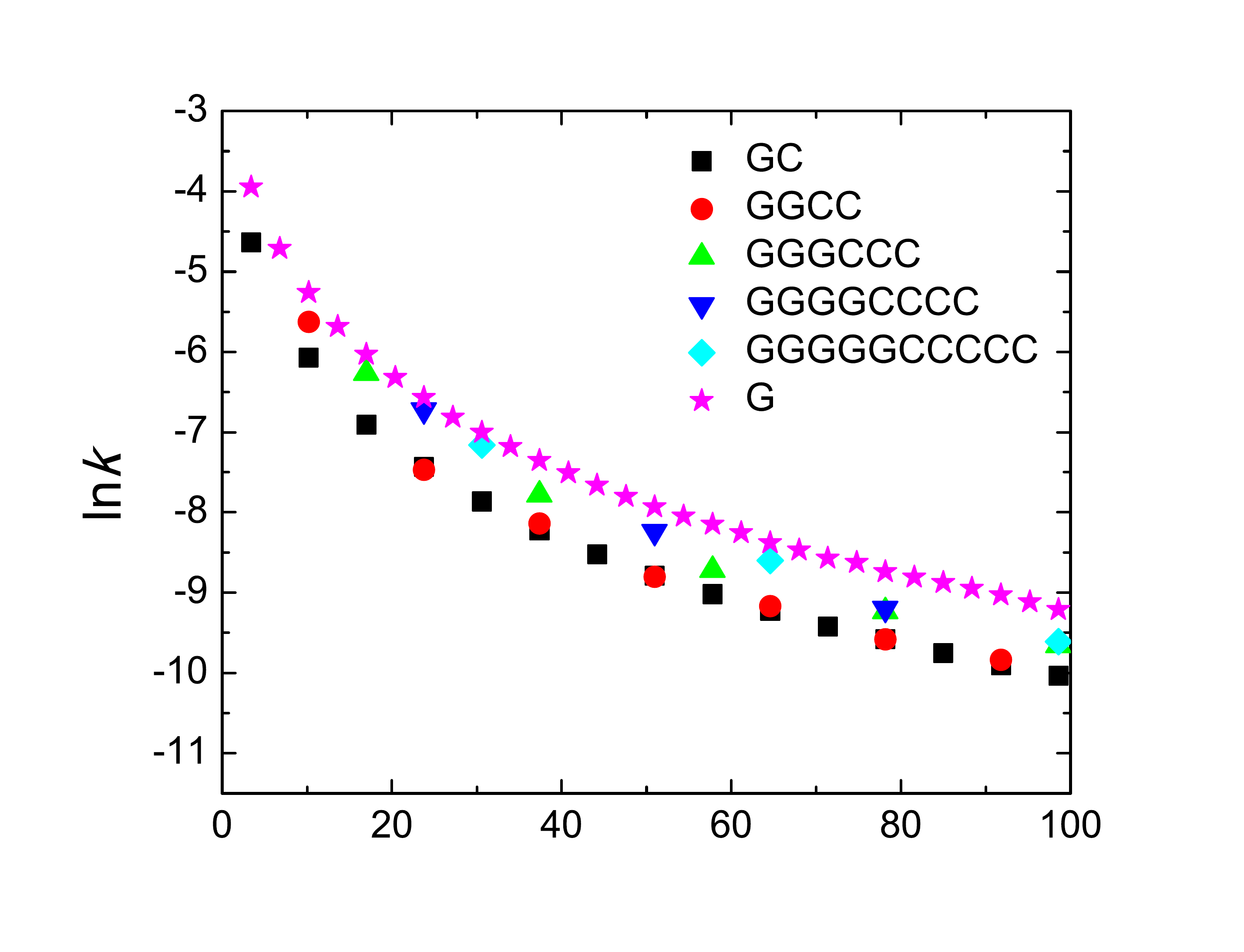}
	\caption{[First line:] $\ln\!k$ as a function of $\ln\!N$ of type I1 (G...), I2 (GC...), I4 (GGCC...), I6  (GGGCCC...), I8 (GGGGCCCC...) and I10 (GGGGGCCCCC...) polymers, for $N$ equal to natural multiples of their $P$, for HOMO (left) and LUMO (right).
		[Second line:] For the same polymers, $\ln\!k$ as a function of $d$, where $d = (N-1)\times 3.4$ {\AA} is the charge transfer distance.}
	\label{fig:kofN-ChV-SM}
\end{figure*}

\begin{figure*}[!h]
	\includegraphics[width=0.45\textwidth]{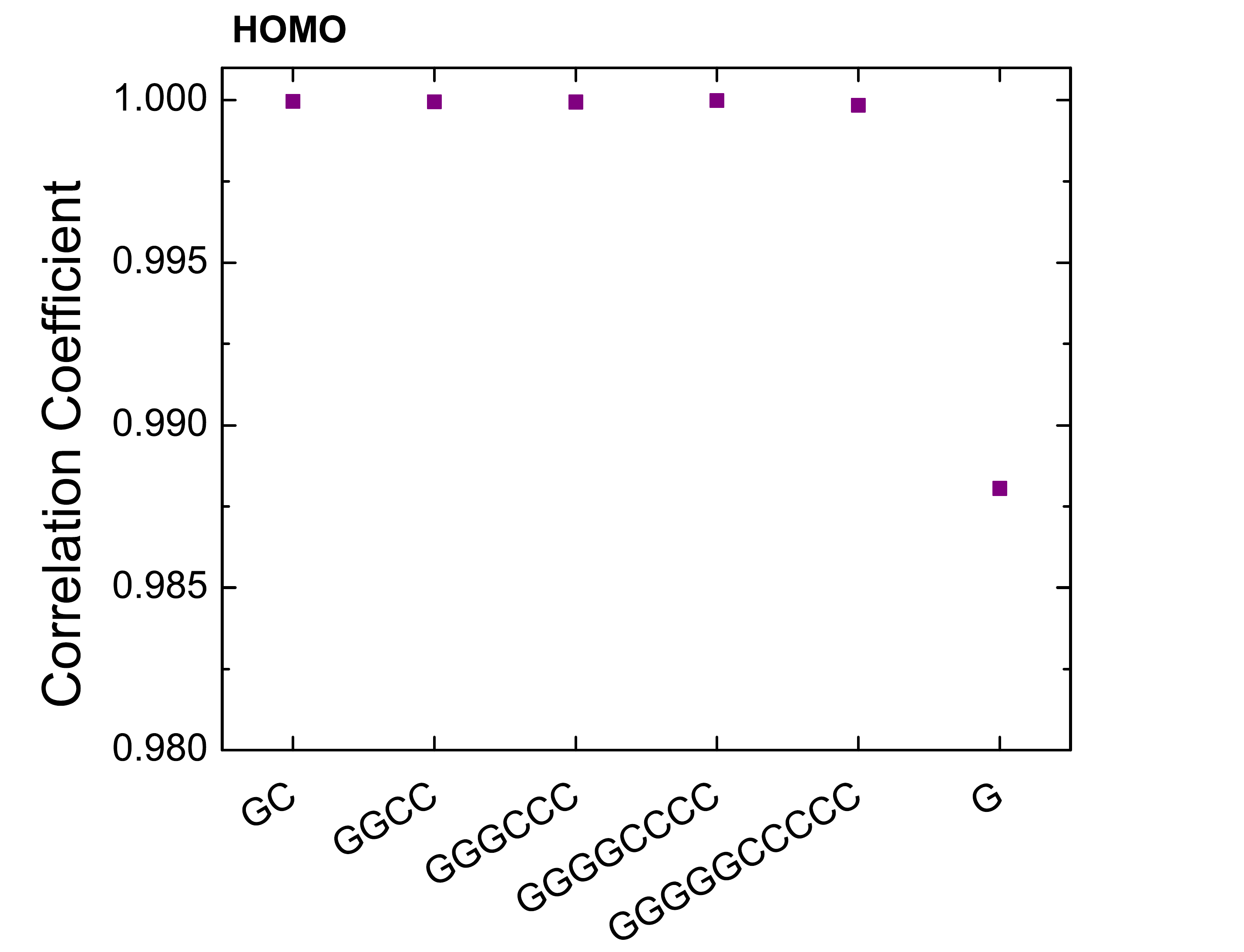}
	\includegraphics[width=0.45\textwidth]{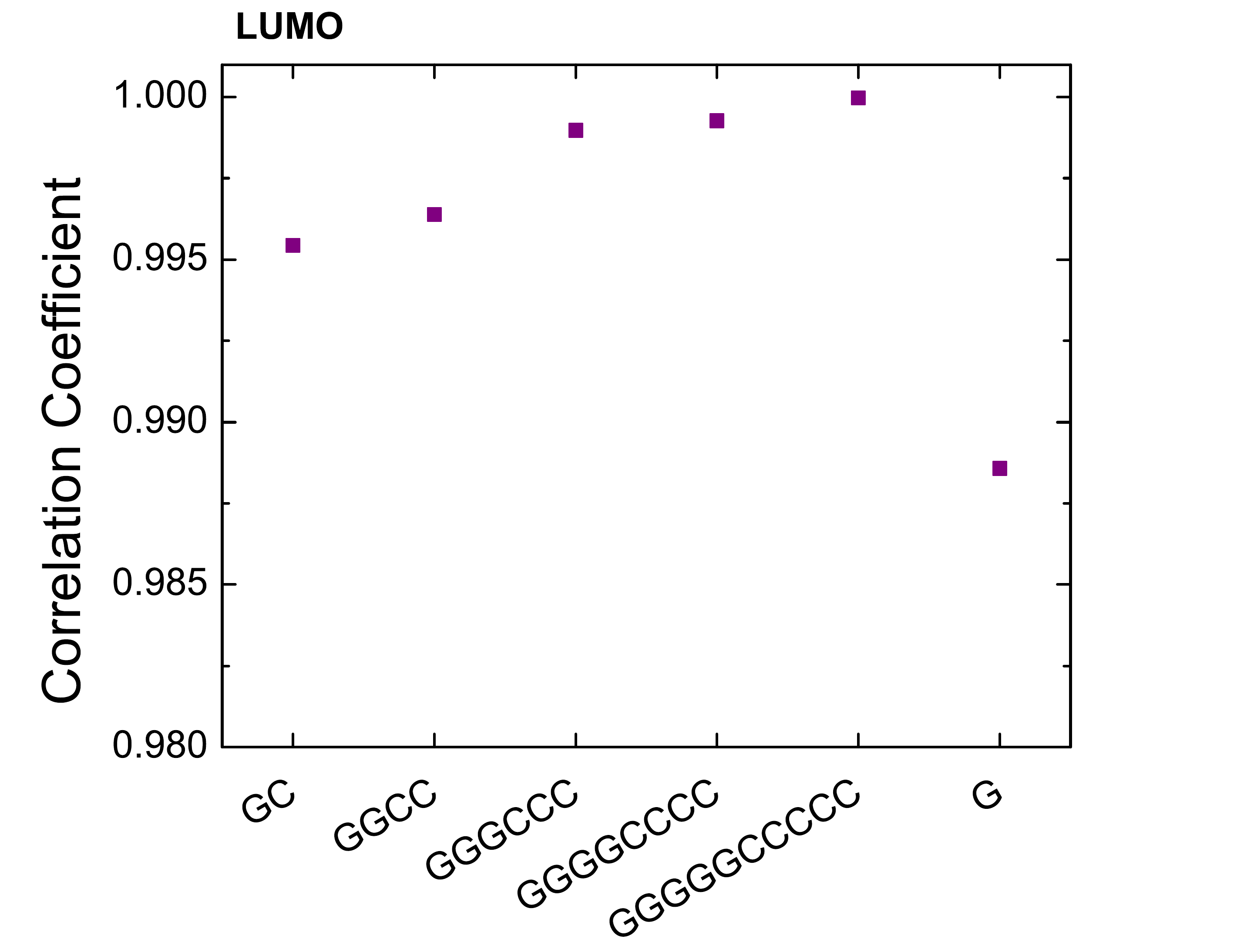}
	\includegraphics[width=0.45\textwidth]{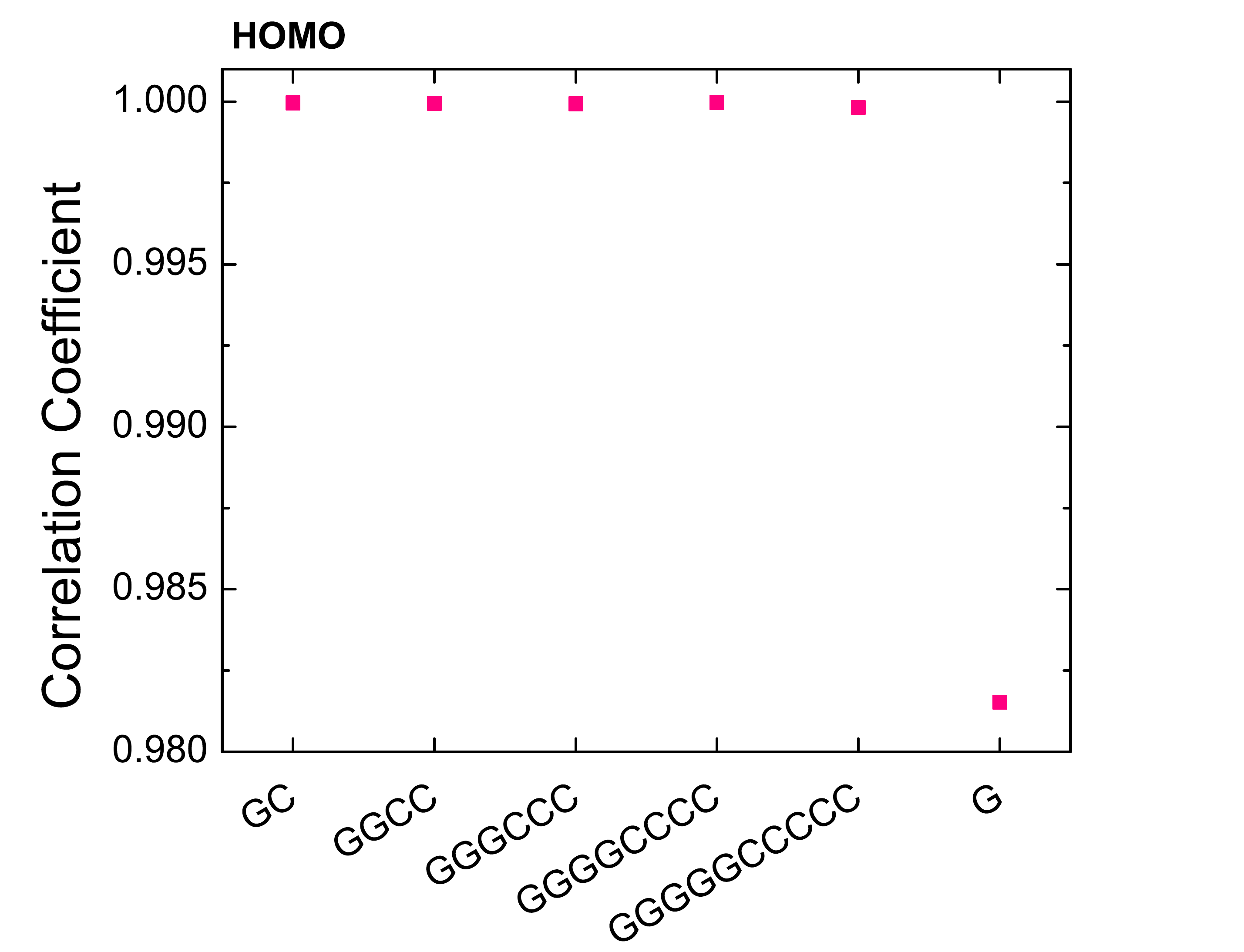}
	\includegraphics[width=0.45\textwidth]{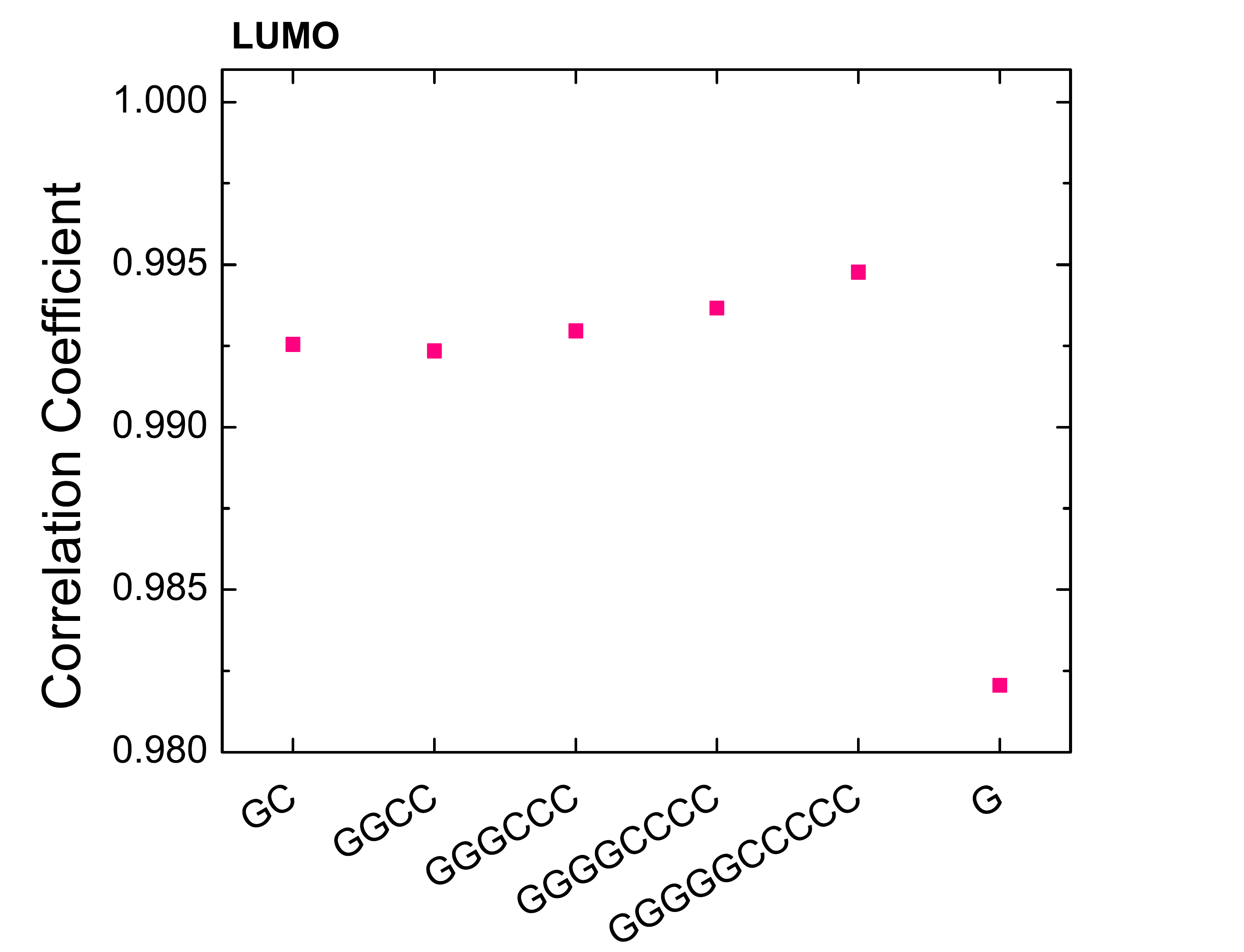}
	\includegraphics[width=0.45\textwidth]{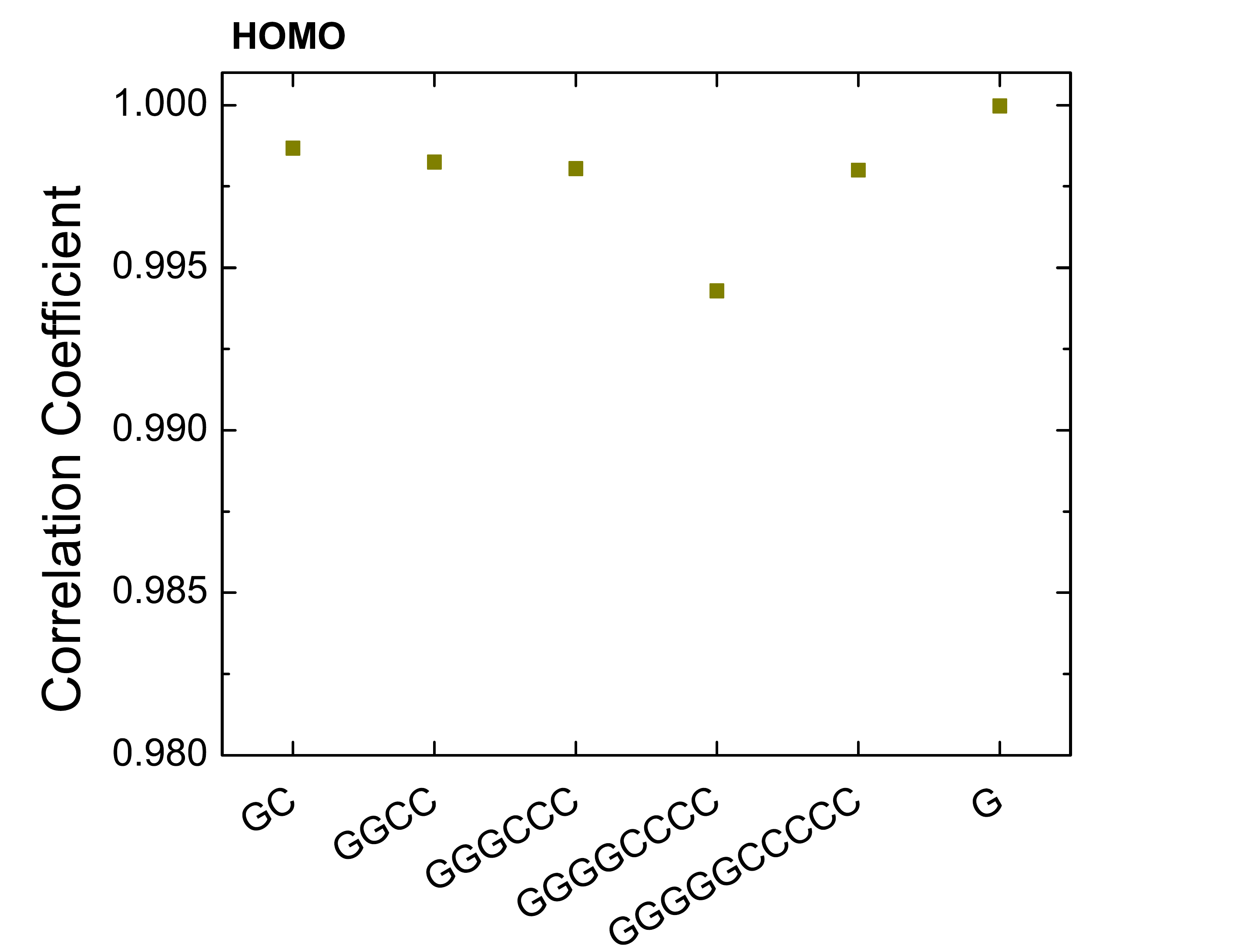}
	\includegraphics[width=0.45\textwidth]{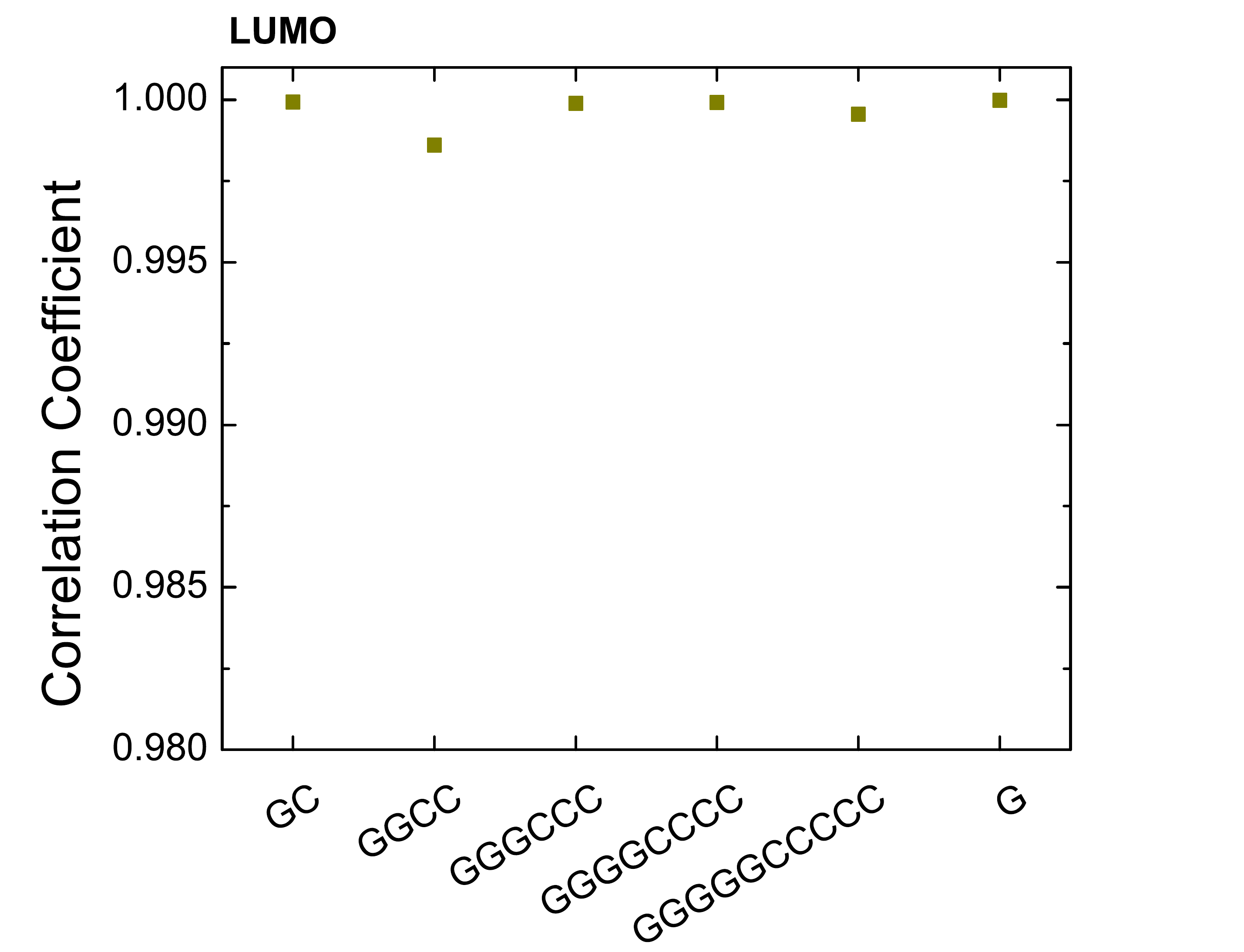}
	\caption{Correlation coefficients for I2 (GC...), I4 (GGCC...), I6 (GGGCCC...), I8 (GGGGCCCC...), I10 (GGGGGCCCCC...) and I1 (G...) polymers, for the exponential fits $k=A+k_{0} \mathrm{e}^{- \beta d}$ (1st line) and $k=k_{0} \mathrm{e}^{- \beta d}$ (2nd line) as well as for the power law fit $k=k_{0}N^{-\eta}$ (3rd line). In all cases, $k$ is from the first to the last monomer, under the condition $N=nP$ and for $N < 40$.}
	\label{fig:CC-ChV}
\end{figure*}

\begin{figure*}[!h]
	\includegraphics[width=0.45\textwidth]{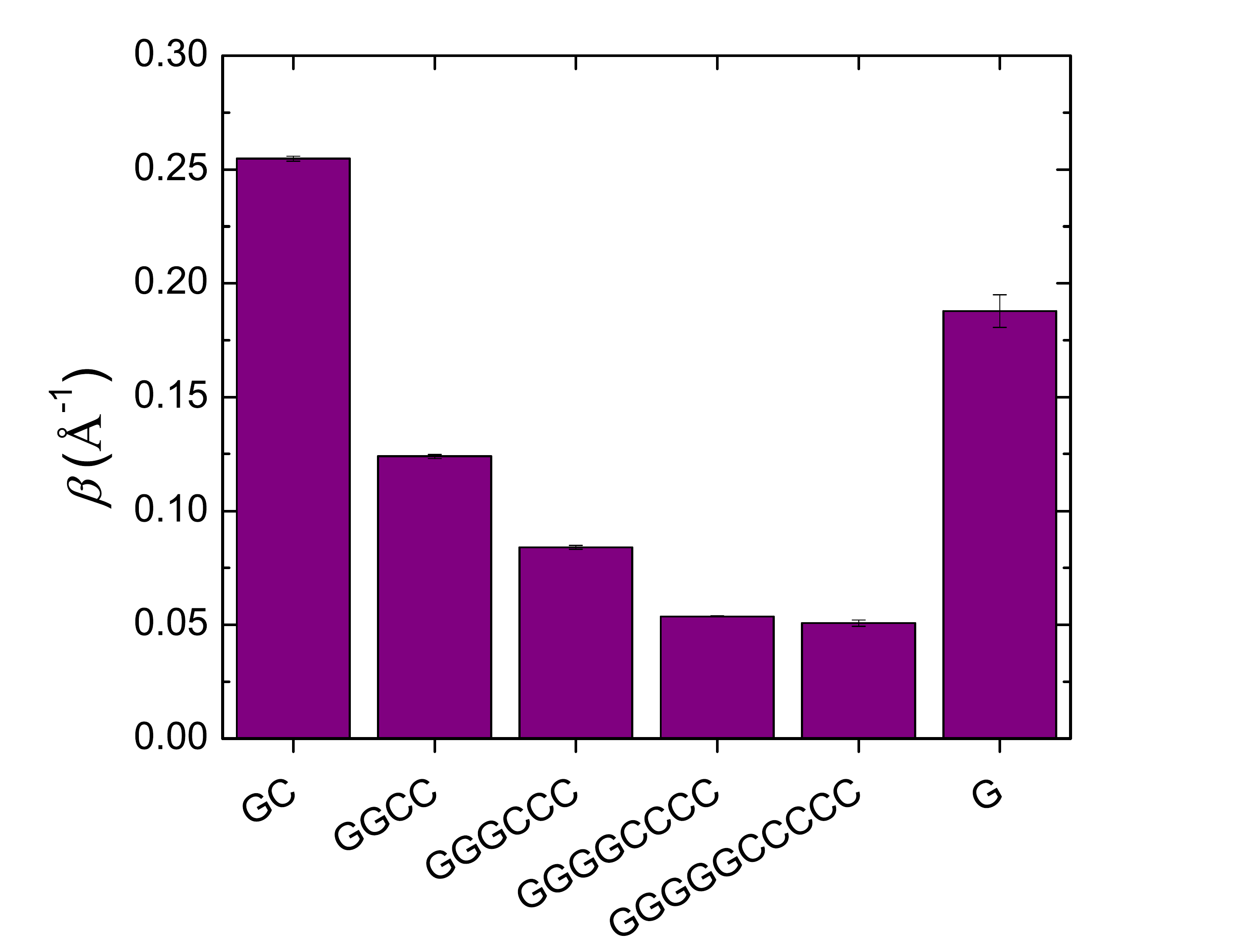}
	\includegraphics[width=0.45\textwidth]{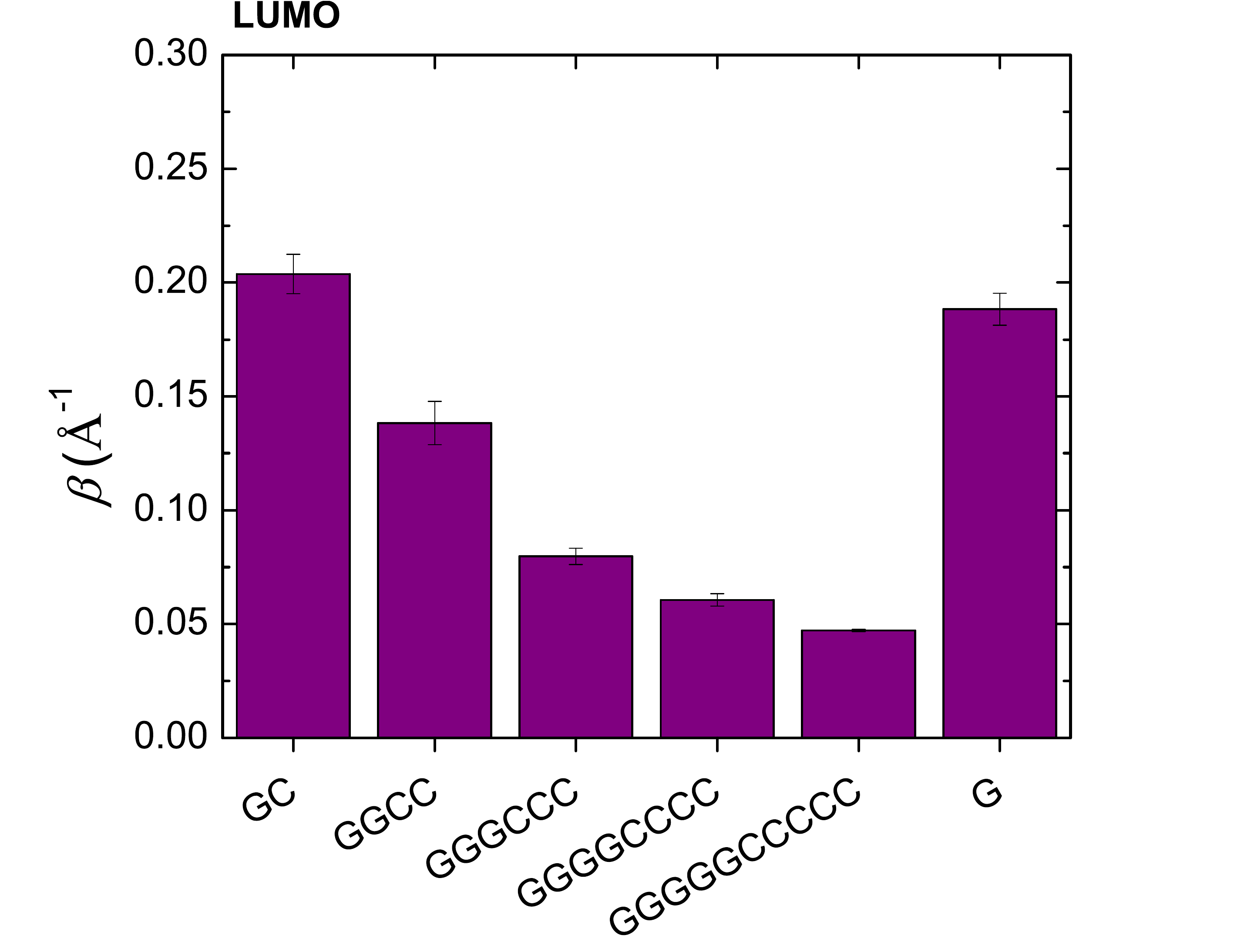}
	\includegraphics[width=0.45\textwidth]{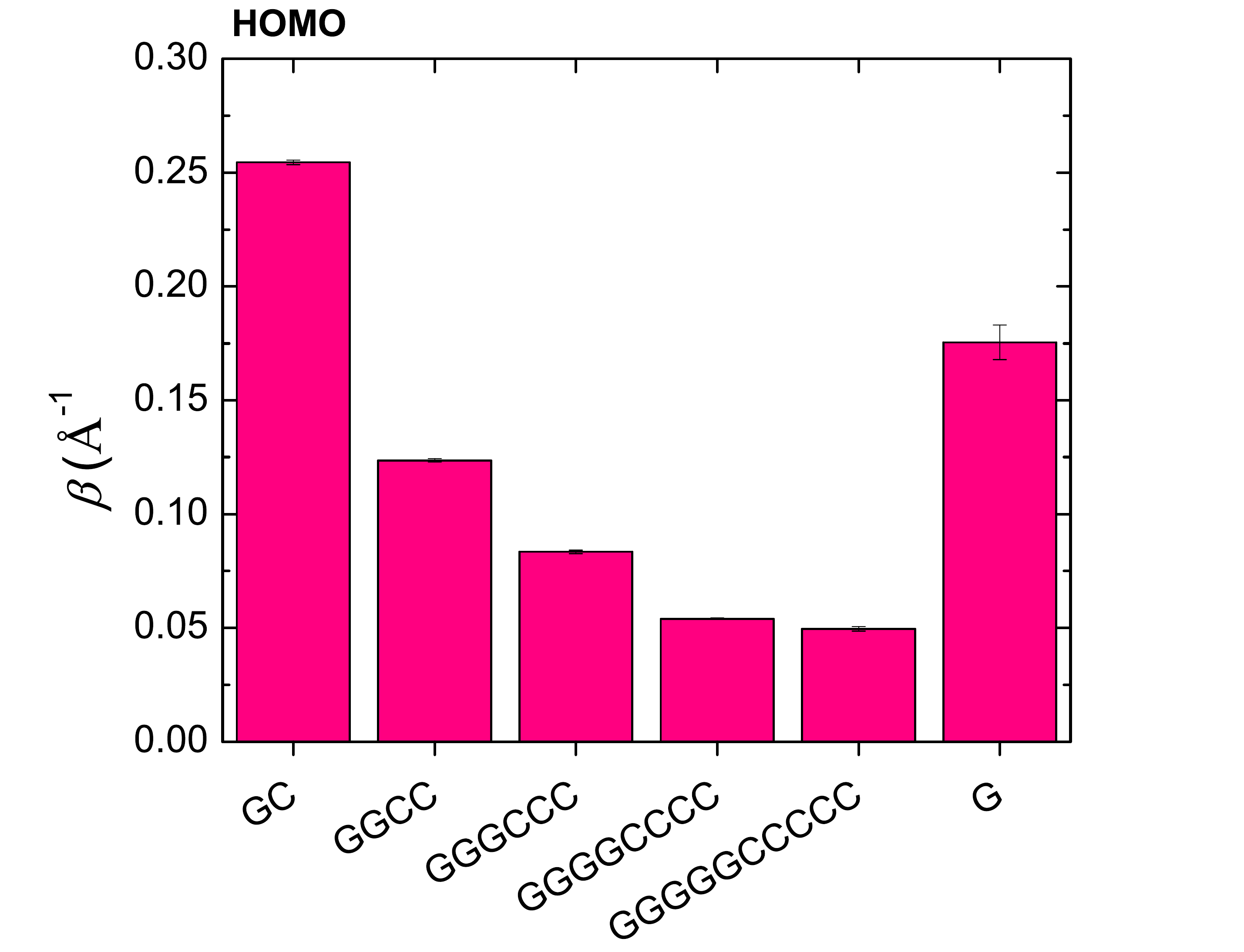}
	\includegraphics[width=0.45\textwidth]{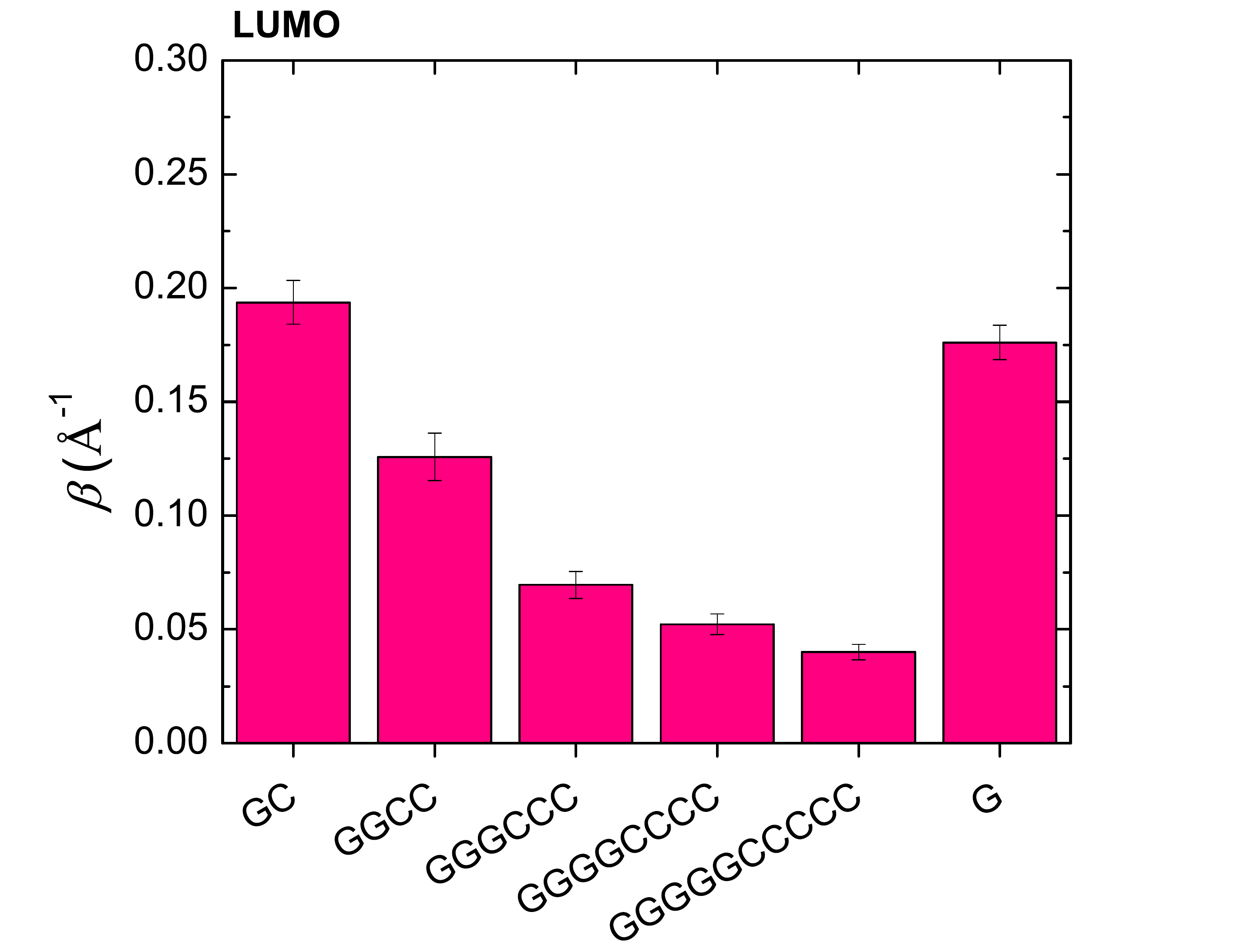}
	\includegraphics[width=0.45\textwidth]{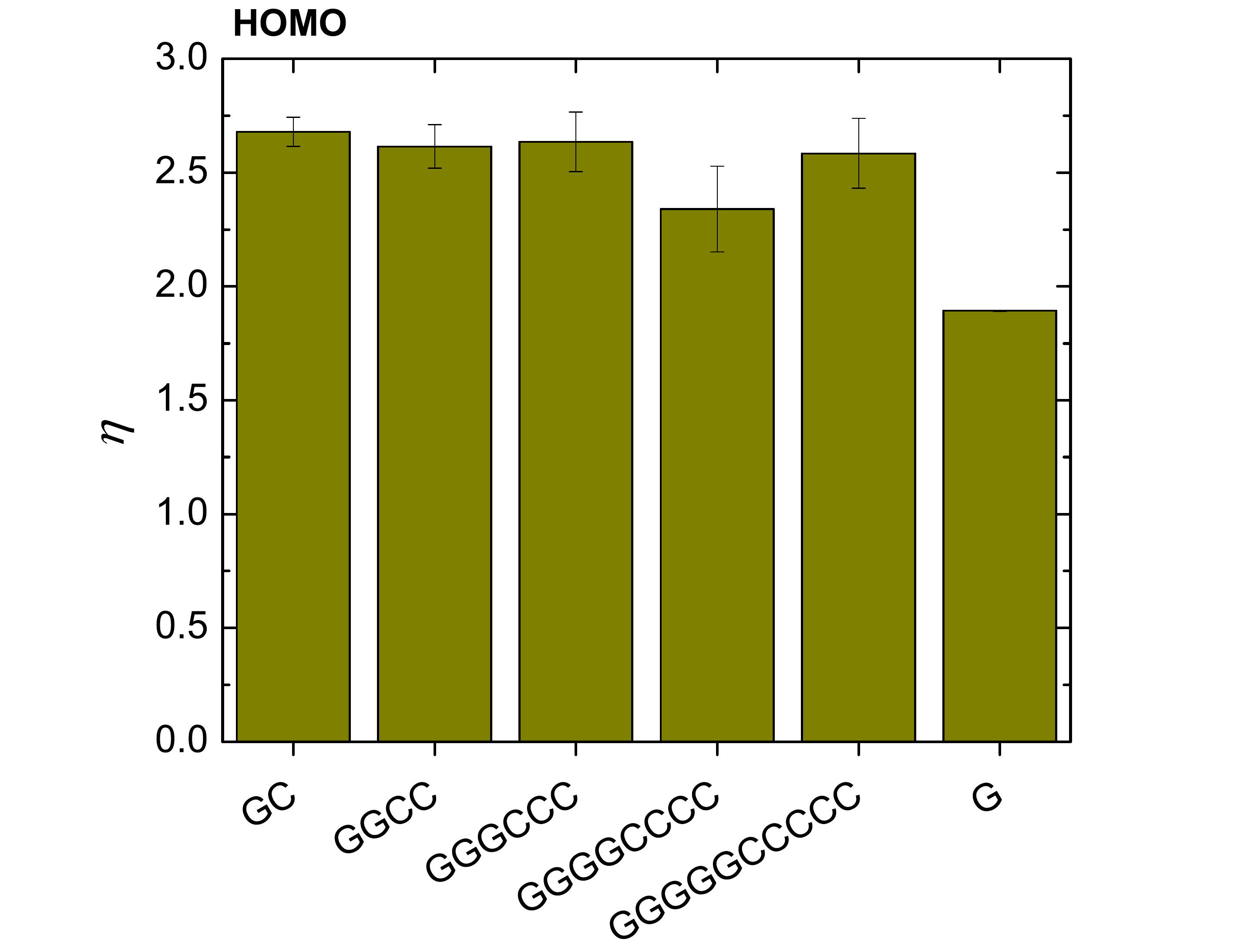}
	\includegraphics[width=0.45\textwidth]{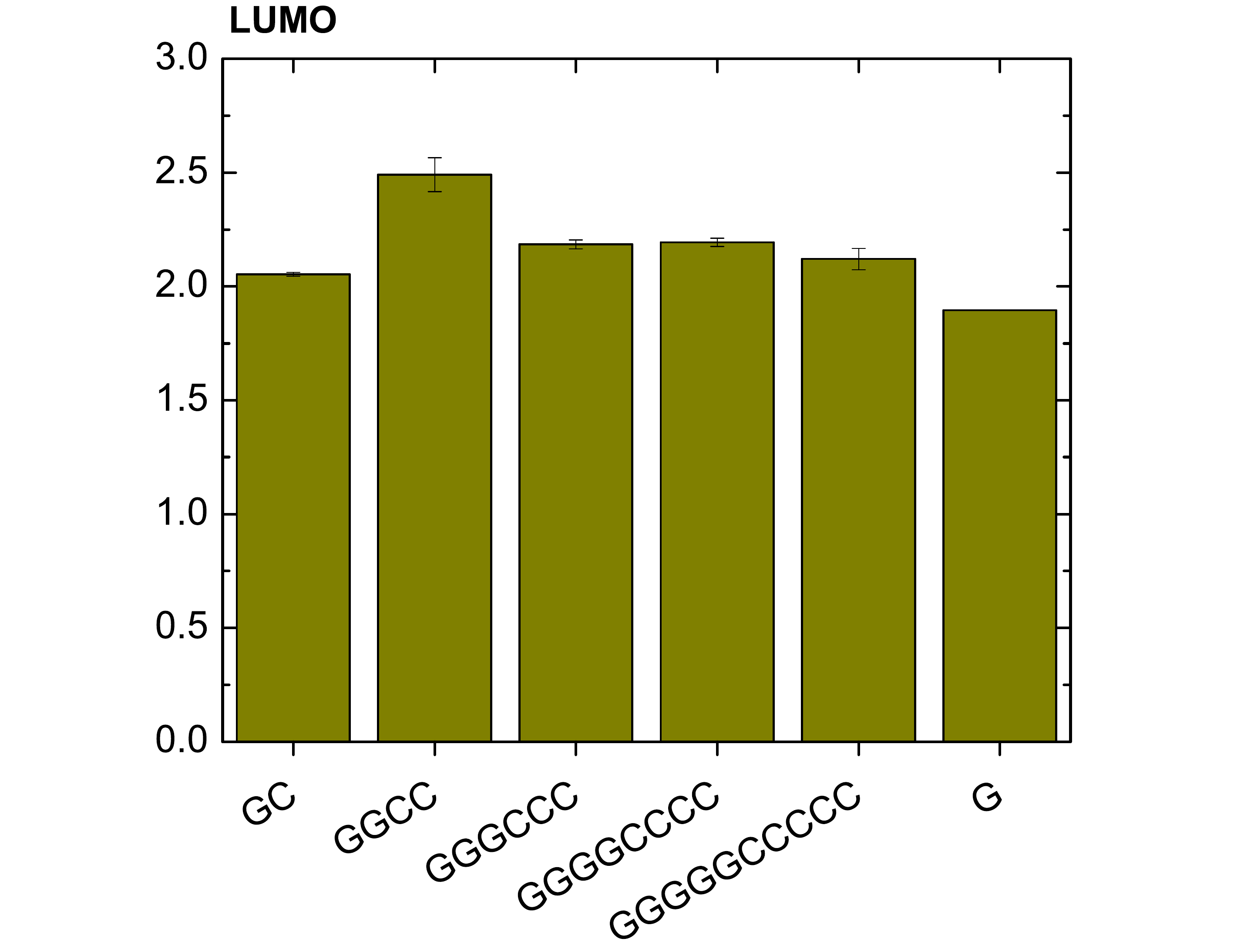}
	\caption{The exponents $\beta$ and $\eta$ for I2 (GC...), I4 (GGCC...), I6 (GGGCCC...), I8 (GGGGCCCC...), I10 (GGGGGCCCCC...) and I1 (G...) polymers, for the exponential fits $k=A+k_{0} \mathrm{e}^{- \beta d}$ (1st line) and $k=k_{0} \mathrm{e}^{- \beta d}$ (2nd line) as well as for the power law fit $k=k_{0}N^{-\eta}$ (3rd line). In all cases, $k$ is from the first to the last monomer, under the condition $N=nP$ and for $N < 40$.}
	\label{fig:betaeta-ChV}
\end{figure*}

\begin{figure*} [h!]
	\includegraphics[width=0.45\textwidth]{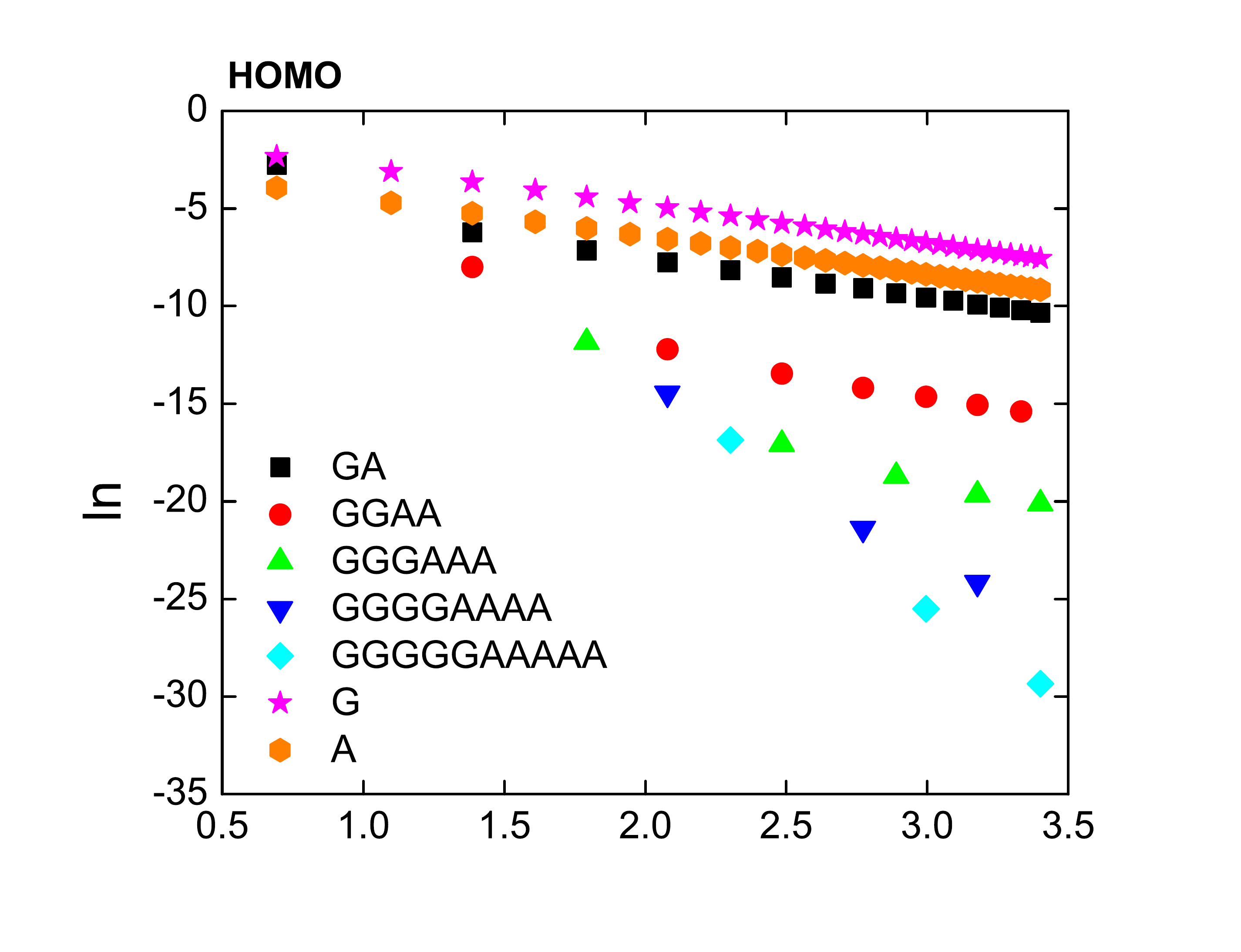}
	\includegraphics[width=0.45\textwidth]{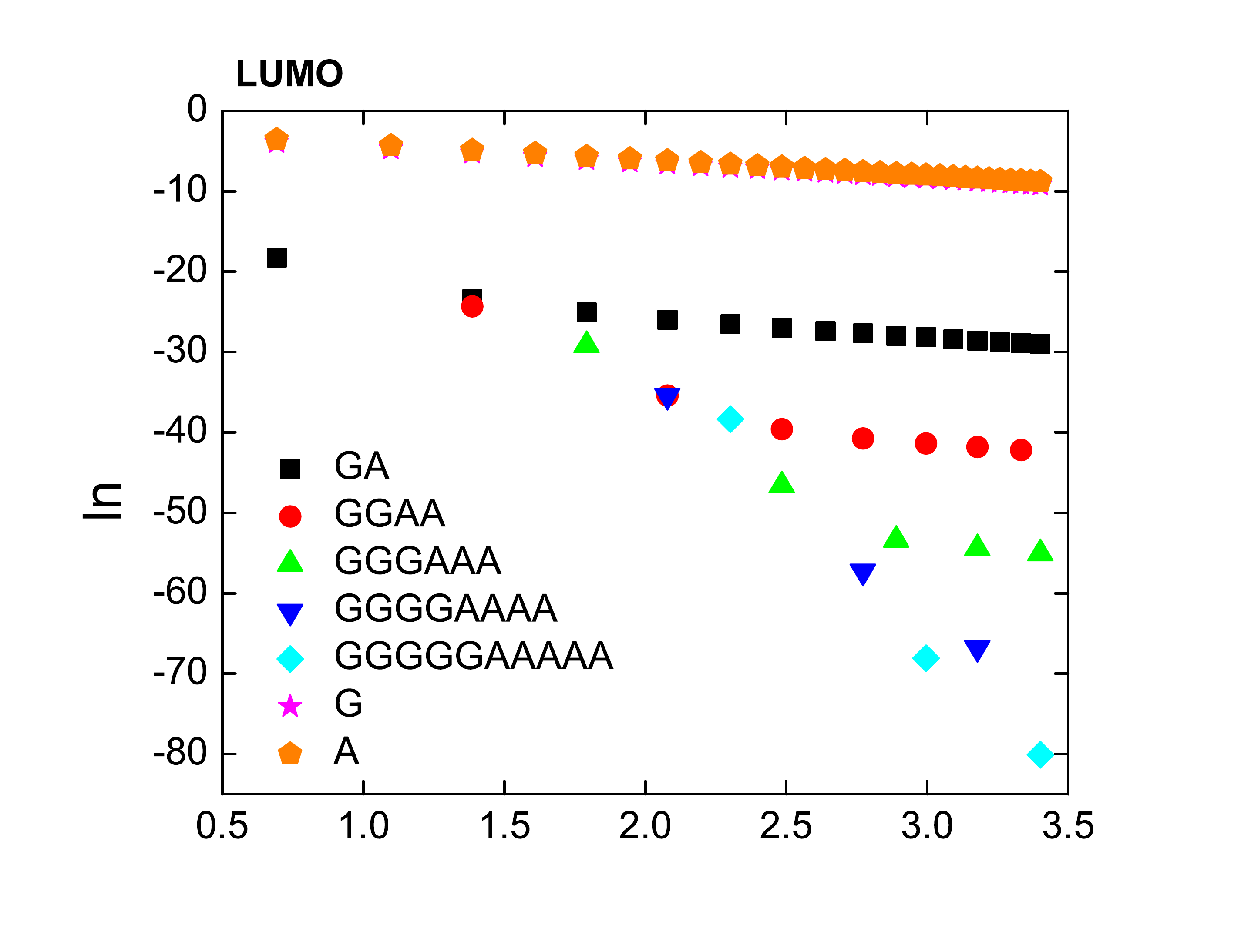}
	\includegraphics[width=0.45\textwidth]{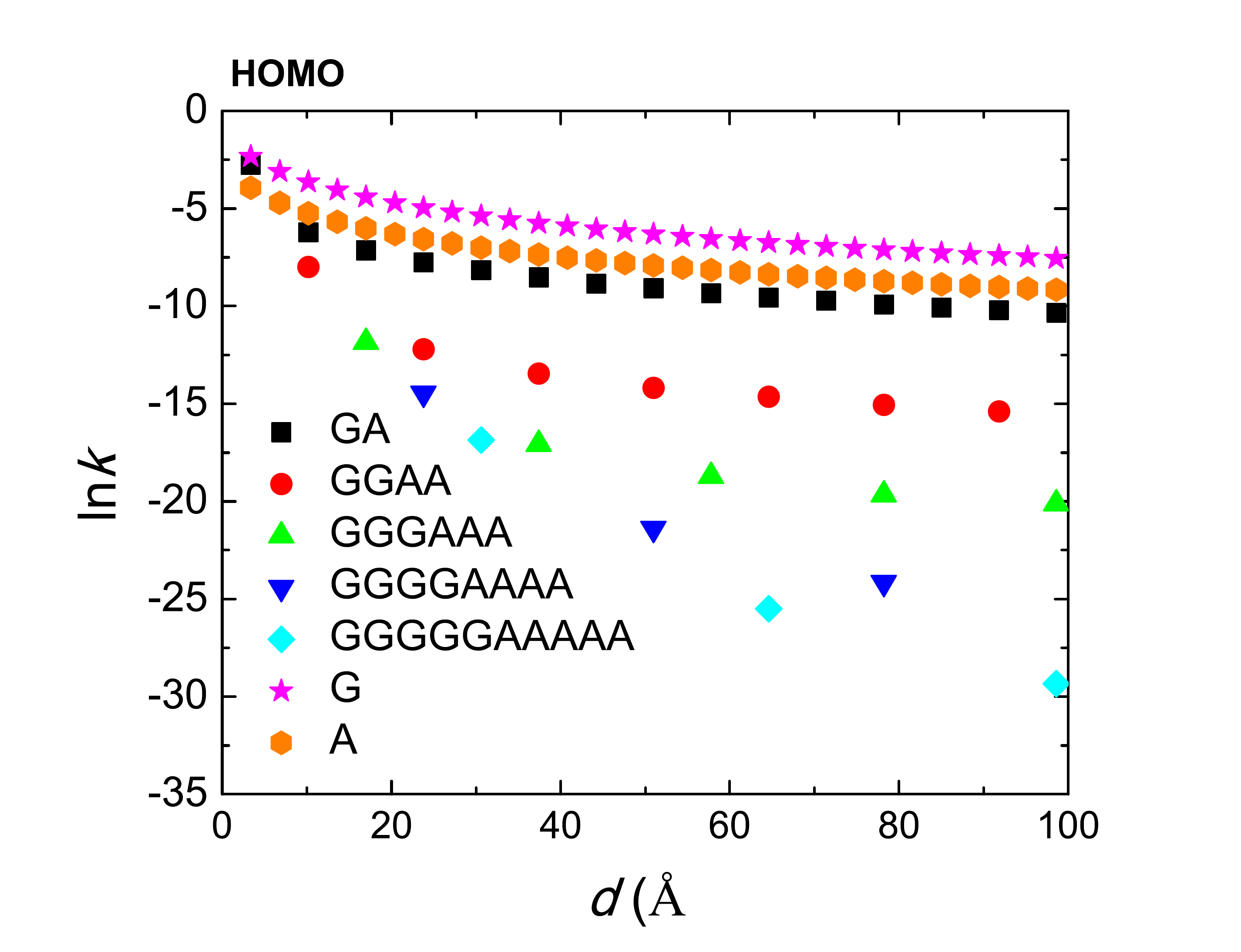}
	\includegraphics[width=0.45\textwidth]{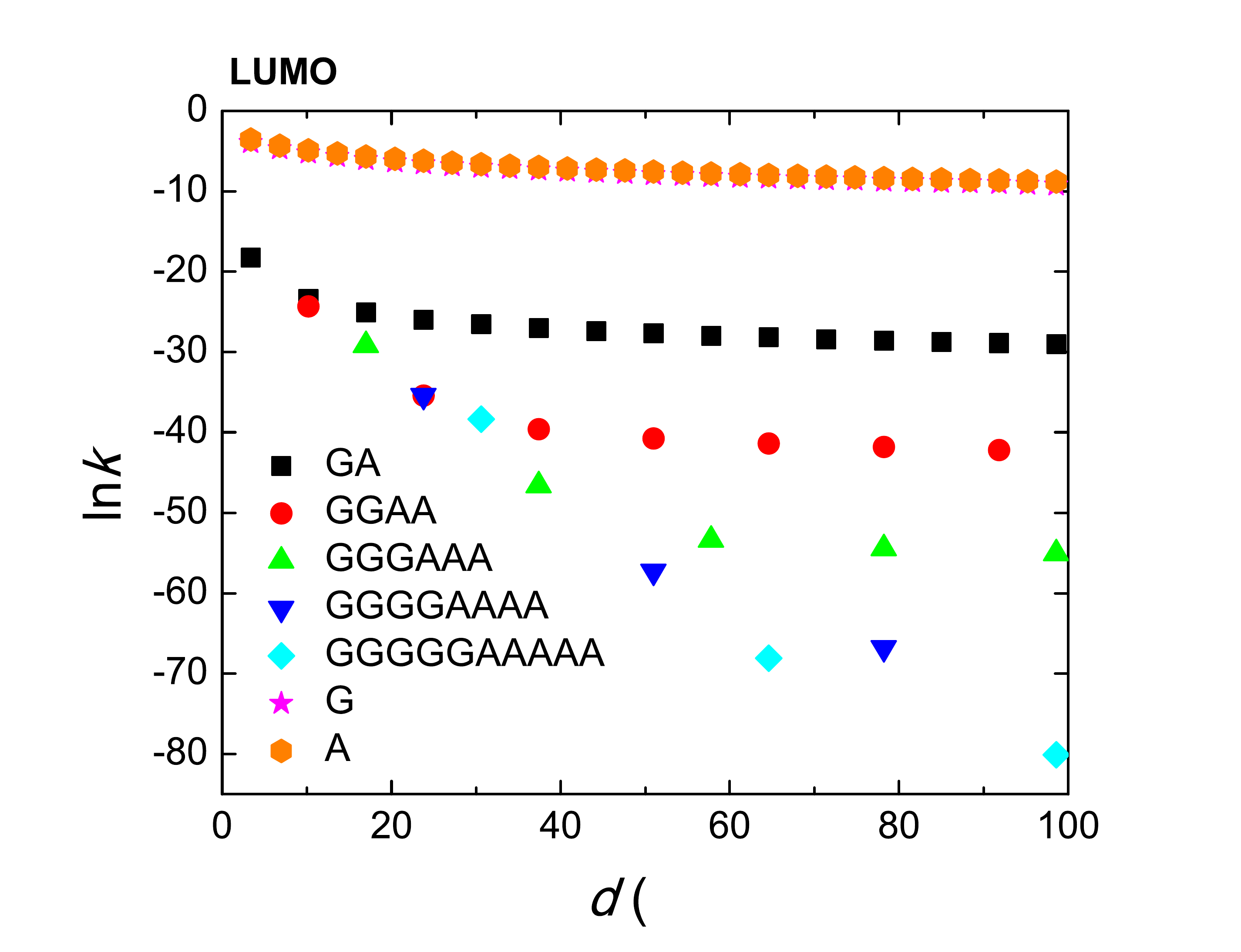}
	\caption{[First line:] $\ln\!k$ as a function of $\ln\!N$ of type I1 (G...), I1 (A...), D2 (GA...), D4 (GGAA...), D6  (GGGAAA...), D8 (GGGGAAAA...) and D10 (GGGGGAAAAA...) polymers, 
		for $N$ equal to natural multiples of their $P$, for HOMO (left) and LUMO (right).
		[Second line:] For the same polymers, $\ln\!k$ as a function of $d$, where $d = (N-1)\times 3.4$ {\AA} is the charge transfer distance.}
	\label{fig:kofN-PMp-SM}
\end{figure*}

\begin{figure*} [h!]
	\includegraphics[width=0.44\textwidth]{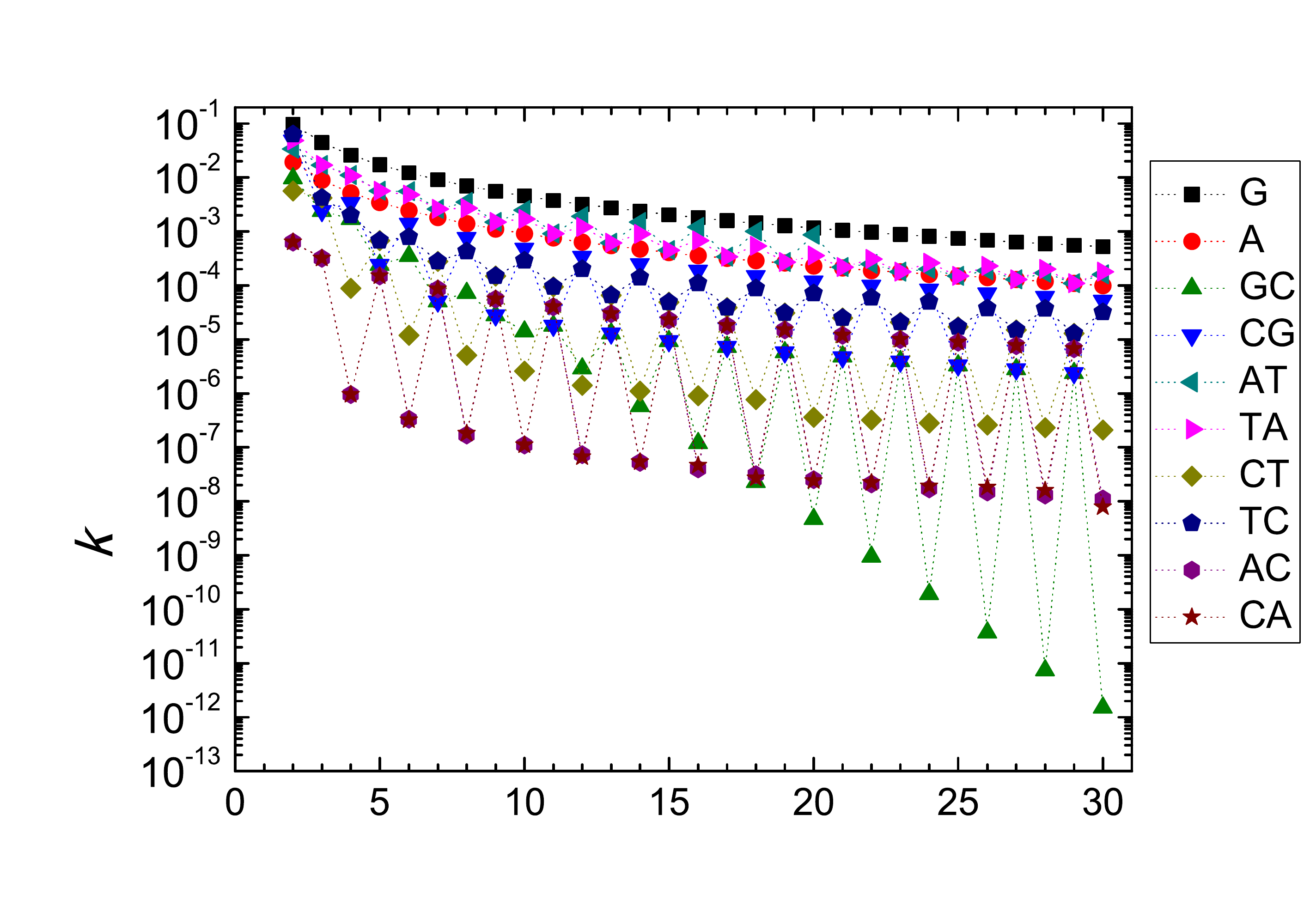}
	\includegraphics[width=0.44\textwidth]{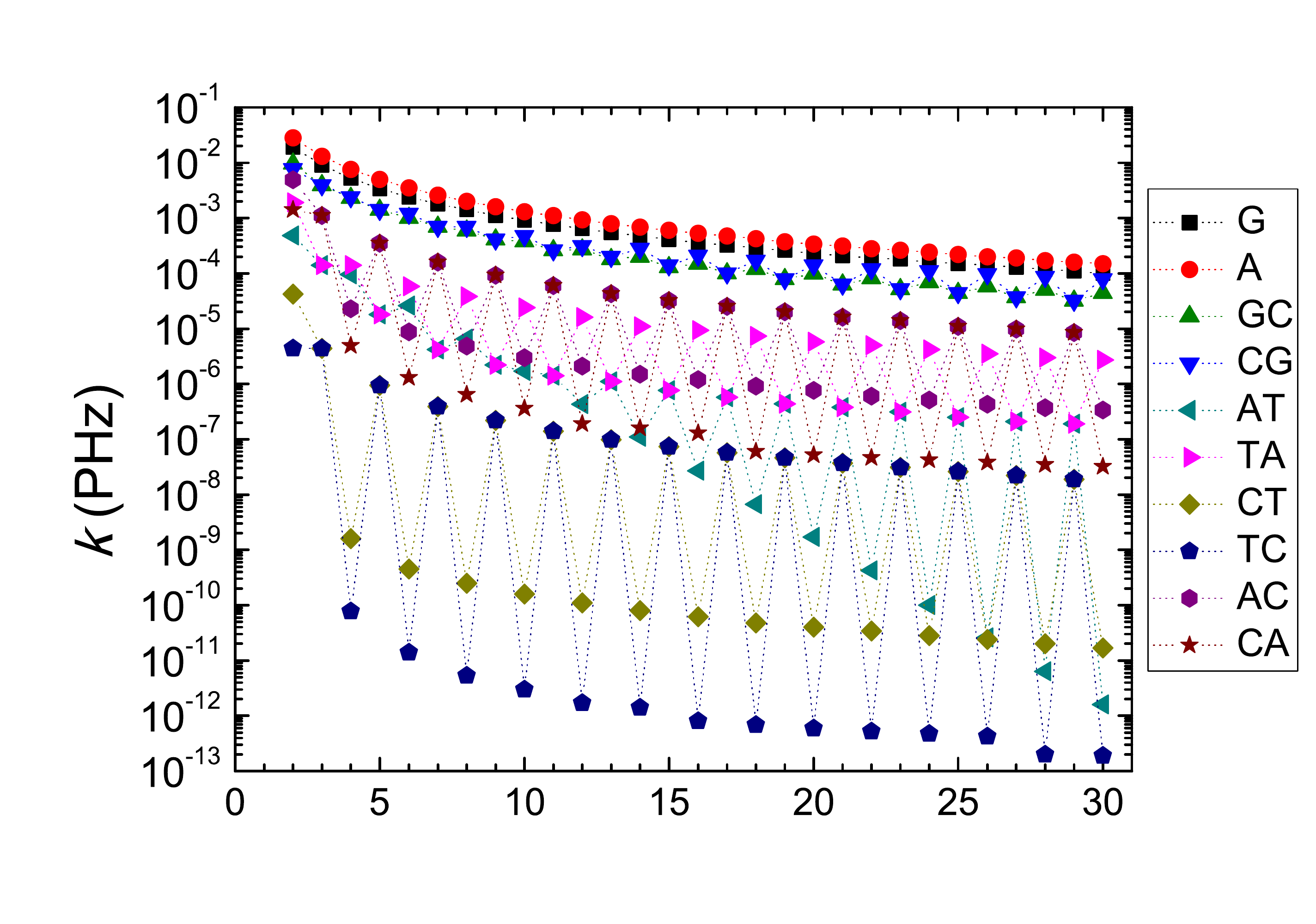}
	\caption{\textit{Pure} mean transfer rate from the first to the last monomer as a function of the number of monomers, $k(N)$, for types I1, I2 and D2 polymers, for HOMO and LUMO, within TBI.
		The frame shows the repetition unit.}
	\label{fig:kofN-TBI}
\end{figure*}

\begin{figure*} [h!]
	\includegraphics[width=0.44\textwidth]{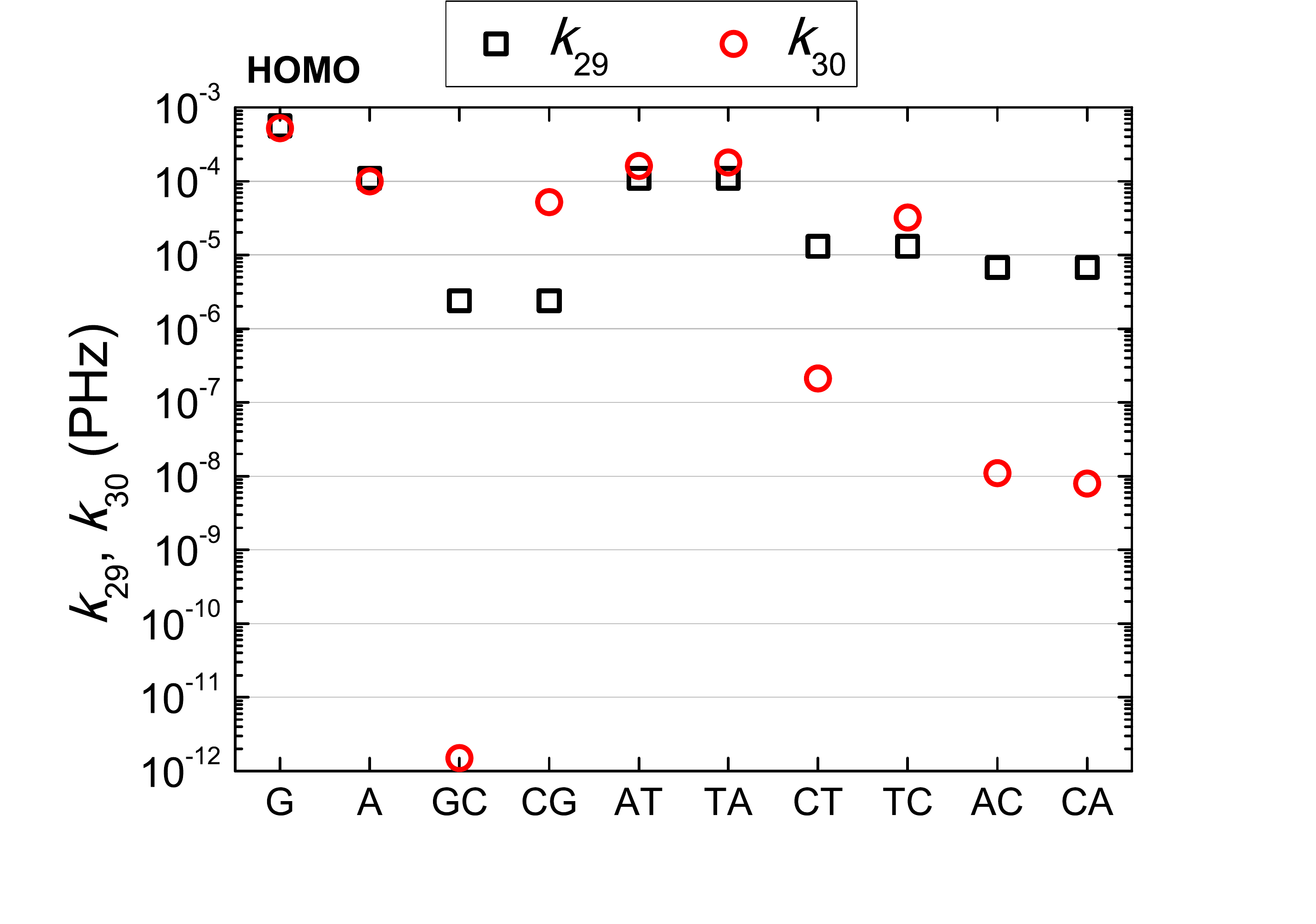}
	\includegraphics[width=0.44\textwidth]{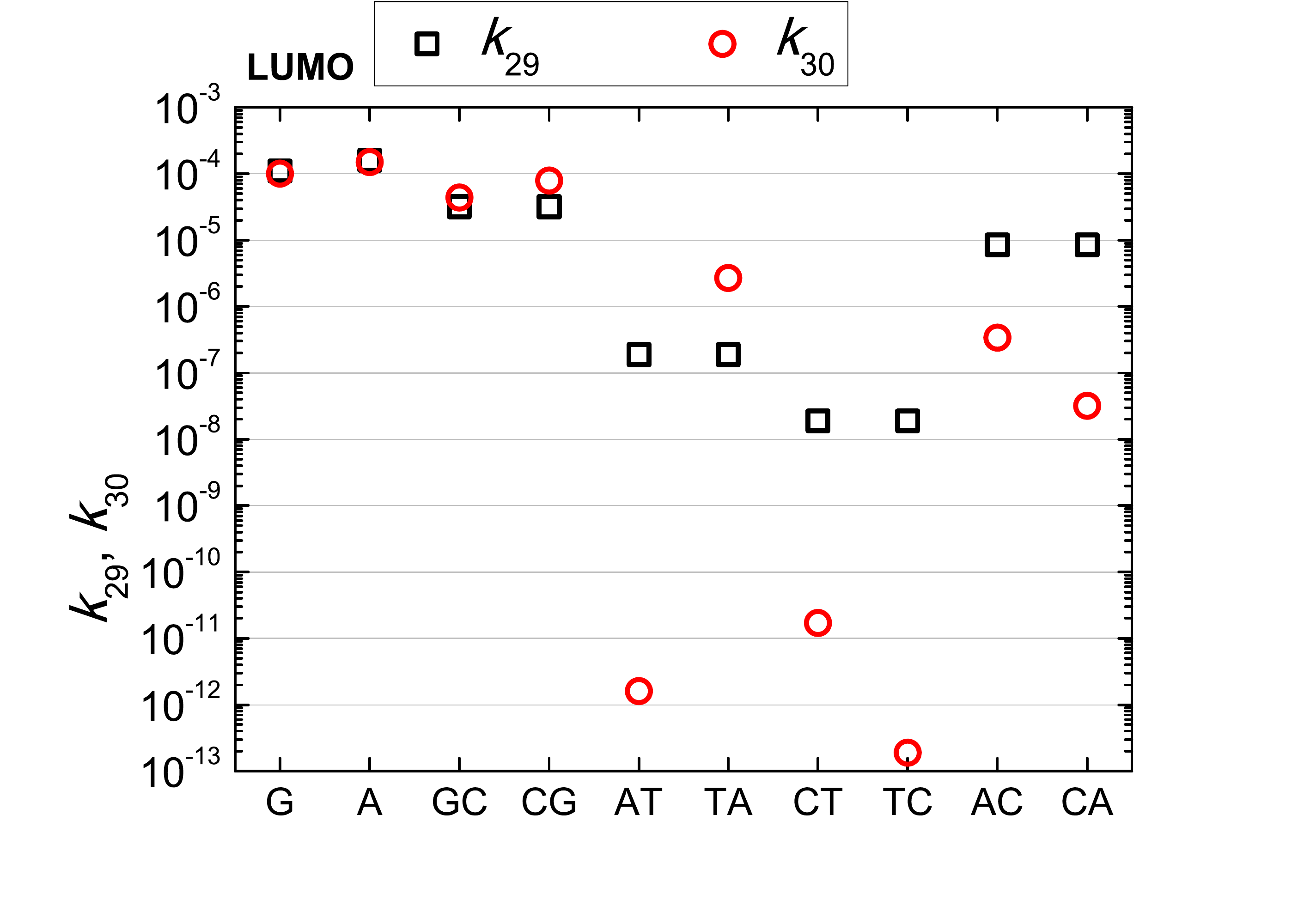}
	\caption{\textit{Pure} mean transfer rate from the first to the 29th ($k_{29}$) and 30th monomer ($k_{30}$), for types I1, I2 and D2 polymers, for HOMO and LUMO, within TBI.
		The horizontal axis shows the repetition unit.}
	\label{fig:k29k30-TBI}
\end{figure*}

\begin{figure*} [h!]
	\includegraphics[width=0.44\textwidth]{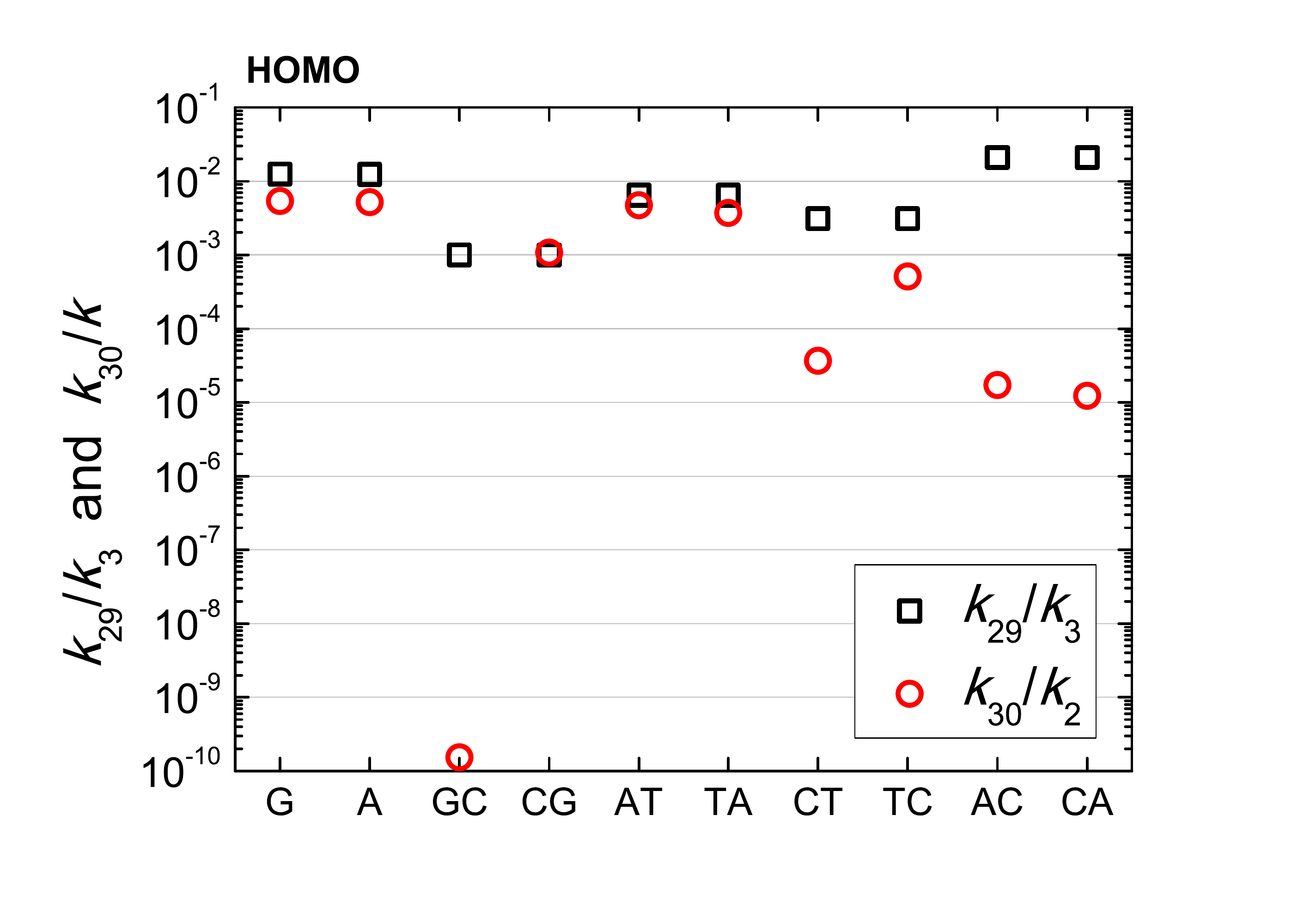}
	\includegraphics[width=0.44\textwidth]{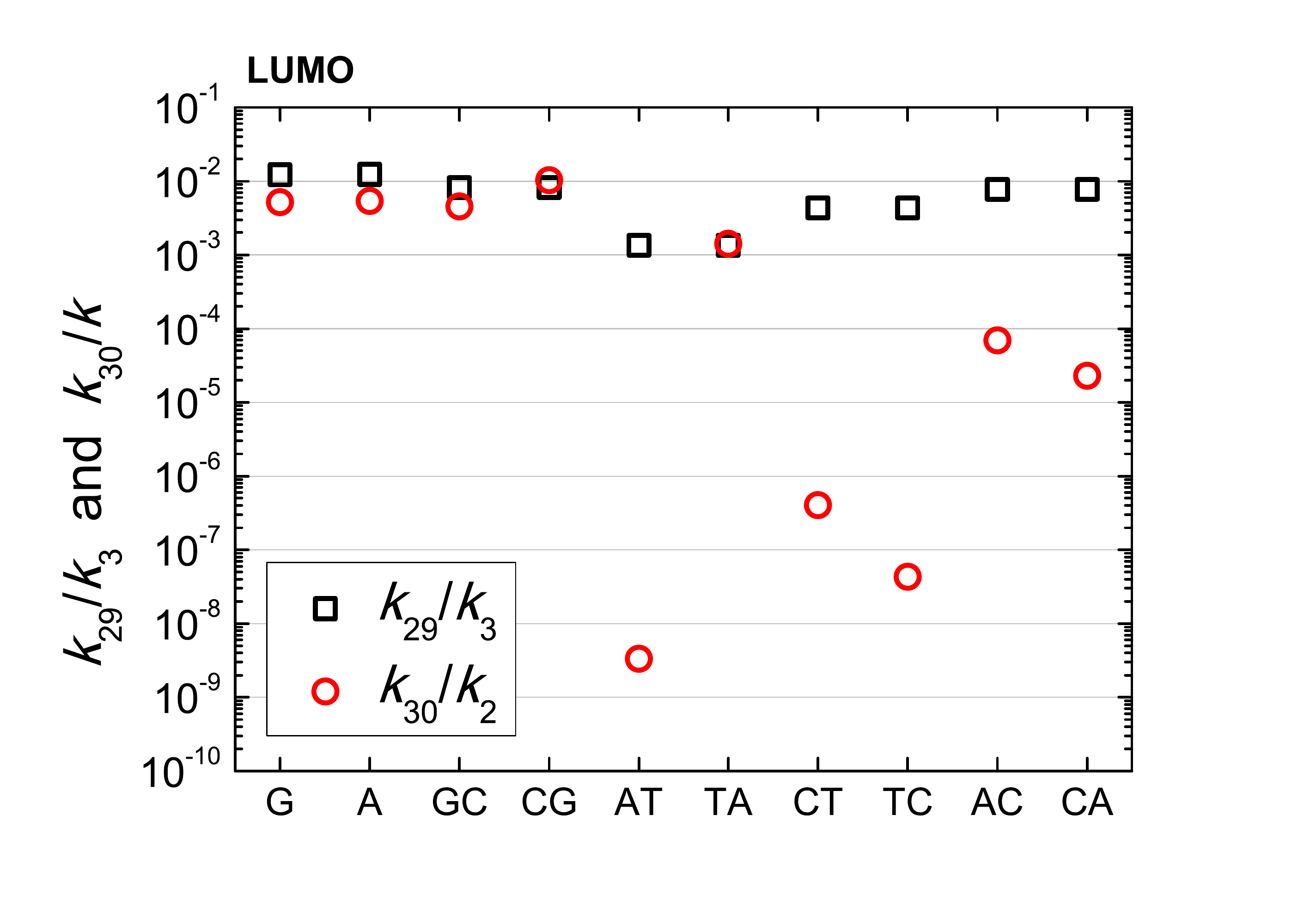}
	\caption{\textit{Pure} mean transfer rate ratio $k_{29}/k_{3}$ and $k_{30}/k_{2}$, for types I1, I2 and D2 polymers, for HOMO and LUMO, within TBI.
		$k_{m}$ is $k$ from the 1st to the $m$-th monomer. The horizontal axis shows the repetition unit.}
	\label{fig:k29or30perk3or2-TBI}
\end{figure*}

\end{widetext}

\end{document}